\documentclass[a4paper]{article}

\usepackage[english]{babel}
\usepackage[a4paper]{geometry}
\usepackage{amsmath}
\usepackage{amssymb}
\usepackage{bm}
\usepackage{graphicx}
\usepackage{url}
\usepackage[T1]{fontenc}
\usepackage{tikz}
\usepackage{authblk}

\title{Semiclassical theory for plasmons in spatially inhomogeneous media}

\author{K. J. A. Reijnders}
\author{T. Tudorovskiy}
\author{M. I. Katsnelson}

\affil{Radboud University, Institute for Molecules and Materials, \\
  Heyendaalseweg 135, 6525 AJ Nijmegen, The Netherlands \\
  \textit{koen.reijnders@ru.nl}, \textit{m.katsnelson@science.ru.nl} }

\date{}                     

\begin{document}
\maketitle

\begin{abstract}

Recent progress in experimental techniques has made the quantum regime in plasmonics accessible. Since plasmons correspond to collective electron excitations, the electron-electron interaction plays an essential role in their theoretical description. Within the Random Phase Approximation, this interaction is incorporated through a system of equations of motion, which has to be solved self-consistently. For homogeneous media, an analytical solution can be found using the Fourier transform, giving rise to Lindhard theory. When the medium is spatially inhomogeneous, this is no longer possible and one often uses numerical approaches, which are however limited to smaller systems.
In this paper, we present a novel semi-analytical approach for bulk plasmons in inhomogeneous media based on the semiclassical (or WKB) approximation, which is applicable when the charge density varies smoothly. By solving the equations of motion self-consistently, we obtain the expressions of Lindhard theory with a spatially varying Fermi wavevector. The derivation involves passing from the operators to their symbols, which can be thought of as classical observables on phase space. In this way we obtain effective (Hamiltonian) equations of motion for plasmons. We then find the quantized energy levels and the plasmon spectrum using Einstein-Brilllouin-Keller quantization.
Our results provide a theoretical basis to describe different setups in quantum plasmonics, such as nanoparticles, quantum dots and waveguides.

\end{abstract}

\vspace*{\baselineskip}


\section{Introduction}
\label{sec:introduction}

A plasma consists of interacting charged particles that are mobile. For classical plasmas, collective excitations of these charged particles have been known since the time of Langmuir~\cite{Tonks29}.
The theory of plasma waves in classical systems was developed by Vlasov~\cite{Vlasov38} and essentially improved by Landau~\cite{Landau46}.
In the early 1950's, Bohm and Pines considered the quantum analog of plasma waves for electrons in metals~\cite{Pines52}. They introduced the concept of plasmons: quasiparticles that represent quantized plasma oscillations~\cite{Vonsovsky89,Giuliani05,Platzman73,Nozieres99}. They considered bulk, or volume, plasmons, which correspond to a collective longitudinal oscillation of all electrons and are defined by a vanishing dielectric function.
Bulk plasmons can be experimentally measured with electron energy loss spectroscopy (EELS)~\cite{Platzman73,Vonsovsky89} and are also relevant for the optical properties of metals and semiconductors~\cite{Kittel05}.

Relatively recently, the field of plasmonics started to develop rapidly, driven by its potential for applications~\cite{Tame13,Fitzgerald16,Fan14}.
An essential part of plasmonics is the transformation of photons to plasmons~\cite{Tame13,Fitzgerald16}. Plasmonic devices that are interesting for applications, for example waveguides~\cite{Tame13}, are therefore unavoidably inhomogeneous.
The operation of these devices typically does not rely on bulk plasmons, but rather on so-called surface plasmon polaritons~\cite{Ritchie73,Vonsovsky89}. These collective oscillations arise in the presence of a surface, and are defined by a different equation than bulk plasmons. 
Surface plasmon polaritons are, by definition, characterized by an exponentially decaying distribution of the electric field on both sides of the surface, the decay length being of the order of the wavelength along the surface.
Moreover, they are lower in energy than bulk plasmons.
Small (metallic) nanoparticles, in which both bulk plasmons and surface plasmon polaritons, also known as localized surface plasmons in this context, can be observed~\cite{Scholl12}, have also been extensively studied~\cite{Tame13,Fitzgerald16,Fan14}.
In the past, the typical wavelength of the plasmons in these systems was much smaller than the de Broglie wavelength of the electrons. One could therefore obtain accurate results with a classical theory.
Nowadays, the quantum limit has been reached in plasmonics~\cite{Tame13,Fitzgerald16,Fan14,Scholl12}, for which classical theories do not suffice.

Different theories have been developed to explain the quantum effects observed in plasmonic systems~\cite{Fitzgerald16,Tame13,Fan14}.
In many cases, the light field can be treated classically, i.e., through Maxwell's equations, and the quantum effects can be incorporated into the matter description. 
One of the more popular approaches is to use the hydrodynamic model~\cite{Fitzgerald16,Eguiluz76,Ciraci13}.
This macroscopic model, based on the hydrodynamic equation of motion, captures some of the non-local physics through the pressure term.
However, the hydrodynamic model leads to an incorrect value for the parameter in front of the non-local term, which should therefore be regarded as a fitting parameter~\cite{Fitzgerald16}.
One of the successes of the hydrodynamic model is its ability to describe the deviations from classical Mie theory~\cite{Mie08} that are found for localized surface plasmons in very small nanoparticles~\cite{Fitzgerald16}.
On the other hand, the hydrodynamic model does not seem very suitable to describe bulk plasmons. Comparing it to the more consistent Random Phase Approximation (RPA), one sees that the hydrodynamic model corresponds to an expansion up to order $q^2$. This implies, among other things, that the hydrodynamic model does not include Landau damping. In certain cases, it therefore leads to qualitatively wrong results in comparison with the RPA~\cite{Ishmukhametov81,Ishmukhametov81b}.

Microscopic models are also sometimes used to describe the quantum effects observed in plasmonic systems. However, these are often tailored to the specific problem at hand, and hard to generalize to different systems.
As an example, we mention the particle-in-a-box model that can be used to describe the resonances in very small nanoparticles~\cite{Genzel75,Scholl12}. In this model, one finds the plasmon resonances by calculating the dielectric function from the discrete energy levels of an infinite spherical well.
Finally, we remark that many different models based on density functional theory are also used~\cite{Tame13,Fitzgerald16,Scholl12}. However, since these approaches are purely numerical in nature, they cannot provide full understanding of all mechanisms that are involved.

In this article, we use a microscopic model based on the RPA to study bulk plasmons in a generic inhomogeneous electron gas. 
We capture the quantum mechanical nature of the matter by starting from the quantum Hamiltonian of the electrons and the Liouville-von Neumann equation for the density operator, and include the classical light field through the Poisson equation.
From the formal point of view, studying a spatially inhomogeneous quantum plasma within this framework is equivalent to studying a Fredholm type integral equation. Although this looks like a well-developed mathematical procedure, it leads to many difficulties in practice. There were several attempts to solve this equation numerically~\cite{Westerhout18,Westerhout21,Jiang21}, but, due to serious computational difficulties, the size of the systems that can be considered is not very large.
A large number of approaches have been developed to simplify the RPA equations~\cite{Fitzgerald16}, including discrete matrix RPA~\cite{Yannouleas92,Yannouleas93}, originally developed in nuclear physics, and local current RPA~\cite{Reinhard90,Brack93}.
However, these methods either heavily rely on computational methods, or simplify the system to a very large degree, and therefore in our opinion do not provide complete understanding.

The idea to use the semiclassical approximation as a bridge between the very well-developed theory of classical plasmas and the theory of quantum plasmas seems to be quite natural. 
One of its great advantages is that it can be used for an arbitrary inhomogeneity, as long as it is sufficiently smooth or the energy is sufficiently high.
The approach already has a pretty long history, being first suggested by Ishmukhametov and collaborators~\cite{Ishmukhametov71,Ishmukhametov75,Ishmukhametov81,Ishmukhametov81b}.
It was originally developed in the context of the atomic plasmon, a hypothetical collective excitation of electrons in atoms, which was hypothesized by Bloch~\cite{Bloch33} and Jensen~\cite{Jensen37}.
After the hypothesis of the atomic plasmon was experimentally disproven~\cite{Verkhovtseva76}, the interest in this direction almost disappeared.
Only recently, the semiclassical approximation was revived in the context of a two-dimensional inhomogeneous plasma~\cite{Torre17}. Nevertheless, all these attempts were heuristic and essentially incomplete.

In this article, we present a systematic derivation, investigation and application of the semiclassical approach in the theory of three-dimensional inhomogeneous quantum plasmas.
In contrast to the previous work~\cite{Ishmukhametov75}, we use a completely rigorous mathematical framework based on the contemporary formulation of the semiclassical approximation in multiple dimensions~\cite{Maslov81,Guillemin77}. More precisely, we make extensive use of the correspondence between the quantum operators and their so-called symbols~\cite{Maslov81,Martinez02,Zworski12,Hormander83}, which can be regarded as classical observables on phase space.
Since we find that the density operator has a non-polynomial symbol, it is an interesting physical example of a so-called pseudodifferential operator.

We derive an effective classical Hamiltonian for bulk plasmons in inhomogeneous three-dimensional systems by applying the semiclassical approximation to the equations of motion that constitute the RPA. This effective classical Hamiltonian corresponds to the dielectric function from Lindhard theory with a spatially varying Fermi wavevector. We also obtain the self-consistent potential that describes the plasmons and show that it has no geometric or Berry phase.

We subsequently investigate the classical motion of the quantum plasmons in phase space by solving Hamilton's equations. We analyze the trajectories, placing particular emphasis on what happens in the vicinity of the classical turning points. We find that the system has two different types of turning points, one of which is connected to the spatial dependence of the plasma frequency and one of which is intimately connected to Landau damping. By investigating both of these turning points in detail, we obtain their Maslov index, which plays an essential role in the Einstein-Brillouin-Keller quantization condition for bound states. With this quantization condition, we obtain the spectrum of the different types of bulk plasmons in the system.

We finally apply our formalism to two physical examples. In our first example, we compute the spectrum of bulk plasmons in a  plasmonic waveguide that is effectively one dimensional.
Physically, this situation corresponds to a slab of a metal or a semiconductor, with a certain charge density, sandwiched between two materials with a higher charge density. However, we need to be somewhat careful here, as the semiclassical approximation is formally only valid when the charge density is sufficiently smooth. A smooth density profile can for instance be created by considering binary alloys, where the composition depends on the position, or by locally doping a semiconductor.
We can also create a plasmonic waveguide by locally modifying the background dielectric constant~\cite{Jiang21}. This modification can be achieved in many ways, for instance by modulating the semiconductor gap using external parameters, e.g. by applying inhomogeneous stress~\cite{Bir74}.

We find that our plasmonic waveguide accomodates two different types of bound states. 
We call the first of these regular bound states and show them in figure~\ref{fig:regular-bound-states-phase-space}. They correspond to bulk plasmons that propagate in the center region of the waveguide, where the plasma frequency is lower than on the sides.
Qualitatively, they look similar to the confined plasmons that were found in numerical calculations of a two-dimensional metal with a spatially varying dielectric environment~\cite{Jiang21}.
We call the second type of bound states Landau-type bound states and show them in figure~\ref{fig:Landau-bound-states-phase-space}. They correspond to bulk plasmons that propagate in the regions where the plasma frequency smoothly changes, that is, in the region between the material on the left (or on the right) and the material in the center. These bound states were previously discussed in Ref.~\cite{Ishmukhametov81b}, but their treatment did not include Landau damping.

In order to avoid possible misunderstanding, we emphasize that the Landau-type bound states are essentially different from the previously mentioned surface plasmons, even though they are localized near the edges of the different materials. The bound states that we find originate entirely from the smooth profile of electron density near the surface~\cite{Ishmukhametov81b}. Mathematically, they are described by the vanishing of the dielectric function, and they are dispersive.
Surface plasmons, on the other hand, are not defined by the vanishing of the dielectric function. They even exist when the boundary between the two materials is sharp, in which case they have a flat dispersion~\cite{Ritchie73,Vonsovsky89}. In the presence of a smooth inhomogeneity near the surface, these surface plasmons acuire a dispersion and become damped~\cite{Blank78}.
A second difference between surface plasmons and the bound states that we find is their degree of localization. The localization of surface plasmons is determined by the wavevector $q_\parallel$ along the surface, and is on the order of $q_\parallel^{-1}$. On the other hand, the localization of our Landau-type bound states is determined by the spatial scale of the profile of the electron density.
Because we only consider bulk plasmons in this paper, we do not consider the additional surface plasmon modes, although a semiclassical theory for these modes is an interesting problem by itself.

In our second example, we consider the bulk plasmons in a system with a spherically symmetric charge density. This can for instance represent a nanoparticle or a plasmonic quantum dot. Although the classical phase space is six dimensional in the case, the integrability of the system is guaranteed by the presence of two additional first integrals, being the square of the total angular momentum and one of its components.
We therefore have three quantization conditions, two of which give rise to the familiar angular quantum numbers~\cite{Maslov81,Curtis04}.
This procedure naturally leads to the Langer substitution~\cite{Langer37,Heading62,Froeman65} in the radial quantization condition, which then defines the spectrum of the bulk plasmons.

For definiteness, we consider two specific radially symmetric setups. In the first setup, we consider an atom and approximate the charge density with the Thomas-Fermi model~\cite{Vonsovsky89,Giuliani05,Lieb81} and the Tietz approximation~\cite{Tietz55,Fluegge94}.
We extend the results of a previous computation~\cite{Ishmukhametov81}, and show that the atomic plasmon indeed does not exist within our model.
In the second setup, we consider a sphere with a fixed radius. we use a parabolic model for the local potential, which corresponds to  a linear electric field inside the sphere. This has been used as a model for the single-particle potential for nucleons in the atomic nucleus~\cite{Mayer55}.
We remark that this model is not very realistic for a spherical nanoparticle, as there the potential is typically constant in the interior and only changes close to the edges~\cite{Toscano15,David14,Lang70}. Our calculations therefore mainly serve as an illustration of our theory. In order to apply it to actual physical systems, one needs to carefully choose the potential.

For this second setup, we compute the spectrum of the bulk plasmons and discuss their physical meaning. We emphasize that our results cannot be compared with the predictions of Mie theory, since the latter is a model for localized surface plasmons.
The theory presented in this article is not yet able to predict the damping of the bulk plasmons in a spherically symmetric potential. Since the bound states in these systems are of Landau type, it is likely that they are damped due to the interaction with the Landau damped region.
For localized surface plasmons, several studies~\cite{Yannouleas92,Li13} have adressed Landau damping within the framework of the RPA. However, since these studies use completely different methodologies, their results cannot be easily transferred to our theory.
We nevertheless consider the determination of the plasmon linewidth a very important problem and will discuss it in a separate publication.

The paper is organized in the following way. Section~\ref{sec:derivation} contains the formal derivation of our semiclassical theory. We start with a more detailed outline of the method in section~\ref{subsec:derivation-outline}, and proceed with the different steps to obtain the effective classical Hamiltonian. We discuss the applicability of the semiclassical approximation in section~\ref{subsec:derivation-applicability-SC}. In order to make section~\ref{sec:derivation} readable for a broad audience, we discuss several technical steps of the derivation in appendix~\ref{app:details-derivation}, where we also briefly review some aspects of pseudodifferential operators.
In section~\ref{sec:investigation}, we analyze the effective classical Hamiltonian for bulk plasmons in inhomogeneous three-dimensional systems.
In the first two subsections, we investigate what happens in the vicinity of the different classical turning points. Several technical aspects of these analyses are discussed in appendices~\ref{app:derivation-Maslov-ind-alternative} and~\ref{app:analytic-cont}.
In section~\ref{subsec:quantization-condition}, we use Einstein-Brillouin-Keller quantization to derive the quantization conditions for the different types of bound states, shown in figures~\ref{fig:regular-bound-states-phase-space} and~\ref{fig:Landau-bound-states-phase-space}. We subsequently discuss how the theory can be applied to spherically symmetric systems in section~\ref{subsec:radial}.
In section~\ref{sec:examples}, we apply our formalism to the two examples discussed above. We numerically calculate the bulk plasmon spectra for a plasmonic waveguide and a spherically symmetric potential and discuss their features.
We present our conclusions and outlook in section~\ref{sec:conclusion}.

We are aware that sections~\ref{sec:derivation} and~\ref{sec:investigation} are somewhat formal and may be considered as cumbersome by some readers. In appendix~\ref{app:prior-derivation}, we therefore review the semiclassical arguments given in Ref.~\cite{Ishmukhametov75}. Although these arguments are not completely rigorous, we hope that they can give the reader a more intuitive idea of the formal transformations presented in section~\ref{sec:derivation}.

We finish this introduction with some words on our notation. Following the practice in the mathematical literature, we do not denote vectors by a bold-faced symbol. The symbol $x$ therefore denotes the position vector and has components $(x_1,\ldots,x_d)$. The quantity $\langle p, x \rangle = \sum_{j=1}^d p_j x_j$ represents the Cartesian inner product on $\mathbb{R}^d$. A star is used to denote complex conjugation, as is common in the physical literature.
We use the convention~(\ref{eq:fourier-transform}) for the Fourier transform, see e.g. Ref.~\cite{Maslov81}, and generally use a bar to denote Fourier transformed quantities. 
Throughout our derivations, the letter $q$ denotes the plasmon momentum. In the mathematical literature, it is more common to denote the momentum by $p$, but this contradicts the physical literature on plasmons, where $p$ is used for the electron momentum over which we integrate to obtain the polarization, see section~\ref{subsec:derivation-outline}. In the physical literature, the letter $q$ often denotes the wavevector of the plasmon. Since the semiclassical approximation considers canonically conjugate position and momentum vectors, we cannot follow this convention in our article. We therefore chose to let the letter $q$ denote the plasmon momentum.
Finally, we work in Gaussian units in our derivations, meaning that the Poisson equation has a factor $4\pi$ on the right-hand side.

\section{Derivation of the induced potential for an inhomogeneous plasma} \label{sec:derivation}

In this section, we derive our semiclassical formalism for inhomogeneous quantum plasmas within the framework of the Random Phase Approximation (RPA).
In section~\ref{subsec:derivation-outline}, we give an outline of the steps in our approach and review the results for the homogeneous quantum plasma. The next three subsections cover the three steps of our approach: obtaining the density operator, computing the induced density and obtaining a self-consistent solution of the Poisson equation. We obtain the effective classical Hamiltonian~(\ref{eq:def-L0-epsilon}), which describes the classical motion of the plasmons in phase space, in section~\ref{subsec:derivation-Poisson-SC}.
In section~\ref{subsec:derivation-induced-potential}, we discuss the connection with classical mechanics in more detail, and obtain our expression~(\ref{eq:solution-sc-potential}) for the induced potential.
We finally discuss the applicability of the semiclassical approximation in section~\ref{subsec:derivation-applicability-SC}.

\subsection{Outline of the approach} \label{subsec:derivation-outline}

Throughout this article, we consider electrons with a quadratic dispersion. When these electrons form a plasma, we have to take the electron-electron interaction into account. 
The RPA provides a powerful framework to capture this interaction. It can be formulated in two different ways. In the first approach, one writes down the Hamiltonian using creation and annihilation operators in Fock space and studies the relevant correlation function using diagrammatic techniques, see. e.g. Ref.~\cite{Giuliani05}.
In this paper, we use a different approach that is equivalent to the first one, and is discussed in, e.g., Ref.~\cite{Vonsovsky89}.
Here, we express the electron-electron interaction 
in the form of an induced (local) potential $V(x,t)$ that is added to the single-electron Hamiltonian.
The total Hamiltonian of the system is then given by
\begin{equation} \label{eq:Ham}
  \hat{H} = \hat{H}_0 + V(x,t) ,
\end{equation}
where $\hat{H}_0$ is the Hamiltonian of the individual electrons, that is, $\hat{H}_0=\hat{p}^2/(2m)$ for a homogeneous plasma. The mass $m$ in this expression corresponds to the effective electron mass.
We can then obtain the induced potential $V(x,t)$, which can be regarded as the wavefunction of the plasmons, using the following self-consistent procedure. First, we compute the effect of $V$ on the density operator by linearizing the Liouville-von Neumann equation. Using this density operator, we can subsequently express the induced density in terms of the induced potential $V$. Finally, the induced potential depends on the induced density through the Poisson equation. These three steps are the starting point of our approach, and we will explain them in more detail below.

Throughout this article, we assume that the Hamiltonian $\hat{H}_0$ does not depend on time. This allows us to write the potential $V(x,t)$ as
\begin{equation}  \label{eq:potential-time-dependence}
  V(x,t) = V(x) \exp\left( -i (E + i \eta) t /\hbar \right) ,
\end{equation}
where $\eta \to 0^+$ is a small parameter which implies that we consider the retarded response function. We discuss the meaning of this parameter in more detail in section~\ref{subsec:damping-homogeneous-systems} and appendix~\ref{app:analytic-cont}.
Alternatively, one may think of $\eta$ as a way to adiabatically switch on the potential. 
Throughout the rest of our derivations, we write $E$ instead of $E+i\eta$, and implicitly assume that $E$ lies slightly above the real axis.

Before we consider inhomogeneous plasmas, let us first review the results for a homogeneous plasma. In the first part of appendix~\ref{app:prior-derivation}, we discuss how the three steps outlined above can be applied to obtain an integro-differential equation for the induced potential $V(x)$ for a three-dimensional plasma, see also Ref.~\cite{Vonsovsky89}.
Since homogeneous systems have translational invariance, we can use the Fourier transform in space to simplify this expression.
Throughout this article, we use the following convention
\begin{equation}
  \begin{aligned} \label{eq:fourier-transform}
    \overline{f}(p) &= \mathcal{F}_{x \to p} g(x) = \frac{e^{-i d \pi/4}}{(2 \pi \hbar)^{d/2}} \int e^{- i \langle p, x \rangle/\hbar} g(x) \, \text{d}x , \\
    g(x) &= \mathcal{F}^{-1}_{p \to x} \overline{g}(p) = \frac{e^{i d \pi/4}}{(2 \pi \hbar)^{d/2}} \int e^{i \langle p, x \rangle/\hbar} \overline{g}(p) \, \text{d}p ,
  \end{aligned}
\end{equation}
where $d$ is the dimensionality of space. As discussed in the introduction, the quantities $x=(x_1,\ldots,x_d)$ and $p=(p_1,\ldots,p_d)$ are $d$-dimensional vectors and their Cartesian inner product is denoted by $\langle p, x \rangle = \sum_{j=1}^d p_j x_j$.
For a homogeneous plasma in three-dimensional space, we therefore write
\begin{equation}  \label{eq:V-fourier-expansion}
  V(x) = \frac{e^{3 i \pi/4}}{(2 \pi \hbar)^{3/2}} \int e^{i \langle q, x \rangle/\hbar} \overline{V}(q) \, \text{d}q ,
\end{equation}
where $q$ is the plasmon momentum.

Inserting expression~(\ref{eq:V-fourier-expansion}) into the integro-differential equation for the induced potential $V(x)$, equation~(\ref{eq:integro-diff-v1}) in appendix~\ref{app:prior-derivation}, we find the secular equation
\begin{equation}  \label{eq:secular-eq-home}
  \overline{\varepsilon}(q, E) \overline{V}(q) = 0 ,
\end{equation}
where $\overline{\varepsilon}(q,E)$ is (the Fourier transform of) the dielectric function.
Within the RPA, the dielectric function (in Gaussian units) is given by
\begin{equation}  \label{eq:dielectric-function-homo}
  \overline{\varepsilon}(q,E) = \varepsilon_b - \overline{v}(q) \overline{\Pi}(q, E) = \varepsilon_b - \frac{4 \pi e^2 \hbar^2}{q^2} \overline{\Pi}(q, E),
\end{equation}
where $\varepsilon_b$ is the (background) dielectric constant and $\overline{v}(q)$ is the Fourier transform of the Coulomb potential. The somewhat unconventional factor $\hbar^2$ arises because we use $q$ to denote the momentum, rather than the wavevector. The function $\overline{\Pi}(q, E)$ represents the polarization or density response function and is given by the Lindhard formula~\cite{Vonsovsky89,Giuliani05}. For a $d$-dimensional system, it is given by
\begin{align}
  \overline{\Pi}(q, E) = \frac{g_s}{(2 \pi \hbar)^d} \int \frac{\rho_0(E_p) - \rho_0(E_{p+q})}{E_p - E_{p+q} + E + i \eta} \text{d}p \label{eq:chi-homogenegous}
\end{align}
where $E_p = p^2/(2m)$ is the energy of a free electron with momentum $p$ and $g_s$ is the spin degeneracy.
Note that we restored the notation $E+i\eta$ in this formula. Since we are dealing with electrons, the equilibrium distribution function $\rho_0(z)$ is given by the Fermi-Dirac distribution
\begin{equation}  \label{eq:Fermi-Dirac-dist}
  \rho_0(z) = \frac{1}{\exp(\beta(z-\mu))+1} ,
\end{equation}
with $\beta=(k_B T)^{-1}$. 
We see that the secular equation~(\ref{eq:secular-eq-home}) admits solutions with $\overline{V}(q) \neq 0$ when $\overline{\varepsilon}(q, E) = 0$.
In this case, we have self-sustained (longitudinal) plasma oscillations, known as plasmons. The equation $\overline{\varepsilon}(q, E) = 0$ defines the dispersion relation $E=E(q)$ of these plasmons, where $E$ is the plasmon energy and $q$ is the plasmon momentum.

\begin{figure}[tb]
  \hfill
  \includegraphics[width=0.5\textwidth]{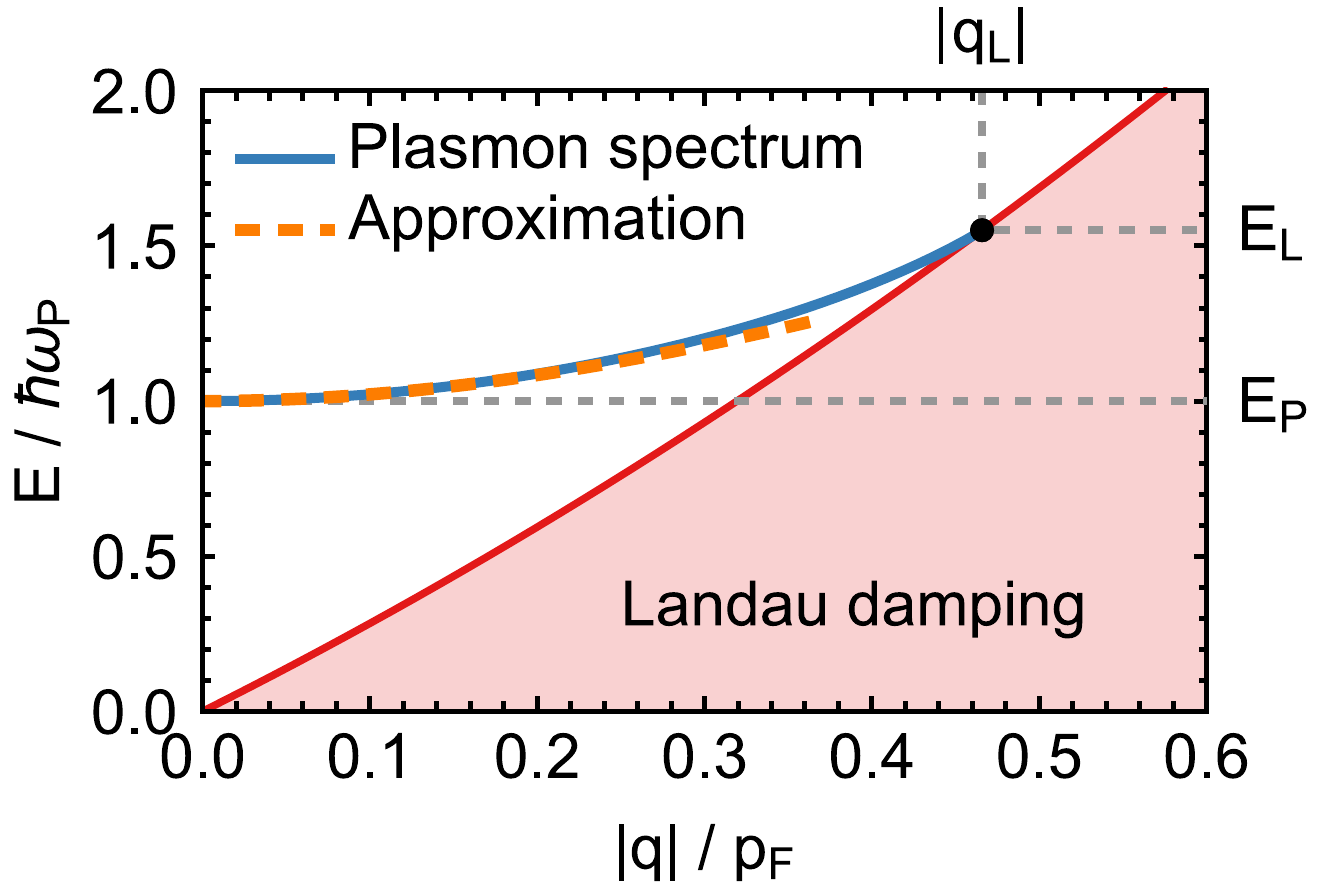}
  \hfill\hfill
  \caption{The dispersion of three-dimensional bulk plasmons for $n^{(0)} = 1/a_0^3$ (solid blue line), and the approximation~(\ref{eq:dispersion-small-q-homo}) (dashed orange line). The shaded red region indicates where Landau damping occurs. We also indicate the plasma energy $E_P$, and the point $(|q_L|, E_L)$, where the plasmon dispersion meets the boundary of the Landau damped region.}
  \label{fig:plasmon-dispersion-homo}
\end{figure}

For a three-dimensional electron plasma at zero temperature,
equation~(\ref{eq:chi-homogenegous}) can be integrated explicitly~\cite{Vonsovsky89,Giuliani05} and we have
\begin{equation}
  \overline{\Pi}(q, E) = \frac{g_s m}{2 \pi^2 \hbar^3} \frac{p_F^2}{|q|} \left( -\frac{|q|}{2 p_F} + \frac{1-\nu_-^2}{4}\log\left(\frac{\nu_- + i\eta + 1 }{\nu_- + i\eta - 1} \right) - \frac{1-\nu_+^2}{4}\log\left(\frac{\nu_+ + i\eta + 1 }{\nu_+ + i\eta - 1} \right) \right) , \label{eq:chi-homogenegous-computed}
\end{equation}
where $\nu_\pm$ is given by
\begin{equation}
  \label{eq:nu-pm}
  \nu_\pm = \frac{m E}{|q| p_F} \pm \frac{|q|}{2 p_F} .
\end{equation}
The Fermi momentum $p_F$ is related to the electron density $n^{(0)}$ by $n^{(0)} = g_s p_F^3/(6 \pi^2 \hbar^3)$ because we consider a homogeneous electron gas with spin degeneracy $g_s$.
It is common to express this electron density in terms of the parameter $r_s$, which is a measure for the distance between electrons and is defined by~\cite{Giuliani05}
\begin{equation} \label{eq:Bohr-radius-rs}
  r_s a_0 = \left( \frac{3}{4 \pi n^{(0)}} \right)^{1/3} , \qquad a_0 = \frac{\hbar^2}{m_e e^2} ,
\end{equation}
where $a_0$ is known as the Bohr radius and $m_e$ is the electron mass.
Although the RPA is formally only valid in the high-density limit $r_s < 1$~\cite{Mahan90,Abrikosov65}, it also works reasonably well for smaller electron densities in practice~\cite{Vonsovsky89,Giuliani05,Platzman73,Nozieres99}.
Quantum Monte Carlo calculations~\cite{Ceperley80} have shown that essentially correlated effects, such as ferromagnetic polarization or Wigner crystallization, which would make RPA qualitatively inappropriate, occur at much higher values of $r_s$ than predicted by naive estimations, on the order of $10^2$.
Equations~(\ref{eq:secular-eq-home}), (\ref{eq:dielectric-function-homo}) and~(\ref{eq:chi-homogenegous-computed}) determine the plasmon dispersion at zero temperature. Since these equations only depend on the length of the momentum vector, we have $E=E(|q|)$.
This dispersion is shown in figure~\ref{fig:plasmon-dispersion-homo} for $n^{(0)}=1/a_0^3$, corresponding to $r_s = (3/(4\pi))^{1/3} \approx 0.62$.

Looking at figure~\ref{fig:plasmon-dispersion-homo}, we immediately see that the plasmon spectrum is bounded from below. This can be shown analytically by expanding the result~(\ref{eq:chi-homogenegous-computed}) for small $|q|$. From this expansion, one finds that the plasmon dispersion can be approximated by~\cite{Vonsovsky89,Giuliani05}
\begin{equation}  \label{eq:dispersion-small-q-homo}
  E^2(q) = \hbar^2 \omega_P^2 + \frac{3 p_F^2}{5 m^2} q^2 + \mathcal{O}(q^4) ,
\end{equation}
where $\omega_P$ is the so-called plasma frequency, defined by
\begin{equation}  \label{eq:plasma-frequency}
  \omega_P^2 = \frac{4 \pi n^{(0)} e^2}{m \varepsilon_b} .
\end{equation}
We can also define the plasma energy $E_P$ by $E_P = \hbar \omega_P$. Equation~(\ref{eq:dispersion-small-q-homo}) and figure~\ref{fig:plasmon-dispersion-homo} then show that we do not have any plasmon modes when $E < E_P$.

The red region shown in figure~\ref{fig:plasmon-dispersion-homo} is defined by $\nu_- < 1$.
We define the energy $E_L$ as the energy where the plasmon dispersion enters this region, i.e., where the blue line meets the red line, as shown in figure~\ref{fig:plasmon-dispersion-homo}. In mathematical terms, this is the point where both equations $\overline{\varepsilon}(q,E)=0$ and $\nu_-=1$ are satisfied. The momentum $|q|$ where this occurs is denoted by $|q_L|$.
Within the red region, the polarization~(\ref{eq:chi-homogenegous-computed}) is complex valued, since $\nu_- < 1$. This leads to a complex-valued dielectric function and implies that a solution $E(q)$ of $\overline{\varepsilon}(q,E)=0$ can only be complex valued beyond $E_L$. In other words, these excitations are damped and have a finite lifetime~\cite{Landau46,Vonsovsky89,Giuliani05}. From a physical point of view, this so-called Landau damping arises because the collective electron excitation transfers energy to incoherent electron-hole pairs, as explained in detail in Ref.~\cite{Vonsovsky89}. 
This transfer of energy is possible when the denominator in expression~(\ref{eq:chi-homogenegous}) vanishes for some value of $p$, that is, $E_p - E_{p+q} + E=0$, while at the same time the numerator does not vanish, i.e., $\rho_0(E_p) -\rho_0(E_{p+q}) \neq 0$.
For $T=0$, the latter requirement means that only transitions from occupied to empty states are allowed, in agreement with the Pauli principle.
One can show that such a value of $p$ exists precisely when $\nu_- < 1$.
We discuss the region where Landau damping occurs in more detail in section~\ref{subsec:Landau-damping-threshold}.

Let us now turn to the case of an inhomogeneous plasma. Since translational invariance is broken in this system, the above derivations based on the Fourier transform no longer give practical results and have to be modified.
From a physical point of view, an inhomogeneous plasma arises for instance when we have a spatially varying electron concentration $n^{(0)}(x)$, as discussed in the introduction. Within the Thomas-Fermi approximation~\cite{Vonsovsky89,Giuliani05,Lieb81}, we can then define a position-dependent Fermi momentum $p_F(x)$ and a local potential $U(x)$ as
\begin{equation}  \label{eq:TF-approx}
  p_F(x) =  \left(\frac{6 \pi^2 \hbar^3}{g_s} n^{(0)}(x)\right)^{1/3}, \qquad U(x) = \mu - \frac{p_F^2(x)}{2 m},
\end{equation}
where $\mu$ is the chemical potential of the system.
We include the local potential $U(x)$ into the Hamiltonian $\hat{H}_0$, that is,
\begin{equation} \label{eq:Ham-0}
  \hat{H}_0 = \frac{\hat{p}^2}{2 m} + U(x) , \qquad \text{with} \; \hat{p} = -i \hbar \frac{\partial}{\partial x} .
\end{equation}
The total Hamiltonian of the system is then still given by equation~(\ref{eq:Ham}).
Alternatively, the inhomogeneity can arise from a spatial variation in the background dielectric constant, i.e., we let $\varepsilon_b = \varepsilon_b(x)$. Such a variation for instance arises when we consider the interface between two different materials, as discussed in the introduction.

Since translational invariance is broken in an inhomogeneous system, we can no longer work with single Fourier modes, that is, we no longer have a dispersion $E(|q|)$.
However, when the typical length scale $\ell$ over which the potential changes is large compared to the electron wavelength, we could say that the system looks ``almost homogeneous'' on a local scale.
In this regime we can use the semiclassical approximation.
Instead of using the Fourier transform, we should therefore use the semiclassical Ansatz 
\begin{equation}  \label{eq:Ansatz-WKB-potential}
  V(x, t) = \exp\left(\frac{i}{\hbar}[ S(x) - E t ] \right) \varphi(x,\hbar) ,
\end{equation}
where $S(x)$ is the classical action and $\varphi(x,\hbar)$ is the amplitude. We come back to the applicability of this approximation in section~\ref{subsec:derivation-applicability-SC}.
Note that when the system is homogeneous, we have $S(x) = \langle q, x \rangle$ and we are considering a single Fourier mode again.

The goal of the next subsections is to obtain an expression for the induced potential $V(x,t)$ in the form of the semiclassical Ansatz~(\ref{eq:Ansatz-WKB-potential}). Our derivation follows the three steps outlined in the beginning of this subsection.
In section~\ref{subsec:derivation-density-matrix}, we express the density operator for a $d$-dimensional system in terms of our semiclassical Ansatz. We then obtain an expression for the induced electron density in terms of this density operator in section~\ref{subsec:derivation-charge-density}. In section~\ref{subsec:derivation-Poisson-SC}, we consider a three-dimensional plasma and solve the Poisson equation to obtain a self-consistent solution for $V(x,t)$. 
In the end, we find that the secular equation~(\ref{eq:secular-eq-home}) for homogeneous systems is replaced by the Hamilton-Jacobi equation~(\ref{eq:Hamilton-Jacobi-epsilon}) when we consider inhomogeneous systems. This equation contains a (local) polarization, which is given by the homogeneous expression~(\ref{eq:chi-homogenegous-computed}) with the substitution $p_F$ to $p_F(x)$.

In order to keep the main text readable for a broad audience, many technical aspects of the derivation are discussed in appendix~\ref{app:details-derivation}.
Nevertheless, our derivations are quite formal in nature, and do not directly appeal to physical intuition. In appendix~\ref{app:prior-derivation}, we therefore review the semiclassical arguments previously given in Ref.~\cite{Ishmukhametov75}. Although these arguments are not completely rigorous, they may give the reader a more intuitive idea of the derivations in the following three subsections.

\subsection{Computation of the density operator} \label{subsec:derivation-density-matrix}

The first step of our analysis is to obtain an expression for the density operator $\hat{\rho}$. We decompose $\hat{\rho}$ as
\begin{equation} \label{eq:rho-decomp}
  \hat{\rho} = \hat{\rho}_0 + \hat{\rho}_1 ,
\end{equation}
where $\hat{\rho}_1$ is the part that is induced by the potential $V(x,t)$. The part $\hat{\rho}_0$ is the equilibrium density operator, and therefore only depends on $\hat{H}_0$.
The density operator obeys the Liouville-von Neumann equation
\begin{equation} \label{eq:liouville-vonNeumann}
  i \hbar \frac{\partial \hat{\rho}}{\partial t} = [ \hat{H}, \hat{\rho} ] .
\end{equation}
Inserting equations~(\ref{eq:Ham}) and~(\ref{eq:rho-decomp}) into this equation, we find
\begin{equation}  \label{eq:liouville-vonNeumann-2}
  i \hbar \left( \frac{\partial \hat{\rho}_0}{\partial t} + \frac{\partial \hat{\rho}_1}{\partial t} \right)
  = [ \hat{H}_0, \hat{\rho}_0 ] + [ \hat{H}_0, \hat{\rho}_1 ] + [ V, \hat{\rho}_0 ] + [ V, \hat{\rho}_1 ] .
\end{equation}
We can consider $\hat{\rho}_1$ as a perturbation to the equilibrium density operator $\hat{\rho}_0$. In the homogeneous case, one can show that $\rho_1$ is proportional to $V$. Although this is somewhat more complicated in the inhomogeneous case, the induced $\hat{\rho}_1$ should still be of the same order as $V$. This implies that the last commutator in equation~(\ref{eq:liouville-vonNeumann-2}) is second order in $V$.
We now demand that equation~(\ref{eq:liouville-vonNeumann-2}) is satisfied for each order of $V$ separately.
Considering the terms that are of zeroth order in the perturbation, we observe that $\partial \hat{\rho}_0/\partial t = 0$, since the density operator in equilibrium is time-independent. We therefore find that the commutator $[ \hat{H}_0, \hat{\rho}_0 ] = 0$. Gathering the terms linear in the perturbation, we obtain
\begin{equation} \label{eq:liouville-vonNeumann-lin}
  i \hbar \frac{\partial \hat{\rho}_1}{\partial t} = [ \hat{H}_0, \hat{\rho}_1 ] + [ V, \hat{\rho}_0 ] ,
\end{equation}
which connects $V$ and $\hat{\rho}_1$.

At this point, we would like to use our Ansatz~(\ref{eq:Ansatz-WKB-potential}) to solve equation~(\ref{eq:liouville-vonNeumann-lin}). However, we also need an Ansatz for $\hat{\rho}_1$. In the homogeneous case, one can show that the matrix elements of $\hat{\rho}_1$ are proportional to the matrix elements of $V$, see e.g. Ref.~\cite{Vonsovsky89}. In the inhomogeneous case, this relation is unfortunately not as simple.
Nevertheless, we can come up with an Ansatz for the operator $\hat{\rho}_1$ by noting that $V(x,t)$ is proportional to $\exp(i[S(x)- E t]/\hbar)$. Considering equation~(\ref{eq:liouville-vonNeumann-lin}), we see that it implies that $\hat{\rho}_1$ is also proportional to this exponent.
Since $\hat{\rho}_1$ is an operator rather than a function, we use
\begin{equation} \label{eq:Ansatz-WKB-rho}
  \hat{\rho}_1 = \exp\left(\frac{i}{\hbar}[ S(x) - E t ] \right) \hat{w} ,
\end{equation}
as Ansatz for $\hat{\rho}_1$. The time-independent operator $\hat{w}$ in this expression needs to be determined from equation~(\ref{eq:liouville-vonNeumann-lin}). We will come back to this Ansatz at the end of this subsection.

Now that we have written down the Ansatzes~(\ref{eq:Ansatz-WKB-potential}) and~(\ref{eq:Ansatz-WKB-rho}), we can insert them into the operator equation~(\ref{eq:liouville-vonNeumann-lin}). Multiplying the entire expression by $\exp(-i[S(x)- E t]/\hbar)$ from the left, we obtain the operator equation
\begin{multline} \label{eq:operator-eq-u}
  E \hat{w} = \exp\left(-\frac{i}{\hbar} S(x) \right) \hat{H}_0 \exp\left(\frac{i}{\hbar} S(x) \right) \hat{w} - \hat{w} \hat{H}_0 + \varphi(x,\hbar) \hat{\rho}_0 \\ - \exp\left(-\frac{i}{\hbar} S(x) \right) \hat{\rho}_0 \exp\left(\frac{i}{\hbar} S(x) \right) \varphi(x,\hbar) .
\end{multline}
This equation contains three unknown quantities, namely the operator $\hat{w}$, the action $S(x)$ and the amplitude $\varphi(x,\hbar)$.
Furthermore, we did not specify the equilibrium part $\hat{\rho}_0$ of the density operator yet, but only stated that it commutes with the Hamiltonian $\hat{H}_0$.

Because equation~(\ref{eq:operator-eq-u}) is an operator equation, we cannot solve it directly. We therefore pass from the operators to their so-called symbols~\cite{Maslov81,Martinez02}. A symbol $a(x,p,\hbar)$ is a function on phase space which corresponds to the quantum operator $\hat{a}$.
If we expand the symbol in powers of $\hbar$, then its leading-order term can be thought of as a classical observable on phase space. For instance, the symbol $H_0(x,p)$ of the operator $\hat{H}_0$ in equation~(\ref{eq:Ham-0}) is simply given by 
\begin{equation} \label{eq:H0-symbol}
  H_0(x,p) = \frac{p^2}{2m} + U(x) .
\end{equation}
When the quantum operator $\hat{a}$ does not contain any products of $x$ and $\hat{p}$, passing from the operator to the symbol simply means replacing the operators $x$ and $\hat{p}$ by the real variables $x$ and $p$.

When the operator $\hat{a}$ contains products of $x$ and $\hat{p}$, we have to determine the operator ordering in order to obtain a unique relation between the quantum operator $\hat{a}$ and its symbol $a(x,p,\hbar)$. Throughout this section, we will use so-called standard quantization, in which the operator $\hat{p}$ acts first, and the operator $x$ acts second. Hence, the quantum operator $\langle x, \hat{p} \rangle$ has symbol $\langle x, p \rangle$ and the quantum operator $\langle \hat{p}, x \rangle$ has symbol $\langle x, p \rangle - i d \hbar$ in $d$-dimensional space. From this simple example, we already see that Hermitian operators may give rise to complex symbols when we use standard quantization. We will come back to this important point in section~\ref{subsec:derivation-induced-potential}. When we consider Weyl quantization instead of standard quantization, Hermitian operators give rise to real symbols and vice versa. However, this choice of quantization makes the calculations much more cumbersome and we therefore use standard quantization throughout this section.

As we will see later on, the functions $a(x,p,\hbar)$ do not have to be polynomial. In general, the operator $\hat{a}$ is therefore a so-called pseudodifferential operator~\cite{Maslov81,Martinez02}. In appendix~\ref{subapp:review-pseudo-diff-operators}, we give a brief review of the correspondence between pseudodifferential operators and their symbols. We discuss how to obtain the symbol of an operator, consider the relation between different quantizations and also review how to compute the symbol of an operator product.

In appendix~\ref{subapp:operator-eq-to-symbol}, we show how to convert equation~(\ref{eq:operator-eq-u}) into an equation for symbols in three steps. We first express the symbols of the operator products in terms of the symbols of the individual operators. We then compute the symbol of the product $\exp(-i S(x)/\hbar) \hat{a} \exp(i S(x)/\hbar)$, see also appendix~\ref{subapp:symbol-action-sandwich}. Finally, we discuss how to determine the symbol of the operator $\hat{\rho}_0$. In order to determine this last symbol, we consider its physical interpretation. In the homogeneous case, discussed in section~\ref{subsec:derivation-outline}, the operator $\hat{\rho}_0$ evaluated on one of the eigenstates becomes the Fermi-Dirac distribution~(\ref{eq:Fermi-Dirac-dist}) evaluated at the energy eigenvalue, i.e. $\rho_0(E_{p})$. In the inhomogeneous case, the leading-order term of the symbol of the operator $\hat{\rho}_0$ equals the Fermi-Dirac distribution evaluated at the symbol $H_0(x,p)$ of the Hamiltonian, as we explain in more detail in appendix~\ref{subapp:operator-eq-to-symbol} and appendix~\ref{subapp:symbol-rho0-functional}.
After performing all steps in the derivation, we obtain an equation which contains the symbol $w(x,p,\hbar)$ of the operator $\hat{w}$ and the symbol $H_0(x,p)$ of the operator $\hat{H}_0$. This equation takes the form of an asymptotic series in $\hbar$, where the zeroth-order term contains the symbols $w(x,p,\hbar)$ and $H_0(x,p)$ and the first-order term also contains their derivatives.
After some calculus, we find that the first two terms of this asymptotic series give rise to the following equation
\begin{align}
  E w(x,p,\hbar) &= H_0\left(x,p+\frac{\partial S}{\partial x} \right) w(x,p,\hbar) - \frac{i\hbar}{2} \sum_{j,k} \frac{\partial^2 H_0}{\partial p_j \partial p_k}\left(x,p+\frac{\partial S}{\partial x}\right) \frac{\partial^2 S}{\partial x_j \partial x_k} w(x,p,\hbar) \nonumber \\
  & \;\; - i\hbar \left\langle \frac{\partial H_0}{\partial p}\left(x,p+\frac{\partial S}{\partial x} \right) , \frac{\partial w}{\partial x}(x,p,\hbar) \right\rangle - w(x,p,\hbar) H_0(x,p) \nonumber \\ 
  & \;\; + i\hbar \left\langle \frac{\partial w}{\partial p}(x,p,\hbar) , \frac{\partial H_0}{\partial x}(x,p) \right\rangle + \varphi(x,\hbar) \rho_0( H_0(x,p)) \nonumber \\
  & \;\; - \frac{i\hbar}{2} \varphi(x,\hbar) \rho_0''(H_0(x,p)) \left\langle \frac{\partial H_0}{\partial p}(x,p) , \frac{\partial H_0}{\partial x}(x,p) \right\rangle \nonumber \\
  & \;\; - \rho_0\left(H_0\left(x,p+\frac{\partial S}{\partial x} \right) \right) \varphi(x,\hbar) \nonumber \\
  & \;\; + \frac{i\hbar}{2} \rho_0''\left(H_0\left(x,p+\frac{\partial S}{\partial x} \right) \right) \left\langle \frac{\partial H_0}{\partial p}\left(x,p+\frac{\partial S}{\partial x} \right) , \frac{\partial H_0}{\partial x}\left(x,p+\frac{\partial S}{\partial x} \right) \right\rangle \varphi(x,\hbar)
  \nonumber \\
  & \;\; + \frac{i\hbar}{2} \rho_0''\left(H_0\left(x,p+\frac{\partial S}{\partial x} \right) \right) \sum_{j,k} \frac{\partial H_0}{\partial p_j} \! \left(x,p+\frac{\partial S}{\partial x} \right) \frac{\partial H_0}{\partial p_k} \! \left(x,p+\frac{\partial S}{\partial x} \right) \frac{\partial^2 S}{\partial x_j \partial x_k} \varphi(x,\hbar) \nonumber \\
  & \;\; + \frac{i\hbar}{2} \rho_0'\left(H_0\left(x,p+\frac{\partial S}{\partial x} \right) \right) \sum_{j,k} \frac{\partial^2 H_0}{\partial p_j \partial p_k}\left(x,p+\frac{\partial S}{\partial x} \right)  \frac{\partial^2 S}{\partial x_j \partial x_k} \varphi(x,\hbar) \nonumber \\
  & \;\; + i\hbar \rho_0'\left(H_0\left(x,p+\frac{\partial S}{\partial x} \right) \right) \left\langle \frac{\partial H_0}{\partial p} \left(x,p+\frac{\partial S}{\partial x} \right) , \frac{\partial \varphi}{\partial x}(x,\hbar) \right\rangle + \mathcal{O}(\hbar^2) . \label{eq:final-operator-eq-u-to-symbol}
\end{align}
As we do not need the terms of $\mathcal{O}(\hbar^2)$ in our analysis, we did not compute them explicitly.

Since we are working in the semiclassical approximation, we would like to solve equation~(\ref{eq:final-operator-eq-u-to-symbol}) order by order in $\hbar$. To this end, we expand the amplitude $\varphi(x,\hbar)$ in orders of $\hbar$, that is,
\begin{equation}  \label{eq:expansion-amplitude}
  \varphi(x,\hbar) = \varphi_0(x) + \hbar \varphi_1(x) + \hbar^2 A_2(x) + \ldots .
\end{equation}
Likewise, we expand the symbol $w(x,p,\hbar)$ of the operator $\hat{w}$ as a power series in $\hbar$, i.e.,
\begin{equation}  \label{eq:expansion-u}
  w(x,p,\hbar) = w_0(x,p) + \hbar w_1(x,p) + \hbar^2 w_2(x,p) + \ldots .
\end{equation}
Finally, we assume that the symbol $H_0(x,p)$ does not depend on $\hbar$, which means that $U(x)$ only has terms of order $\hbar^0$.
Inserting the expansions~(\ref{eq:expansion-amplitude}) and~(\ref{eq:expansion-u}) into equation~(\ref{eq:final-operator-eq-u-to-symbol}), we can construct a solution for $w_0(x,p)$ and $w_1(x,p)$ by equating all terms of the same order in $\hbar$.

Gathering the terms of order $\hbar^0$, we find that
\begin{multline}  \label{eq:relation-u0-A0-eq}
  E w_0(x,p) = H_0\left(x,p+\frac{\partial S}{\partial x} \right) w_0(x,p) - w_0(x,p) H_0(x,p) + \varphi_0(x) \rho_0( H_0(x,p)) \\ - \rho_0\left(H_0\left(x,p+\frac{\partial S}{\partial x} \right) \right) \varphi_0(x) .
\end{multline}
Solving this equation for $w_0(x,p)$, we obtain a relation between the principal part $w_0(x,p)$ of the symbol and the amplitude $\varphi_0(x)$, namely
\begin{equation}  \label{eq:relation-u0-A0}
  w_0(x,p) = \frac{\rho_0( H_0(x,p)) - \rho_0\left(H_0\left(x,p+\frac{\partial S}{\partial x} \right)\right)}{H_0(x,p) - H_0\left(x,p+\frac{\partial S}{\partial x} \right) + E} \varphi_0(x) \equiv  \zeta\left(x,p,\frac{\partial S}{\partial x}\right) \varphi_0(x) .
\end{equation}
In the last equality we have introduced $\zeta(x,p,q)$, given by
\begin{equation} \label{eq:def-eta}
  \zeta(x,p,q) = \frac{\rho_0( H_0(x,p)) - \rho_0\left(H_0\left(x,p+q \right)\right)}{H_0(x,p) - H_0\left(x,p+q \right) + E} .
\end{equation}
We immediately note the similarities between $\zeta(x,p,q)$ and the integrand in the Lindhard formula~(\ref{eq:chi-homogenegous}) for the homogeneous case.

In the same way, we can collect all terms of order $\hbar^1$. After extensive algebraic manipulations, which are shown in full detail in appendix~\ref{subapp:operator-eq-to-symbol}, we find that the subprincipal part $w_1(x,p)$ of the symbol can be expressed as
\begin{multline}
  w_1(x,p) = 
  \zeta\left(x,p,\frac{\partial S}{\partial x}\right) \varphi_1(x) 
  - \frac{i}{2} \sum_j \frac{\partial^2 \zeta}{\partial q_j \partial x_j}\left(x,p,\frac{\partial S}{\partial x}\right) \varphi_0(x)
  + \sum_j \frac{\partial B_{3,j}}{\partial p_j}\left(x,p,\frac{\partial S}{\partial x}\right) \varphi_0(x) \\
  - i \left\langle \frac{\partial \zeta}{\partial q}\left(x, p,\frac{\partial S}{\partial x}\right), \frac{\partial \varphi_0}{\partial x} \right\rangle
  - \frac{i}{2} \sum_{j,k} \frac{\partial^2 \zeta}{\partial q_j \partial q_k}\left(x,p,\frac{\partial S}{\partial x}\right) \frac{\partial^2 S}{\partial x_j \partial x_k} \varphi_0(x) .
  \label{eq:u1-final}
\end{multline}
where $B_{3,j}(x,p,q)$ is defined by
\begin{multline}
  B_{3,j}(x,p,q) = 
  - \frac{i}{2} \frac{\partial H_0}{\partial x_j}(x,p) \frac{\rho_0'( H_0(x,p))}{H_0(x,p) - H_0\left(x,p+q \right) + E} \\
  + \frac{i}{2} \frac{\partial H_0}{\partial x_j}(x,p) \frac{\rho_0( H_0(x,p)) - \rho_0\left(H_0\left(x,p+q \right)\right)}{(H_0(x,p) - H_0\left(x,p+q \right) + E)^2} .
  \label{eq:def-B3}
\end{multline}
Equations~(\ref{eq:relation-u0-A0}) and~(\ref{eq:u1-final}) give us the principal part $w_0(x,p)$ and subprincipal part $w_1(x,p)$ of the symbol of the operator $\hat{w}$, which was defined in equation~(\ref{eq:Ansatz-WKB-rho}).
As we already anticipated, the symbols $w_0(x,p)$ and $w_1(x,p)$ are not polynomial in $p$, but have a more complicated structure.
Moreover, we observe that the principal symbol $w_0(x,p)$ is proportional to $\varphi_0(x)$. This shows the similarity to the homogeneous case, where the matrix elements of $\hat{\rho}_1$ are proportional to the matrix elements of $V$, as we previously discussed.

\subsection{Computation of the induced electron density}  \label{subsec:derivation-charge-density}

The second step of our analysis is to compute the induced electron density $n(x,t)$ using the density matrix $\hat{\rho}_1$ that we obtained in the previous subsection.
As the electron density operator is given by
\begin{equation}
  \hat{n}(x) = \delta(x-x') ,
\end{equation}
the density that is induced by the potential $V(x,t)$ is given by
\begin{equation}  \label{eq:def-density}
  n(x,t) = g_s \text{Tr}(\delta(x-x')\hat{\rho}_1) ,
\end{equation}
where $g_s$ is the spin degeneracy. We remark that $n(x,t)$ represents the density of electrons, and not the charge density. The latter can be obtained by multiplying both sides by the charge $-e$ of an electron.

In order to compute the density $n(x,t)$, we first have to understand how to calculate the trace of a pseudodifferential operator. 
In appendix~\ref{subapp:trace-general}, we review why the trace of a pseudodifferential operator is given by the integral of its symbol over phase space.
Using our previous expression~(\ref{eq:Ansatz-WKB-rho}) for the density matrix, we therefore obtain
\begin{align}
  n(x,t) &= \frac{g_s}{(2\pi \hbar)^d} \int \delta(x-x') \exp\left(\frac{i}{\hbar}[ S(x') - E t ] \right) w(x',p,\hbar) \, \text{d}x' \text{d}p  \label{eq:charge-density-integral-expression-general} \\
  &= \frac{g_s}{(2\pi \hbar)^d} \exp\left(\frac{i}{\hbar}[ S(x) - E t ] \right) \int \Big(w_0(x,p) +\hbar w_1(x,p) + \mathcal{O}(\hbar^2) \Big) \, \text{d}p , 
  \label{eq:charge-density-integral-expression-specific}
\end{align}
where $d$ is the dimensionality of space and $w(x,p)$ is the symbol (within standard quantization) of $\hat{w}$. 
Although this looks like a simple computation, we swept a few subtle points under the rug. We discuss these in detail in appendix~\ref{subapp:trace-precise}, and briefly summarize them here.

First of all, we did not verify whether the trace is actually well-defined, i.e. whether the operator is trace-class. Loosely speaking, this translates to the question whether the integral~(\ref{eq:charge-density-integral-expression-general}) converges. 
Intuitively, we expect this to be the case. First, the delta function is zero outside of a single point. Second, the Fermi-Dirac function is zero for large $|p|$ when we consider zero temperature and exponentially decays when we consider finite temperatures.
In appendix~\ref{subapp:trace-precise}, we prove in detail that the operator is trace class. Throughout this proof, we assume that $U(x)$ and its derivatives are bounded at the point $x$ under consideration, and that $\partial S/\partial x$ is bounded as well.
A second aspect concerns the delta function, which is of course not an element of the Hilbert space $L^2(\mathbb{R}^d)$ over which we take the trace. We should therefore formally consider a limiting procedure, as we discuss in appendix~\ref{subapp:trace-precise}.
Finally, the requirements discussed in appendix~\ref{subapp:trace-precise} imply that the symbol cannot have any poles. This demand is somewhat problematic in our case, as $\zeta(x,p,\frac{\partial S}{\partial x}) \varphi_0(x)$ has simple poles at the points where $H_0(x,p) - H_0\left(x,p+\frac{\partial S}{\partial x}\right) + E = 0$, see equation~(\ref{eq:def-eta}). The presence of this pole is intimately related to Landau damping, and we discuss its implications in detail in section~\ref{subsec:Landau-damping-threshold}. For the moment, we may assume that it does not present any significant complications.

We therefore continue our exposition by computing the various terms in equation~(\ref{eq:charge-density-integral-expression-specific}). The leading-order term $n_0(x,t)$ is obtained by considering the terms in the integrand proportional to $\hbar^0$. With the help of expression~(\ref{eq:relation-u0-A0}) for $w_0(x,p)$, we obtain
\begin{align}
  n_0(x,t) 
    &= \frac{g_s}{(2\pi \hbar)^d} e^{\frac{i}{\hbar}( S(x) - E t )} \int w_0(x,p) \text{d}p 
    = \frac{g_s}{(2\pi \hbar)^d} e^{\frac{i}{\hbar}( S(x) - E t )} \int \zeta\left(x,p,\frac{\partial S}{\partial x} \right) \varphi_0(x) \text{d}p 
    \nonumber \\
    &= e^{\frac{i}{\hbar}( S(x) - E t )} \Pi_0\left(x,\frac{\partial S}{\partial x} \right) \varphi_0(x) ,
    \label{eq:n0}
\end{align}
where we have introduced the local polarization $\Pi_0(x,q)$ in the last equality. It is given by
\begin{equation} \label{eq:def-polarization}
  \Pi_0(x,q) = \frac{g_s}{(2\pi \hbar)^d} \int \zeta(x,p,q) \text{d}p 
  = \frac{g_s}{(2\pi \hbar)^d} \int \frac{\rho_0( H_0(x,p)) - \rho_0\left(H_0\left(x,p+q \right)\right)}{H_0(x,p) - H_0\left(x,p+q \right) + E} \text{d}p .
\end{equation}
Comparing this expression to the Lindhard formula~(\ref{eq:chi-homogenegous}) for a homogeneous plasma, we observe that both expressions are almost identical. The difference is that we now have an additional dependence on $x$ because we are considering an inhomogeneous plasma. This dependence enters in the local polarization $\Pi_0(x,q)$ through the symbol $H_0(x,p)$.

In a similar way, we now consider the subleading term $n_1(x,t)$, obtained by gathering the terms of order $\hbar^1$ in the integrand in equation~(\ref{eq:charge-density-integral-expression-specific}). We have
\begin{equation}
  n_1(x,t) 
    = \frac{g_s}{(2\pi \hbar)^d} e^{\frac{i}{\hbar}( S(x) - E t )} \int w_1(x,p) \text{d}p ,
\end{equation}
where $w_1(x,p)$ is given by equation~(\ref{eq:u1-final}). We first focus on the term $\sum_j \frac{\partial B_{3,j}}{\partial p_j} \varphi_0$ in $w_1$, which is in fact the divergence (with respect to $p$) of the $d$-dimensional vector field $B_3 \varphi_0$. By Gauss' theorem, the integral of this divergence over a $d$-dimensional volume equals the surface integral of the vector field over its boundary. By a proof similar to the one given in appendix~\ref{subapp:trace-precise}, one can easily show that the vector field $B_3$ vanishes at infinity. Hence, the integral of $\sum_j \frac{\partial B_{3,j}}{\partial p_j} \varphi_0$ vanishes.
When integrating the other terms in $w_1$, we can commute the derivatives with respect to $q$ with the integral over $p$. We then obtain an expression for $n_1(x,t)$ in terms of the polarization, i.e.,
\begin{multline}
  n_1(x,t) 
    = e^{\frac{i}{\hbar}( S(x) - E t )} \Bigg( \Pi_0\left(x,\frac{\partial S}{\partial x} \right) \varphi_1(x) 
    - \frac{i}{2} \sum_{j,k} \frac{\partial^2 \Pi_0}{\partial q_j \partial q_k}\left(x,\frac{\partial S}{\partial x}\right) \frac{\partial^2 S}{\partial x_j \partial x_k} \varphi_0(x) \\
    - i \left\langle \frac{\partial \Pi_0}{\partial q}\left(x,\frac{\partial S}{\partial x}\right), \frac{\partial \varphi_0}{\partial x} \right\rangle
    - \frac{i}{2} \sum_j \frac{\partial^2 \Pi_0}{\partial q_j \partial x_j}\left(x,\frac{\partial S}{\partial x}\right) \varphi_0(x) 
    \Bigg) .
    \label{eq:n1}
\end{multline}
We remark that this term has a very particular structure which commonly appears in the semiclassical approximation and will prove to be essential in the following section, where we construct a self-consistent solution of the Poisson equation.
In fact, the specific structure of equation~(\ref{eq:n1}) implies that we can introduce a pseudodifferential operator $\hat{\Pi}$ that represents the polarization. We discuss this is more detail in appendix~\ref{subapp:pd-pol-Ham}, as it is not essential for the development of our formalism but can provide a deeper insight for the interested reader.

We obtain the induced electron density $n(x,t)$ by adding the leading and subleading term. We have $n(x,t) = n_0(x,t) + \hbar n_1(x,t) + \mathcal{O}(\hbar^2)$, where $n_0(x,t)$ and $n_1(x,t)$ are given by expressions~(\ref{eq:n0}) and~(\ref{eq:n1}), respectively.

\subsection{Self-consistent solution of the Poisson equation}  \label{subsec:derivation-Poisson-SC}

In the third step of our analysis, we use the Poisson equation to compute the potential $V(x,t)$ induced by the electron density $n(x,t)$ that we computed in the previous subsection. Although our previous computations were performed for an arbitrary number of dimensions $d$, we now specialize to the three-dimensional case. In particular, we consider a three-dimensional plasma, so $n(x,t)$ depends on all three components of the vector $x=(x_1,x_2,x_3)$. Because the density $n(x,t)$ depends on $V(x,t)$, we have to solve the Poisson equation self-consistently in order to obtain a solution for the induced potential $V(x,t)$.

Since we also allow the background dielectric constant $\varepsilon_b$ to depend on $x$, the Poisson equation reads
\begin{equation}  \label{eq:Poisson-epsilon}
  \sum_j \frac{\partial}{\partial x_j} \left( \varepsilon_b(x) \frac{\partial V(x,t)}{\partial x_j} \right) = -4 \pi e^2 n(x, t) .
\end{equation}
Substituting our Ansatz~(\ref{eq:Ansatz-WKB-potential}) for the potential $V(x,t)$ on the left-hand side and our expressions~(\ref{eq:n0}) and~(\ref{eq:n1}) for the induced density $n(x,t)=n_0(x,t)+\hbar n_1(x,t)$ on the right-hand side, we obtain
\begin{multline}  \label{eq:Poisson-with-Ansatz-epsilon}
  \frac{1}{\hbar^2} \exp\left(\frac{i}{\hbar}[ S(x) - E t ]\right) 
  \Bigg( - \varepsilon_b(x) \big( \varphi_0(x) + \hbar \varphi_1(x) \big) \left\langle \frac{\partial S}{\partial x} , \frac{\partial S}{\partial x} \right\rangle
  + 2 i \hbar \varepsilon_b(x) \left\langle \frac{\partial \varphi_0}{\partial x} , \frac{\partial S}{\partial x} \right\rangle  
  \\
  \hspace*{4cm} 
  + i \hbar \varepsilon_b(x) \varphi_0(x) \sum_j \frac{\partial^2 S}{\partial x_j \partial x_j} 
  + i \hbar \varphi_0(x) \left\langle \frac{\partial \varepsilon_b}{\partial x} , \frac{\partial S}{\partial x} \right\rangle
  + \mathcal{O}(\hbar^2)
  \Bigg) \\
  = -4 \pi e^2 \exp\left(\frac{i}{\hbar}[ S(x) - E t ] \right) 
  \Bigg( \Pi_0\left(x,\frac{\partial S}{\partial x} \right) \left( \varphi_0(x) + \hbar \varphi_1(x) \right)
  - i \hbar \left\langle \frac{\partial \Pi_0}{\partial q}\left(x,\frac{\partial S}{\partial x}\right), \frac{\partial \varphi_0}{\partial x} \right\rangle
  \\
  - \frac{i \hbar}{2} \sum_j \frac{\partial^2 \Pi_0}{\partial q_j \partial x_j}\left(x,\frac{\partial S}{\partial x}\right) \varphi_0(x)
  - \frac{i \hbar}{2} \sum_{j,k} \frac{\partial^2 \Pi_0}{\partial q_j \partial q_k}\left(x,\frac{\partial S}{\partial x}\right) \frac{\partial^2 S}{\partial x_j \partial x_k} \varphi_0(x)
  + \mathcal{O}(\hbar^2) 
  \Bigg) .
\end{multline}
In the homogeneous case, we already saw in equation~(\ref{eq:dielectric-function-homo}) that we had a somewhat unconventional factor of $\hbar^2$ because we consider the plasmon momentum instead of the plasmon wavevector. In the inhomogeneous case, the local plasmon momentum is given by $q = \frac{\partial S}{\partial x}$. This means that when we want to solve equation~(\ref{eq:Poisson-with-Ansatz-epsilon}) order by order in $\hbar$, we have to equate the term proportional to $\hbar^{-2}$ on the left-hand side to the term proportional to $\hbar^{0}$ on the right-hand side. We come back to this in section~\ref{subsec:derivation-applicability-SC}, where we discuss the applicability of the semiclassical approximation and introduce proper dimensionless parameters.
Gathering the leading-order terms and bringing all terms to the same side, we find
\begin{equation}  \label{eq:comm-rel-h0}
  \frac{1}{\hbar^2} \Bigg( \varepsilon_b(x) \left\langle \frac{\partial S}{\partial x} , \frac{\partial S}{\partial x} \right\rangle 
  - 4 \pi e^2 \hbar^2 \Pi_0\left(x,\frac{\partial S}{\partial x} \right) \Bigg)
  \varphi_0(x) \exp\left(\frac{i}{\hbar}[ S(x) - E t ]\right) = 0 .
\end{equation}
Since we are looking for a solution with a non-vanishing wavefunction, the amplitude $\varphi_0(x) \neq 0$ and we find
\begin{equation}  \label{eq:Hamilton-Jacobi-epsilon}
  \varepsilon_b(x) \left\langle \frac{\partial S}{\partial x} , \frac{\partial S}{\partial x} \right\rangle - 4 \pi e^2 \hbar^2 \Pi_0\left(x,\frac{\partial S}{\partial x} \right) = 0.
\end{equation}
This so-called Hamilton-Jacobi equation~\cite{Maslov81,Arnold89} determines the action $S(x)$.
By introducing an effective classical Hamiltonian $L_0(x,q)$, given by
\begin{equation}  \label{eq:def-L0-epsilon}
  L_0(x,q) = \varepsilon_b(x) \langle q , q \rangle - 4 \pi e^2 \hbar^2 \Pi_0\left(x,q \right) ,
\end{equation}
we can also write the Hamilton-Jacobi equation as $L_0(x, \partial S/\partial x) = 0$.
The effective classical Hamiltonian $L_0$ is a very important object, since it describes the motion of the quantum plasmons in classical phase space. We discuss this motion in some more detail in the next subsection. In section~\ref{sec:investigation}, we analyze $L_0$ in detail and show the phase space trajectories for several examples.

Let us now pause here for a moment to compare our results with the homogeneous case. In the previous subsection, we already noticed that the polarization~(\ref{eq:def-polarization}) is very similar to the Lindhard formula~(\ref{eq:chi-homogenegous}). 
We can now compare the effective classical Hamiltonian $L_0(x,q)$ with the full dielectric function~$\overline{\varepsilon}(q,E)$ for the homogenenous case, see expression~(\ref{eq:dielectric-function-homo}). Their structure is once again very similar, apart from a factor of $q^2$. The key difference is the dependence on $x$, which manifests itself in the background dielectric constant $\varepsilon_b(x)$ and in the aforementioned polarization $\Pi_0(x,q)$ through the symbol $H_0(x,p)$ of the Hamiltonian.
When we recognize that $\varphi_0(x) e^{\frac{i}{\hbar}(S(x) - E t)}$ is the leading-order term of our Ansatz~(\ref{eq:Ansatz-WKB-potential}), we also see that equation~(\ref{eq:comm-rel-h0}) has the same structure as equation~(\ref{eq:secular-eq-home}) for the homogeneous case, $\overline{\varepsilon}(q,E) \overline{V}(q) = 0$.

Since we specialized to a three-dimensional plasma in this subsection, we can compute the polarization explicitly for the case of zero temperature. In this way, we clearly see how the spatial dependence enters in $L_0(x,p)$.
As in the homogeneous case~\cite{Giuliani05}, we split the fraction into two parts and perform the substitution $p+q \to p$ in the second part.
\begin{align}
  \Pi_0(x,q) 
  &= \frac{g_s}{(2\pi \hbar)^3} \int \frac{\rho_0( H_0(x,p)) - \rho_0\left(H_0\left(x,p+q \right)\right)}{H_0(x,p) - H_0\left(x,p+q \right) + E} \text{d}p \nonumber \\
  &= \frac{g_s}{(2\pi \hbar)^3} \left( \int \frac{\rho_0( H_0(x,p))}{H_0(x,p) - H_0\left(x,p+q \right) + E} \text{d}p - \int \frac{\rho_0\left(H_0\left(x,p \right)\right)}{H_0(x,p-q) - H_0\left(x,p \right) + E} \text{d}p \right)
\end{align}
For zero temperature, the Fermi-Dirac distribution $\rho_0(x,p)$ is a step function and is only non-zero when $H_0(x,p) < \mu$. When we consider the definition~(\ref{eq:TF-approx}) of the position-dependent Fermi momentum $p_F(x)$ and the definition~(\ref{eq:H0-symbol}) of $H_0(x,p)$, we see that the condition $H_0(x,p) < \mu$ is equivalent to $|p| < p_F(x)$. We therefore have
\begin{equation}
  \rho_0(H_0(x,p)) = \left\{
    \begin{aligned} &1, \qquad && |p| < p_F(x) \\ & 0, && |p| > p_F(x) \end{aligned} \right.
\end{equation}
Our second observation is that the differences $H_0(x,p) - H_0\left(x,p+q \right)$ and $H_0(x,p-q) - H_0\left(x,p \right)$ do not depend on $x$, as the dependence on $U(x)$ drops out. Because of these observations, we can compute the polarization $\Pi_0(x,q)$ in the same way as in the homogeneous case, by introducing polar coordinates and performing the integrals. For the details of the computation, we refer to Ref.~\cite{Giuliani05}. We obtain the final result by replacing $p_F$ by $p_F(x)$ in expression~(\ref{eq:chi-homogenegous-computed}) and $\nu_\pm$ by $\nu_\pm(x)$. The only difference between the polarization for the homogeneous case and the inhomogeneous case is therefore we now have a position-dependent Fermi momentum $p_F(x)$, instead of a constant $p_F$.
Inserting our result for $\Pi_0(x,q)$ into expression~(\ref{eq:def-L0-epsilon}) for $L_0(x,q)$, we find
\begin{multline}
  L_0(x,q) = 
  \varepsilon_b(x) q^2 
  - 4 \pi e^2 \hbar^2 \frac{g_s m}{2 \pi^2 \hbar^3} \frac{p_F^2(x)}{|q|} \Bigg( -\frac{|q|}{2 p_F(x)} + \frac{1-\nu_-^2(x)}{4}\log\left(\frac{\nu_-(x) + 1 }{\nu_-(x) - 1} \right) 
  \\ - \frac{1-\nu_+^2(x)}{4}\log\left(\frac{\nu_+(x) + 1 }{\nu_+(x) - 1} \right) 
  \Bigg)  ,
  \label{eq:L0-computed-epsilon}
\end{multline}
where $\nu_\pm(x) = \frac{m E}{|q| p_F(x)} \pm \frac{|q|}{2 p_F(x)}$, in the same way as in definition~(\ref{eq:nu-pm}). This function determines the action $S(x)$ through the Hamilton-Jacobi equation $L_0(x,\partial S/\partial x) = 0$.

We now continue our analysis of the Poisson equation~(\ref{eq:Poisson-epsilon}). Gathering the terms of subleading order in expression~(\ref{eq:Poisson-with-Ansatz-epsilon}) and rearranging them, we arrive at
\begin{align}  \label{eq:Poisson-with-Ansatz-epsilon-subleading}
  \frac{\hbar}{\hbar^2} \exp\left(\frac{i}{\hbar}[ S(x) - E t ]\right) 
  \Bigg( 
  & -i \left\langle 2 \varepsilon_b(x) \frac{\partial S}{\partial x} - 4 \pi e^2 \hbar^2 \frac{\partial \Pi_0}{\partial q}\left(x,\frac{\partial S}{\partial x}\right) , \frac{\partial \varphi_0}{\partial x} \right\rangle
  \nonumber \\
  & -\frac{i}{2} \sum_{j,k} \Bigg( 2 \varepsilon_b(x) \delta_{jk} - 4 \pi e^2 \hbar^2 \sum_{j,k} \frac{\partial^2 \Pi_0}{\partial q_j \partial q_k}\left(x,\frac{\partial S}{\partial x}\right) \Bigg) \frac{\partial^2 S}{\partial x_j \partial x_k} \varphi_0(x)
  \nonumber \\
  & -\frac{i}{2} \Bigg( 2 \left\langle \frac{\partial \varepsilon_b}{\partial x} , \frac{\partial S}{\partial x} \right\rangle - 4 \pi e^2 \hbar^2 \sum_j \frac{\partial^2 \Pi_0}{\partial q_j \partial x_j}\left(x,\frac{\partial S}{\partial x}\right) \Bigg) \varphi_0(x)
  \nonumber \\
  & + \Bigg( \varepsilon_b(x) \left\langle \frac{\partial S}{\partial x} , \frac{\partial S}{\partial x} \right\rangle - 4 \pi e^2 \hbar^2 \Pi_0\left(x,\frac{\partial S}{\partial x} \right) \Bigg) \varphi_1(x) 
  \Bigg) = 0.
\end{align}
Using our definition~(\ref{eq:def-L0-epsilon}) of $L_0(x,q)$, we can also write this as
\begin{multline}
  \label{eq:comm-rel-h1}
  \frac{1}{\hbar} \exp\left(\frac{i}{\hbar}[ S(x) - E t ]\right) 
  \Bigg( 
  -i \left\langle \frac{\partial L_0}{\partial q}\left(x,\frac{\partial S}{\partial x}\right) , \frac{\partial \varphi_0}{\partial x} \right\rangle
  - \frac{i}{2} \sum_{j,k} \frac{\partial^2 L_0}{\partial q_j \partial q_k}\left(x,\frac{\partial S}{\partial x}\right)  \frac{\partial^2 S}{\partial x_j \partial x_k} \varphi_0(x)
  \\
  - \frac{i}{2} \sum_j \frac{\partial^2 L_0}{\partial q_j \partial x_j}\left(x,\frac{\partial S}{\partial x}\right) \varphi_0(x)
  + L_0\left(x, \frac{\partial S}{\partial x} \right) \varphi_1(x)
  \Bigg) = 0.
\end{multline}
We observe that this expression has the same structure as our previous result~(\ref{eq:n1}) for $n_1(x,t)$. We discuss this structure in more detail in appendix~\ref{subapp:pd-pol-Ham} and explain why it allows us to introduce a pseudodifferential operator $\hat{L}$.

Equations~(\ref{eq:Hamilton-Jacobi-epsilon}) and~(\ref{eq:comm-rel-h1}) provide the self-consistent solution to the Poisson equation~(\ref{eq:Poisson-epsilon}) that we were looking for. They determine the action $S(x)$ and the amplitude $\varphi_0(x)$ in the semiclassical Ansatz~(\ref{eq:Ansatz-WKB-potential}). In the next subsection, we discuss how to solve these equations.

\subsection{Solution for the induced potential} \label{subsec:derivation-induced-potential}

The structure of the equations that we obtained in the previous subsection is typical for the semiclassical approximation. We can therefore follow the standard semiclassical scheme to solve them. In this subsection, we summarize this scheme and show how to obtain both $S(x)$ and $\varphi_0(x)$ in the semiclassical Ansatz~(\ref{eq:Ansatz-WKB-potential}).
For a more extensive discussion of the method and its background, we refer to Refs.~\cite{Maslov81,Reijnders18}.

In the previous subsection, we established that the action $S(x)$ satisfies the Hamilton-Jacobi equation $L_0(x,\partial S/\partial x)=0$, with $L_0(x,q)$ given by equation~(\ref{eq:def-L0-epsilon}). According to the theory of classical mechanics, see e.g. Ref.~\cite{Arnold89}, the Hamilton-Jacobi equation is equivalent to the system of Hamilton equations
\begin{equation} \label{eq:Hamilton}
  \frac{\text{d} x}{\text{d} t} =  \frac{\partial L_0}{\partial q}, \qquad
  \frac{\text{d} q}{\text{d} t} = -\frac{\partial L_0}{\partial x} .
\end{equation}
Given a set of initial conditions $\alpha$, we can integrate these equations to obtain the trajectories $( X(t,\alpha), Q(t,\alpha) )$ in the six-dimensional phase space. We can then determine the action from these solutions by computing
\begin{equation}
  \label{eq:def-action}
  S(x) = \int_{x_0}^x \langle Q, \text{d} X \rangle ,
\end{equation}
where we integrate from an initial point $x_0$ to the point $x$. In practice, this line integral is often performed along the trajectories.
When going from the Hamilon-Jacobi equation to the system of Hamilton's equations, we lift the problem from the so-called configuration space, where the points are parametrized by the position vector $x$, to phase space, where the points are parametrized by $(x,q)$, the position and the momentum vector~\cite{Maslov81}.

In many problems, the initial conditions $\alpha$ parameterize a surface $\Lambda^2$ in phase space. For instance, when we consider a three-dimensional scattering problem, $\alpha$ typically parameterizes the two-dimensional wavefront at an initial time $t_0$. 
We have two additional conditions on $\Lambda^2$~\cite{Maslov81}. First of all, the surface $\Lambda^2$ parameterized by $\alpha$ should be contained in a level set of $L(x,q)$. For the Hamilton-Jacobi equation~(\ref{eq:Hamilton-Jacobi-epsilon}), we therefore require that $L\big(X(t=0,\alpha),Q(t=0,\alpha)\big)=0$. Second, the surface $\Lambda^2$ should be a so-called isotropic manifold. This means that the inner product $\langle Q(t=0,\alpha) , \text{d} X(t=0,\alpha) \rangle$ has to equal an exact differential $\text{d} S_0(\alpha)$, which ensures that the action~(\ref{eq:def-action}) is locally path-independent.
In more mathematical terms, see e.g. Refs.~\cite{Arnold89,Maslov81}, one can say that the symplectic form $\text{d} x \wedge \text{d}q$ has to vanish on $\Lambda^2$. This is, in turn, equivalent to vanishing of the Lagrange brackets, see e.g. Ref.~\cite{Reijnders18}.

The time evolution of all points on the surface $\Lambda^2$ generates a three-dimensional surface $\Lambda^3$. One can prove that $\Lambda^3$ is also an isotropic manifold, and because its dimension equals half the dimension of phase space it is known as a Lagrangian manifold~\cite{Maslov81,Arnold89}. The points on $\Lambda^3$ can be parametrized by the time $t$ and the parameters $\alpha$. We can therefore introduce the Jacobian
\begin{equation}
  \label{eq:def-jacobian}
  J(t,\alpha) =  \det \left( \frac{\partial X(t,\alpha)}{\partial (t,\alpha) } \right) .
\end{equation}
The solution of the Hamilton-Jacobi equation exists and is unique as long as the Jacobian $J(t,\alpha)$ is nonzero~\cite{Maslov81}. 
In this case, the function $x=X(t,\alpha)$ is invertible, and we can project the Lagrangian manifold onto the configuration space with coordinates $x$. With a slight abuse of notation, we can then also write $J(x)$.

We have thus solved the Hamilton-Jacobi equation by computing the trajectories generated by the system of Hamilton equations. Our next step is to essentially simplify equation~(\ref{eq:comm-rel-h1}) along these trajectories in order to find a solution for $\varphi_0(x)$. Our first observation is that the term with $\varphi_1(x)$ vanishes along the trajectories, since $L_0(x,\partial S/\partial x)=0$ by the Hamilton-Jacobi equation. Our result for $\varphi_0$ therefore reads
\begin{multline}
  \label{eq:transport}
  -i \left\langle \frac{\partial L_0}{\partial q}\left(x,\frac{\partial S}{\partial x}\right) , \frac{\partial \varphi_0}{\partial x} \right\rangle
  - \frac{i}{2} \sum_{j,k} \frac{\partial^2 L_0}{\partial q_j \partial q_k}\left(x,\frac{\partial S}{\partial x}\right)  \frac{\partial^2 S}{\partial x_j \partial x_k} \varphi_0(x) \\
  - \frac{i}{2} \sum_j \frac{\partial^2 L_0}{\partial q_j \partial x_j}\left(x,\frac{\partial S}{\partial x}\right) \varphi_0(x)
  = 0 .
\end{multline}
This equation is known as the transport equation. As we explain in appendix~\ref{subapp:pd-pol-Ham}, it straightforwardly appears when one applies the semiclassical approximation to a partial differential equation. It is however quite curious that it also appears in our multi-step procedure involving multiple differential equations.

Because of the specific structure of the transport equation~(\ref{eq:transport}), it can be essentially simplified. To this end, we study the time-evolution of the Jacobian along the trajectories of the Hamiltonian system.
Using the Liouville formula~\cite{Maslov81,Arnold89}, one can show that
\begin{equation}
  \label{eq:liouville-formula}
  \frac{\text{d}}{\text{d} t} \log J = \sum_{j,k} \frac{\partial^2 L_0}{\partial q_j \partial q_k}\left(x,\frac{\partial S}{\partial x}\right) \frac{\partial^2 S}{\partial x_j \partial x_k} + \sum_j \frac{\partial^2 L_0}{\partial x_j \partial q_j}\left(x,\frac{\partial S}{\partial x}\right) .
\end{equation}
Along the trajectories, we also have~\cite{Maslov81}
\begin{equation}
  \left\langle \frac{\partial L_0}{\partial q}\left(x,\frac{\partial S}{\partial x}\right) , \frac{\partial \varphi_0}{\partial x} \right\rangle = \left\langle \frac{\text{d} x}{\text{d} t} , \frac{\partial \varphi_0}{\partial x} \right\rangle 
  = \frac{\text{d} \varphi_0}{\text{d} t} .
\end{equation}
When we insert these results into equation~(\ref{eq:transport}), we find
\begin{equation}
  \label{eq:amplitude-eq-pre-final}
  -i \frac{\text{d} \varphi_0}{\text{d} t}
  - \frac{i}{2} \left(\frac{\text{d}}{\text{d} t} \log J \right) \varphi_0
  = 0 .
\end{equation}
It is important to note that the term containing $\sum_j \partial^2 L_0/\partial x_j \partial q_j \varphi_0$ has dropped out of the equation.
We note that this is fairly non-trivial. As we explain in somewhat more detail in appendix~\ref{subapp:pd-pol-Ham}, one generally has a term proportional to $(L_1(x,q) + (i/2) \sum_j \partial^2 L_0/\partial x_j \partial q_j) \varphi_0$ in the above equation. When there are no non-trivial geometric phases, or Berry phases, in the system, we have $L_1(x,q) = -(i/2) \sum_j \partial^2 L_0/\partial x_j \partial q_j$ and this term vanishes. This is exactly what happens in our system. Equation~(\ref{eq:amplitude-eq-pre-final}) therefore shows that the induced potential $V(x,t)$ has no geometric or Berry phase.

Our next step is to introduce a new amplitude $A_0(x)$, defined by $A_0 = \varphi_0 \sqrt{J}$. Equation~(\ref{eq:amplitude-eq-pre-final}) can then be written as
\begin{equation}
  \label{eq:amplitude-eq-final}
  \frac{1}{\sqrt{J}} \frac{\text{d} A_0}{\text{d} t}
  = 0 .
\end{equation}
This equation has the obvious solution $A_0(x) = A_0^0$.

Putting everything together, we find that the leading-order term of the asymptotic solution for the induced potential $V(x,t)$ is given by
\begin{equation}  \label{eq:solution-sc-potential}
  V(x,t) = \frac{A_0^0}{\sqrt{J(x)}} \exp\left(\frac{i}{\hbar}[ S(x) - E t ] \right) ,
\end{equation}
where $S(x)$ is the action defined by expression~(\ref{eq:def-action}) through the Hamiltonian system~(\ref{eq:Hamilton}). 
Note that we should formally write $S(r)=S(x,q)$ instead of $S(x)$, since the action is defined on a trajectory, which is in turn parametrized by the coordinates $(t,\alpha)$. However, as long as the Jacobian $J(t,\alpha)$ is nonzero, we can project the Langrangian manifold onto the configuration space and write $S(x)$.

We thus obtain the asymptotic solution~(\ref{eq:solution-sc-potential}), which is self-consistent and takes the electron-electron interaction into account through the RPA. We further investigate its properties in section~\ref{sec:investigation}, after we discuss the applicability of the semiclassical approximation in the next subsection.

\subsection{Applicability of the semiclassical approximation}  \label{subsec:derivation-applicability-SC}

Throughout the derivations that we presented in this section, our main guideline was the comparison with the homogeneous case. In this subsection, we identify the proper dimensionless semiclassical parameters that underlie our derivations and justify our arguments.

We start our considerations by identifying the relevant scales in the problem. The first of these is the length scale $\ell$ of a typical change in the external potential $U(x)$ and/or in the background dielectric constant $\varepsilon_b(x)$. The second scale is the typical value of the Fermi momentum $p_F(x)$, which we denote by $p_0$. Finally, we denote the typical value of the background dielectric constant $\varepsilon_b(x)$ by $\varepsilon_{b0}$.
Note that we do not have to consider the typical value of the plasmon momentum $q$ as a separate parameter. Since we are interested in a quantum plasma, $q$ should be of the same order of magnitude as $p_0$. If, instead, $q$ were much smaller than $p_0$, we would be dealing with a classical plasma.

Following these considerations, we introduce a dimensionless spatial coordinate $\tilde{x} = x/\ell$ and dimensionless momenta $\tilde{q}=q/p_0$ and $\tilde{p}_F(\tilde{x})=p_F(x)/p_0$.
We also introduce the dimensionless semiclassical parameter $h$, given by
\begin{equation}
  \label{eq:def-h}
  h = \frac{\hbar}{p_0 \ell} = \frac{1}{2\pi} \frac{\lambda_\text{el}}{\ell} ; \qquad \lambda_\text{el} = 2 \pi \frac{\hbar}{p_0} ,
\end{equation}
where $\lambda_\text{el}$ is the electron wavelength. This parameter $h$ is the proper dimensionless parameter in which we perform the asymptotic expansion. A first indication for this can be found by examining the exponent in the semiclassical Ansatz~(\ref{eq:Ansatz-WKB-potential}). We have
\begin{equation}
  \frac{1}{\hbar} S(x) = \frac{1}{\hbar} \int_{x_0}^x \langle Q, \text{d} X \rangle
    = \frac{p_0 \ell}{\hbar} \int_{x_0}^x \left\langle \frac{Q}{p_0} , \frac{\text{d} X}{\ell} \right\rangle
    = \frac{1}{h} \tilde{S}(\tilde{x}) ,
\end{equation}
where we used expression~(\ref{eq:def-action}) in the second equality, and defined the dimensionless action~$\tilde{S}(\tilde{x})$ in the last equality.

Let us continue to examine other dimensionless quantities.
Because we consider electron-electron interaction, another important length scale in our problem is the Thomas-Fermi screening length, given by 
\begin{equation}  \label{eq:def-TF-screening-length}
  \lambda_\text{TF} = \sqrt{\frac{\pi \hbar^3 \varepsilon_{b0}}{4 m e^2 p_0}} 
  = \sqrt{\frac{\pi}{4} \frac{m_e}{m} \frac{\hbar a_0}{p_0} \varepsilon_{b0}} 
  = a_0 \sqrt{\frac{\pi}{4} \frac{m_e}{m} \frac{\hbar}{p_0 a_0} \varepsilon_{b0}} ,
\end{equation}
where $a_0$ is the Bohr radius, defined by
\begin{equation}
  a_0 = \frac{\hbar^2}{m_e e^2} ,
\end{equation}
in Gaussian units, with $m_e$ the electron mass. We therefore introduce a second dimensionless parameter $\kappa$,
\begin{equation}  \label{eq:def-mu}
  \kappa = \frac{1}{2\pi} \frac{\lambda_\text{TF}}{\ell} .
\end{equation}
the ratio of the Thomas-Fermi screening length and the length scale $\ell$.
Subsequently, we make the energy dimensionless with $p_0^2/(2m)$, i.e., $\tilde{E}=2 m E/p_0^2$, where $m$ is the effective electron mass. Of course, one can freely choose the numerical factor here, but using the free-electron energy seems elegant. The background dielectric constant $\varepsilon_b(x)$ has no dimension, and we set $\tilde{\varepsilon}_b(\tilde{x}) = \varepsilon_b(x)/\varepsilon_{b0}$.
Finally, it is clear from the first term in expression~(\ref{eq:L0-computed-epsilon}) that $L_0(x,q)$ has dimension $q^2$. We therefore introduce the dimensionless $\tilde{L}_0$ by $\tilde{L}_0 = L_0/(\varepsilon_{b0} p_0^2)$.

We can now rewrite expression~(\ref{eq:L0-computed-epsilon}) using these dimensionless quantities. We find
\begin{multline}
  \tilde{L}_0(\tilde{x},\tilde{q}) = 
  \tilde{\varepsilon}_b(\tilde{x}) \tilde{q}^2 
  - \frac{g_s}{8\pi^2} \frac{h^2}{\kappa^2} \frac{\tilde{p}_F^2(\tilde{x})}{|\tilde{q}|} \Bigg( -\frac{|\tilde{q}|}{2 \tilde{p}_F(\tilde{x})} + \frac{1-\tilde{\nu}_-^2(\tilde{x})}{4}\log\left(\frac{\tilde{\nu}_-(\tilde{x}) + 1 }{\tilde{\nu}_-(\tilde{x}) - 1} \right) 
  \\ - \frac{1-\tilde{\nu}_+^2(\tilde{x})}{4}\log\left(\frac{\tilde{\nu}_+(\tilde{x}) + 1 }{\tilde{\nu}_+(\tilde{x}) - 1} \right) 
  \Bigg)  ,
  \label{eq:L0-epsilon-dimless}
\end{multline}
where
$
  \tilde{\nu}_\pm(\tilde{x}) = \frac{\tilde{E}}{2 |\tilde{q}| \tilde{p}_F(\tilde{x})} \pm \frac{|\tilde{q}|}{2 \tilde{p}_F(\tilde{x})}
$
and we have used that 
\begin{equation} \label{eq:ratio-h-mu}
  \frac{h^2}{\kappa^2} = \frac{\lambda_\text{el}^2}{\lambda_\text{TF}^2} 
  = 16 \pi \frac{m e^2}{p_0 \hbar \varepsilon_{b0}}
  = 16 \pi \frac{m}{m_e} \frac{1}{\varepsilon_{b0}} \frac{\hbar}{p_0 a_0}.
\end{equation}
Expression~(\ref{eq:L0-epsilon-dimless}) explains, from a formal point of view, why we combined terms with different orders of $\hbar$ when we derived equation~(\ref{eq:comm-rel-h0}). Now that we have introduced proper dimensionless parameters, we see that the terms coming from the polarization contain the ratio $h/\kappa$ of our two dimensionless parameters, instead of only the dimensionless semiclassical parameter $h \ll 1$.
In contrast to $h$, the quotient $h/\kappa$ is not small in typical applications, as we will see in section~\ref{sec:examples}.
In fact, the prefactor $\frac{g_s}{8\pi^2} \frac{h^2}{\kappa^2}$ should be of order one, as the second term in expression~(\ref{eq:L0-epsilon-dimless}) should cancel the first in order to satisfy the Hamilton-Jacobi equation $L_0(x,\partial S/\partial x)=0$. When $r_s$ is of order one, we can show this explicitly using expression~(\ref{eq:Bohr-radius-rs}).
This argument provides another indication that $h$ is the proper dimensionless semiclassical parameter.

At this point, we discuss a few other dimensionless quantities. 
We can define a dimensionless electron concentration $\tilde{n}^{(0)}(x)$ as 
\begin{equation}
  \tilde{n}^{(0)}(x) = n^{(0)}(x) \ell^3 
    = \frac{g_s}{6 \pi^2} \frac{p_0^3 \ell^3}{\hbar^3} \frac{p_F^3(x)}{p_0^3} 
    = \frac{g_s}{6 \pi^2} \frac{\tilde{p}_F^3(\tilde{x})}{h^3} ,
\end{equation}
where we used our previous definition~(\ref{eq:TF-approx}) in the second equality.
Anticipating the results in the next section, we also define a dimensionless plasma energy $\tilde{E}_P(\tilde{x})$, analogous to the plasma energy $E_P$ for the homogeneous case, which was defined in the text below equation~(\ref{eq:plasma-frequency}). We have
\begin{equation}  \label{eq:plasma-energy-dimless}
  \tilde{E}_P^2(\tilde{x}) = \frac{(2 m)^2}{p_0^4} \hbar^2 \omega_P^2(x) 
  = \frac{(2 m)^2}{p_0^4} \hbar^2 \frac{4 \pi n^{(0)}(x) e^2}{m \varepsilon_b(x)}
  = \frac{g_s}{6\pi^2} 16 \pi \frac{m e^2}{p_0 \hbar\varepsilon_{b0}} \frac{\varepsilon_{b0}}{\varepsilon_b(x)} \frac{p_F^3(x)}{p_0^3}
  = \frac{g_s}{6\pi^2} \frac{h^2}{\kappa^2} \frac{\tilde{p}_F^3(\tilde{x})}{\tilde{\varepsilon}_b(\tilde{x})} ,
\end{equation}
where we used equations~(\ref{eq:TF-approx}) and~(\ref{eq:ratio-h-mu}) in the third and fourth equalities, respectively.
We can then also define a dimensionless plasma frequency $\tilde{\omega}_P(\tilde{x})$ by $\tilde{E}_P(\tilde{x}) = h \tilde{\omega}_P(\tilde{x})$.

We obtain our last indication that $h$ is the proper dimensionless semiclassical parameter by considering equation~(\ref{eq:comm-rel-h1}). We would like to show that the terms in this expression are one order of $h$ larger than the terms in equation~(\ref{eq:comm-rel-h0}). Looking at the latter expression, we observe that its dimensions are equal to the dimensions of $L_0 \varphi_0/\hbar^2$. The terms in equation~(\ref{eq:comm-rel-h1}) all have the same units, and from the third term we see their dimension is equal to the dimension of $L_0 \varphi_0/(q x \hbar)$. Introducing our dimensionless units, and disregarding $\varphi_0$ which is equally present in both expressions, we find $\varepsilon_{b0} p_0^2 \tilde{L}_0/\hbar^2$ for the leading order and $\hbar/(p_0 \ell) \times \varepsilon_{b0} p_0^2 \tilde{L}_0/(\hbar^2\tilde{q}\tilde{x})$ for the subleading order. As the quantities with a tilde are dimensionless and of order one, we see that the terms of subleading order are a factor of $h=\hbar/(p_0\ell)$ smaller than the terms of leading order.

In this subsection, we have thus shown that our semiclassical approximation is an asymptotic expansion in the dimensionless semiclassical parameter $h$. This approximation is, strictly speaking, valid when $h \ll 1$. This means that the electron wavelength $\lambda_\text{el}$ should be (much) smaller than the typical length scale $\ell$ of spatial variations of $n^{(0)}(x)$ and $\varepsilon_b(x)$. We also require that the typical screening length is such that quantum effects play a role, as measured by our second dimensionless parameter $\kappa$.

For future reference, we finish this subsection with a brief discussion on the relation between our dimensionless units and atomic units. By setting $\ell=a_0$, we measure all distances in Bohr radii, which is the unit of length in atomic units. Similarly, we can set $p_0 = \hbar/a_0$ to obtain the momenta in atomic units. The dimensionless semiclassical parameter $h$ is then equal to unity, and $\kappa= (4\sqrt{\pi m/m_e})^{-1} = (4\sqrt{\pi m_\text{at}})^{-1}$, where $m_\text{at}$ is the effective electron mass in atomic units.
The energy $E_\text{at}$ in atomic units is measured in terms of the Hartree energy $E_H=\hbar^2/(m_e a_0^2)$, and is therefore related to the dimensionless energy $\tilde{E}$ by $\tilde{E} = 2 m E/p_0^2 = 2 (m/m_e) (E/E_H) = 2 m_\text{at} E_\text{at}$.

\section{Analysis of the semiclassical equations}  \label{sec:investigation}

In the previous section, we constructed the classical Hamiltonian $L_0(x,q)$ for the quantum plasmons, given by expression~(\ref{eq:def-L0-epsilon}). We also obtained the asymptotic solution~(\ref{eq:solution-sc-potential}) for the induced potential $V(x,t)$.
In this section, we analyze these expressions in detail. Our main goal is to obtain the quantization condition for bound states, which determines the spectrum of the bulk plasmons in our system.
The key ingredients for this quantization condition are the periodic trajectories of the system and the Maslov indices, which are related to the scattering phases at the classical turning points.

A classical turning point lies at the boundary of a classically allowed and a classically forbidden region. In a classically allowed region, the equation $L_0(x,q)=0$ has solutions with real $q$ for a given (fixed) value of the energy; in the classically forbidden region this is not the case. A bound state, which has a periodic trajectory, arises when a classically allowed region is surrounded by two classically forbidden regions.

In section~\ref{subsec:derivation-outline} and figure~\ref{fig:plasmon-dispersion-homo}, we saw that the plasmon spectrum of a homogeneous system is bounded from below by $E_P$ and from above by $E_L$. In an inhomogeneous system, the charge density $n^{(0)}(x)$ becomes spatially varying, and we can think of the quantities $E_P(x)$ and $E_L(x)$ as dependent on position, as we explain in more detail in sections~\ref{subsec:simple-turning-point} and~\ref{subsec:Landau-damping-threshold}. This generalization implies that we are in the classically allowed region whenever $E_P(x) < E < E_L(x)$. Hence, we can reach a classically forbidden region in two different ways: either because the energy becomes equal to $E_P(x)$, or because the energy becomes equal to $E_L(x)$.

We therefore start our analysis of the semiclassical equations by considering what happens at these points. We present our considerations for $E = E_P(x)$ in section~\ref{subsec:simple-turning-point} and for $E = E_L(x)$ in section~\ref{subsec:Landau-damping-threshold}. In these subsections, we focus on the case of an effectively one-dimensional waveguide.
After our analysis of the classical turning points, we construct the periodic trajectories in phase space in section~\ref{subsec:quantization-condition} and write down the quantization condition.
We find that there are two different types of bound states, which we call regular bound states and Landau-type bound states. They are shown in figures~\ref{fig:regular-bound-states-phase-space} and~\ref{fig:Landau-bound-states-phase-space}, respectively.
The energies of these bound states depend on the momentum $q_\parallel$ along the waveguide, which leads to a plasmon spectrum.
The final subsection~\ref{subsec:radial} considers what happens for spherically symmetric potentials.

\subsection{Simple turning points}  \label{subsec:simple-turning-point}

In section~\ref{subsec:derivation-outline}, we reviewed the plasmon dispersion for a homogeneous system, which is shown in figure~\ref{fig:plasmon-dispersion-homo}. In particular, we saw that the plasmon spectrum is bounded from below by $E_P = \hbar \omega_P$, defined by expression~(\ref{eq:plasma-frequency}): there are no propagating plasmons for $E<E_P$. Moreover, one can derive the dispersion relation~(\ref{eq:dispersion-small-q-homo}) for small $|q|$ by expanding the Lindhard function~(\ref{eq:chi-homogenegous-computed}) for small $|q|$ and subsequently solving the equation~$\overline{\varepsilon}(q,E)=0$ for $E$, see e.g. Refs~\cite{Vonsovsky89,Giuliani05}.

For an inhomogeneous plasma, we are dealing with the classical Hamiltonian $L_0(x,q)$ given by expression~(\ref{eq:def-L0-epsilon}), or, equivalently, by expression~(\ref{eq:L0-computed-epsilon}). As we discussed in section~\ref{subsec:derivation-Poisson-SC}, the main difference between this Hamiltonian and the expression for the homogeneous case is that $p_F$ is replaced by $p_F(x)$ and $\varepsilon_b$ by $\varepsilon_b(x)$.
For small $|q|$, we can therefore expand $L_0(x,q)$ as one does for the homogeneous case. We find
\begin{equation}  \label{eq:L0-E-expansion-small-q-Ham}
\begin{aligned}
  L_0(x,q) 
    &= \varepsilon_b(x) q^2 \left( 1 -\frac{2 g_s}{3\pi} \frac{e^2 p_F^3(x)}{m \hbar \varepsilon_b(x) E^2} - \frac{2 g_s}{5 \pi} \frac{e^2 p_F^5(x)}{m^3 \hbar \varepsilon_b(x) E^4} q^2 \right) + \mathcal{O}(q^6) \\
    &= \varepsilon_b(x) q^2 \left( 1 - \frac{E_P^2(x)}{E^2} - \frac{3 p_F^2(x)}{5 m^2} \frac{E_P^2(x)}{E^4} q^2 \right) + \mathcal{O}(q^6) ,
\end{aligned}
\end{equation}
where we have introduced the plasma energy $E_P(x)$, defined by
\begin{equation}  \label{eq:plasma-energy}
  E_P^2(x) = \hbar^2 \frac{4 \pi n^{(0)}(x) e^2}{m \varepsilon_b(x)} = \frac{2 g_s e^2}{3 \pi m \hbar} \frac{p_F^3(x)}{\varepsilon_b(x)} ,
\end{equation}
cf. expression~(\ref{eq:plasma-frequency}). Using the expansion~(\ref{eq:L0-E-expansion-small-q-Ham}), we can solve the equation $L_0(x,q)=0$ for the energy $E$. When we expand the result up to order $q^2$, we obtain
\begin{equation}  \label{eq:L0-E-expansion-small-q}
  E^2 = E_P^2(x) + \frac{3 p_F^2(x)}{5 m^2} q^2 + \mathcal{O}(q^4), 
\end{equation}
which is equivalent to expression~(\ref{eq:dispersion-small-q-homo}). In dimensionless units, the plasma energy $\tilde{E}_P(\tilde{x})$ is given by expression~(\ref{eq:plasma-energy-dimless}), and we have
\begin{equation}  \label{eq:L0-E-expansion-small-q-dimless}
  \tilde{E}^2 = \tilde{E}_P^2(\tilde{x}) + \frac{12}{5} \tilde{p}_F^2(\tilde{x}) \tilde{q}^2 +\mathcal{O}(\tilde{q}^4).
\end{equation}
For a given energy $E$, the classically allowed region now consists of all points $x$ for which the equation~$L_0(x,q)=0$ has a solution for real $q$. From the expansion~(\ref{eq:L0-E-expansion-small-q}) it is clear that such a solution certainly does not exist when $E<E_P(x)$.

\begin{figure}[!tb]
  \hfill
  \begin{tikzpicture}
    \node at (0cm,0cm)
    {\includegraphics[width=0.95\textwidth]{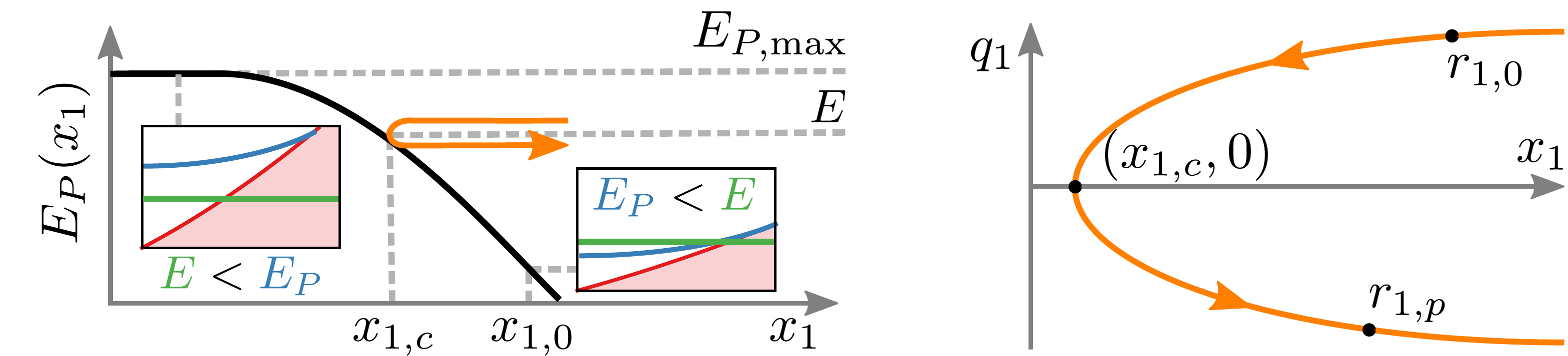}};
    \node at (-6.85cm,1.35cm) {(a)}; 
    \node at (1.3cm,1.35cm) {(b)};
  \end{tikzpicture}
  \hfill\hfill
  \caption{(a) Schematic representation of the spatially varying plasma energy $E_P(x_1)$ (solid black line), given by expression~(\ref{eq:plasma-energy}). 
    The classical turning point $x_{1,c}$ separates the classically allowed ($x_1 > x_{1,c}$) and the classically forbidden ($x_1 < x_{1,c}$) region. The plasmon cannot propagate when $E<E_P(x_1)$, as indicated in the inset, which shows the spectrum of a homogeneous quantum plasma with the parameters $n^{(0)}$ and $\varepsilon_b$ given by their values at the point $x_1$, cf. figure~\ref{fig:plasmon-dispersion-homo}.
    (b) The trajectory in phase space corresponding to the propagating plasmon shown by the orange arrow in panel (a). Since the velocity is parallel to $-q_1$, the incoming wave lies on the upper leaf of the Lagrangian manifold, as indicated by the arrows. The point $x_{1,c}$ is a fold point, with $q_1^2 \propto x_1 - x_{1,c}$.
  }
  \label{fig:simple-turning-point-phase-space}
\end{figure}

In what follows, we investigate what happens near the classical turning point $x_c$ that separates the classically allowed and the classically forbidden region.
To make our discussion more concrete, we consider a charge density $n^{(0)}(x_1)$ and a background dielectric constant $\varepsilon_b(x_1)$ that only depend on the first coordinate $x_1$. Specifically, we consider the situation sketched in figure~\ref{fig:simple-turning-point-phase-space}(a), where the plasmon energy $E_P(x_1)$ has a maximal value $E_{P,\text{max}}$ and decreases as $x_1$ increases. We can infer from definition~(\ref{eq:plasma-energy}) that this decrease can be caused by a decrease of $p_F(x_1)$, and therefore by a decrease of the charge density $n^{(0)}(x_1)$, or by an increase of $\varepsilon_b(x_1)$, or by a combination of the two.
Since the effective Hamiltonian $L_0(x,q)$ does not depend on $x_2$ and $x_3$ in this case, it is immediately clear from Hamilton's equations~(\ref{eq:Hamilton}) that $q_2$ and $q_3$ are constant. We may therefore define $q_\parallel = \sqrt{q_2^2+q_3^2}$.
Furthermore, the action $S(x)$ is separable and can be written as
\begin{equation}  \label{eq:action-separated-cartesian}
  S(x) = \int_{x_{1,0}}^{x_1} Q_1(X_1) \text{d}X_1 + q_2 x_2 + q_3 x_3 .
\end{equation}
where, following the notation introduced in section~\ref{subsec:derivation-induced-potential}, $(X_1(t),Q_1(t))$ represents the solution of Hamilton's equations.
Our problem is therefore effectively one-dimensional. Characterizing what happens near the turning point $x_{1,c}$, which separates the classically allowed from the classically forbidden region, therefore means understanding how $q_1$ depends on $x_1$ near $x_{1,c}$.

We start with the simplest case, where $q_\parallel$ equals zero. 
With the help of equation~(\ref{eq:L0-E-expansion-small-q}), we can understand where the classically allowed and the classically forbidden region lie. When $E > E_{P}(x_1)$, this equation has real solutions $q_1(x_1)$, which means that we are in the classically allowed region. When $E < E_{P}(x_1)$, the solutions are purely imaginary, which means that we are in the classically forbidden region. 
We conclude that the point $x_{1,c}$, which separates the two regions, is defined by $E=E_P(x_{1,c})$. As can be seen from equation~(\ref{eq:L0-E-expansion-small-q}), the turning point $x_{1,c}$ corresponds to $q_1=0$.
The difference between the two different regions is schematically indicated in figure~\ref{fig:simple-turning-point-phase-space}(a): the two insets show the spectrum for a homogeneous system with the parameters $n^{(0)}$ and $\varepsilon_b$ at those points.
Note that, in general, a turning point $x_{1,c}$ exists when $E<E_{P,\text{max}}$ and does not exist when $E>E_{P,\text{max}}$.

In order to quantitatively understand what happens in the vicinity of the point $x_{1,c}$, we may assume that $E_P(x_1)$ can be approximated by a linear function in the vicinity of $E_P(x_{1,c})$, see also figure~\ref{fig:simple-turning-point-phase-space}(a).
In other words, we perform a first-order Taylor expansion of equation~(\ref{eq:L0-E-expansion-small-q}) at $x_{1,c}$, which yields
\begin{equation}  \label{eq:L0-E-expansion-small-q-Taylor}
  E^2 \approx E_P^2(x_{1,c}) + 2 E_P(x_{1,c}) E_P'(x_{1,c}) (x_1 - x_{1,c}) + \frac{3 p_F^2(x_{1,c})}{5 m^2} q_1^2 .
\end{equation}
Since $x_{1,c}$ is defined by $E=E_P(x_{1,c})$, we easily find
\begin{equation}  \label{eq:L0-E-expansion-small-q-Taylor-fold}
  q_1^2 = - \frac{10 m^2}{3} \frac{E_P(x_{1,c})}{p_F^2(x_{1,c})} E_P'(x_{1,c}) (x_1 - x_{1,c}) + \mathcal{O}\left((x_1-x_{1,c})^2\right) .
\end{equation}
This behavior is illustrated in figure~\ref{fig:simple-turning-point-phase-space}(b), where we show the motion of the plasmon in phase space $(x_1,q_1)$. Expression~(\ref{eq:L0-E-expansion-small-q-Taylor-fold}) shows that $x_{1,c}$ is a so-called fold point~\cite{Poston78,Arnold82}, since $q_1^2 \propto x_1 - x_{1,c}$.
When we look at the curve in the $(x_1,q_1)$-plane, shown in figure~\ref{fig:simple-turning-point-phase-space}(b), we see that it indeed looks like a fold.
This curve is of course a projection of the full Lagrangian manifold $\Lambda^3$, which we discussed in section~\ref{subsec:derivation-induced-potential}, onto the coordinates $(x_1,q_1)$.

We remark that, in equation~(\ref{eq:L0-E-expansion-small-q-Taylor}), we performed the Taylor expansion of $E_P^2(x_1)$ up to first order, and the expansion of $p_F^2(x_1)$ up to zeroth order. Although this may seem inconsistent, it is justified by the conclusion that $q^2 \propto x_1 - x_{1,c}$, as it shows that $(x_1 - x_{1,c}) q^2$ contributes to the correction terms in expression~(\ref{eq:L0-E-expansion-small-q-Taylor-fold}) and not to the leading-order term. Similarly, higher-order terms in the expansion~(\ref{eq:L0-E-expansion-small-q}) only contribute to the correction terms.

In the same way, we can consider the case when $q_\parallel \neq 0$. When $q_\parallel$ is small, we can still use the expansion~(\ref{eq:L0-E-expansion-small-q}) for small $q_1$. We have
\begin{equation}  \label{eq:L0-E-expansion-small-q-Taylor-finite-qp-first}
  E^2 = E_P^2(x_1) + \frac{3 p_F^2(x_1)}{5 m^2} (q_1^2 + q_\parallel^2) + \mathcal{O}(q^4),
\end{equation}
and we see that the point $x_{1,c}$ is defined by the equation
\begin{equation}  \label{eq:L0-E-expansion-small-q-Taylor-finite-qp}
  E^2 = E_P^2(x_{1,c}) + \frac{3 p_F^2(x_{1,c})}{5 m^2} q_\parallel^2 .
\end{equation}
We subsequently perform a Taylor expansion of $E_P(x_1)$ and $p_F(x_1)$ around $x_{1,c}$ up to first order. After some calculus we find
\begin{equation}  \label{eq:L0-E-expansion-small-q-Taylor-fold-finite-qp}
  \begin{aligned}
  q_1^2 
    &= - \frac{2 E_P(x_{1,c}) E_P'(x_{1,c}) + \frac{6}{5m^2} p_F(x_{1,c}) p_F'(x_{1,c}) q_\parallel^2}{\frac{3}{5m^2} p_F^2(x_{1,c}) + \frac{6}{5m^2} p_F(x_{1,c}) p_F'(x_{1,c}) (x_1 - x_{1,c}) } (x_1 - x_{1,c}) + \mathcal{O}\left((x_1 - x_{1,c})^2\right) \\
    &= - \left(\frac{10 m^2}{3} \frac{E_P(x_{1,c})}{p_F^2(x_{1,c})} E_P'(x_{1,c}) + \frac{2 p_F'(x_{1,c})}{p_F(x_{1,c})} q_\parallel^2 \right)(x_1 - x_{1,c}) + \mathcal{O}\left((x_1 - x_{1,c})^2\right) ,
  \end{aligned}
\end{equation}
which indeed reduces to our previous expression~(\ref{eq:L0-E-expansion-small-q-Taylor-fold}) when $q_\parallel=0$.
This result shows that $x_{1,c}$ is still a fold point when $q_\parallel$ is small, since $q_1^2 \propto x_1 - x_{1,c}$. 
Hence, the momentum $q_1(x_1)$ is purely real when $E > E_{P}(x_1)$, and purely imaginary when $E < E_{P}(x_1)$.
When $q_\parallel$ is not small, we cannot use the expansion~(\ref{eq:L0-E-expansion-small-q}). However, we can still consider the full expression for $L_0(x,q)$ and define $x_{1,c}$ by $L_0(x_{1,c},q_\parallel)=0$ for a given energy $E$. One can then verify numerically that we still have $q_1^2 \propto x_1 - x_{1,c}$ around $x_{1,c}$.

To understand the direction of the arrows in figure~\ref{fig:simple-turning-point-phase-space}(b), we compute the velocity in the vicinity of $x_{1,c}$.
When $q_\parallel$ is small, the expansion~(\ref{eq:L0-E-expansion-small-q-Ham}) and Hamilton's equations~(\ref{eq:Hamilton}) lead to
\begin{align}
  \frac{\text{d} x_1}{\text{d} t} 
  &= \frac{\partial L_0}{\partial q_1}
  = \frac{\partial L_0}{\partial |q|} \frac{\partial |q|}{\partial q_1} \nonumber \\
  &\approx \frac{q_1}{|q|} \left( \varepsilon_b(x) (2 |q|) \left( 1 - \frac{E_P^2(x)}{E^2} - \frac{3 p_F^2(x)}{5 m^2} \frac{E_P^2(x)}{E^4} q^2 \right) -  \varepsilon_b(x) q^2 \frac{3 p_F^2(x)}{5 m^2} \frac{E_P^2(x)}{E^4} (2|q|) \right) \nonumber \\
  &\approx - 2 q_1 q^2 \varepsilon_b(x) \frac{3 p_F^2(x)}{5 m^2} \frac{E_P^2(x)}{E^4} ,
    \label{eq:velocity-momentum-opposite}
\end{align}
where we used that $L_0(x,q)=0$ to cancel the first term in the penultimate expression.
We therefore conclude that the velocity is parallel to $-q_1$, which corresponds to the direction of the arrows in figure~\ref{fig:simple-turning-point-phase-space}(b). In particular, the velocity vanishes when $q_1=0$, that is, at $x_{1,c}$.
The point $x_{1,c}$ is called a turning point, because the plasmon turns around there:
it comes in with positive momentum $q_1$, is decelerated and is then reflected with negative momentum $q_1$. Turning points for which $q_1^2 \propto x_1 - x_{1,c}$ are known as simple turning points.

The point $x_{1,c}$ is also a so-called singular point~\cite{Maslov81,Poston78,Arnold82}. To explain what this means, we have to consider the projection of the Lagrangian manifold, shown in figure~\ref{fig:simple-turning-point-phase-space}(b), onto the $x_1$-axis. We immediately see that the projection of the entire manifold is two-to-one. However, when we consider a single point $r_{1,0}=(x_{1,0},q_{1,0})$ in phase space, we can generally find a neighborhood for which the projection onto the $x_1$-axis is (locally) one-to-one. The only point for which this is impossible is precisely the fold point $(x_1,q_1) = (x_{1,c},0)$. By the implicit function theorem, this is equivalent to the vanishing of the Jacobian $J$, defined in equation~(\ref{eq:def-jacobian}) and discussed in section~\ref{subsec:derivation-induced-potential}.

To see how this works in more detail, we compute the Jacobian explicitly. Since $L_0(x,q)$ does not depend on $x_2$ and $x_3$, our problem is effectively one-dimensional and $q_2$ and $q_3$ are constant. The plane $\Lambda^2$ of initial conditions can be parametrized by the initial values of $x_2$ and $x_3$. We therefore have
\begin{equation}  \label{eq:Lambda-2-1d}
  \Lambda^2 = \{ (x,q), \quad x_1 = x_{1,0} , \; x_2 = \alpha_1, \; x_3 = \alpha_2, \; q_1 = q_{1,0}, \; q_2 = q_{2,0}, \; q_3 = q_{3,0} \},
\end{equation}
where $q_\parallel^2 = q_{2,0}^2+q_{3,0}^2$ and the energy $E$ is determined by $L_0(x,q) = 0$. Since $L_0(x,q)$ does not depend on $x_2$ and $x_3$, it assumes the same value on all points of $\Lambda^2$.
From a physical point of view, it is therefore obvious that $X_1(t,\alpha_1,\alpha_2)$ does not depend on $\alpha_1$ or $\alpha_2$. This can be proven using the so-called variational system, which follows from Hamilton's equations~(\ref{eq:Hamilton}):
\begin{equation}
  \begin{aligned}
    \frac{\text{d}}{\text{d} t} \left( \frac{\partial X_i}{\partial \alpha_j} \right) 
      = \frac{\partial}{\partial \alpha_j} \left( \frac{\text{d} X_i}{\text{d} t} \right) 
      &= \sum_k \frac{\partial^2 L_0}{\partial q_i \partial x_k} \frac{\partial X_k}{\partial \alpha_j} + \sum_k \frac{\partial^2 L_0}{\partial q_i \partial q_k} \frac{\partial Q_k}{\partial \alpha_j} , \\
    \frac{\text{d}}{\text{d} t} \left( \frac{\partial Q_i}{\partial \alpha_j} \right) 
      = \frac{\partial}{\partial \alpha_j} \left( \frac{\text{d} Q_i}{\text{d} t} \right) 
      &= -\sum_k \frac{\partial^2 L_0}{\partial x_i \partial x_k} \frac{\partial X_k}{\partial \alpha_j} - \sum_k \frac{\partial^2 L_0}{\partial x_i \partial q_k} \frac{\partial Q_k}{\partial \alpha_j} .
  \end{aligned}
\end{equation}
Considering that several derivatives are zero, and taking the initial conditions~(\ref{eq:Lambda-2-1d}) into account, one finds that the solution of this system of differential equations is given by
\begin{equation}
  \frac{\partial Q_i}{\partial \alpha_j} = 0, \; \frac{\partial X_1}{\partial \alpha_j} = 0, \;
  \frac{\partial X_2}{\partial \alpha_1} = 1, \; \frac{\partial X_2}{\partial \alpha_2} = 0, \;
  \frac{\partial X_3}{\partial \alpha_1} = 0, \; \frac{\partial X_3}{\partial \alpha_2} = 1.  
\end{equation}
The Jacobian is therefore given by
\begin{equation}  \label{eq:jacobian-1d-calculated}
  J = \det 
    \begin{pmatrix}
      \frac{\text{d} X_1}{\text{d} t} & \frac{\partial X_1}{\partial \alpha_1} & \frac{\partial X_1}{\partial \alpha_2} \\
      \frac{\text{d} X_2}{\text{d} t} & \frac{\partial X_2}{\partial \alpha_1} & \frac{\partial X_2}{\partial \alpha_2} \\
      \frac{\text{d} X_3}{\text{d} t} & \frac{\partial X_3}{\partial \alpha_1} & \frac{\partial X_3}{\partial \alpha_2}
    \end{pmatrix}
    = \det 
    \begin{pmatrix}
      \frac{\partial L_0}{\partial q_1} & 0 & 0 \\
      \frac{\partial L_0}{\partial q_2} & 1 & 0 \\
      \frac{\partial L_0}{\partial q_3} & 0 & 1
    \end{pmatrix}
    = \frac{\partial L_0}{\partial q_1} 
    = \frac{\partial L_0}{\partial |q|} \frac{\partial |q|}{\partial q_1}
    = \frac{q_1}{|q|} \frac{\partial L_0}{\partial |q|} .
\end{equation}
One could say that the Jacobian factorizes because the problem is effectively one-dimensional. 
Expression~(\ref{eq:jacobian-1d-calculated}) shows that the Jacobian indeed vanishes at $x_{1,c}$, since $q_1$ vanishes at that point. It is good to note that we are actually dealing with a plane of singular points, since the parameters $\alpha_1$ and $\alpha_2$ are arbitrary. A connected set of singular points is known as a caustic~\cite{Maslov81,Poston78,Arnold82}.

It is actually not that surprising that we have a fold point, as the general theory of caustics and singularities, see e.g. Refs~\cite{Poston78,Arnold82}, states that the fold point is the generic singularity in (effectively) one-dimensional problems. The word generic in this statement means that any other singularity will become a fold when one slightly changes the parameters. When the singular points in a system do not change type upon slight changes in the parameters, one says that the system is in general position.
Within the classification of caustics, a fold point is also called a singularity of type $A_2$. 

Since the Jacobian $J(x)$ vanishes at the fold point $x_{1,c}$, the asymptotic solution~(\ref{eq:solution-sc-potential}) diverges at the fold point. It is therefore no longer a good approximation to the true solution of the original differential equations.
It is well known that the leading-order term of the asymptotic solution near a fold point can be expressed in terms of the Airy function~\cite{Griffiths05,Maslov81,Berry72,Reijnders18,Chester57,Berry21,Anikin19}.
For many differential operators, one can explicitly construct this solution using the Maslov canonical operator~\cite{Maslov81,Reijnders18,Anikin19}. The key observation in this construction is that, although the projection of the Lagrangian manifold onto the $x_1$-axis is no longer one-to-one in the vicinity of $x_{1,c}$, the projection onto the $q_1$-axis is one-to-one. Unfortunately, the construction of this asymptotic solution requires a new and lengthy derivation in our case, since we are dealing with the system of equations discussed in section~\ref{sec:derivation}. We will therefore discuss this asymptotic solution in a separate publication. We remark that its importance is far greater than just the fold point, as it can also be used to construct asymptotic solutions near different types of singular points, such as the cusp point $A_3$. The latter naturally occurs when we consider potentials that depend on two spatial coordinates, as it is one of the generic singularities in effectively two-dimensional problems. We also remark that the vanishing of the Jacobian also leads to mathematical problems with the definition of the trace in section~\ref{subsec:derivation-charge-density}, as explained in appendix~\ref{subapp:trace-precise} below expression~(\ref{eq:operator-trace-lo-bound}).

For many applications, however, one does not need an explicit solution near the turning point. 
In fact, if we want to determine the quantization condition for bound states, we only have to determine the scattering phase that is acquired upon passing the turning point.
This scattering phase can be expressed in terms of the so-called Maslov index $\mu(r_{0},r)$ of the path between the starting point $r_0=(x_0,q_0)$ and the final point $r=(x,q)$ in phase space~\cite{Maslov81}.
It is important to realize that this index depends on the point in phase space, and not just on the coordinate $x$.
The index can be included in the asymptotic solution~(\ref{eq:solution-sc-potential}) by writing
\begin{equation}  \label{eq:solution-sc-potential-index}
  V(x,t) = \frac{A_0^0}{\sqrt{|J(x)|}} \exp\left( - \frac{i \pi}{2} \mu(r_0,r) \right) \exp\left(\frac{i}{\hbar}[ S(x) - E t ] \right) .
\end{equation}
From a practical point of view, the Maslov index regulates the analytic continuation of $\sqrt{J(x)}$, since the Jacobian~(\ref{eq:jacobian-1d-calculated}) changes sign when we pass the turning point $x_{1,c}$.
There are various equivalent definitions of the index~\cite{Maslov81,Arnold67,Dobrokhotov03}, but for our one-dimensional problem the most convenient definition of the Maslov index $\mu(r_{1,0},r_{1,p})$ of the path between the starting point $r_{1,0}$ and the point $r_{1,p}$ is~\cite{Maslov81}
\begin{equation}   \label{eq:maslov-index-calc}
  \mu(r_{1,0},r_{1,p}) = \text{inerdex} \,\frac{\partial x_1}{\partial q_1}(r_{1,0}) - \text{inerdex} \,\frac{\partial x_1}{\partial q_1}(r_{1,p}) ,
\end{equation}
where $\text{inerdex}\, A$ denotes the number of negative eigenvalues of the matrix $A$. For our one-dimensional example, the Lagrangian manifold is shown in figure~\ref{fig:simple-turning-point-phase-space}(b). 
Since we start with an incoming wave and $\text{d}x_1/\text{d}t \propto - q_1$, the starting point $r_{1,0}$ lies on the upper leaf of the Lagrangian manifold, as indicated in figure~\ref{fig:simple-turning-point-phase-space}(b).
Similarly, a point $r_{1,p}$ on the trajectory of the reflected wave lies on the lower leaf. By inspecting the sign of the derivative $\partial x_1/\partial q_1$ at the different points, we see from expression~(\ref{eq:maslov-index-calc}) that $\mu(r_{1,0},r_{1,p}) = -1$.
Although a rigorous proof of this result, as mentioned before, requires a representation of the solution in the vicinity of the singular point, we emphasize that the Maslov index is a geometric object that only depends on the geometry of the Lagrangian manifold~\cite{Maslov81,Arnold67}. This means that the index is always the same for a fold point and does not depend on the specific details of the problem.

In one-dimensional problems, one can also take an alternative approach to determining the scattering phase and the Maslov index. In this approach, pioneered by Zwaan~\cite{Zwaan29}, we derive this quantity by analytically continuing the asymptotic solution~(\ref{eq:solution-sc-potential}) in the complex plane. A detailed account of this approach was given in Refs.~\cite{Heading62,Froeman65}, see also Refs.~\cite{Berry72,Reijnders13}.
Although the method is only proven for the one-dimensional Schr\"odinger equation in the literature, we can apply the same idea to our asymptotic solution~(\ref{eq:solution-sc-potential}), since it has exactly the same structure.
We use this method to confirm that the Maslov index $\mu(r_{1,0},r_{1,p})$ equals $-1$. Since the derivation is quite technical, we present it in appendix~\ref{app:derivation-Maslov-ind-alternative}.

In this section, we investigated in detail what happens near one type of boundary between a classically allowed and a classically forbidden region. By investigating the effective Hamiltonian, we were able to clarify the structure of the Lagrangian manifold and identify that we are dealing with a simple turning point. This allowed us to compute the scattering phase, expressed in terms of the Maslov index, that is acquired when passing the turning point.
Although we presented our considerations for a decreasing plasmon energy $E_P(x)$ and charge density $n^{(0)}(x)$, they also apply to an increasing plasmon energy. In that case we find the same value of the Maslov index, namely $\mu(r_{1,0},r_{1,p}) = -1$.

\subsection{Investigation of the Landau damping threshold}  \label{subsec:Landau-damping-threshold}

In the previous subsection, we considered what happens in the vicinity of a simple turning point, where $E$ is close to $E_P(x)$. We found that $q_1^2 \propto x_1 - x_{1,c}$ and that the Maslov index $\mu(r_{1,0},r_{1,p})$ equals $-1$. In this subsection, we investigate the second type of turning point in our system, which is more complicated and arises when $E$ is close to $E_L(x)$. In terms of the dispersion for a homogeneous system, shown in figure~\ref{fig:plasmon-dispersion-homo}, this means that we hit the upper boundary of the plasmon spectrum. We first carefully set the scene in section~\ref{subsec:Landau-turning-point-setting} and define the Landau-type turning point $x_L$. In section~\ref{subsec:no-solutions-real-energies}, we then prove that there are no solutions $q_1(x_1)$, neither real nor complex, of $L_0(x,q,E)=0$ for $x>x_{L}$. In section~\ref{subsec:damping-homogeneous-systems}, we take a closer look at what happens in homogeneous systems when $E>E_L$ and study the complex-energy solutions of the secular equation. Finally, in section~\ref{subsec:complex-energies-consequences-inhomogeneous}, we discuss what our findings imply for the value of the Maslov index.

\subsubsection{Definition of the Landau-type turning point}
\label{subsec:Landau-turning-point-setting}

In section~\ref{subsec:derivation-outline}, we discussed the plasmon dispersion for a homogeneous system. As indicated in figure~\ref{fig:plasmon-dispersion-homo}, the equation~$\overline{\varepsilon}(q,E)=0$ has no solutions for real values of $E$ when we are above $E_L$.
Before we consider what happens in the vicinity of the point $E_L$, let us define it more precisely.
To this end, we first define $|q_L(E)|$ as the momentum for which $\nu_- = 1$. For $|q|>|q_L(E)|$, we have $\nu_- < 1$ and the logarithm in equation~(\ref{eq:chi-homogenegous-computed}) becomes complex. In figure~\ref{fig:plasmon-dispersion-homo}, the line that indicates the boundary of the red region corresponds exactly to $|q_L(E)|$. The second criterion to define $E_L$ is that it satisfies the secular equation $\overline{\varepsilon}=0$. For a homogeneous system, the quantity $E_L$ is therefore defined by $\overline{\varepsilon}(q_L(E_L),E_L) = 0$.

For an inhomogeneous system, we have to consider the classical Hamiltonian $L_0(x,q,E)$, given by expression~(\ref{eq:L0-computed-epsilon}), where we now explicitly included the energy dependence in our notation. The main difference between $L_0(x,q,E)$ and
the dielectric function $\overline{\varepsilon}(q,E)$ for a homogeneous system is that $p_F$ is replaced by $p_F(x)$ and $\varepsilon_b$ by $\varepsilon_b(x)$.
We can therefore define the quantities $|q_L|$ and $E_L$ for each point $x$. 
In other words, the equation $\nu_-(x)=1$ implicitly defines $|q_L(x,E)|$. The energy $E_L(x)$ is also implicitly defined:
\begin{equation}  \label{eq:E-L-def}
  L_0(x,q_L(x,E_L),E_L) = 0 \; \text{ defines } E_L(x).
\end{equation}
For a given point $x$, we do not have any real energy solutions beyond $E_L(x)$, as can be concluded in analogy with the homogeneous case. When we instead consider a fixed energy $E$, we can define the point $x_L$ by the condition $E_L(x_L) = E$. An equivalent definition of the point $x_L$ can be obtained from the momentum. If $|q(x)|$ is the solution of $L_0(x,q,E)= 0$ for a given energy $E$, then the point $x_L$ is defined by the condition $|q(x_L)| = |q_L(x_L,E)|$. The equivalence of these definitions comes from the fact that both equations~$\nu_-(x) = 1$ and $L_0(x,q,E)=0$ are satisfied in the two cases. 
We will see later on that the behavior near the point $x_L$ is very different from the behavior near the simple turning point $x_{1,c}$ from the previous subsection. Because of its special nature, which is related to Landau damping, we call the point $x_{L}$ a Landau-type turning point.

Before investigating what happens in the vicinity of the Landau-type turning point $x_L$, we briefly point out the relation with section~\ref{subsec:derivation-charge-density} and appendix~\ref{subapp:trace-precise}. In these sections, we mentioned that $\zeta(x,p,q)$ has simple poles at the points where $H_0(x,p) - H_0\left(x,p+q\right) + E = 0$ and discussed its mathematical implications. More precisely, we found that the symbol $\rho_1(x,p)$ formally does not give rise to a trace-class operator when these poles lie inside the region where $\rho_0( H_0(x,p)) - \rho_0\left(H_0\left(x,p+q \right)\right)$ is non-zero. We now see how this relates to physics: these issues arise precisely when single-particle excitations are possible, that is, when $x > x_L$. For $x > x_L$, the integral in expression~(\ref{eq:def-polarization}) for the polarization contains a simple pole, which leads to a complex outcome. This is precisely the same as saying that the logarithm in expression~(\ref{eq:L0-epsilon-dimless}) becomes complex for $x > x_L$. We may therefore also say that we investigate the physical consequences of the pole in this section.

\begin{figure}[!tb]
    \hfill
  \begin{tikzpicture}
    \node at (0cm,0cm)
    {\includegraphics[width=0.95\textwidth]{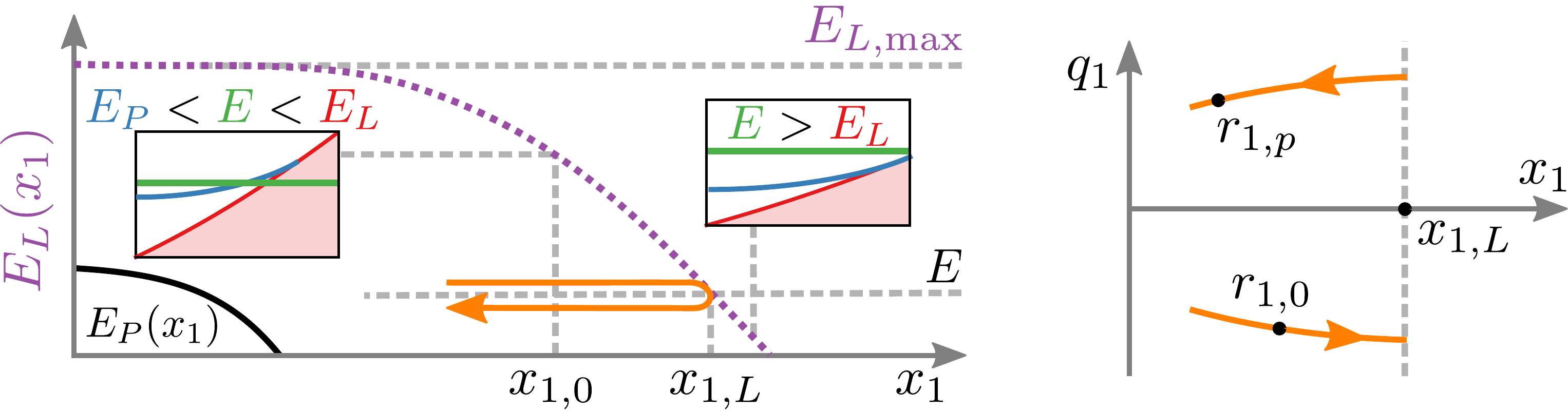}};
    \node at (-6.85cm,1.6cm) {(a)}; 
    \node at (2.2cm,1.6cm) {(b)};
  \end{tikzpicture}
  \hfill\hfill
  \caption{
    (a) Schematic representation of the spatial variation of $E_L(x_1)$ (dotted purple line), defined by equation~(\ref{eq:E-L-def}). For completeness, we also show the spatially varying plasma energy $E_P(x_1)$ (solid black line), given by expression~(\ref{eq:plasma-energy}). The plasmon can propagate in the region where $E_P(x_1) < E < E_L(x_1)$, i.e., for $x_1<x_{1,L}$. This is indicated by the orange arrow and in the insets, which show the spectrum of a homogeneous plasma with the parameters $n^{(0)}$ and $\varepsilon_b$ given by their values at the respective points, cf. figures~\ref{fig:plasmon-dispersion-homo} and~\ref{fig:simple-turning-point-phase-space}.
    (b) The phase space trajectory corresponding to the orange arrow in panel (a). At the Landau-type turning point $x_{1,L}$, there is a jump from the lower leaf of the Lagrangian manifold to the upper leaf, and the momentum discontinuously changes from a negative to a positive value.
  }
  \label{fig:Landau-turning-point-phase-space}
\end{figure}

To make our discussion more concrete, we consider the situation sketched in figure~\ref{fig:Landau-turning-point-phase-space}(a), where $E_L(x_1)$ monotonically decreases and only depends on the first coordinate $x_1$.
As for $E_P(x_1)$ in the previous subsection, this decrease in $E_L(x_1)$ may either arise from a decrease in $n^{(0)}(x_1)$, or from an increase in $\varepsilon_b(x_1)$, or from a combination of both. As in the previous subsection, the other two components $q_2$ and $q_3$ of the momentum are constant, and we define $q_\parallel=\sqrt{q_2^2+q_3^3}$. The action is once again separable as indicated in expression~(\ref{eq:action-separated-cartesian}). Our problem is thus effectively one-dimensional, and $x_L$ becomes $x_{1,L}$.
Since the function $E_L(x_1)$ is defined by $L_0(x_1,q_L(x_1,E_L),E_L) = 0$, where $q_L(x_{1,L},E)$ is defined by $\nu_-(x_1)=1$, it does not depend on $q_\parallel$. This means that the point $x_{1,L}$, defined by $E=E_L(x_{1,L})$ does not depend on $q_\parallel$. Note, however, that $q_1(x_{1,L})$ does depend on $q_\parallel$, since it is defined by $q_1^2(x_{1,L}) + q_\parallel^2 = |q_L(x_{1,L},E)|^2$. This is in marked difference with the simple turning point, since $x_{1,c}$ does depend on $q_\parallel$.
In section~\ref{subsec:simple-turning-point}, we obtained the Maslov index for a simple turning point by studying the structure of the Lagrangian manifold in the vicinity of the turning point. We should therefore study the behavior of $q_1(x_1)$ in the vicinity of $x_{1,L}$.

\subsubsection{Non-existence of solutions $q_1(x_1)$ for real energies $E$}
\label{subsec:no-solutions-real-energies}

In this subsection, we study the behavior of $q_1(x_1)$ in the vicinity of $x_{1,L}$ for real energies. We first prove that, for real $E$, there are no solutions $q_1(x_1)$ of $L_0(x,q,E)=0$ for $x_1>x_{1,L}$. We then discuss its implications.
Although the proof may seem somewhat technical, its main ingredient is the behavior of the function $z \log z$ in the complex plane around the origin. To make the notation a bit more readable, we do not explicitly indicate the energy dependence $E$ in what follows.

In the first step of the proof, we expand $L_0(x,q)$ in the vicinity of $|q_L(x_1)|$ for a given energy $E$. When we translate this to the spectrum depicted in figure~\ref{fig:plasmon-dispersion-homo}, we can say that we expand near the red line that delimits the Landau damping region.
We define the variable $\delta q$ by $\delta q = |q_L(x_1)| - |q|$. Since $\nu_-(x_1)=1$ when $|q| = |q_L(x_1)|$, we can approximate $\nu_-(x_1)-1 \approx c_1(x_1) (|q_L(x_1)|-|q|) = c_1(x_1) \delta q$ for $|q|$ close to $|q_L(x_1)|$. Note that $c_1(x_1)>0$, as $\nu_-(x)-1>0$ for $|q|<|q_L(x_1)|$.
The effective Hamiltonian $L_0(x,q)$ contains a term with $(\nu_-(x_1)-1) \log (\nu_-(x_1)-1)$, which near the point $q_L(x_1)$ behaves as $c_1(x_1) \delta q (\log \delta q + \log c_1(x_1))$. The other terms in $L_0(x,q)$ do not show singular behavior, and we can use a first-order Taylor expansion to approximate their behavior. In the vicinity of $|q| = |q_L(x_1)|$, the effective Hamiltonian $L_0(x,q)$ therefore behaves as
\begin{equation} \label{eq:L0-approx-qL}
  L_0(x,q) = c_2(x_1) + c_3(x_1) \delta q \log \delta q + c_4(x_1) \delta q,
\end{equation}
for certain real functions $c_j(x_1)$.
A direct computation shows that we can eliminate the term $c_4(x_1) \delta q$ by introducing the new variable $z = \exp(c_4(x_1)/c_3(x_1)) \delta q$. We find
\begin{equation}  \label{eq:L0-approx-qL-reduced}
  L_0(x,q) = c_2(x_1) + c_5(x_1) z \log z,
\end{equation}
where $c_5(x_1) = c_3(x_1) \exp(-c_4(x_1)/c_3(x_1))$.
At this point, we specialize to the situation where $x_1$ is close to $x_{1,L}$. Since we have a well-defined plasmon at $x_{1,L}$, with momentum $|q_L(x_{1,L})|$, we have $L_0(x_{1,L}, q_L(x_{1,L})) = 0$. This implies that $c_2(x_{1,L}) = 0$, and that $c_2(x_1)$ is small for $x_1$ close to $x_{1,L}$, i.e.,  $c_2(x_1) \approx c_6(x_1- x_{1,L})$.

In the second step of the proof, we therefore consider the solutions of the equation
\begin{equation}
  z \log z = - c_2(x_1)/c_5(x_1) \equiv \lambda,
\end{equation}
where $\lambda$ is real and small. Allowing the variable $z$ to be complex means that we allow $q_1^2(x_1)+q_\parallel^2$ to be complex. In other words, we consider complex $q_1(x_1)$. Using the polar form $z=r e^{i\phi}$ and separating real and imaginary parts, we obtain the set of equations
\begin{equation} \label{eq:singularity-near-qL}
  \begin{aligned}
    & r \cos\phi \log r - r \phi \sin\phi = \lambda \\
    & r \sin\phi \log r + r \phi \cos\phi = 0 .
  \end{aligned}
\end{equation}
The variable $\phi$ is constrained to the interval $(-\pi,\pi]$. This is a consequence of the fact that we are discussing the retarded response function, cf. the discussion below expression~(\ref{eq:potential-time-dependence}), since it implies that we have to consider the principal branch of the logarithm. We discuss this point in more detail in appendix~\ref{app:analytic-cont}.

Let us start with the case $x_1<x_{1,L}$, where the above set of equations should have a solution for real $z$, see also figure~\ref{fig:Landau-turning-point-phase-space}.
From the definition of $z$, we have $|q| = |q_L(x_1)| - r e^{i \phi} \exp(-c_4(x_1)/c_3(x_1))$. Since we have a well-defined plasmon, the solution satisfies $|q|<|q_L(x_1)|$, meaning that $r>0$ and $\phi=0$.
Indeed, the second equation~(\ref{eq:singularity-near-qL}) is automatically satisfied for $\phi=0$, whilst the first one reduces to $r \log r = \lambda$. We previously established that $\lambda$ is small, and proportional to $x_1- x_{1,L}$ in first order. We now see that the constant of proportionality has to be positive, since $r \log r < 0$ for small $r$.

For $x_1>x_{1,L}$, the solution with $\phi=0$ ceases to exist, since the left-hand side of the first equation~(\ref{eq:singularity-near-qL}) is negative, whereas the right-hand side is positive. The first equation then implies that $r$ has to be small, since $\lambda$ is small. 
Because we established that $\phi \neq 0$, we can divide by it and rewrite the second equation~(\ref{eq:singularity-near-qL}) as
\begin{equation}
  \frac{\tan \phi}{\phi} = -\frac{1}{\log r}
\end{equation}
For small $r$, the right-hand side is positive and small. The left-hand side, however, lies between 1 and infinity for $\phi \in (-\pi/2,\pi/2)$ and between negative infinity and zero for $\phi \in (-\pi,-\pi/2)$ or $\phi \in (\pi/2,\pi)$. This implies that the system of equations~(\ref{eq:singularity-near-qL}) does not have any solutions for $x_1>x_{1,L}$, which means that there are no solutions $q_1(x_1)$ of $L_0(x,q,E)=0$ for $x_1>x_{1,L}$. 
Note that this proof does not rule out solutions that cannot be continuously connected to $|q_L(x_{1,L})|$, i.e. that appear as a sudden jump in the value of $|q|$. We therefore verified numerically that these do not exist.

The above discussion shows that we have a classically allowed region for $x_1< x_{1,L}$, where there are real solutions $q_1(x_1)$ for real energies $E$. We can call the region $x_1 > x_{1,L}$ classically forbidden, since there are no solutions $q_1(x_1)$, neither real nor complex, for real energies $E$. At the point $x_{1,L}$, the momentum $q_1$ does not vanish, but attains a finite value. In figure~\ref{fig:Landau-turning-point-phase-space}(b), we show the curve in phase space (the Lagrangian manifold) that corresponds to this situation. It is important to note that this graph differs dramatically from the phase space curve for a simple turning point, depicted in figure~\ref{fig:simple-turning-point-phase-space}(b). Looking at the induced potential~(\ref{eq:solution-sc-potential-index}), we see that the Jacobian $J$ does not vanish at $x_{1,L}$, in stark contrast to what happens at the simple turning point, as discussed in the previous subsection.

Qualitatively, we could say that what happens at $x_{1,L}$ looks somewhat similar to scattering by an infinite potential barrier for a particle governed by the Schr\"odinger equation. For an infinite potential barrier, we also do not have any solutions, neither real nor complex, beyond the point $x_\text{inf}$ where the potential becomes infinite. Moreover, the momentum can also be finite at $x_\text{inf}$. To satisfy the boundary conditions at this point, the incoming wave and the reflected wave have to add up to zero, meaning that the reflected wave equals $e^{\pm i \pi}$ times the incoming wave.
Exploiting this analogy with scattering by an infinite potential barrier, and considering our asymptotic solution~(\ref{eq:solution-sc-potential-index}), we may therefore speculate that the plasmon is completely reflected at the point $x_{1,L}$ and that the Maslov index $\mu(r_{1,0},r_{1,p})$ equals $-2$.
In the phase space curve depicted in figure~\ref{fig:Landau-turning-point-phase-space}(b), this leads to a jump from the lower leaf of the Lagrangian manifold, with negative $q_1$, to the upper leaf of the Lagrangian manifold, with positive $q_1$, at the point $x_{1,L}$.

\subsubsection{Complex energies and damping in homogeneous systems}
\label{subsec:damping-homogeneous-systems}

Unfortunately, the situation seems to be more complicated than the simple analysis from the previous subsection suggests.
As we mentioned in section~\ref{subsec:derivation-outline}, a homogeneous system admits complex energy solutions beyond the point $(|q_L|, E_L)$. These excitations have a finite lifetime due to Landau damping~\cite{Landau46,Vonsovsky89,Giuliani05}: the collective excitation transfers energy to incoherent electron-hole excitations, that is, to single-particle excitations.
Before we consider what this means for an inhomogeneous system, let us first discuss these damped excitations in more detail for a homogeneous system.

From an experimental perspective, damping manifests itself as a broadening of the plasmon resonances. In electron energy loss spectroscopy (EELS), one measures the dynamic structure factor $S(q,E)$, which is proportional to $\text{Im}[-1/\overline{\varepsilon}(q,E)]$, see e.g. Refs.~\cite{Vonsovsky89,Giuliani05}.
When $\overline{\varepsilon}(q,E)$ has a real root, that is, when $E=E(q)$ satisfies the plasmon dispersion with $E_P<E<E_L$, the dynamic structure factor becomes proportional to a delta function. It is nonzero exactly at the point where $\overline{\varepsilon}(q,E)$ vanishes, i.e. $\text{Im}[-1/\overline{\varepsilon}(q,E)] \propto \delta(\overline{\varepsilon}(q,E))$.
When $E>E_L$, this peak acquires a finite width, which increases with increasing energy. We can use the position of the maximum as a measure for the plasmon dispersion $E(q)$ and half of the full width at half maximum (FWHM) as a measure for the damping $\gamma(q)$.

From a theoretical perspective, it seems more appropriate to consider the roots of the dielectric function. Indeed, equation~(\ref{eq:dielectric-function-homo}) suggests that we should consider $\overline{\varepsilon}(q,z)=0$ for complex energies $z=E-i\gamma$ when $E>E_L$.
Based on our previous considerations, we expect the solutions of this equation to correspond to damped plasmons.
Because $V(x,t) \propto \exp(-i(E-i \gamma) t/\hbar) = \exp(-\gamma t/\hbar) \exp(-i E t/\hbar)$, this implies that the roots should lie in the lower half plane, i.e. $\gamma>0$.
This intuition is confirmed by the proof in Ref.~\cite{Bonitz94}, which shows that the equation $\overline{\varepsilon}(q,E-i\gamma)=0$ does not have any roots for $\gamma<0$, i.e. in the upper half plane, for any $|q|$ and an arbitrary isotropic distribution function $\rho_0(E_p)$. The induced potential $V(x,t)$ therefore cannot become exponentially large over time and there are no plasma instabilities.

However, our expression~(\ref{eq:dielectric-function-homo}) for the dielectric function, with polarization~(\ref{eq:chi-homogenegous}), is only valid in the upper half plane, and not in the lower half plane.
This is a consequence of causality. As pointed out by Landau~\cite{Landau46}, the system initially has an equilibrium distribution $\rho_0$ and only becomes polarized after an external perturbation is applied at time $t_0$. This implies that the polarization, which is a retarded response function, has to vanish for $t<t_0$.
By Cauchy's theorem, this is equivalent to the requirement that $\overline{\Pi}(q,z)$ is analytic for $z$ in the upper half plane. This result is known as the Titchmarsh theorem, see e.g. Ref.~\cite{Giuliani05}, which additionally states that $\overline{\Pi}(q,E)$ satisfies the Kramers-Kronig relations. This is why we said that $E$ lies slightly above the real axis in section~\ref{subsec:derivation-outline}, and wrote $E+i\eta$.
Consequentially, we first have to consider the analytic continuation of $\overline{\Pi}(q,z)$ and $\overline{\varepsilon}(q,z)$ into the lower half plane before we can study its roots.

For the classical plasma, where $\rho_0$ is the Maxwell-Boltzmann distribution, this analytic continuation can be performed directly and was studied by Landau~\cite{Landau46}. A key ingredient in his derivation is that, after a suitable choice of coordinates, the problem effectively becomes one dimensional. It also relies on the observation that the Maxwell-Boltzmann distribution is isotropic and retains its functional form when we integrate over perpendicular directions. Since the latter does not hold for the Fermi-Dirac distribution, one cannot apply the technique developed by Landau to a three-dimensional quantum plasma. Note, however, that it can be applied to a one-dimensional quantum plasma, see Ref.~\cite{Bonitz94b}.

Nevertheless, one can perform the analytic continuation of $\overline{\varepsilon}(q,z)$ for a three-dimensional quantum plasma by considering the spectral representation, as shown in Ref.~\cite{Bonitz93}. In Ref.~\cite{Hamann20}, this technique was applied to a plasma at finite temperature, and the roots of the analytical continuation $\check{\varepsilon}(q,z)$ were studied.
In appendix~\ref{app:analytic-cont}, we show how this formalism can be applied to the case of zero temperature.
We first review the methods discussed in Refs.~\cite{Bonitz93,Hamann20} and subsequently consider their application for finite temperatures and for zero temperature. At finite temperature, we find that the analytic continuation $\check{\varepsilon}(q,z)$ is not unique due to the presence of branch cuts in the spectral function. Moreover, we find that for some choices of the branch cuts, the equation $\check{\varepsilon}(q,z)=0$ does not have solutions for sufficiently low temperatures. 
We then present a physical argument to choose the most suitable analytic continuation, namely that $\check{\varepsilon}(q,z)$ should have a well-defined limit for $T \to 0$. In this respect, we understand the case of zero temperature as a limit of the finite temperature case. We construct the analytic continuation $\check{\varepsilon}(q,z)$ for zero temperature and numerically extract the dispersion $E$ and the damping $\gamma$ from the position of the root, setting $z=E-i\gamma$. Finally, we analytically prove that, at zero temperature, the equation $\check{\varepsilon}(q,z) = 0$ has roots for $|q|>|q_L(E_L)|$, i.e., for $E > E_L$.

\begin{figure}[tb]
  \hfill
  \begin{tikzpicture}
    \node at (0cm,0cm)
    {\includegraphics[width=0.45\textwidth]{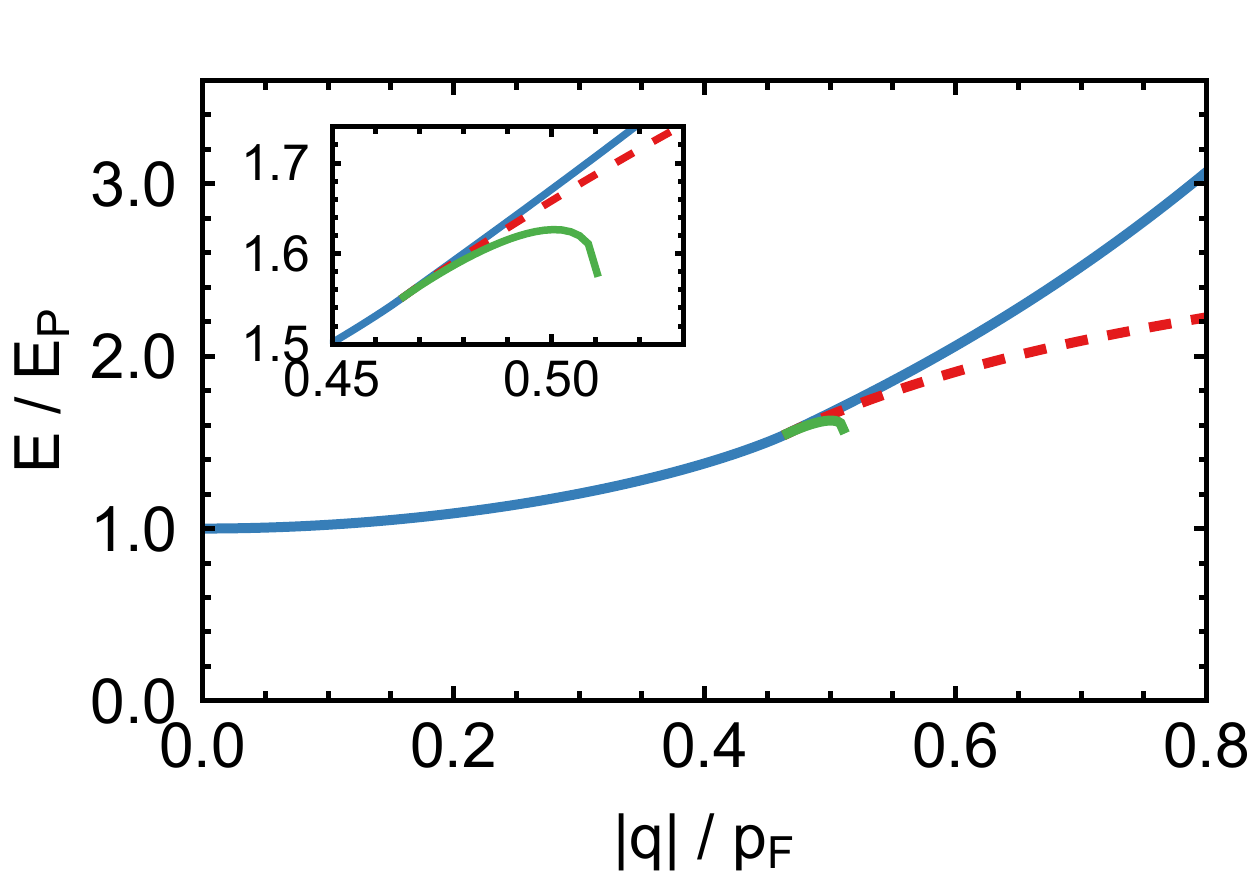}};
    \node at (-3.3cm,1.85cm) {(a)};
    
    \node at (7.1cm,0cm)
    {\includegraphics[width=0.45\textwidth]{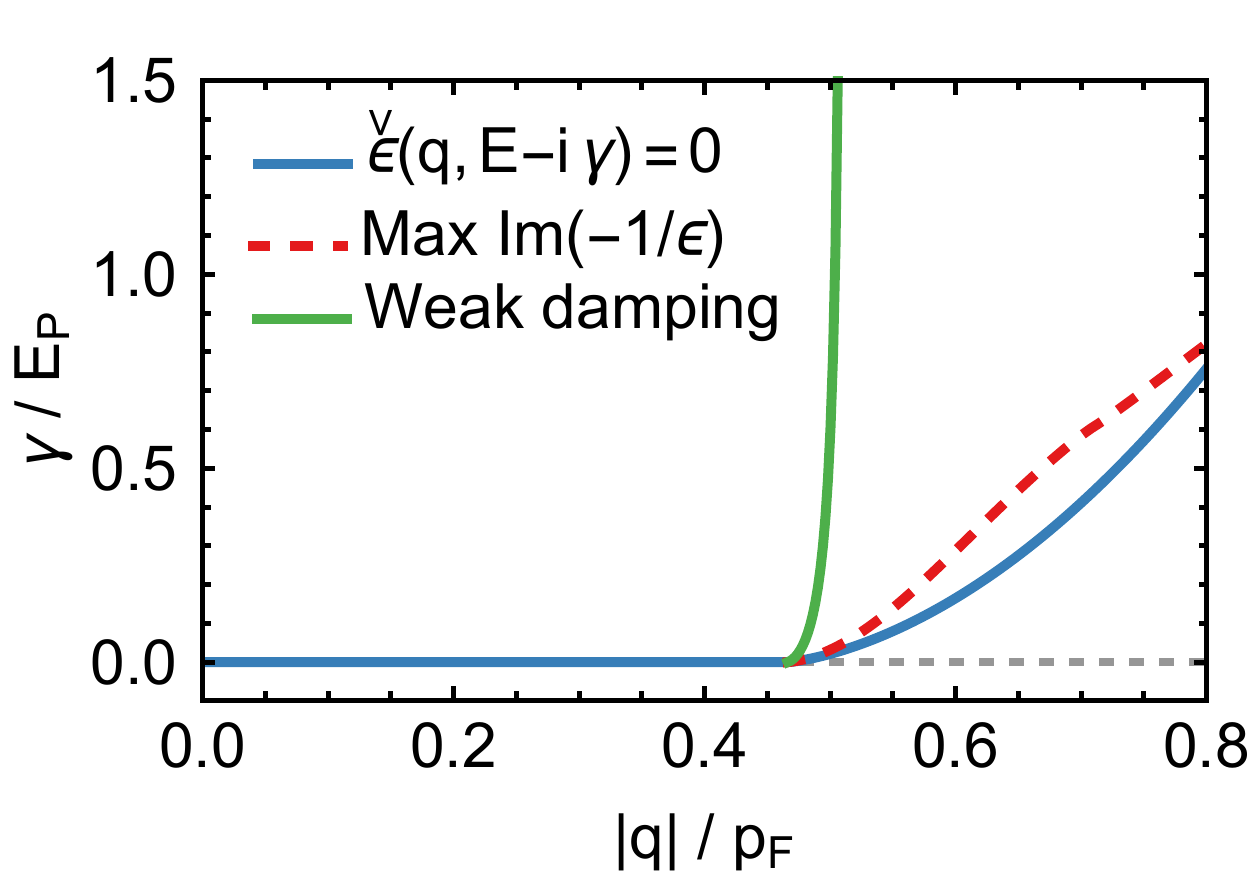}}; 
    \node at (3.8cm,1.85cm) {(b)};
  \end{tikzpicture}
  \hfill\hfill
  \caption{(a) The plasmon dispersion $E$ and (b) the plasmon damping $\gamma$ as a function of $|q|$ for $n^{(0)} = 1/a_0^3$. The solid blue line shows the results obtained from the roots of the analytic continuation $\check{\varepsilon}(q,E-i\gamma)$. The results extracted from the maximum of $\text{Im}[-1/\overline{\varepsilon}(q,E)]$ are shown by the dashed red line. The solid green line shows the weak damping approximation. The inset in panel (a) shows a closeup of the region where damping sets in, and shows the limited validity of the weak damping approximation. The dashed grey line in panel (b) indicates zero and serves as a guide to the eye.}
  \label{fig:dispersion-damping-homo}
\end{figure}

Figure~\ref{fig:dispersion-damping-homo} shows the dispersion $E$ and damping $\gamma$ for the two methods we discussed, for $n^{(0)} = 1/a_0^3$. The solid blue line shows the results obtained from solving $\check{\varepsilon}(q,E-i\gamma)=0$, whilst the dashed red line shows the results obtained from analyzing the peak of $\text{Im}[-1/\overline{\varepsilon}(q,E)]$. For finite temperatures, the qualitative difference between these two quantities is discussed in Ref.~\cite{Hamann20}: a sharp peak of $\text{Im}[-1/\overline{\varepsilon}(q,E)]$ can be called a collective mode, whereas a broad peak consists of collective effects as well as single-particle effects. Whether we call the modes with $|q|>|q_L(E_L)|$ plasmons or single-particle excitations is partically semantic, as the division between the two is not absolute.
For $|q|$ just above $|q_L(E_L)|$, we may nevertheless say that we are dealing with a damped plasmon, since the peak of $\text{Im}[-1/\overline{\varepsilon}(q,E)]$ is narrow. When $|q|$ is much larger than $|q_L(E_L)|$, the damping is large, leading to a broad peak which has more single-particle character.

A third approach to obtain the dispersion and damping is the so-called weak damping approximation~\cite{Vonsovsky89,Hamann20}. Here, one assumes that $\gamma$ is small, and performs a first-order Taylor expansion of $\overline{\varepsilon}(q,E-i\gamma)=0$. Setting both the real and imaginary parts to zero, one arrives at
\begin{equation}  \label{eq:small-damping}
  \text{Re} \, \overline{\varepsilon}(q,E) = 0 , \qquad
  \gamma = \frac{\text{Im} \, \overline{\varepsilon}(q,E)}{\frac{\partial}{\partial E} \text{Re} \, \overline{\varepsilon}(q,E)} .
\end{equation}
The dispersion $E$ and damping $\gamma$ obtained from these equations are shown by the solid green lines in figure~\ref{fig:dispersion-damping-homo}. We immediately see that this approximation performs very poorly. First, the damping quickly becomes very large as we increase $|q|$, which means that we can no longer neglect the higher-order terms in the Taylor expansion. Second, the equation $\text{Re} \, \overline{\varepsilon}(q,E) = 0$ does not have any solutions when $|q|$ is only slightly larger than $|q_L(E_L)|$, see also the inset in figure~\ref{fig:dispersion-damping-homo}(a). 
For an extensive discussion of this approximation and its extensions, we refer to Ref.~\cite{Hamann20}.
We finally remark that the weak damping approximation implicitly relies on the analytic continuation, since the equation $\overline{\varepsilon}(q,z)=0$ does not have any solutions for complex $z$ when $|q|>|q_L(E_L)|$, as shown in appendix~\ref{app:analytic-cont}.

We have thus established that, for homogeneous systems, there are damped plasmon modes for $|q|>|q_L(E_L)|$. We can understand these modes in various ways, namely by studying the peak of $\text{Im}[-1/\overline{\varepsilon}(q,E)]$, or by studying the complex roots of the analytic continuation $\check{\varepsilon}(q,z)$. For our study, the latter description seems most relevant.

\subsubsection{Consequences for an inhomogeneous plasma}
\label{subsec:complex-energies-consequences-inhomogeneous}

In section~\ref{subsec:no-solutions-real-energies}, we considered real energies and used the analogy with scattering by an infinite potential barrier for Schr\"odinger electrons to speculate that the Maslov index $\mu(r_{1,0},r_{1,p})$ equals $-2$.
However, our findings in the previous subsection imply that we should consider complex energies whenever the trajectory in phase space includes a Landau-type turning point. Our previous considerations therefore do not seem entirely accurate.
From a physical point of view, considering a complex energy means that the plasmon acquires damping, which leads to broadening of the resonances corresponding to bound states. This seems sensible, as it implies that the plasmon loses some energy when it hits the Landau-type turning point, where it interacts with incoherent electron-hole excitations.

A rigorous semiclassical analysis of this process however requires more advanced techniques, because of two reasons. 
The first reason for this is that we jump from one leaf of the Lagrangian manifold to another at the Landau-type turning point. Using more mathematical terms, we can say that the Lagrangian manifold does not have one of the canonical forms~\cite{Arnold82} in the vicinity of the turning point. This jump is not unique to our problem, and a class of these problems was studied in Refs.~\cite{Nazaikinskii14,Anikin18,Dobrokhotov21}. In these papers, the authors constructed an asymptotic solution in the vicinity of the turning point by modifying the standard semiclassical construction in a way tailored to the problem at hand, and were able to obtain the Maslov index.
The second reason is that, as we mentioned before, we probably have to consider complex energies when our system includes a Landau-type turning point. We may therefore have to consider the complex phase space, at least in the vicinity of the turning point, which also makes the analysis more complicated.

We can illustrate these two points by considering the small damping approximation.
Although we saw in the previous subsection that this approximation does not perform well outside of a small neighborhood of the point $(|q_L(E_L)|,E_L)$, we can still use it to gain some qualitative understanding of what is going on.
When we are beyond the point $x_{1,L}$, the polarization $\Pi_0(x,q,E)$ in the effective classical Hamiltonian $L_0(x,q,E)$ is complex. For $x_1$ (very) close to $x_{1,L}$, the imaginary part of $\Pi_0(x,q,E)$ is comparable to the terms of subleading order, i.e., it is of order $\hbar$. This implies that $\text{Re} \, \Pi_0(x,q,E)$ should be included in the leading-order term of the asymptotic series, whereas $\text{Im} \, \Pi_0(x,q,E)$ should be included in the subleading-order term. In other words, equation~(\ref{eq:def-L0-epsilon}) should be replaced by the real part of this expression. Similarly, the imaginary part of $\Pi_0$ should be added to the terms in expression~(\ref{eq:comm-rel-h1}), which gives rise to an additional term in the amplitude $\varphi_0(x)$.

Since we use the leading-order term to construct the Lagrangian manifold, the above considerations imply that we can construct the Lagrangian manifold with $\text{Re} \, L_0(x,q,E)$. This means that the Lagrangian manifold extends slightly beyond the point $x_{1,L}$ in figure~\ref{fig:Landau-turning-point-phase-space}. However, as is clear from figure~\ref{fig:dispersion-damping-homo}, this does not change the essential conclusion, namely that there are no solutions $q_1(x_1)$ beyond a certain point $x_{1,LC}$. In fact, this was explicitly shown in Ref.~\cite{Ishmukhametov81b}, where the phase space trajectories were constructed based on $\text{Re} \, L_0(x,q,E)$. On the one hand, we can use this observation to speculate once again that the Maslov index $\mu(r_{1,0},r_{1,p})$ equals $-2$, as the induced potential $V(x,t)$ has to vanish at the point $x_{1,LC}$. On the other hand, it also means that we have to extend the integration interval to the point $x_{1,LC}$. Moreover, we have not taken the additional term in the amplitude $\varphi_0(x)$, which encodes the plasmon damping, into account in these arguments. Since the interval $(x_{1,L},x_{1,LC})$ becomes larger as the length $\ell$ of the potential decrease becomes larger, we may speculate that the damping is larger for smoother potentials.

We conclude that a rigorous semiclassical analysis of the Landau-type turning point is not at all straightforward, and postpone it to a separate paper. In the remainder of this paper, we take $x_{1,L}$ as the classical turning point, where the plasmon is reflected. 
Based on our previous arguments based on the small damping approximation, it seems that this approximation is most accurate when the interval $(x_{1,L},x_{1,LC})$ is not too long, that is, when the potential does not change too smoothly. We do not take $x_{1,LC}$ as the classical turning point, since the small damping approximation does not perform well outside a small neighborhood of $x_{1,L}$, see figure~\ref{fig:dispersion-damping-homo}. We therefore believe that it would not lead to more accurate results.
We thus assert that the momentum abruptly changes from a negative value to a positive value at the point $x_{1,L}$, leading to a jump in the phase space curve shown in figure~\ref{fig:Landau-turning-point-phase-space}(b).
Furthermore, we say that the asymptotic solution corresponding to the reflected plasmon equals $\exp(i\pi\delta/2)$ times the one for the incoming plasmon. This corresponds to a Maslov index $\mu(r_{1,0},r_{1,p}) = -\delta$.
In our numerical calculations in section~\ref{sec:examples}, we set $\delta$ equal to two, based on our previous arguments. We do not make any explicit statements about the damping in the remainder of this article, as it is beyond the scope of our present analysis.

\subsection{Quantization condition for bound plasmons}  \label{subsec:quantization-condition}

In the previous two subsections, we considered what happens at the boundary of a classically allowed and a classically forbidden region. By considering the trajectories of the Hamiltonian system in the vicinity of a classical turning point, we constructed the Lagrangian manifold in phase space. We analyzed its structure and obtained a value for the Maslov index for the two types of turning points in our system.
In this section, we consider the quantization condition for bound states. A bound state, which has a periodic trajectory, arises when a classically allowed region is surrounded by two classically forbidden regions. 
This means that we can find these bound states by combining the results from the previous two subsections. In this subsection, we show how this leads to two different types of bound states: regular bound states and Landau-type bound states. We derive the quantization conditions for both of these types, and thereby obtain the spectrum of the different types of bulk plasmons in the system.

\begin{figure}[p]
  \hfill
  \begin{tikzpicture}
    \node at (0cm,0cm)
    {\includegraphics[width=0.95\textwidth]{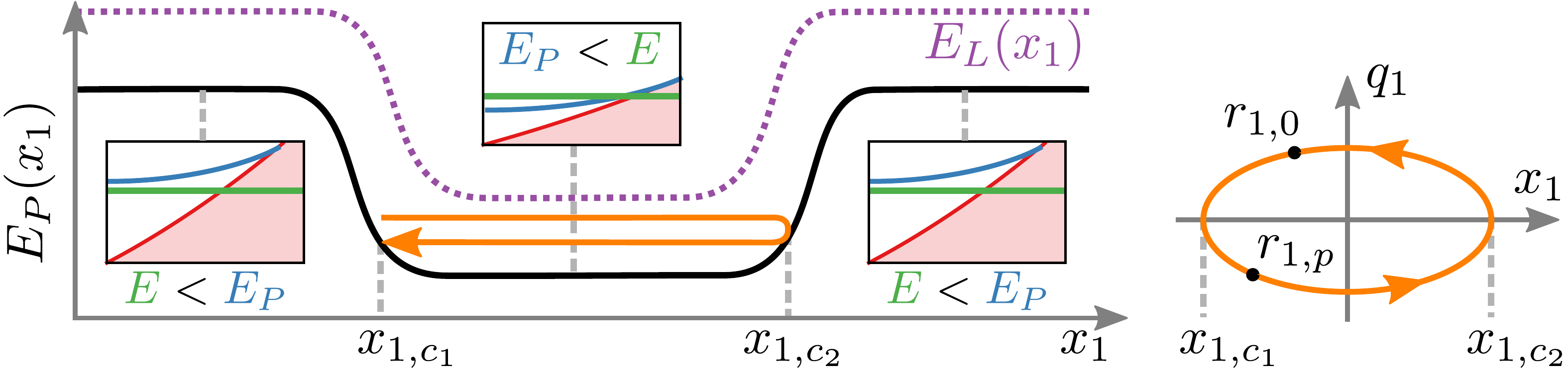}};
    \node at (-6.85cm,1.45cm) {(a)}; 
    \node at (3.5cm,1.45cm) {(b)};
  \end{tikzpicture}
  \hfill\hfill
  \caption{(a) Schematic representation of the regular bound states for $q_\parallel=0$. The solid black line indicates the spatial dependence of $E_P(x_1)$, given by expression~(\ref{eq:plasma-energy}), and the dotted purple line shows the spatial variation of $E_L(x_1)$, defined by equation~(\ref{eq:E-L-def}). Regular bound states arise when we have two simple turning points $x_{1,c_1}$ and $x_{1,c_2}$, which satisfy $E_P(x_{1,c_j}) = E$. These states therefore propagate in the entire center region, as indicated by the orange arrow, and satisfy $E_{P,\text{min}}<E<E_{L,\text{min}}$. Similar to figures~\ref{fig:simple-turning-point-phase-space} and~\ref{fig:Landau-turning-point-phase-space}, the insets show the spectrum of a homogeneous plasma with the parameters given by their values at the indicated point. 
    (b) The periodic trajectory in phase space corresponding to a regular bound state, with the arrow indicating the direction of propagation. This trajectory is obtained by solving the equation $L_0(x_1,\sqrt{q_1^2+q_\parallel^2},E) = 0$. At the two simple turning points $x_{1,c_1}$ and $x_{1,c_2}$, the momentum $q_1$ vanishes and the velocity equals zero, cf. equation~(\ref{eq:velocity-momentum-opposite}).
  }
  \label{fig:regular-bound-states-phase-space}
\end{figure}

Similar to the previous two subsections, we consider a situation where the electron density $n^{(0)}(x_1)$ and background dielectric constant $\varepsilon_b(x_1)$ only depend on the first coordinate $x_1$. This implies that the momenta $q_2$ and $q_3$ are constant, and we define the momentum parallel to the interface as $q_\parallel = \sqrt{q_2^2 + q_3^2}$. This allows us to study the effectively one-dimensional plasmonic waveguide discussed in the introduction. 
Specifically, we consider the situation shown in figure~\ref{fig:regular-bound-states-phase-space}(a), where the position-dependent plasma energy $E_P(x_1)$ has a minimum in the center. This minimum can arise from a decrease in $n^{(0)}(x_1)$, an increase in  $\varepsilon_b(x_1)$, or a combination of both. Although we stick to a schematic representation in this section, we already mention that we will consider specific profiles in the next section, see expressions~(\ref{eq:charge-density-tanh-numerics}) and~(\ref{eq:background-epsilon-tanh-numerics}) in section~\ref{subsec:examples-1d-waveguide-numerics}.
Physically, one can think of this setup as a slab of a metal or a semiconductor sandwiched between two materials with a higher charge density, although we should keep it mind that the density profile this creates will probably not be very smooth. A much smoother density profile could for instance be created by locally doping a semiconductor.

\begin{figure}[p]
  \hfill
  \begin{tikzpicture}
    \node at (0cm,0cm)
    {\includegraphics[width=0.95\textwidth]{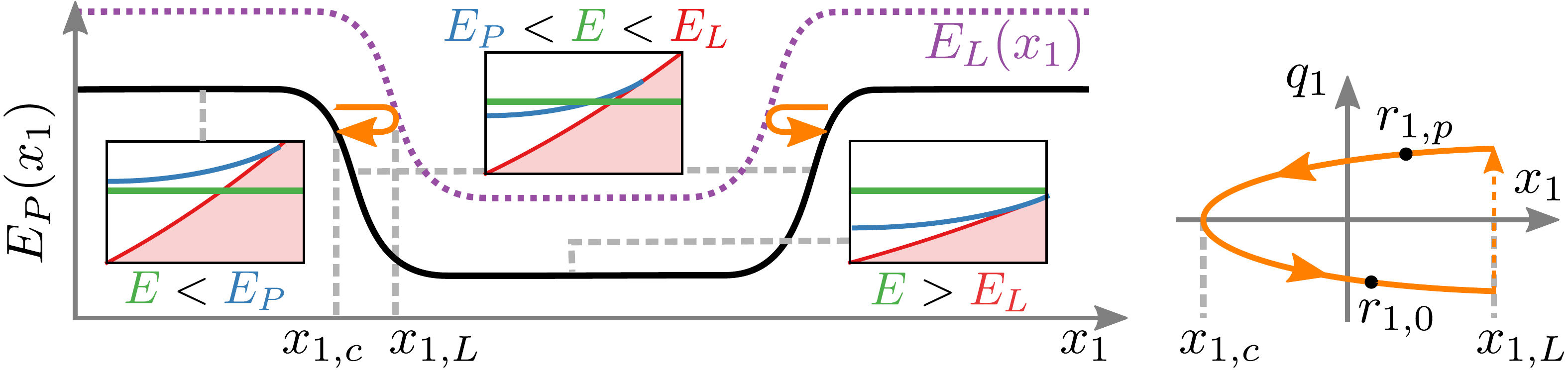}};
    \node at (-6.85cm,1.45cm) {(a)}; 
    \node at (3.5cm,1.45cm) {(b)};
  \end{tikzpicture}
  \hfill\hfill
  \caption{(a) Schematic representation of the Landau-type bound states for $q_\parallel=0$. The solid black line indicates the spatial dependence of $E_P(x_1)$, given by expression~(\ref{eq:plasma-energy}), and the dotted purple line shows the spatial variation of $E_L(x_1)$, defined by equation~(\ref{eq:E-L-def}). 
    Landau-type bound states arise when we have one simple turning point $x_{1,c}$, satisfying $E_P(x_{1,c}) = E$, and one Landau-type turning point, satisfying $E_L(x_{1,L}) = E$. These states therefore propagate in a narrow region near the decrease (left) or increase (right) of $E_P(x_1)$, as indicated by the two orange arrows, and satisfy $E_{L,\text{min}} < E < E_{P,\text{max}}$.
    As before, the insets show the spectrum of a homogeneous plasma with the parameters given by their values at the indicated point. 
    (b) The periodic trajectory in phase space corresponding to a regular bound state, with the arrow indicating the direction of propagation. This trajectory is obtained by solving the equation $L_0(x_1,\sqrt{q_1^2+q_\parallel^2},E) = 0$. 
    The Lagrangian manifold is smooth at the simple turning point $x_{1,c}$. At the Landau-type turning point $x_{1,L}$, we jump from the lower leaf of the Lagrangian manifold to the upper leaf, as indicated by the dashed orange line with the arrow. 
  }
  \label{fig:Landau-bound-states-phase-space}
\end{figure}

Using the results from the previous two subsections, we now derive that our waveguide hosts two different types of bound states, which we call regular bound states and Landau-type bound states.
A regular bound state arises when we have a closed trajectory with two simple turning points $x_{1,c_1}$ and $x_{1,c_2}$.
For $q_\parallel=0$, these simple turning points are defined by $E_P(x_{1,c_j}) = E$, see equation~(\ref{eq:L0-E-expansion-small-q}).
A regular bound state therefore propagates in the entire center region of the waveguide, as shown schematically in figure~\ref{fig:regular-bound-states-phase-space}(a). Consequentially, it satisfies $E_{P,\text{min}} < E < E_{L,\text{min}}$ when $q_\parallel = 0$, where $E_{P,\text{min}} =  \text{min}_{x_1} E_P(x_1)$ and $E_{L,\text{min}} = \text{min}_{x_1} E_L(x_1)$.
The corresponding periodic trajectory in phase space can be obtained by solving the equation $L_0(x_1,\sqrt{q_1^2+q_\parallel^2},E) = 0$ and is shown in figure~\ref{fig:regular-bound-states-phase-space}(b). Note that the velocity is parallel to $-q_1$, as shown in equation~(\ref{eq:velocity-momentum-opposite}). Moreover, the velocity vanishes at the simple turning points $x_{1,c_1}$ and $x_{1,c_2}$, as does the momentum $q_1$.

A Landau-type bound state arises when we have a closed trajectory with one simple turning point and one Landau-type turning point. For $q_\parallel = 0$, the simple turning point $x_{1,c}$ is defined by $E_P(x_{1,c}) = E$, and the Landau-type turning point $x_{1,L}$ is defined by $E=E_L(x_{1,L})$, see equation~(\ref{eq:E-L-def}) and section~\ref{subsec:Landau-damping-threshold}.
A Landau-type bound state therefore propagates in a narrow region near the decrease (left) or increase (right) of $E_P(x_1)$, as indicated schematically in figure~\ref{fig:Landau-bound-states-phase-space}(a). For $q_\parallel=0$, a Landau-type bound state can arise when the inequality $E_{L,\text{min}} < E < E_{P,\text{max}}$ is satisfied, where $E_{P,\text{max}} = \text{max}_{x_1} E_P(x_1)$. 
Note that a Landau-type bound state cannot exist in the center of the waveguide, since we have $E>E_L(x_1)$ there.
Figure~\ref{fig:Landau-bound-states-phase-space} shows the trajectory in phase space corresponding to the Landau-type bound state, obtained by solving the equation $L_0(x_1,\sqrt{q_1^2+q_\parallel^2},E) = 0$. In accordance with the results of section~\ref{subsec:Landau-damping-threshold}, it includes a jump from negative momentum to positive momentum at $x_{1,L}$.

As we already mentioned in the introduction, the Landau-type bound states should not be regarded as surface plasmons. Although they propagate in a narrow region near the potential increase or decrease, as shown in figure~\ref{fig:Landau-bound-states-phase-space}(a), their spatial scale is defined by the length scale $\ell$ of the change in the potential. Surface plasmons, instead, have a spatial scale determined by $q_\parallel^{-1}$. Moreover, they satisfy a different equation. Thus, our Landau-type bound states are not surface plasmons, but instead correspond to localized bulk plasmons.

When $q_\parallel \neq 0$, the above pictures do not change qualitatively. There are some quantitative changes, since, e.g., $E_{P,\text{min}}$ becomes dependent on $q_\parallel$, as can be inferred from equation~(\ref{eq:L0-E-expansion-small-q-Taylor-finite-qp}) and section~\ref{subsec:simple-turning-point}. Since the precise definitions of the energy boundaries are quite technical for $q_\parallel \neq 0$, we discuss them in section~\ref{subsec:examples-1d-waveguide}, where we also present our numerical results. Although we tacitly assumed that $E_{L,\text{min}} < E_{P,\text{max}}$ in the above discussion, this need not be the case. In that situation, the Landau-type bound states only appear at finite $q_\parallel$, as we also discuss in section~\ref{subsec:examples-1d-waveguide}.

We can now write down the quantization conditions for the two types of bound states. As we previously saw in figure~\ref{fig:regular-bound-states-phase-space}(b), a regular bound state has a periodic trajectory. 
When we start at an arbitrary point $r_{1,0}$ in phase space and follow the trajectory until we return at the same point, we should end up with the same value for the induced potential $V$. When we are not close to any of the turning points, the asymptotic solution for $V$ is given by expression~(\ref{eq:solution-sc-potential-index}).
It is important that we regard this asymptotic solution as a function of the phase space coordinate $r=(x,q)$, since our requirement that it is single-valued only holds in phase space. This can be viewed as a consequence of the procedure we performed in section~\ref{subsec:derivation-induced-potential}, where we lifted the problem from configuration space to phase space in passing from the Hamilton-Jacobi equation to the system of Hamilton's equations.
We should therefore also write $S(r)$ instead of $S(x)$, since our projection of the phase space onto the coordinate space is two-to-one, see figure~\ref{fig:regular-bound-states-phase-space}(b). Physically, these two contributions simply correspond to the right-moving and left-moving solutions.
The condition that the induced potential has to be single-valued then naturally leads to the quantization condition~\cite{Maslov81,Berry72,Guillemin77}
\begin{equation}  \label{eq:quantization-condition-general}
  \frac{1}{\hbar} \oint \langle Q, \text{d} X \rangle - \frac{\pi}{2} \mu(r_{1,0},r_{1,0}) = - 2 n \pi,
\end{equation}
where $n$ can in principle be an arbitrary integer, and the minus sign is chosen for convenience, as we will see shortly. We use the notation $\oint$ to indicate that the initial point $r_{1,0}$ is irrelevant.
This quantization condition is known under several names, including the Bohr-Sommerfeld quantization condition and Einstein-Brillouin-Keller quantization.
We remark that the above expression does not reflect the most general form of the quantization condition. As we noted in section~\ref{subsec:derivation-induced-potential}, the induced potential $V$ does not have a geometric or Berry phase. If this were the case, then the quantization condition would include an additional term reflecting this phase factor, see e.g. Refs.~\cite{Maslov81,Guillemin77,Reijnders18}. However, since we do not have a geometric or Berry phase in this problem, equation~(\ref{eq:quantization-condition-general}) is the correct quantization condition in our case.

For a simple turning point, the Maslov index $\mu(r_{1,0},r_{1,p})$ equals $-1$, as shown in section~\ref{subsec:simple-turning-point}. For a regular bound state, shown in figure~\ref{fig:regular-bound-states-phase-space}(b), we therefore have $\mu(r_{1,0},r_{1,0})=-2$ for our periodic trajectory. Furthermore, since the velocity is parallel to $-q_1$ for our bound plasmons, we can write
\begin{equation}  \label{eq:S-tot-negative}
  -S_\text{tot} \equiv \oint \langle Q, \text{d} X \rangle = - \left| \oint \langle Q, \text{d} X \rangle \right| = - 2 \left| \int_{x_{1,c_1}}^{x_{1,c_2}} \langle Q, \text{d} X \rangle \right| ,
\end{equation}
where the last equality follows from the symmetry in $q_1$, and the points $x_{1,c_1}$ and $x_{1,c_2}$ are the left and right turning points, see figure~\ref{fig:regular-bound-states-phase-space}(b).
Our quantization condition for regular bound states therefore reads
\begin{equation}  \label{eq:quantization-cond-regular-bound}
  \frac{S_\text{tot}}{2 \hbar} = \frac{1}{\hbar} \left| \int_{x_{1,c_1}}^{x_{1,c_2}} \langle Q, \text{d} X \rangle \right| = \left(n +\frac{1}{2} \right) \pi ,
\end{equation}
where $n$ is now a non-negative integer.

The quantization condition for Landau-type bound states can be derived in almost the same way. The only difference is the occurence of the special Landau-type turning point, see also figure~\ref{fig:Landau-bound-states-phase-space}(b). As discussed in the previous subsection, we postulate that the Maslov index $\mu(r_{1,0},r_{1,p})$ equals $-\delta$ for this turning point. The total Maslov index of the periodic trajectory is therefore given by $\mu(r_{1,0},r_{1,0}) = -(1 + \delta)$.
Thus, the quantization condition for Landau-type bound states reads
\begin{equation}  \label{eq:quantization-cond-Landau-bound}
  \frac{S_\text{tot}}{2 \hbar} = \frac{1}{\hbar} \left| \int_{x_{1,c}}^{x_{1,L}} \langle Q, \text{d} X \rangle \right| = \left(n + \frac{1+\delta}{4} \right) \pi ,
\end{equation}
where we use $\delta=2$ for our numerical calculations in section~\ref{sec:examples}.

Finally, we briefly discuss the asymptotic solution for the induced potential $V(x,t)$. When we consider a certain point $x$ in the configuration space, there are two points $r=(x,q)$ in phase space that are projected onto this point. As shown in figures~\ref{fig:regular-bound-states-phase-space}(b) and~\ref{fig:Landau-bound-states-phase-space}(b), these correspond to positive and negative values of $q_1$, that is, to left-moving and right-moving waves, respectively. The total asymptotic solution $V(x,t)$ is now given as the sum of the two asymptotic solutions constructed for these two points. As long as we are not in the vicinity of a turning point, we can use our formula~(\ref{eq:solution-sc-potential-index}) to express this asymptotic solution. In the vicinity of the turning points, we need different expressions to adequately represent the asymptotic solution, as we briefly discussed in section~\ref{subsec:simple-turning-point}. These constructions, including the arguments regarding the projection, can be made more precise by within the framework of the so-called Maslov canonical operator~\cite{Maslov81,Guillemin77}. However, we do not discuss this topic in more detail in this paper.

We have thus identified two types of bound states in a one-dimensional waveguide: regular bound states and Landau-type bound states. Moreover, we derived the quantization condition for both of these types.
Before we consider some numerical examples in section~\ref{subsec:examples-1d-waveguide}, we first discuss the bound states in a spherically symmetric potential in the next subsection.

\subsection{Spherically symmetric problems}  \label{subsec:radial}

In the previous subsection, we considered the bound states for an effectively one-dimensional waveguide. In this subsection, we discuss the quantization condition for bound states in a spherically symmetric system. First, we briefly review the transformation to spherical coordinates for our effective classical Hamiltonian $L_0(|x|,|q|)$, using the framework of canonical transformations. We then discuss the semiclassical quantization conditions.
It is important to note that our treatment is different from the approach that is usually taken for the Schr\"odinger equation with a spherically symmetric potential. In the latter problem, one typically uses separation of variables to obtain a radial and an angular differential equation. The angular equation is then solved exactly, giving rise to the spherical harmonics and the angular quantum numbers~\cite{Griffiths05}. One finally writes down a semiclassical quantization condition for the radial equation~\cite{Berry72, Heading62,Froeman65}.
We cannot use the same approach in this problem, since we are dealing with a classical Hamiltonian $L_0(|x|,|q|)$ and not with a differential equation. We therefore have to consider the trajectories in six-dimensional phase space. This leads to three semiclassical quantization conditions, which correspond to the three spherical coordinates $\phi$, $\theta$ and $\varrho$.
The same approach was previously used to study the energy levels of the hydrogen atom, see e.g. Ref.~\cite{Curtis04}, that is, to quantize the orbits in the Kepler problem. Interestingly, it leads to the same angular quantum numbers as the angular differential equation. For the Schr\"odinger equation, the radial quantization condition obtained in our approach corresponds to the quantization condition for the radial differential equation with the Langer substitution~\cite{Langer37,Heading62,Froeman65}, see also Ref.~\cite{Curtis04}. In what follows, we discuss the various steps and their implications in detail.

We introduce conventional spherical coordinates $(\varrho,\theta,\phi)$, which are related to the Cartesian coordinates $(x_1,x_2,x_3)$ by the transformation
\begin{equation} \label{eq:coord-cartesian-in-spherical}
  x_1 = \varrho \sin\theta \cos\phi , \qquad 
  x_2 = \varrho \sin\theta \sin\phi , \qquad 
  x_3 = \varrho \cos\theta
\end{equation}
These formulas describe the relation between the two coordinate systems, but do not directly indicate the relation between the corresponding canonical momenta.
Since a coordinate transformation is a special case of a canonical transformation, the relation between the canonical momenta $(q_\rho,q_\theta,q_\phi)$ and their Cartesian counterparts can be derived from the requirement that the fundamental Poisson brackets are invariant. In other words, we require that the relations
\begin{equation} \label{eq:Poisson-fundamental}
  \{ x_i, x_j \} = 0, \quad \{ q_i , q_j \} = 0, \quad \{ x_i, q_j \} = \delta_{ij} , \qquad \text{where }
  \{ a, b \} = \left\langle \frac{\partial a}{\partial x}, \frac{\partial b}{\partial q} \right\rangle 
  - \left\langle \frac{\partial a}{\partial q}, \frac{\partial b}{\partial x} \right\rangle ,
\end{equation}
are valid in both coordinate systems.

A particular convenient way to express canonical transformations is through the use of generating functions~\cite{Arnold89,Goldstein02}.
Coordinate transformations can be expressed with the generating function $F_2$, which depends on the old coordinates $(x_1,x_2,x_3)$ and the new momenta $(q_\varrho,q_\theta,q_\phi)$. In our case, it is given by the expression~\cite{Goldstein02}
\begin{equation}  \label{eq:generating-spherical-one}
  F_2(x_1,x_2,x_3,q_\varrho,q_\theta,q_\phi) = \varrho(x_1,x_2,x_3) q_\varrho + \theta(x_1,x_2,x_3) q_\theta + \phi(x_1,x_2,x_3) q_\phi ,
\end{equation}
where $\varrho(x_1,x_2,x_3)$ expresses the coordinate transformation~(\ref{eq:coord-cartesian-in-spherical}), and the same holds for $\theta$ and $\phi$.
When using generating function $F_2$, the relation between the new coordinates and the old coordinates is defined by~\cite{Arnold89,Goldstein02}
\begin{equation}  \label{eq:S2-coordinate-transform}
  \varrho = \frac{\partial F_2}{\partial q_\varrho} = \varrho(x_1,x_2,x_3), \quad
  \theta = \frac{\partial F_2}{\partial q_\theta} = \theta(x_1,x_2,x_3), \quad
  \phi = \frac{\partial F_2}{\partial q_\phi} = \phi(x_1,x_2,x_3) .
\end{equation}
This is exactly as we desired and corresponds to the coordinate transformation. The relation between the new momenta and the old momenta is defined by
\begin{equation}  \label{eq:p-in-spherical}
  \begin{aligned}
    q_1 &= \frac{\partial \varrho}{\partial x_1} q_\varrho + \frac{\partial \theta}{\partial x_1} q_\theta + \frac{\partial \phi}{\partial x_1} q_\phi 
      = \sin\theta\cos\phi \, q_\varrho + \frac{\cos\theta\cos\phi}{\varrho} q_\theta - \frac{\sin\phi}{\varrho \sin\theta} q_\phi, \\
    q_2 &= \frac{\partial \varrho}{\partial x_2} q_\varrho + \frac{\partial \theta}{\partial x_2} q_\theta + \frac{\partial \phi}{\partial x_2} q_\phi 
      = \sin\theta\sin\phi \, q_\varrho + \frac{\cos\theta\sin\phi}{\varrho} q_\theta + \frac{\cos\phi}{\varrho \sin\theta} q_\phi, \\
    q_3 &= \frac{\partial \varrho}{\partial x_3} q_\varrho + \frac{\partial \theta}{\partial x_3} q_\theta + \frac{\partial \phi}{\partial x_3} q_\phi 
      = \cos\theta \, q_\varrho - \frac{\sin\theta}{\varrho} q_\theta ,
  \end{aligned}
\end{equation}
where the second equalities follow after some calculus.
With these relations, we can easily derive the familiar relation
\begin{equation}  \label{eq:squared-momenta}
  q_1^2 + q_2^2 + q_3^3 = q_\varrho^2 + \frac{q_\theta^2}{\varrho^2} + \frac{q_\phi^2}{\varrho^2\sin^2\theta}
\end{equation}
for the square of the momentum vector.

In spherical coordinates, our effective classical Hamiltonian $L_0(|x|,|q|)$ therefore reads
\begin{equation}  \label{eq:Hamiltonian-spherical}
  L_0(|x|,|q|) = L_0\left(\varrho, \sqrt{q_\varrho^2 + \frac{q_\theta^2}{\varrho^2} + \frac{q_\phi^2}{\varrho^2\sin^2\theta}} \right)
\end{equation}
Since this Hamiltonian has rotational symmetry, we can immediately identify two constants of motion~\cite{Goldstein02}. The first one of these is $\mathcal{L}_3 = (x \times q)_3 = x_1 q_2 - x_2 q_1$, the third component of the angular momentum. The second constant of motion is $\mathcal{L}^2$, the square of the total angular momentum. 
Because we study the equation $L_0(|x|,|p|)=0$, the classical Hamiltonian itself is also a constant of motion. We have therefore found three constants of motion, also called first integrals.

Conserved quantities have a vanishing Poisson bracket with the classical Hamiltonian, since $\frac{\text{d} a}{\text{d} t} = \{ a, L_0 \}$ has to be zero.
Moreover, using the relation $\{ \mathcal{L}_i, \mathcal{L}_j \} = \sum_k \epsilon_{ijk} \mathcal{L}_k$, where $\epsilon_{ijk}$ is the Levi-Cività symbol, one can show that the Poisson bracket of $\mathcal{L}^2$ and $\mathcal{L}_3$ also vanishes.
We therefore have the relations
\begin{equation}
  \{ \mathcal{L}_3 , L_0 \} = 0 ,  \quad \{ \mathcal{L}^2 , L_0 \} = 0 , \quad \{ \mathcal{L}^2 , \mathcal{L}_3 \} = 0 .
\end{equation}
One may also say that our first integrals are in involution. Since they are also independent, the system is completely integrable~\cite{Arnold89,Maslov81,Goldstein02}.

Our next goal is to construct a solution of the Hamilton-Jacobi equation, or, in other words, to determine the total action $S(\varrho,\theta,\phi)$. We first note that the coordinate $\phi$ is cyclic: the classical Hamiltonian $L_0(\varrho, |q|)$ does not depend on it~\cite{Goldstein02}.
Since we have expressed our coordinate transformation as a canonical transformation, Hamilton's equations have the same form in both coordinate systems. We therefore immediately see that $\frac{\text{d} q_\phi}{\text{d} t} = -\frac{\partial H_0}{\partial \phi} = 0$, and $q_\phi$ is constant. Expressions~(\ref{eq:p-in-spherical}) show that $q_\phi = x_1 q_2 - x_2 q_1 = \mathcal{L}_3$, of which we previously concluded that it is constant. We can therefore separate the action as~\cite{Goldstein02}
\begin{equation}  \label{eq:action-separation-1}
  S(\varrho,\theta,\phi) = S_{\varrho\theta}(\varrho,\theta) + S_\phi(\phi) , \qquad S_\phi = \int_0^{\phi} q_\phi \text{d}\phi' = - \mathcal{L}_3 \phi ,
\end{equation}
where the minus sign comes from the fact that the velocity is opposite to the momentum, see the discussion in section~\ref{subsec:simple-turning-point}.
The coordinate $\theta$ is not cyclic, but it is separable~\cite{Goldstein02}: the dependence of $L_0$ on $\theta$ and $q_\theta$ can be expressed by the function $f(\theta,q_\theta) = q_\theta^2 + q_\phi^2/\sin^2\theta$, which does not involve $\varrho$ and $q_\varrho$, see expression~(\ref{eq:Hamiltonian-spherical}). In this case, one can prove that the action is completely separable~\cite{Goldstein02}, that is,
\begin{equation}  \label{eq:action-separation-2}
  S(\varrho,\theta,\phi) = S_{\varrho}(\varrho) + S_\theta(\theta) + S_\phi(\phi) ,
\end{equation}
and that the function $f(\theta,q_\theta)$ is constant. After some calculus, one finds that $q_\theta^2 + \mathcal{L}_3^2/\sin^2\theta = \mathcal{L}^2$, of which we indeed previously concluded that it is constant.
We therefore have
\begin{equation}
  S_\theta(\theta) = \int_{\theta_0}^\theta q_\theta(\theta') \text{d}\theta' 
    = \int_{\theta_0}^\theta \sqrt{\mathcal{L}^2 - \frac{\mathcal{L}_3^2}{\sin^2\theta'}} \text{d}\theta' , \qquad
  S_\varrho(\varrho) = \int_{\varrho_0}^{\varrho} q_\varrho(\varrho') \text{d}\varrho' ,
\end{equation}
where $\varrho_0$ and $\theta_0$ are (arbitrary) reference points.

The next step in the construction of the quantization conditions for bound states is to describe the Lagrangian manifold $\Lambda^3$ in phase space. For the angular coordinates, we can follow the arguments given in Ref.~\cite{Maslov81}.
Since $q_\theta^2 + \mathcal{L}_3^2/\sin^2\theta = \mathcal{L}^2$, where $\mathcal{L}^2$ and $\mathcal{L}_3$ are constant, the motion in the polar angle $\theta$ is described by an ellipse in phase space. There are two classical turning points, which are given by $(\theta,q_\theta)=(\theta_0,0)$ and $(\theta,q_\theta)=(\pi-\theta_0,0)$, where $\theta_0 = \arcsin(|\mathcal{L}_3|/|\mathcal{L}|)$.
We subsequently consider the projection of the Lagrangian manifold onto the coordinates $(\phi,q_\phi)$, which describe the motion of the azimuthal angle. The momentum $q_\phi=\mathcal{L}_3$ is constant, and the coordinate satisfies $0\leq \phi < 2\pi$. We therefore do not have any classical turning points, and the motion is a rotation. Thus, the projection is diffeomorphic to the one-dimensional circle $S^1$. Since the projection onto $(\theta,q_\theta)$ is also diffeomorphic to a circle, the projection of $\Lambda^3$ onto the angular part is diffeomorphic to the two-dimensional torus $T^2=S^1 \times S^1$.

The motion in the radial coordinate $\varrho$ is described by the effective classical Hamiltonian $L_0(\varrho, \sqrt{q_\varrho^2+\mathcal{L}^2/\varrho^2})$. The periodic trajectories that describe the bound states have two turning points, at the coordinates $\varrho_c$ and $\varrho_L$, with $\varrho_c<\varrho_L$. The analysis of these turning points is slightly more complicated than the analysis performed in sections~\ref{subsec:simple-turning-point} and~\ref{subsec:Landau-damping-threshold}, because the term $\mathcal{L}^2/\varrho^2$ in the total momentum depends on position. Before, this term was equal to $q_\parallel^2$ and was therefore constant.
In what follows, we only consider the case where the charge density monotonously decreases and goes to zero as the radius becomes large. When the charge density is not monotonously decreasing, we may end up in a situation where there are more than two turning points. A proper description of this system should include tunneling, which we do not discuss in this paper.

In section~\ref{subsec:simple-turning-point}, we saw that $q_\parallel$ effectively raises the plasma energy, see expression~(\ref{eq:L0-E-expansion-small-q-Taylor-finite-qp-first}). Since the term $\mathcal{L}^2/\varrho^2$ diverges for $\varrho\to 0$, the effective plasma energy also keeps increasing as $\varrho$ becomes smaller. Eventually, we reach the point where the energy $E$ becomes equal to this effective plasma energy. This qualitatively explains why we always have a simple turning point at $\varrho_c$. Performing calculations similar to those in section~\ref{subsec:simple-turning-point}, one can also show this explicitly.

As we increase the radius beyond $\varrho_c$, the charge density decreases and the term $\mathcal{L}^2/\varrho^2$ also decreases in size. The radial momentum $q_\varrho$ therefore becomes larger, and eventually the total momentum $\sqrt{q_\varrho^2+\mathcal{L}^2/\varrho^2}$ reaches the Landau-damping threshold. The second turning point $\varrho_L$ is therefore a Landau-type turning point, and the analysis performed in section~\ref{subsec:Landau-damping-threshold} is applicable. The projection of the Lagrangian manifold $\Lambda^3$ onto the radial part $(\varrho,q_\varrho)$ therefore looks precisely like figure~\ref{fig:Landau-bound-states-phase-space}(b).

In order to construct the quantization conditions, we have to choose three closed, linearly independent, paths $\gamma_j$ on the Lagrangian manifold $\Lambda^3$. In order for our asymptotic solution~(\ref{eq:solution-sc-potential-index}) for the induced potential $V(x,t)$ to be well-defined, that is, single-valued, we have to satisfy the conditions
\begin{equation}
  \frac{1}{\hbar} \oint_{\gamma_j} \langle Q, \text{d} X \rangle - \frac{\pi}{2} \mu_j = 2 \pi n_j ,
\end{equation}
where $\mu_j$ is the Maslov index of the closed path $\gamma_j$. Since the final results are independent of the choice of $\gamma_j$, see e.g. Ref.~\cite{Maslov81}, we may choose paths that correspond to the different coordinates. In this way, we obtain three quantization conditions, all of which have to be satisfied simultaneously, see also Refs.~\cite{Maslov81,Curtis04}.

As we previously discussed, the motion in the azimuthal angle $\phi$ is a rotation. This means that there are no turning points, and we have $\mu_\phi=0$. The first quantization condition therefore reads
\begin{equation}  \label{eq:quantization-phi-first}
  - \frac{S_{\phi,\text{tot}}}{\hbar} 
    \equiv \frac{1}{\hbar} \oint_{\gamma_\phi} q_\phi \, \text{d}\phi' 
    = - \frac{1}{\hbar} \int_{0}^{2\pi} \mathcal{L}_3 \, \text{d}\phi'
    = - 2 \pi n_\phi .
\end{equation}
As in the previous subsection, we included a minus sign in the definition of $n_\phi$ for convenience.
When we consider the projection of $\Lambda^3$ onto $(\theta,q_\theta)$, we have two simple turning points, at $(\theta_0,0)$ and $(\pi-\theta_0,0)$. Because the momentum and the velocity are in opposite directions, see also section~\ref{subsec:simple-turning-point} and expression~(\ref{eq:velocity-momentum-opposite}), the Maslov index equals minus two, and the second quantization condition becomes
\begin{equation}  \label{eq:quantization-theta-first}
  - \frac{S_{\theta,\text{tot}}}{\hbar}
    \equiv \frac{1}{\hbar} \oint_{\gamma_\theta} q_\theta \, \text{d}\theta' 
    = - \frac{2}{\hbar} \int_{\theta_0}^{\pi-\theta_0} \sqrt{\mathcal{L}^2 - \frac{\mathcal{L}_3^2}{\sin^2\theta'}} \text{d}\theta' 
    = - 2 \pi n_\theta - \pi .
\end{equation}
We finally construct the radial quantization condition. 
As in equation~(\ref{eq:S-tot-negative}), we have
\begin{equation}  \label{eq:quantization-r-first}
  - S_{\varrho,\text{tot}}
  \equiv \oint_{\gamma_r} q_\varrho(\varrho') \text{d}\varrho' 
  = - 2 \left| \int_{\varrho_c}^{\varrho_L} q_\varrho(\varrho') \text{d}\varrho' \right| .
\end{equation}
Since we have one simple turning point and one Landau-type turning point, the Maslov index equals $-(1+\delta)$. Using the same manipulations as in the previous subsection, we therefore conclude that
\begin{equation} \label{eq:quantization-r-first-quant}
  \frac{S_{\varrho,\text{tot}}}{\hbar} = \frac{2}{\hbar} \left| \int_{\varrho_c}^{\varrho_L} q_\varrho(\varrho') \text{d}\varrho' \right| = 2 \pi n_\varrho + \frac{\pi(1+\delta)}{2} ,
\end{equation}
where $n_\varrho$ should be a non-negative integer.

We now proceed to simplify the angular quantization conditions, in the same way as in Ref.~\cite{Curtis04}. From equation~(\ref{eq:quantization-phi-first}), we immediately find that
\begin{equation}  \label{eq:quantization-phi}
  \frac{1}{\hbar} 2 \pi \mathcal{L}_3 = 2 \pi n_\phi , \quad \text{or } \mathcal{L}_3 = n_\phi \hbar .
\end{equation}
The integral in the second quantization condition can also be computed explicitly~\cite{Goldstein02,Curtis04}. After some calculus, one arrives at
\begin{equation}  \label{eq:quantization-theta}
  S_{\theta,\text{tot}} = 2 \pi (|\mathcal{L}| - |\mathcal{L}_3|) = 2 \pi \left(n_\theta + \frac{1}{2}\right) \hbar ,
    \quad \text{or }  |\mathcal{L}| - |\mathcal{L}_3| = \left( n_\theta + \frac{1}{2} \right) \hbar .
\end{equation}
Since we require that $|\mathcal{L}_3| < |\mathcal{L}|$, we have $n_\theta \geq 0$. Combining the conditions~(\ref{eq:quantization-phi}) and~(\ref{eq:quantization-theta}), we find that
\begin{equation}
  |\mathcal{L}| 
    = |\mathcal{L}_3| + \left( n_\theta + \frac{1}{2} \right) \hbar
    = \left( n_\theta + |n_\phi| + \frac{1}{2} \right) \hbar .
\end{equation}
Because $|\mathcal{L}|$ should not be negative, we also have $n_\theta+|n_\phi| \geq 0$.

Keeping the analogy with the spherical harmonics in mind, one can introduce the new quantum numbers $l$ and $m_l$ by $l = n_\theta+|n_\phi|$ and $m_l = n_\phi$. The condition $n_\theta+|n_\phi| \geq 0$ then automatically translates to $l \geq 0$. The condition $n_\theta \geq 0$ instead becomes $l-|m_l| \geq 0$, which means that $|m_l| \leq l$, or $-l \leq m_l \leq l$. These are exactly the conditions on the angular quantum numbers that we are used to from the spherical harmonics. We thus have
\begin{equation}  \label{eq:quantization-conditions-angular}
  \mathcal{L}_3 = m_l \hbar, \qquad |\mathcal{L}| = \left(l+\frac{1}{2}\right) \hbar .
\end{equation}
The second expression is slightly different from the quantization obtained from the spherical harmonics, as we have $\mathcal{L}^2/\hbar^2 = (l+\tfrac{1}{2})^2$ instead of $\mathcal{L}^2/\hbar^2 = l(l+1)$. The difference between these two expressions becomes small for $l\gg 1$, that is, in the deep semiclassical limit. In fact, the replacement of $l(l+1)$ by $(l+\tfrac{1}{2})^2$ corresponds to the so-called Langer substitution in the semiclassical approximation~\cite{Langer37,Heading62,Froeman65}. This substitution often leads to better results in problems with spherical symmetry.

It is good to note that the quantization conditions~(\ref{eq:quantization-conditions-angular}) and~(\ref{eq:quantization-r-first-quant}) lead to a degeneracy: all energy levels with the same values of $n_\varrho$ and $l$ are degenerate, regardless of the value of $m_l$. This degeneracy comes from the spherical symmetry and can already be observed for the Laplace operator on the sphere~\cite{Maslov81,Goldstein02}. 
However, the degeneracy is not as rich as the degeneracy observed in the hydrogen atom, where energy levels with different quantum number $l$ are also degenerate. The latter degeneracy is is only present for a quadratic Hamiltonian with a $1/r$ potential, and can be explained by noting that there is an additional conserved vector, known as the Runge-Lenz vector~\cite{Goldstein02}.

We have thus obtained the quantization conditions for a spherically symmetric system. Renaming $n_\varrho$ to $n$, we find that the radial quantization condition reads
\begin{equation}  \label{eq:quantization-r-final}
   \frac{S_{\varrho,\text{tot}}}{2\hbar} = \frac{1}{\hbar} \left| \int_{\varrho_c}^{\varrho_L} q_\varrho(\varrho') \, \text{d}\varrho' \right| = \left( n + \frac{1+\delta}{4} \right) \pi ,
\end{equation}
where $n$ is a non-negative integer and $q_\varrho(\varrho)$ is the solution of 
\begin{equation}  \label{eq:L0-root-radial-quantizated-angular}
  L_0\left(\varrho, \sqrt{q_\varrho^2 + \frac{\left(l + \frac{1}{2} \right)^2 \hbar^2}{\varrho^2} } \right) = 0 .
\end{equation}
The total angular momentum is quantized as $|\mathcal{L}| = \left(l+\frac{1}{2}\right) \hbar$, and its third component as $\mathcal{L}_3 = m_l \hbar$. 
The energy levels $E_{n,l}$ are $(2 l+1)$-fold degenerate.

\section{Numerical examples}  \label{sec:examples}

In this section, we illustrate our theory with a few numerical examples.
In section~\ref{subsec:examples-1d-waveguide}, we compute the spectrum of bulk plasmons in a plasmonic waveguide that is effectively one dimensional. Using the quantization conditions obtained in section~\ref{subsec:quantization-condition}, we obtain both the regular bound states and the Landau-type bound states.
We consider two spherically symmetric potentials in section~\ref{subsec:examples-radial}. In section~\ref{subsec:examples-radial-atomic}, we take a new look at the problem of the atomic plasmon within the Thomas-Fermi model with the Tietz approximation. We consider a parabolic potential in section~\ref{subsec:examples-radial-parabolic}. 
Using the quantization conditions derived in section~\ref{subsec:radial}, we obtain the spectrum of bulk plasmons for both cases.

\subsection{One-dimensional waveguide} \label{subsec:examples-1d-waveguide}

In this section, we consider a one-dimensional waveguide. In terms of physics, one can for instance think of this system as a slab of a metal or a semiconductor, with a certain charge density, sandwiched between two materials with a higher charge density.
We start by discussing the dependence of the bound states on $q_\parallel$ in section~\ref{subsec:examples-1d-waveguide-dependence-qpar}, and establish for which energies regular bound states and Landau-type bound states can occur. In section~\ref{subsec:examples-1d-waveguide-numerics}, we consider the numerical implementation and compute the spectrum for several different parameters, indicative of metals and semiconductors.

\subsubsection{Dependence of the bound states on $q_\parallel$}
\label{subsec:examples-1d-waveguide-dependence-qpar}

In section~\ref{subsec:quantization-condition}, we obtained the quantization conditions for the two types of bound states in an effectively one-dimensional waveguide, shown in figure~\ref{fig:regular-bound-states-phase-space}(a). We discussed the bounds on the spectrum when $q_\parallel=0$, but did not consider explicitly how these boundaries change when $q_\parallel$ is not equal to zero. In this subsection, we establish these bounds.

For non-zero $q_\parallel$, we can define the quantity $E_{P}(x_1,q_{\parallel})$ as the solution of $L_0(x_1,q_\parallel,E) = 0$, cf. the discussion in section~\ref{subsec:simple-turning-point}. 
When $E > E_{P}(x_1,q_{\parallel})$, we are in the classically allowed region, where plasmons can propagate. Likewise, the region with $E < E_{P}(x_1,q_{\parallel})$ is classically forbidden. In the context of our waveguide, we can therefore view $E_{P}(x_1,q_{\parallel})$ as the effective plasma energy. As in section~\ref{subsec:simple-turning-point}, we can then define a classical turning point $x_{1,c}$ by $E = E_{P}(x_{1,c},q_{\parallel})$, and we have $q_{1}(x_{1,c}) = 0$.
The discussion in section~\ref{subsec:simple-turning-point} shows that $x_{1,c}$ is always a simple turning point, regardless of the value of $q_\parallel$. Expression~(\ref{eq:L0-E-expansion-small-q-Taylor-finite-qp}) gives an explicit formula for $E_{P}(x_1,q_{\parallel})$ when $q_\parallel$ is small.

We can now define the minimal and maximal values of $E_{P}(x_1,q_{\parallel})$ by
\begin{equation}
  E_{P,\text{min}}(q_\parallel) = \min_{x_1} E_{P}(x_1,q_{\parallel}) , \qquad
  E_{P,\text{max}}(q_\parallel) = \max_{x_1} E_{P}(x_1,q_{\parallel}) .
\end{equation}
There are no bound states for $E < E_{P,\text{min}}(q_\parallel)$, since there is no classically allowed region in this case. For $E > E_{P,\text{max}}(q_\parallel)$, the system does not have simple turning points and we are in the continuum spectrum. Bound states can therefore exist when $E_{P,\text{min}}(q_\parallel) < E < E_{P,\text{max}}(q_\parallel)$.

To make the distinction between regular bound states and Landau-type bound states, we have to consider the quantity $E_L(x_1)$, defined in section~\ref{subsec:Landau-turning-point-setting}. As previously discussed, $E_L(x_1)$ does not depend on $q_\parallel$, since the position of the Landau-type turning point is determined by the (length of the) total momentum vector, and not by one of its components. The quantities $E_{L,\text{min}} = \text{min}_{x_1} E_L(x_1)$ and $E_{L,\text{max}} = \text{max}_{x_1} E_L(x_1)$ therefore do not depend on $q_\parallel$, see also section~\ref{subsec:quantization-condition}.

\begin{figure}[tb]
  \hfill
  \begin{tikzpicture}
    \node at (0cm,0cm)
    {\includegraphics[width=0.45\textwidth]{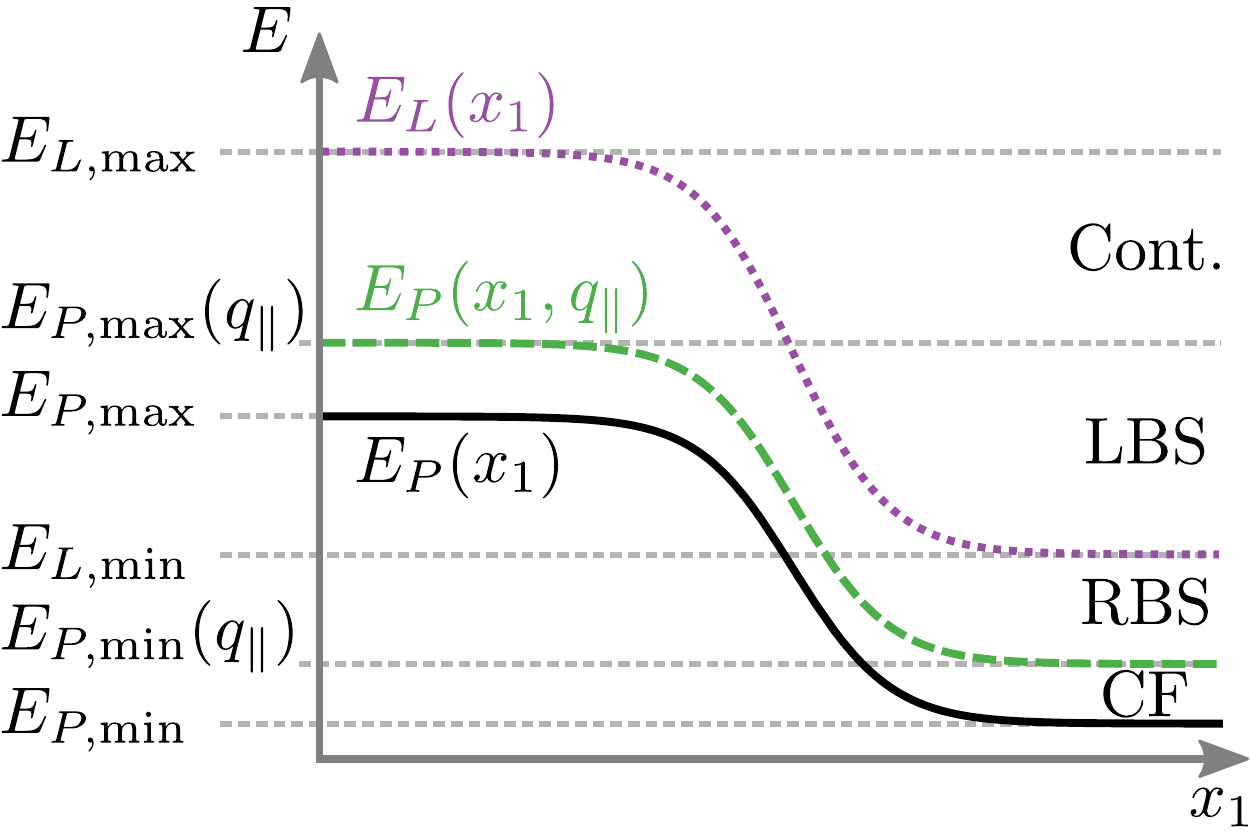}};
    \node at (-3.3cm,2.0cm) {(a)};
    
    \node at (7.1cm,0cm)
    {\includegraphics[width=0.45\textwidth]{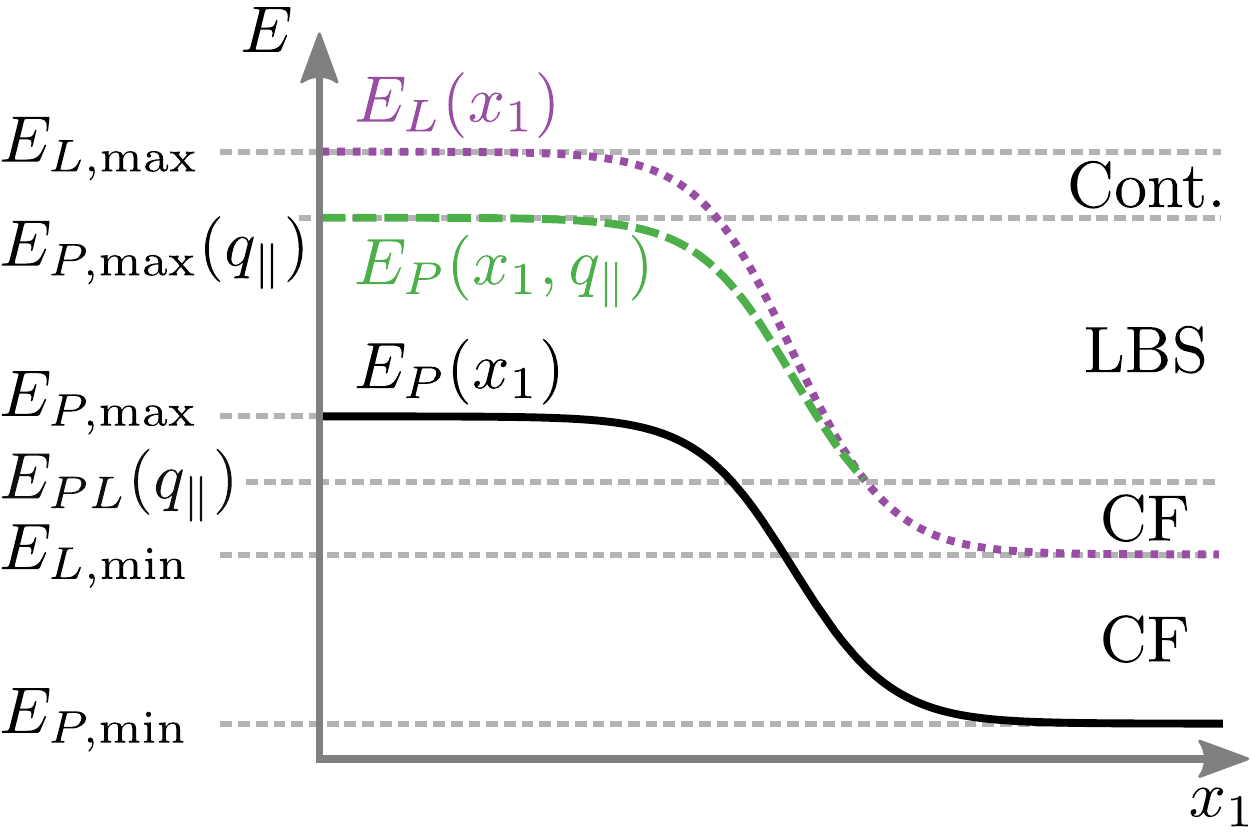}}; 
    \node at (3.8cm,2.0cm) {(b)};
  \end{tikzpicture}
  \hfill\hfill
  \caption{Spatial dependence of the plasma energy $E_P(x_1)$ (solid black line), given by expression~(\ref{eq:plasma-energy}), the energy $E_L(x_1)$ (dotted purple line), defined by equation~(\ref{eq:E-L-def}), and the energy $E_P(x_1,q_\parallel)$ (dashed green line), defined in this section. We only show the left-hand side of our waveguide; the complete waveguide also includes the right-hand side, where these functions increase, as in figures~\ref{fig:regular-bound-states-phase-space}(a) and~\ref{fig:Landau-bound-states-phase-space}(a).
  Panel (a) shows the dependencies for $q_\parallel < q_{L,\text{min}}$. In analogy with figure~\ref{fig:regular-bound-states-phase-space}(a), we see that regular bound states (RBS) can occur in the region $E_{P,\text{min}}(q_\parallel) <  E_{L,\text{min}}$, as indicated by the dashed horizontal grey lines. Landau-type bound states (LBS) occur for $E_{L,\text{min}} < E < E_{P,\text{max}}(q_\parallel)$, since we have one simple turning point and one Landau-type turning point in this energy range, cf. figure~\ref{fig:Landau-bound-states-phase-space}(a). We have a continuum spectrum (Cont.) for $E_{P,\text{max}}(q_\parallel) < E < E_{L,\text{max}}$, and the region between $E_{P,\text{min}}$ and $E_{P,\text{min}}(q_\parallel)$ is classically forbidden (CF).
  Panel (b) shows what happens for $q_{L,\text{min}} < q_\parallel < q_{L,\text{max}}$. There are no regular bound states in this case. Landau-type bound states can occur when $E_{PL}(q_\parallel) < E < E_{P,\text{max}}(q_\parallel)$, where $E_{PL}(q_\parallel)$ denotes the point where $E_P(x_1,q_\parallel)$ and $E_L(x_1)$ cross.
}
  \label{fig:charge-density-delimiters}
\end{figure}

In figure~\ref{fig:charge-density-delimiters}(a), we have schematically drawn $E_P(x)$, $E_P(x,q_\parallel)$ and $E_L(x)$ for the left-hand side or our waveguide, assuming that $E_{L,\text{min}} < E_{P,\text{max}}$. Of course, the complete waveguide also includes the right-hand side, where these functions increase, as in figures~\ref{fig:regular-bound-states-phase-space}(a) and~\ref{fig:Landau-bound-states-phase-space}(a). 
We immediately see that $E_P(x,q_\parallel)$ lies between $E_P(x)$ and $E_L(x)$.
Comparing figure~\ref{fig:charge-density-delimiters}(a) with figures~\ref{fig:regular-bound-states-phase-space}(a) and~\ref{fig:Landau-bound-states-phase-space}(a), we can understand where the regular bound states and the Landau-type bound states occur. Alternatively, we can repeat the arguments given in section~\ref{subsec:quantization-condition}.
We conclude that regular bound states occur when $E_{P,\text{min}}(q_\parallel) < E < E_{L,\text{min}}$ and that Landau-type bound states occur when $E_{L,\text{min}} < E < E_{P,\text{max}}(q_\parallel)$. 
We note that we also have a continuous spectrum for $E_{P,\text{max}}(q_\parallel) < E < E_{L,\text{max}}$, and that we are in the Landau damped region for $E > E_{L,\text{max}}$.
For $E_{L,\text{min}} > E_{P,\text{max}}$, the picture is slightly different. With the help of a figure similar to figure~\ref{fig:charge-density-delimiters}(a), one can show that, for $q_\parallel=0$, we have regular bound states and a continuous spectrum in this case. The Landau-type bound states only appear beyond the value of $q_\parallel$ that satisfies $E_{L,\text{min}} = E_{P,\text{max}}(q_\parallel)$.

Let us now look more closely at the dependence on $q_\parallel$.
Since we are in the Landau-damped region whenever $q_\parallel > q_L(x,E)$, we do not have bound states for all values of $q_\parallel$.
We therefore define two additional quantities, namely
\begin{equation}
  q_{L,\text{min}} = \min_{x_1} q_L(x_1,E_L(x_1)) , \qquad q_{L,\text{max}} = \max_{x_1} q_L(x_1,E_L(x_1)) .
\end{equation}
When $q_\parallel < q_{L,\text{min}}$, we have $E_{P,\text{min}}(q_\parallel) < E_{L,\text{min}}$. This implies that both regular bound states and Landau-type bound states can occur. It corresponds to the situation shown in figure~\ref{fig:charge-density-delimiters}(a). When $q_{L,\text{min}} < q_\parallel < q_{L,\text{max}}$, we are in the situation shown in figure~\ref{fig:charge-density-delimiters}(b). Here, the functions $E_{P}(x_1,q_{\parallel})$ and $E_L(x_1)$ cross at a certain point. We denote the energy at which this crossing occurs by $E_{PL}(q_\parallel)$. In this case, regular bound states cannot exist, and the system can only host Landau-type bound states, which can occur for $E_{PL}(q_\parallel) < E < E_{P,\text{max}}(q_\parallel)$.
When $q_\parallel > q_{L,\text{max}}$, we are in the Landau-damped region for all values of $x_1$. A schematic representation of the different regions of the spectrum is shown in figure~\ref{fig:dispersion-metal-regions}(b).

\subsubsection{Numerical implementation}
\label{subsec:examples-1d-waveguide-numerics}

\begin{figure}[tb]
  \hfill
  \begin{tikzpicture}
    \node at (0cm,0cm)
    {\includegraphics[width=0.45\textwidth]{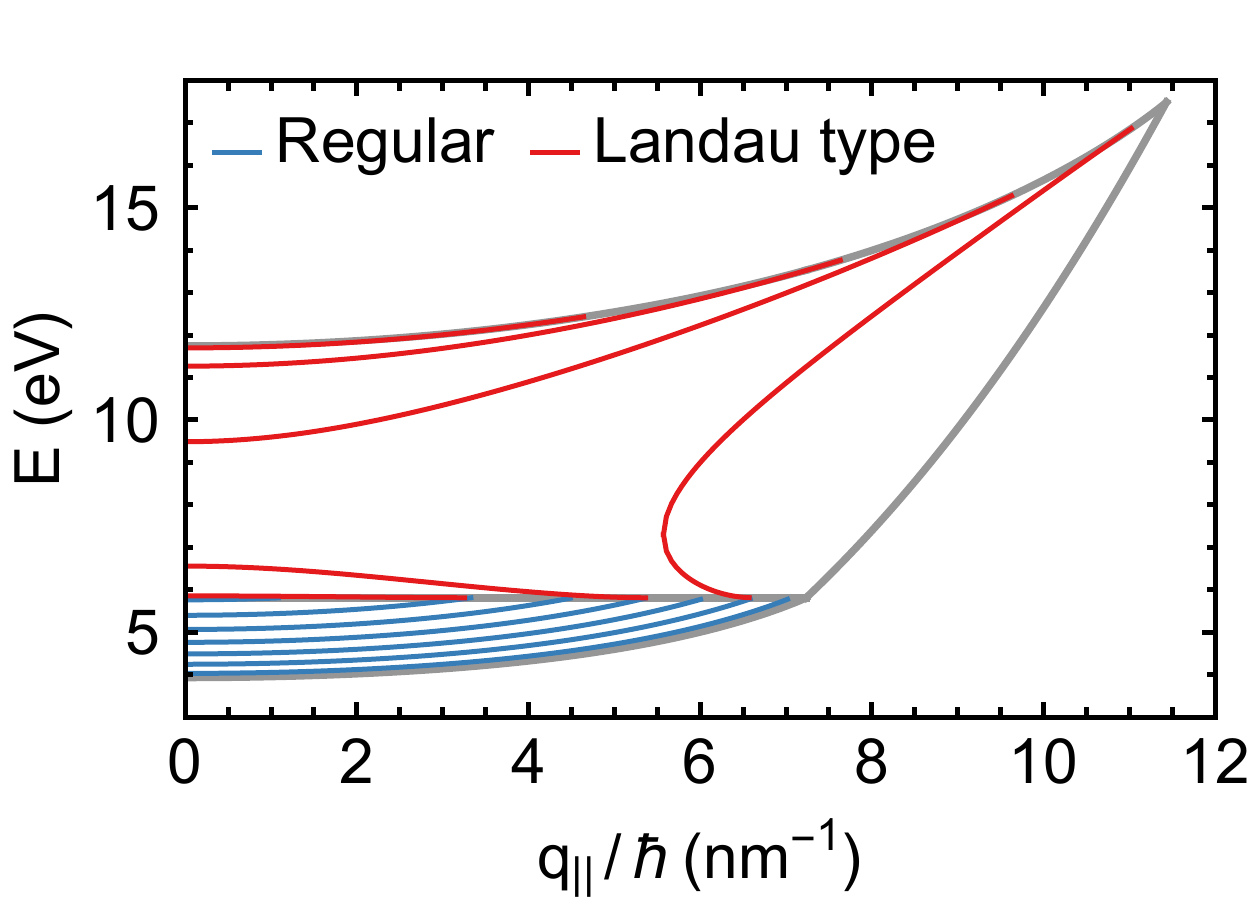}};
    \node at (-3.3cm,1.85cm) {(a)};
    
    \node at (7.1cm,0cm)
    {\includegraphics[width=0.45\textwidth]{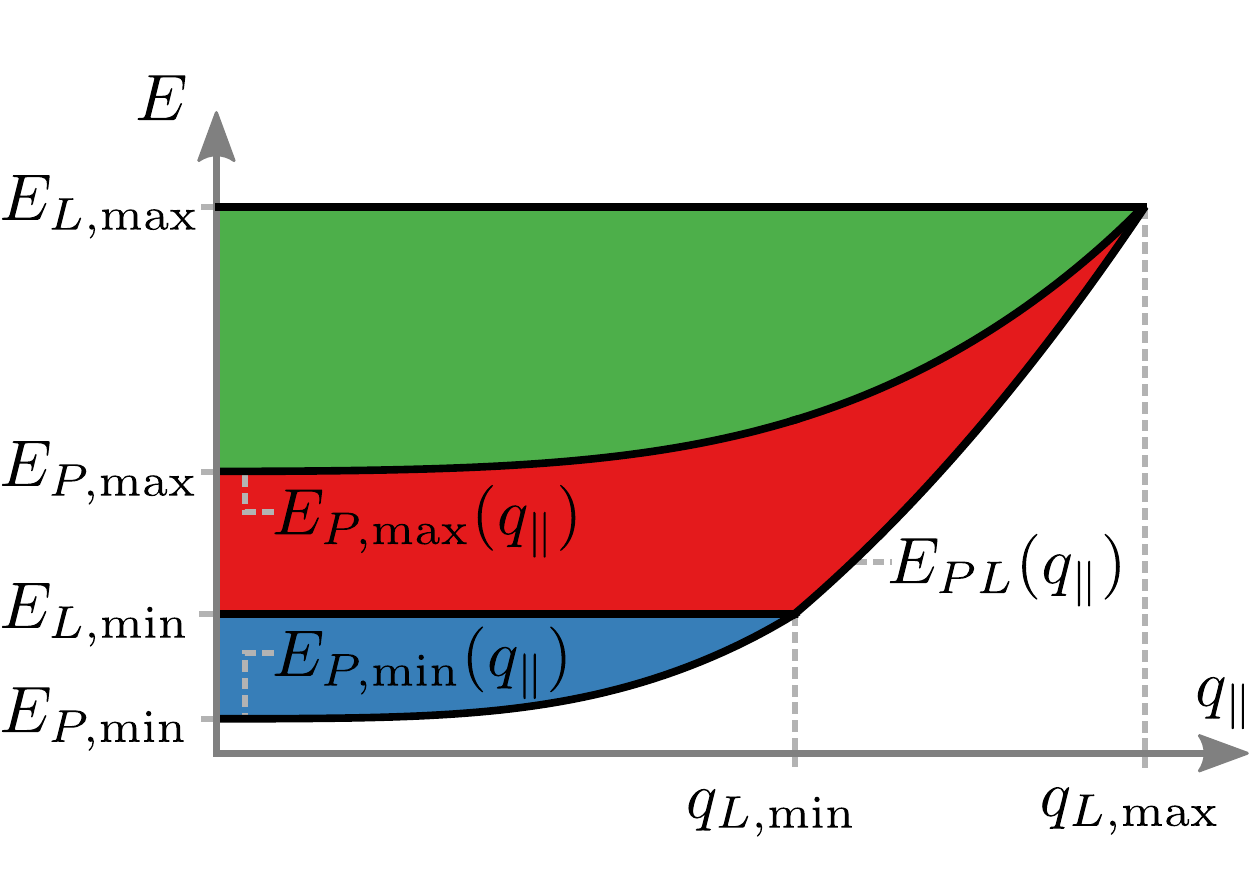}}; 
    \node at (3.8cm,1.85cm) {(b)};
  \end{tikzpicture}
  \hfill\hfill
  \caption{(a) Spectrum of bulk plasmons as a function of the wavevector $q_\parallel/\hbar$ for the charge density~(\ref{eq:charge-density-tanh-numerics}), with parameters indicative of a metal, namely $n^{(0)}_{\text{max}} = 10^{23}$~cm$^{-3}$, $n^{(0)}_{\text{min}} = 10^{22}$~cm$^{-3}$, $\ell_m = 5$~nm, $\ell_c=1$~nm, $m=m_e$, $g_s=2$ and $\varepsilon_b(x_1)=1$. The blue lines denote the regular bound states, and the red lines denote the Landau-type bound states. (b) Schematic representation of the various parts of the spectrum, indicating the regions where regular bound states can occur (blue), where Landau-type bound states can occur (red), and where the spectrum is continuous (green). Also indicated are the energies considered in figure~\ref{fig:charge-density-delimiters}, as well as the values $q_{L,\text{min}}$ and $q_{L,\text{max}}$.}
  \label{fig:dispersion-metal-regions}
\end{figure}

Before presenting our numerical results, we briefly discuss some details of the implementation in Wolfram Mathematica~\cite{Mathematica}. 
We use dimensionless parameters to make sure that all quantities are of order one, which improves numerical stability. Since the effective Hamiltonian $\tilde{L}_0(\tilde{x},\tilde{q})$ exhibits many oscillations for small $q$, which arise from numerical inaccuracies, we replace it by its Taylor expansion up to order $\tilde{q}^6$ for $\tilde{q}<0.12$. The difference between the function value and the approximation at the cutoff point is typically on the order of $10^{-8}$.
We obtain $q_1(x_1)$ by finding the root of $\tilde{L}_0(\tilde{x}_1,\tilde{q})/\tilde{q}^2$. It turns out that the greatest numerical stability is reached by choosing our initial guesses for the root close to $q_L(x_1,E)$. We subsequently compute the action by integrating $q_1(x_1)$ over the relevant interval. We explicitly check that there is no significant difference between the results obtained from the built-in continuum integration and from the Riemann sum constructed by evaluating $q_1(x_1)$ on a grid.
We finally compute the bound states from the quantization conditions~(\ref{eq:quantization-cond-regular-bound}) and~(\ref{eq:quantization-cond-Landau-bound}), with $\delta=2$, using root finding. We explicitly check that all roots are converged to the desired precision, on the order of $10^{-7}$.

In figure~\ref{fig:dispersion-metal-regions}(a), we show the spectrum of bound states for a one-dimensional waveguide with charge density
\begin{equation}  \label{eq:charge-density-tanh-numerics}
  n^{(0)}(x_1) = n^{(0)}_{\text{max}} - \frac{n^{(0)}_{\text{max}} - n^{(0)}_{\text{min}}}{2} \left( \tanh\left( \frac{x_1 + \ell_m/2}{\ell_c} \right)- \tanh\left( \frac{x_1 - \ell_m/2}{\ell_c} \right) \right) ,
\end{equation}
which leads to the variation of $E_P(x_1)$ schematically shown in figure~\ref{fig:regular-bound-states-phase-space}(a). The length scales $\ell_m$ and $\ell_c$ measure, respectively, the width of the center region and the length scale of the increase and decrease.
For the example shown in figure~\ref{fig:dispersion-metal-regions}(a), we use $n^{(0)}_{\text{max}} = 10^{23}$~cm$^{-3}$ and $n^{(0)}_{\text{min}} = 10^{22}$~cm$^{-3}$, corresponding to two metals. The length scales are given by $\ell_m = 5$~nm and $\ell_c=1$~nm, and we set $m=m_e$, $g_s=2$ and $\varepsilon_b(x_1)=1$. Setting $\ell = 2.5 \ell_c$ as a measure for the typical length scale, $p_0 = \hbar (6 \pi^2 n^{(0)}_{\text{max}}/g_s)^{1/3}$ and $\varepsilon_{b0}=1$, we obtain $h \approx 0.028$ and $\kappa \approx 0.0034$. The factor $\frac{g_s}{8\pi^2} \frac{h^2}{\kappa^2}$ equals approximately $1.7$ in this case, in accordance with the demands formulated in section~\ref{subsec:derivation-applicability-SC}. We are therefore well inside the semiclassical regime.

\begin{figure}[tb]
  \hfill
  \begin{tikzpicture}
    \node at (0cm,0cm)
    {\includegraphics[width=0.45\textwidth]{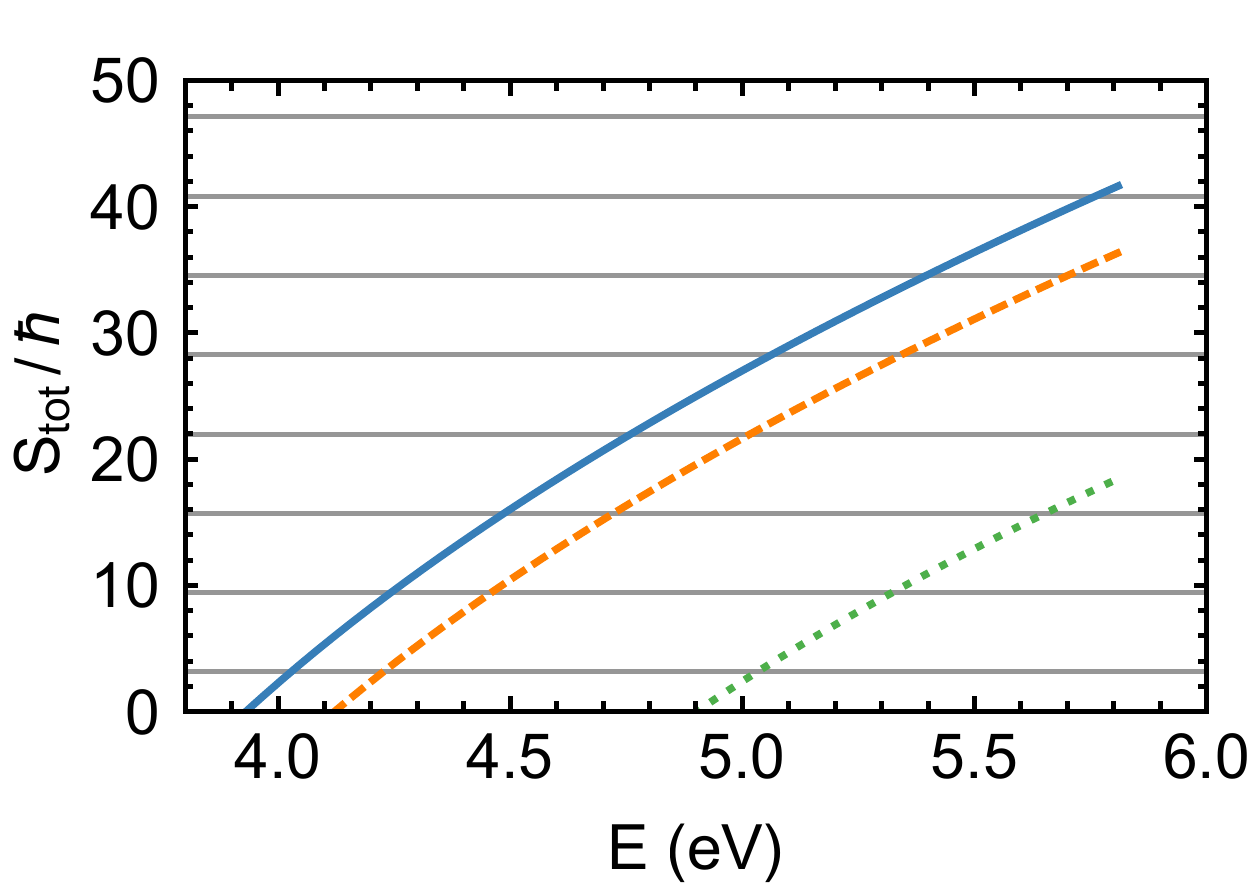}};
    \node at (-3.3cm,1.85cm) {(a)};
    
    \node at (7.1cm,0cm)
    {\includegraphics[width=0.45\textwidth]{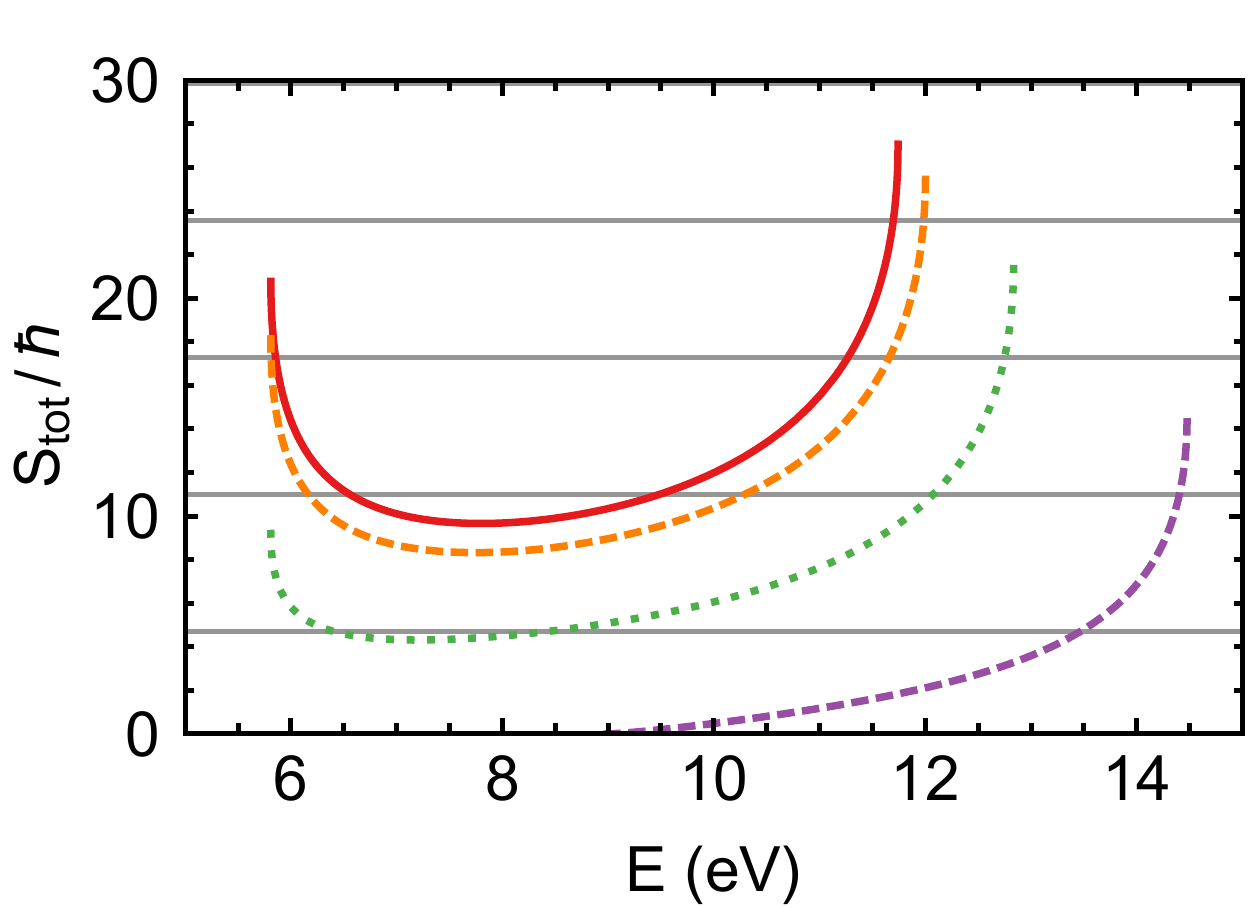}}; 
    \node at (3.8cm,1.85cm) {(b)};
  \end{tikzpicture}
  \hfill\hfill
  \caption{(a) The action $S_\text{tot}$ for regular bound states for $q_\parallel = 0$ (solid blue line), $q_\parallel = 0.4 q_{L,\text{min}}$ (dashed orange line) and $q_\parallel = 0.8 q_{L,\text{min}}$ (dotted green line). The action behaves as expected, increasing with increasing energy and decreasing with increasing $q_\parallel$. 
  (b) The action $S_\text{tot}$ for Landau-type bound states for $q_\parallel = 0$ (solid red line), $q_\parallel = 0.4 q_{L,\text{min}}$ (dashed orange line), $q_\parallel = 0.8 q_{L,\text{min}}$ (dotted green line) and $q_\parallel = 1.2 q_{L,\text{min}}$ (dashed purple line). The action behaves differently, as it has a minimum and sharply increases near the edges. Moreover, for a given $q_\parallel$, it does not necessarily become zero at a certain energy.
  The grey horizontal lines show the thresholds (a) $2(n+1/2)\pi$ and (b) $2(n+3/4)\pi$. The crossing points of these lines with the action indicate the energies of the bound states.
  }
  \label{fig:actions-regular-Landau-metal}
\end{figure}

The regular bound states shown in figure~\ref{fig:dispersion-metal-regions}(a) have a familiar shape: their energy increases with increasing $q_\parallel$ and their quantum number $n$ increases for increasing energy. This pattern can be explained by taking a closer look at the action $S_\text{tot}$ along a periodic trajectory, shown in figure~\ref{fig:actions-regular-Landau-metal}(a). We observe that the action monotonically increases with energy, and decreases as $q_\parallel$ increases. The number of states therefore decreases with increasing $q_\parallel$, in agreement with figure~\ref{fig:dispersion-metal-regions}(a).

The Landau-type bound states, on the other hand, exhibit a rather unconventional dispersion, with a new state emerging at finite $q_\parallel$. To explain this dispersion, we once again consider the action along a closed trajectory, shown in figure~\ref{fig:actions-regular-Landau-metal}(b). For $q_\parallel < q_{L,\text{min}}$, we see that it has a minimum and sharply increases near the edges. We can explain this shape by noting that the action $S_\text{tot}$ is determined by both the size of $q_1$ and the length of the integration interval. Since $q_1$ does not vanish at $x_{1,L}$, the length of the integration interval plays a key role here. Looking at figure~\ref{fig:charge-density-delimiters}(a), we see that this interval is largest near $E_{L,\text{min}}$ and $E_{P,\text{max}}(q_\parallel)$, and decreases in size as we move away from these points.
This leads to the minimum in the action, and the sharp increase near the edges.
Figure~\ref{fig:actions-regular-Landau-metal}(b) shows that, for $q_\parallel = 0$, we have two states with $n=1$, two with $n=2$ and one with $n=3$. When $q_\parallel$ increases, the action decreases. Eventually, the minimum of the action crosses the threshold for a bound state with $n=0$, which explains why a new state emerges at finite $q_\parallel$ in the spectrum depicted in figure~\ref{fig:dispersion-metal-regions}(a). For $q_\parallel>q_{L,\text{min}}$, the action is zero at $E=E_{PL}(q_\parallel)$, since the integration interval has zero length at that point, see figure~\ref{fig:charge-density-delimiters}(b). The action subsequently increases for increasing energy, leading to the spectrum of Landau-type bound states shown in figure~\ref{fig:dispersion-metal-regions}(a).

\begin{figure}[tb]
  \hfill
  \begin{tikzpicture}
    \node at (0cm,0cm)
    {\includegraphics[width=0.45\textwidth]{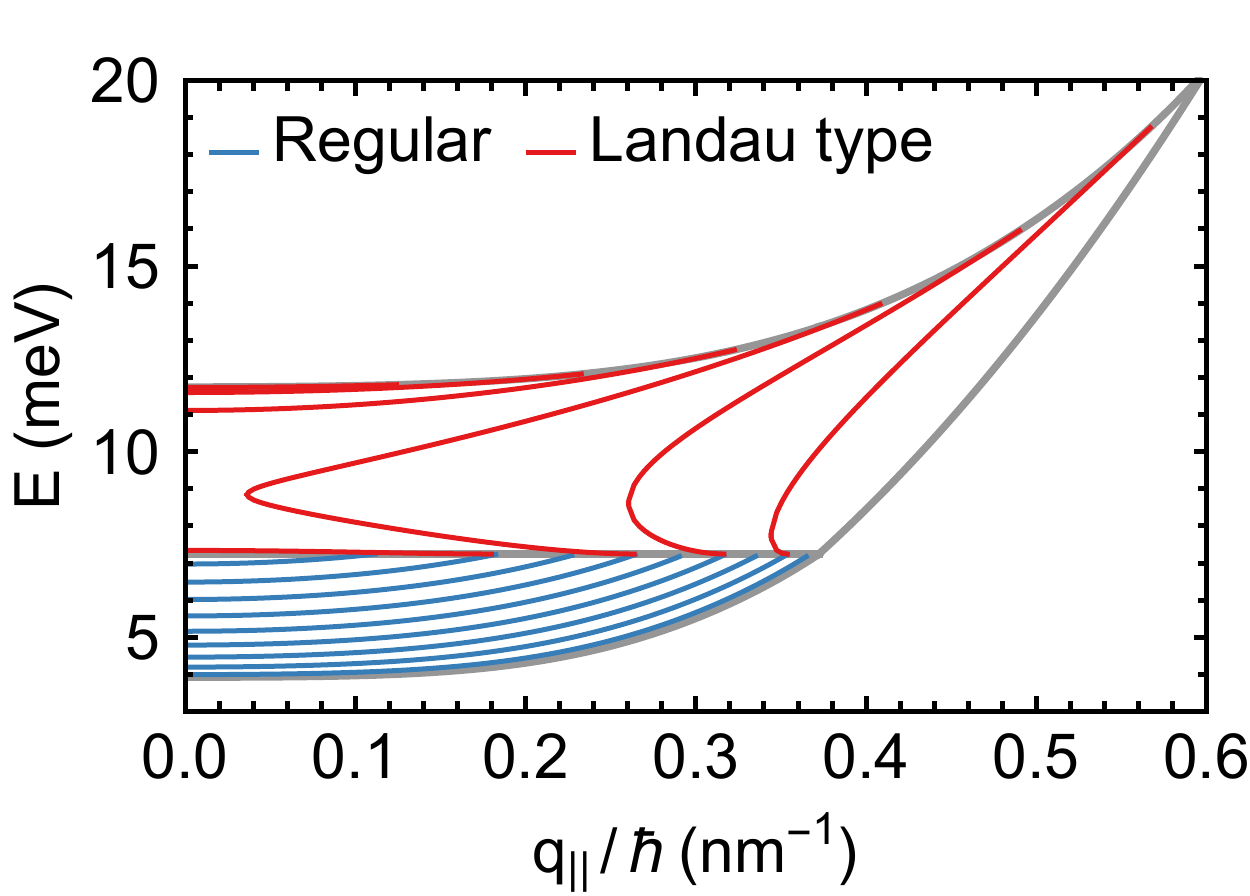}};
    \node at (-3.3cm,1.85cm) {(a)};
    
    \node at (7.1cm,0cm)
    {\includegraphics[width=0.45\textwidth]{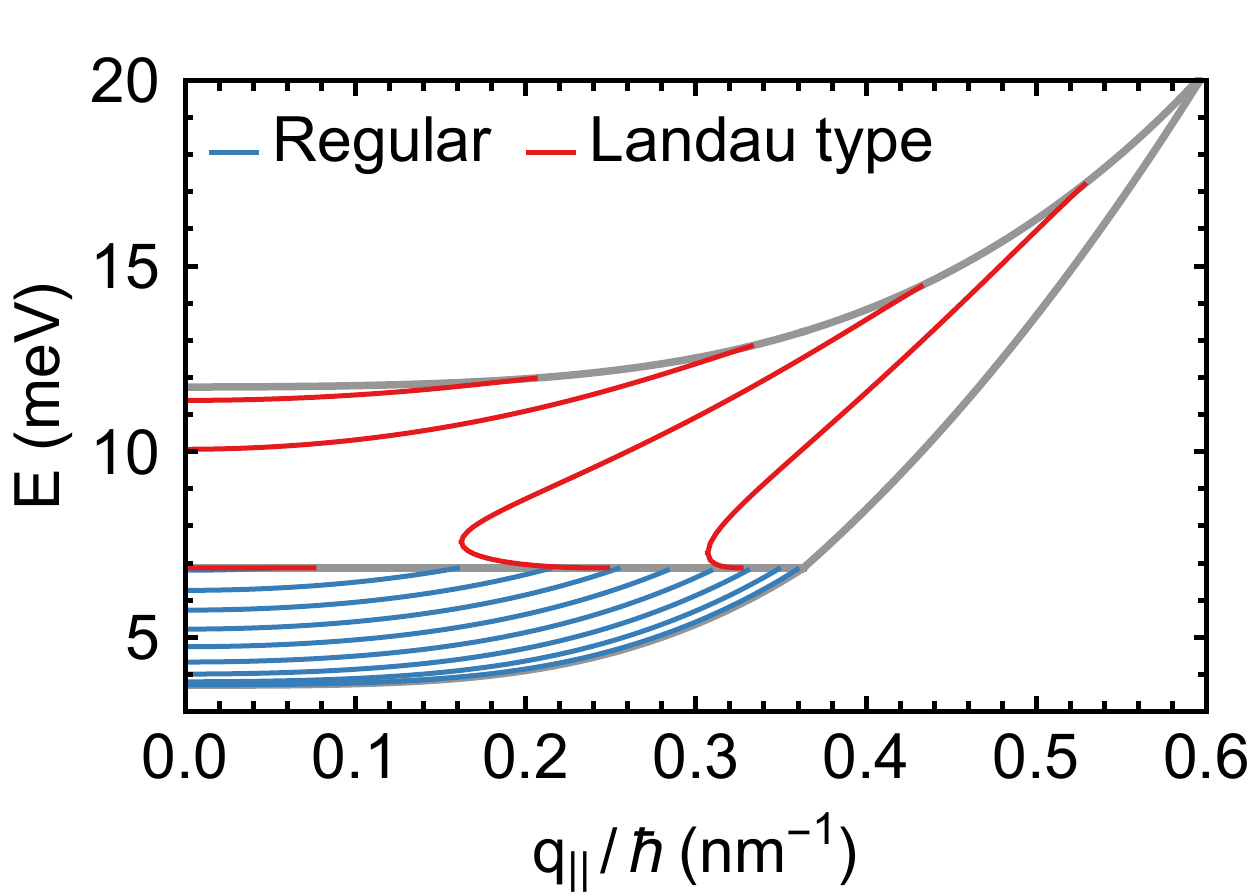}}; 
    \node at (3.8cm,1.85cm) {(b)};
    
    \node at (0cm,-4.65cm)
    {\includegraphics[width=0.45\textwidth]{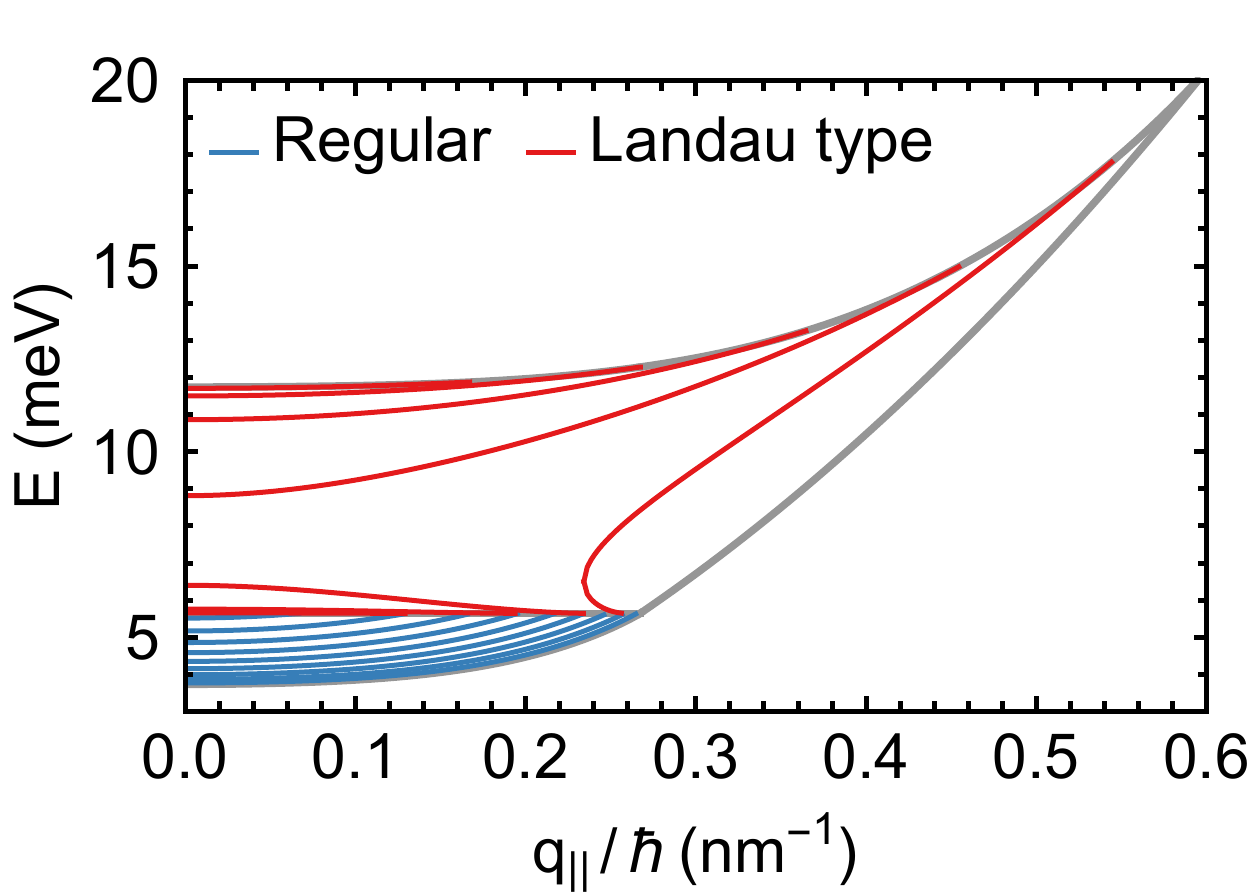}};
    \node at (-3.3cm,-2.8cm) {(c)};
    
    \node at (7.1cm,-4.65cm)
    {\includegraphics[width=0.45\textwidth]{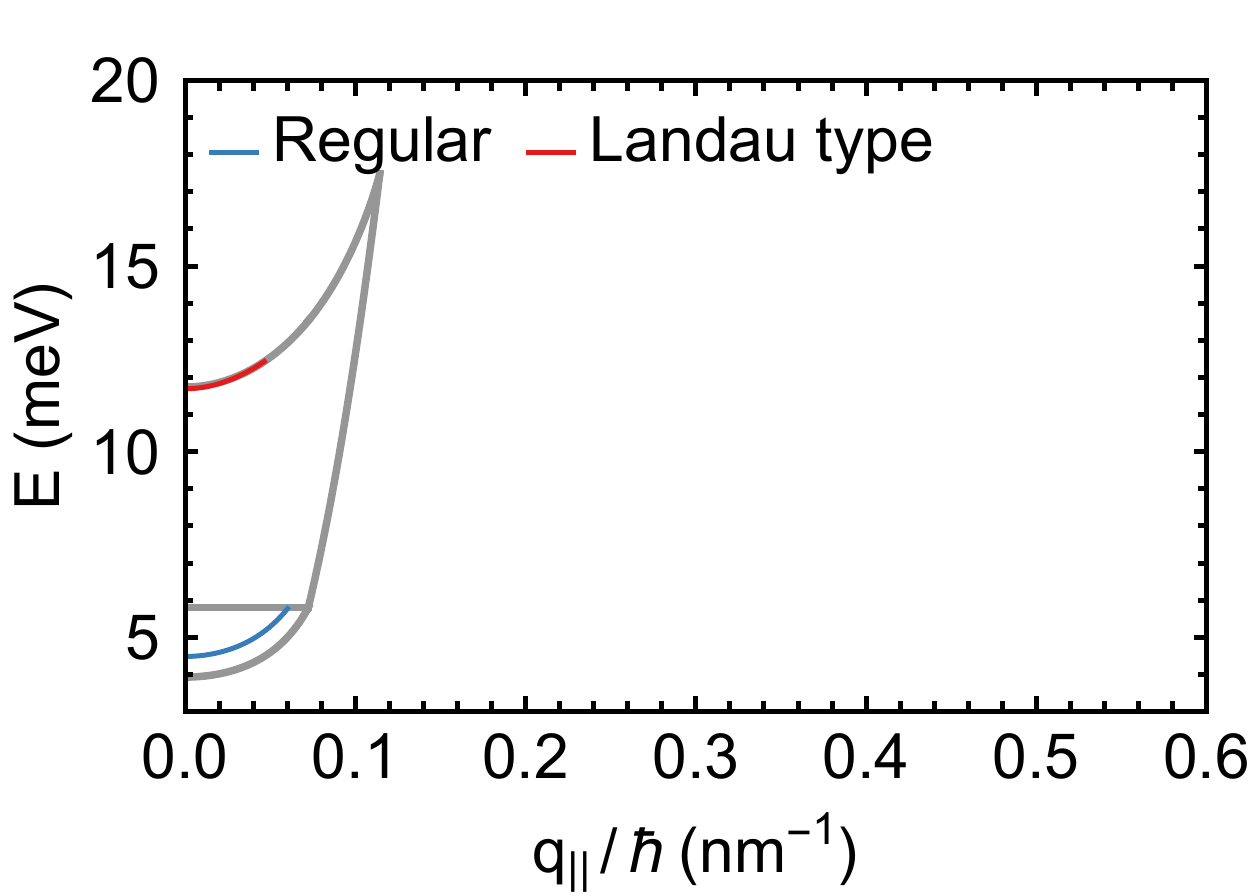}}; 
    \node at (3.8cm,-2.8cm) {(d)};
  \end{tikzpicture}
  \hfill\hfill
  \caption{Spectrum of bulk plasmons as a function of the wavevector $q_\parallel/\hbar$, with parameters indicative of a semiconductor. (a) Result for the spatially varying charge density~(\ref{eq:charge-density-tanh-numerics}), with $n^{(0)}_{\text{max}} = 10^{17}$~cm$^{-3}$, $n^{(0)}_{\text{min}} = 10^{16}$~cm$^{-3}$, $\ell_m = 100$~nm, $\ell_c=20$~nm, $m=m_e$, $g_s=2$ and $\varepsilon_b(x_1)=1$. (b) Spectrum obtained when the hyperbolic tangent in the spatially varying charge density is replaced by the spline~(\ref{eq:spline-function-charge-density}). All other parameters are kept the same. (c) Result for the spatially varying background dielectric constant~(\ref{eq:background-epsilon-tanh-numerics}), with $\varepsilon_{b,\text{min}}=1$, $\varepsilon_{b,\text{max}}=10$, $n^{(0)}(x_1)=10^{17}$~cm$^{-3}$, $\ell_m = 100$~nm, $\ell_c=20$~nm, $m=m_e$, $g_s=2$ and $\varepsilon_b(x_1)=1$. (d) Spectrum obtained for the spatially varying charge density considered in panel (a), except here we use $m = 0.1 m_e$, and $\varepsilon_b(x_1) = 10$.
  }
  \label{fig:dispersion-sc-epsilon-mass-spline}
\end{figure}
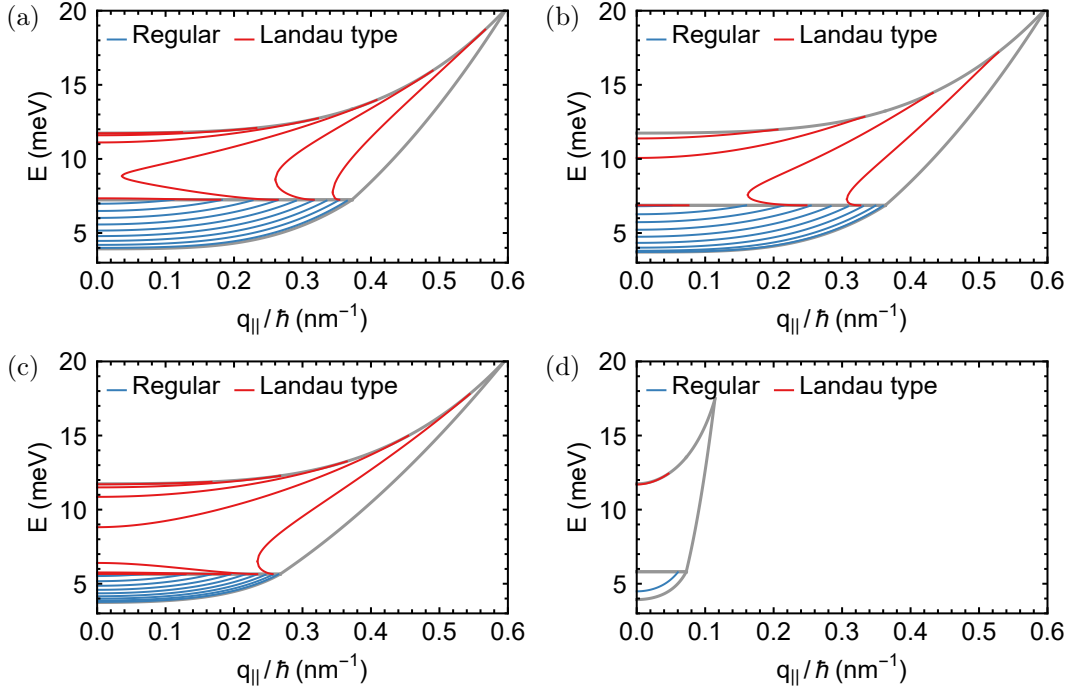

Instead of considering metals, we may also consider semiconductors. As discussed in the introduction, we can create a smooth density profile in these materials by local doping. We can also locally modify the background dielectric constant $\varepsilon_b(x)$ by modulating the semiconductor gap using external parameters. In figure~\ref{fig:dispersion-sc-epsilon-mass-spline}, we show the plasmon spectrum for different values of the parameters. Since the main idea of these calculations is to illustrate how the different parameters influence the spectrum, not all of these parameter choices represent realistic semiconductors.

In figure~\ref{fig:dispersion-sc-epsilon-mass-spline}(a), we show our results for a semiconductor with the charge density~(\ref{eq:charge-density-tanh-numerics}). We set $n^{(0)}_{\text{max}} = 10^{17}$~cm$^{-3}$, $n^{(0)}_{\text{min}} = 10^{16}$~cm$^{-3}$, $\ell_m = 100$~nm, $\ell_c=20$~nm, $m=m_e$, $g_s=2$ and $\varepsilon_b(x_1)=1$. This leads to the dimensionless parameters $h=0.14$ and $\kappa=0.0017$. Although the ratio $h/\kappa$ is much larger than for the parameters considered in figure~\ref{fig:dispersion-metal-regions}(a), the spectrum looks quite similar. In figure~\ref{fig:dispersion-sc-epsilon-mass-spline}(b), we use the same parameters, but a slightly different charge density, in which the hyperbolic tangent is replaced by the function
\begin{equation}  \label{eq:spline-function-charge-density}
  f(x_1) = 2 g\left(\frac{4 x_1}{15} + \frac{1}{2} \right) - 1, \qquad 
  g(x_1) = \left\{ \begin{array}{ll} 0, & x_1 < 0 \\ 6 x_1^5 - 15 x_1^4 + 10 x_1^3, \; & 0 \leq x_1 \leq 1 \\ 1, & x_1 > 1 \end{array} \right. .
\end{equation}
The function $g(x_1)$ is smooth at $x_1=0$ and $x_1=1$ and also has vanishing first and second derivatives at those points. 
The function $f(x_1)$ has the same derivative as the hyperbolic tangent at $x=0$, but reaches its minimal and maximal values much faster.
The increase of the action near $E_{L,\text{min}}$ and $E_{P,\text{max}}$ is therefore much less sharp.
As a result, there are less states near the edges of the spectrum in figure~\ref{fig:dispersion-sc-epsilon-mass-spline}(b).

Figure~\ref{fig:dispersion-sc-epsilon-mass-spline}(c) shows our results for a constant charge density $n^{(0)}=10^{17}$~cm$^{-3}$ and varying background dielectric constant
\begin{equation}  \label{eq:background-epsilon-tanh-numerics}
  \varepsilon_b(x_1) = \varepsilon_{b,\text{min}} + \frac{\varepsilon_{b,\text{max}} - \varepsilon_{b,\text{min}}}{2} \left( \tanh\left( \frac{x_1 + \ell_m/2}{\ell_c} \right)- \tanh\left( \frac{x_1 - \ell_m/2}{\ell_c} \right) \right) ,
\end{equation}
with $\varepsilon_{b,\text{min}}=1$ and $\varepsilon_{b,\text{max}}=10$. All other parameters correspond to those in figure~\ref{fig:dispersion-sc-epsilon-mass-spline}(a). The Thomas-Fermi screening length~$\lambda_{TF}$, when viewed as a function of the coordinate $x_1$, 
behaves somewhat differently for the systems in figures~\ref{fig:dispersion-sc-epsilon-mass-spline}(a) and~\ref{fig:dispersion-sc-epsilon-mass-spline}(c), since a tenfold decrease in $n^{(0)}$ leads to an increase of $10^{1/6}$ in $\lambda_{TF}$, whilst a tenfold increase in $\varepsilon_b$ leads to an increase of $10^{1/2}$ in $\lambda_{TF}$. Nevertheless, the spectrum does not change very much.

On the other hand, the spectrum changes significantly when we consider the same charge density as in figure~\ref{fig:dispersion-sc-epsilon-mass-spline}(a), but set $m=0.1 m_e$ and $\varepsilon_b(x_1)=10$. The plasmon spectrum for this system is shown in figure~\ref{fig:dispersion-sc-epsilon-mass-spline}(d). 
We can explain the difference between figures~\ref{fig:dispersion-sc-epsilon-mass-spline}(a) and~~\ref{fig:dispersion-sc-epsilon-mass-spline}(d) by considering the parameter $\kappa$. As can be inferred from expression~(\ref{eq:def-TF-screening-length}), the parameter $\kappa$ is ten times larger for the system shown in figure~\ref{fig:dispersion-sc-epsilon-mass-spline}(d). Note that, on the other hand, the dimensionless parameter $h$ and the plasma energy~$E_P$ are exactly the same for both systems.
We remark that the same spectrum as in figure~\ref{fig:dispersion-sc-epsilon-mass-spline}(d), when considered entirely in terms of dimensionless parameters, is obtained for $n^{(0)}_{\text{max}} = 10^{23}$~cm$^{-3}$, $n^{(0)}_{\text{min}} = 10^{22}$~cm$^{-3}$, $\ell_m = 1$~nm, $\ell_c=0.2$~nm, $m=m_e$, $g_s=2$ and $\varepsilon_b(x_1)=1$, which leads to the same values for $h$ and $\kappa$.
We also remark that the parameters used to construct figure~\ref{fig:dispersion-sc-epsilon-mass-spline}(d) are likely closer to those of realistic semiconductors than those considered in figure~\ref{fig:dispersion-sc-epsilon-mass-spline}(a).

We conclude that both dimensionless parameters $h$ and $\kappa$ influence the number of bound states in a one-dimensional plasmonic waveguide. Decreasing $h$ leads to a larger number of bound states, and increasing $\kappa$ leads to a smaller number of bound states.

\subsection{Spherically symmetric potentials}
\label{subsec:examples-radial}

In this section, we consider two different spherically symmetric potentials. We first consider the possibility of an atomic plasmon in section~\ref{subsec:examples-radial-atomic}, and then consider a parabolic potential in section~\ref{subsec:examples-radial-parabolic}.

\subsubsection{The atomic plasmon}
\label{subsec:examples-radial-atomic}

Atoms are composed of a positively charged nucleus, and a negatively charged electron cloud. If an atomic plasmon existed, this electron cloud would exhibit a collective oscillation. Since a sufficiently large amount of electrons is needed to form a collective excitation, this is more likely to happen for heavier atoms. As mentioned in the introduction, the atomic plasmon was hypothesized by Bloch~\cite{Bloch33} and Jensen~\cite{Jensen37}. Later on, experiments showed that this collective excitation does not exist~\cite{Verkhovtseva76}.

In order to show that this collective oscillation does not exist in our model, we describe the electron density around the atom with the Thomas-Fermi model~\cite{Vonsovsky89,Giuliani05,Lieb81}. We use the Tietz approximation~\cite{Tietz55,Fluegge94} for the Thomas-Fermi function, as was done in Ref.~\cite{Ishmukhametov81}. Within this approximation, the spherically symmetric Fermi momentum is given by
\begin{equation}
  p_F(\varrho) = \frac{\hbar}{a_0} \sqrt{\frac{2 Z}{\varrho/a_0}} \frac{1}{1 + \left( \frac{2 Z}{9} \right)^{1/3} \frac{\varrho}{a_0}} ,
\end{equation}
where $Z$ is the atomic number.
Our numerical implementation follows the approach discussed in section~\ref{subsec:examples-1d-waveguide-numerics}, with a few minor differences. Due to the nature of the problem, we use atomic units. Since the effective dimensionless parameter $h$ is equal to one with this choice, the formal criterion for the applicability of the semiclassical approximation implies that the action should be sufficiently large, i.e., $S_{\varrho,\text{tot}}/h \gg 1$.

\begin{figure}[tb]
  \hfill
  \begin{tikzpicture}
    \node at (0cm,0cm)
    {\includegraphics[width=0.45\textwidth]{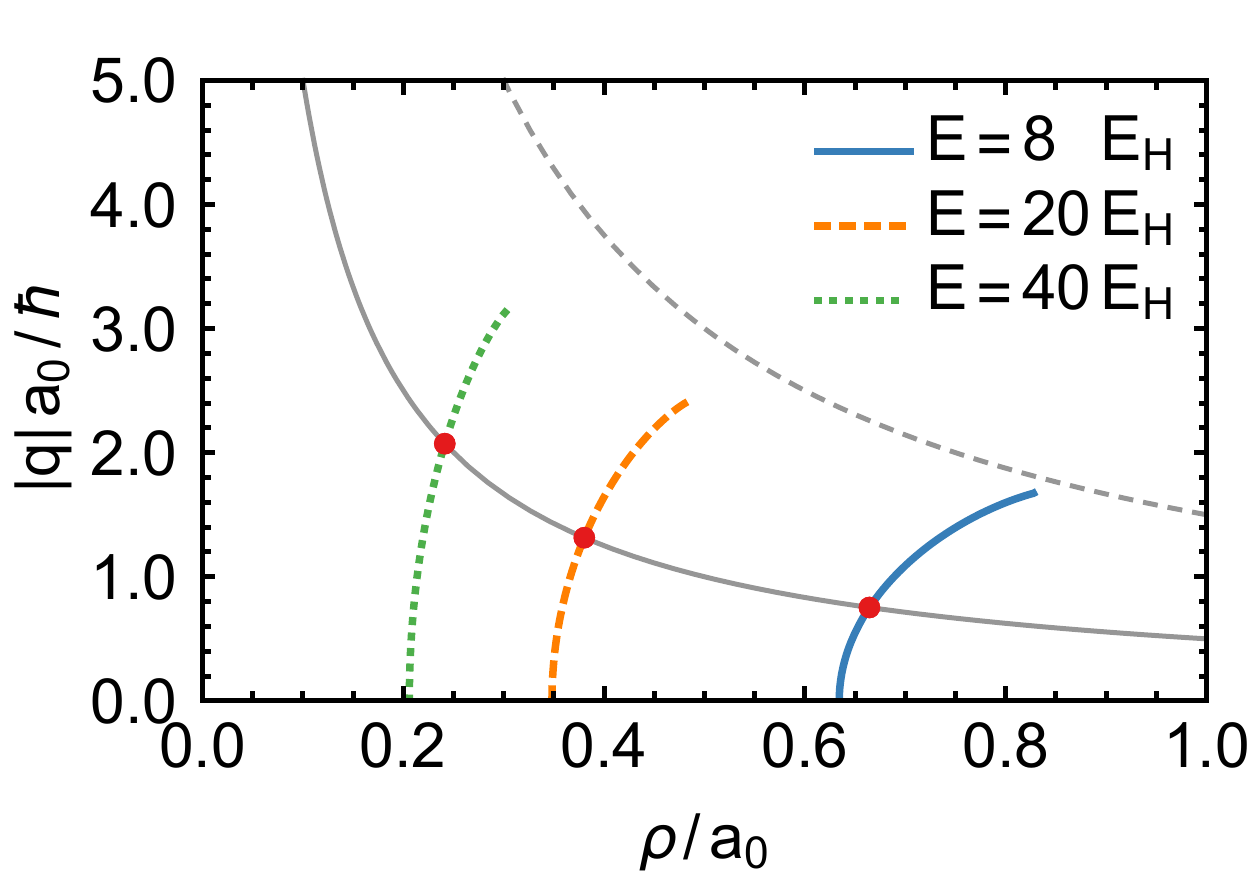}};
    \node at (-3.3cm,1.85cm) {(a)};
    
    \node at (7.1cm,0cm)
    {\includegraphics[width=0.45\textwidth]{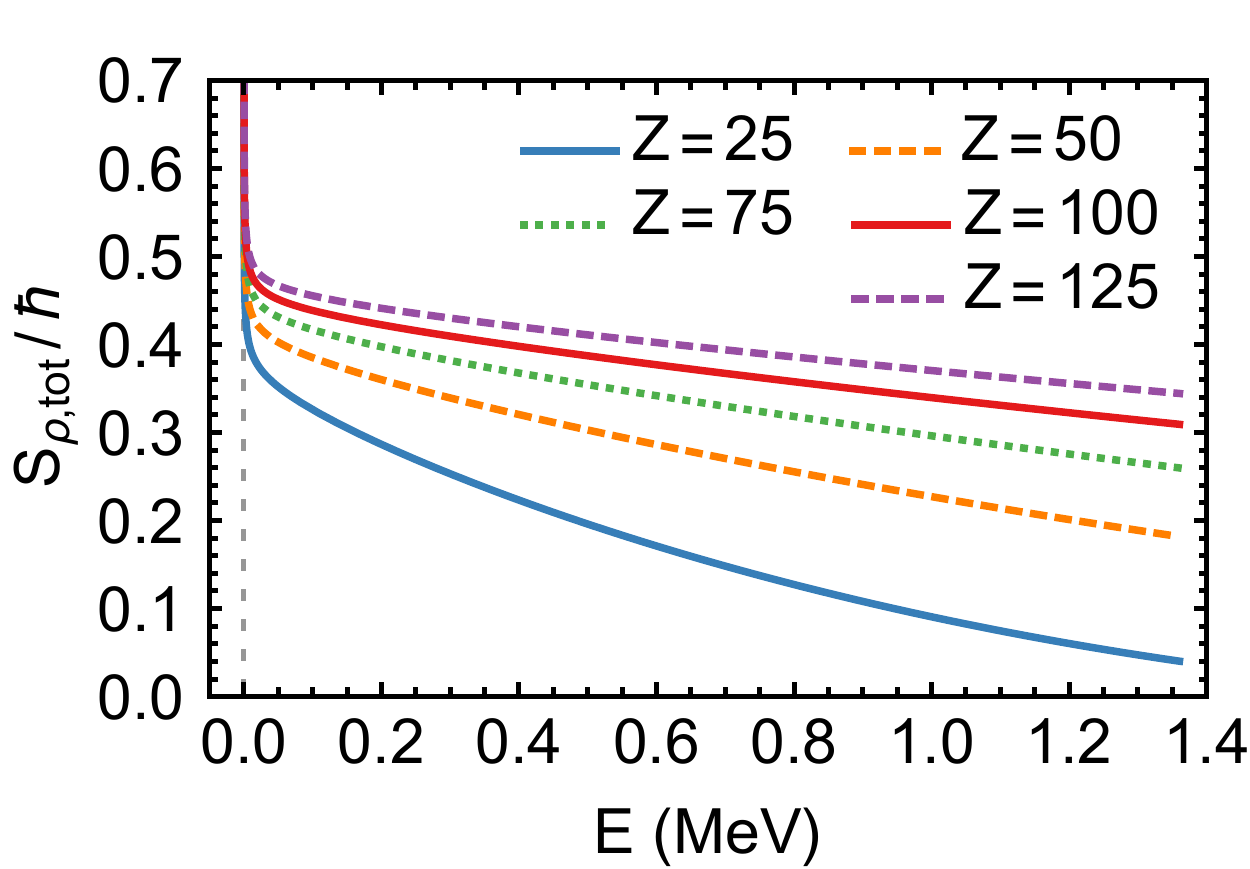}};
    \node at (3.8cm,1.85cm) {(b)};
  \end{tikzpicture}
  \hfill\hfill
  \caption{(a) The total momentum $|q| a_0/\hbar$ as a function of the radial coordinate $\varrho/a_0$ for three different energies (solid blue line, dashed orange line and dotted green line). We also show the centrifugal term $(l + \frac{1}{2} )^2 a_0^2/\varrho^2$ for $l=0$ (solid grey line) and $l=1$ (dashed grey line). The red dot indicates the simple turning point $\varrho_c$, where $|q|= \hbar / (2 \varrho_c)$.
    (b) The radial action $S_{\varrho,\text{tot}}$ as a function of the energy for different atomic numbers $Z$. The vertical, dashed grey line indicates zero and serves as a guide to the eye. Although the action steeply increases for small energies, our approximations are not valid there. We therefore do not reach the threshold for the existence of an energy level. 
  }
  \label{fig:q-dep-action-atomic-plasmon}
\end{figure}

We start by computing $q_\varrho(\varrho)$ numerically in two different ways, and check that they lead to the same result.
The first way is to determine the root of expression~(\ref{eq:L0-root-radial-quantizated-angular}), that is, we solve $L_0\big(\varrho, \sqrt{q_\varrho^2 + (l + \frac{1}{2} )^2 \hbar^2/\varrho^2 } \big) = 0$. The second way is to solve the equation $L_0(\varrho,|q|)=0$ for the total momentum $|q|$ and then to determine $q_\varrho$ from $|q|^2 = q_\varrho^2(\varrho) + (l + \frac{1}{2} )^2 \hbar^2/\varrho^2$. 
This second method is illustrated in figure~\ref{fig:q-dep-action-atomic-plasmon}(a), where we show the total momentum $|q|$ as a function of the radius $\varrho$, as well as the lines $(l + \frac{1}{2} )^2 \hbar^2/\varrho^2$ for $l=0$ and $l=1$.
The figure shows that, for the energies under consideration, the radial momentum $q_\varrho(\varrho)$ is only real for $l=0$. 
It turns out that this situation is typical, that is, setting $l=1$ often does not lead to a real value for the radial momentum.
The red dots in figure~\ref{fig:q-dep-action-atomic-plasmon}(a) indicate the simple turning points $\varrho_c$, at which $q_\varrho = 0$ for $l=0$. We also have Landau-type turning points $\varrho_L$ at the end of the different curves. As discussed in section~\ref{subsec:radial}, we have $\varrho_c < \varrho_L$.

Our next step is to compute the action $S_{\varrho,\text{tot}}$ along a closed trajectory, given by expression~(\ref{eq:quantization-r-first}). Figure~\ref{fig:q-dep-action-atomic-plasmon}(b) shows this action for $l=0$, as a function of the energy $E$, for a wide range of atomic numbers. Although the action sharply increases when the energy becomes very small, this does not have any physical meaning, since our approximations are not valid in this regime. First of all, the Thomas-Fermi approximation is no longer valid in this case, since the electron density becomes too small~\cite{Lieb81}. Second, the Tietz approximation is no longer valid, since the electron density should decrease exponentially at large distances rather than polynomially. Excluding these small energies from our considerations, we see that $S_{\varrho,\text{tot}} \leq 0.6$. We therefore cannot satisfy the radial quantization condition~(\ref{eq:quantization-r-final}), neither for $\delta=2$ nor for $\delta=1$. The action for $l=1$, when it exists, is always smaller than the action for $l=0$. 
We therefore confirm the conclusion of Ref.~\cite{Ishmukhametov81} that the atomic plasmon does not exist within our model.

We finally remark that the values of the action are formally too small to use the semiclassical approximation. On the other hand, despite this formal criterion, the semiclassical approximation usually also works reasonably well for low-lying states. Furthermore, given that the action is much smaller than the threshold for a bound state, we can safely draw the conclusion that the atomic plasmon does not exist in our model, in agreement with the experimental results~\cite{Verkhovtseva76}.

\subsubsection{Parabolic potential}
\label{subsec:examples-radial-parabolic}

In our second example of a spherically symmetric system, we consider a parabolic potential $U(\varrho)$. 
This potential decreases from its maximal value at $\varrho=0$ to zero at $\varrho =R$ and equals zero beyond this point.
As we mentioned in the introduction, this potential has been used as a model for the single-particle potential for nucleons in the atomic nucleus~\cite{Mayer55}. Loosely speaking, one may think of a spherical nanoparticle with radius $R$. Unfortunately, a spherical potential does not provide a realistic model of a spherical nanoparticle, as there the potential is typically constant in the interior and only changes close to the edges~\cite{Toscano15,David14,Lang70}. The calculations in this subsection therefore mainly serve to illustrate what kind of bound states one can expect in a spherically symmetric system.
Since the Fermi momentum $p_F(\varrho)$ is related to the square root of the potential $U(\varrho)$, see equation~(\ref{eq:TF-approx}), we write
\begin{equation}  \label{eq:pF-radial-parabolic}
  p_F(\varrho) = \begin{cases} p_{F,\text{max}} \sqrt{1-\varrho^2/R^2} \, , & \varrho < R  \\ 0 \, , & \varrho \geq R \end{cases} .
\end{equation}
The parameter $p_{F,\text{max}}$ corresponds to the Fermi momentum in the center and is related to the maximal charge density $n^{(0)}_\text{max}$ by $p_{F,\text{max}} =  (6 \pi^2 \hbar^3 n^{(0)}_\text{max} / g_s)^{1/3}$.

\begin{figure}[!tb]
  \hfill
  \begin{tikzpicture}
    \node at (0cm,0cm)
    {\includegraphics[width=0.45\textwidth]{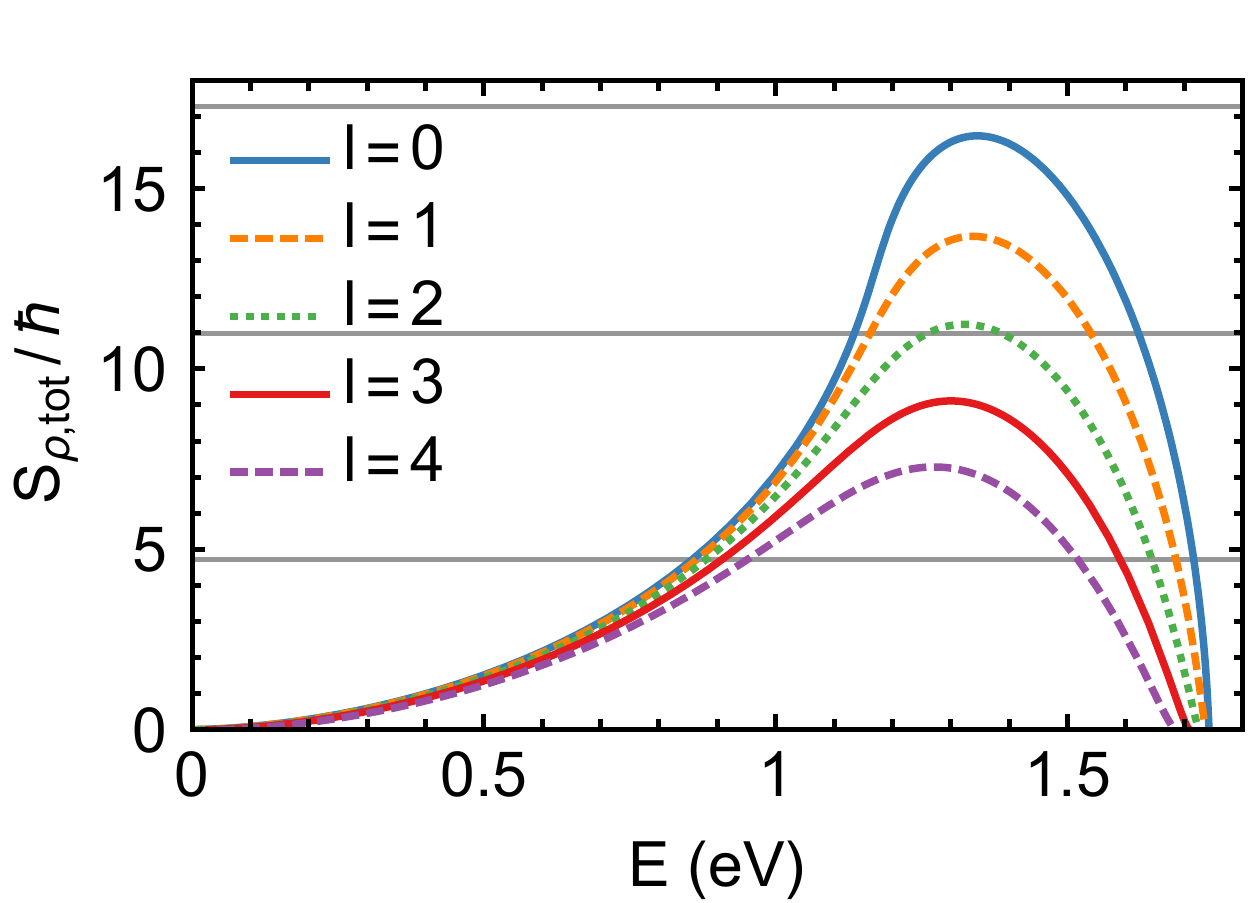}};
  \end{tikzpicture}
  \hfill\hfill
  \caption{The action $S_{\varrho,\text{tot}}$ as a function of the energy $E$ for the position-dependent Fermi momentum~(\ref{eq:pF-radial-parabolic}) with $n^{(0)}_\text{max} = 10^{21}$~cm$^{-3}$ and $R=5$~nm. We set $\varepsilon_b(\varrho)=1$, $m=m_e$ and $g_s=2$. We show different values of the angular quantum number $l$, and observe that the maximum decreases with increasing $l$. The grey horizontal lines indicate the values of $2(n+3/4)\pi$, and therefore show where the energy levels $E_{n,l}(R)$ are located.}
  \label{fig:action-spherical-parabolic}
\end{figure}

\begin{figure}[!tb]
  \hfill
  \begin{tikzpicture}
    \node at (0cm,0cm)
    {\includegraphics[width=0.45\textwidth]{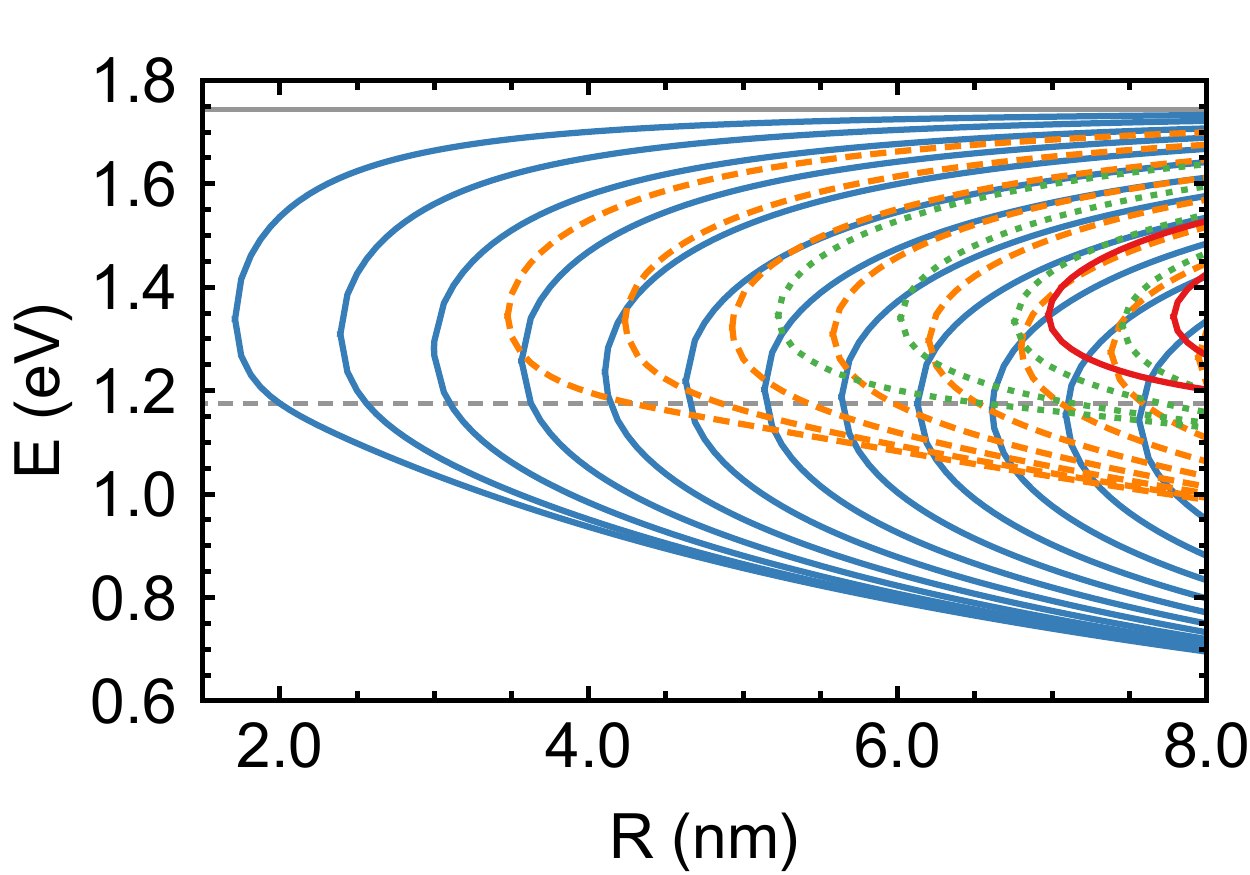}};
    \node at (-3.3cm,1.85cm) {(a)};
    
    \node at (7.1cm,0cm)
    {\includegraphics[width=0.45\textwidth]{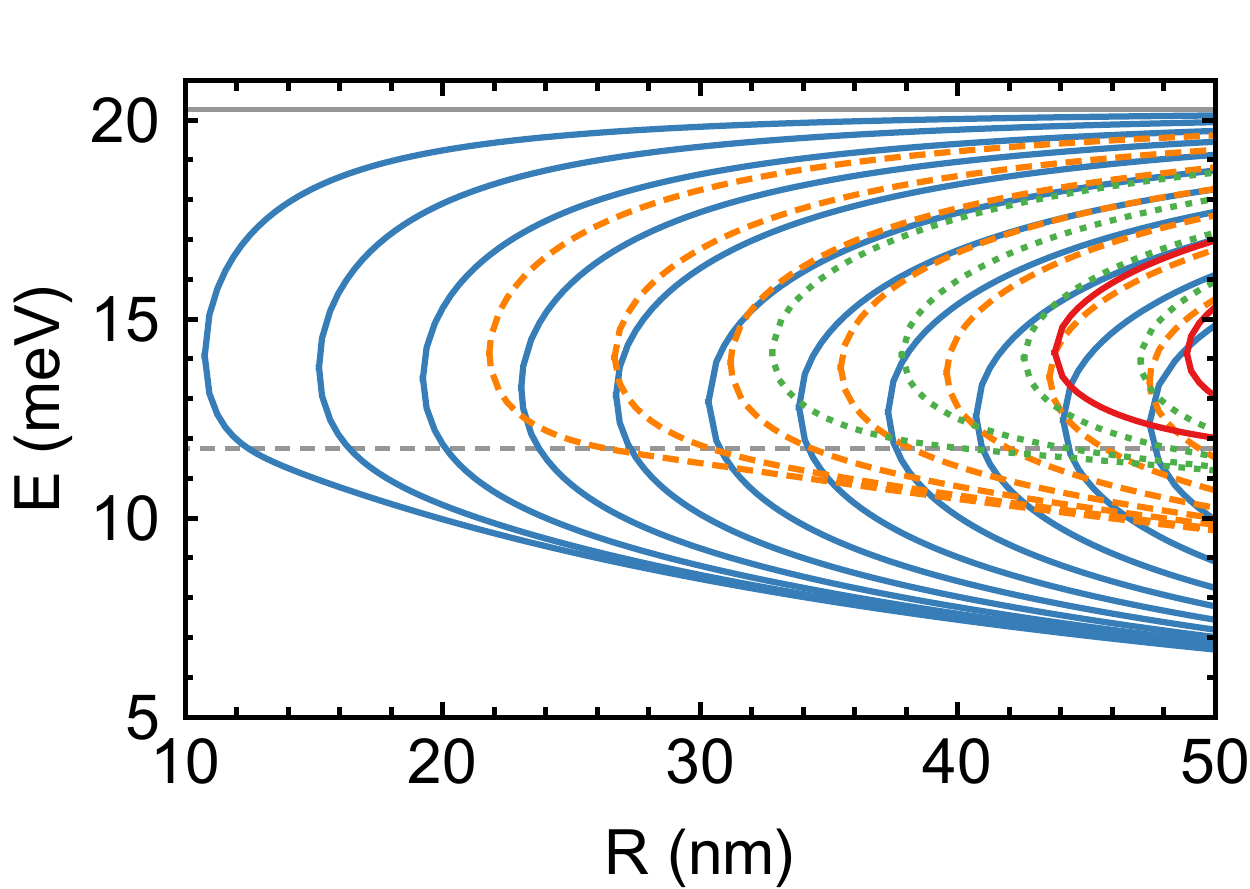}}; 
    \node at (3.8cm,1.85cm) {(b)};
  \end{tikzpicture}
  \hfill\hfill
  \caption{The energy levels $E_{n,l}$ as a function of the radius $R$ for (a) a metallic charge density $10^{21}$~cm$^{-3}$ and (b) a semiconductor charge density $10^{17}$~cm$^{-3}$. We set $\varepsilon_b(\varrho)=1$, $m=m_e$ and $g_s=2$ in both cases. Lines with different colors correspond to different values of the radial quantum number $n$, where the solid blue line corresponds to $n=0$, the dashed orange line to $n=1$, the dotted green line to $n=2$ and the solid red line to $n=4$. Different lines with the same color correspond to different values of the angular quantum number $l$, where the left-most line corresponds to $l=0$, the second from the left to $l=1$, and so forth. The solid grey horizontal line indicates the energy $E_L$ corresponding to $n^{(0)}_\text{max}$. The dashed grey horizontal line indicates the plasma energy $E_P$ corresponding to $n^{(0)}_\text{max}$.}
  \label{fig:dispersion-spherical-parabolic}
\end{figure}

We utilize the numerical implementation discussed in the previous subsection, using dimensionless units instead of atomic units.
In this way we compute the action $S_{\varrho,\text{tot}}$ as a function of the energy $E$. Figure~\ref{fig:action-spherical-parabolic} shows our result for a metallic charge density $n^{(0)}_\text{max} = 10^{21}$~cm$^{-3}$ and a radius $R=5$~nm, which corresponds to $h=0.065$ and $\kappa=0.0037$. We observe that the action has a maximum, which decreases as the angular quantum number $l$ increases. According to the radial quantization condition~(\ref{eq:quantization-r-final}), we can then determine the energy levels $E_{n,l}(R)$ by computing the intersections with the lines $2(n+3/4)\pi$, where we have once again set $\delta$ equal to two. 
As for the Landau-type bound states in an effectively one-dimensional waveguide, we find that there are two roots for each value of $n$ and $l$. We believe that the difference between these two roots can be understood by considering their localization. The root with the largest energy is located closer to the center of the sphere than the state with the smallest energy. 
Moreover, the state with the largest energy is located above $E_{P,\text{max}}$, the plasma energy corresponding to $n^{(0)}_\text{max}$. It therefore seems to be similar to the bulk states that arise in a spherical nanoparticle with a constant charge density. On the other hand, the state with the smallest energy can also occur at an energy lower than $E_{P,\text{max}}$, depending on the parameters. It therefore seems to arise from the decay of the potential near the edge.

Figure~\ref{fig:dispersion-spherical-parabolic} shows the energy levels $E_{n,l}(R)$ for the parabolic potential~(\ref{eq:pF-radial-parabolic}). We consider two different charge densities, corresponding to typical values for metals and semiconductors. In panel (a), where $n^{(0)}_\text{max} = 10^{21}$~cm$^{-3}$, the dimensionless parameters are equal to $h=0.065$ and $\kappa=0.0037$ at $R=5$~nm. In panel (b), where $n^{(0)}_\text{max} = 10^{17}$~cm$^{-3}$, the dimensionless parameters are equal to $h=0.28$ and $\kappa=0.0034$ at $R=25$~nm. The charge density indicative of a semiconductor thus requires larger radii, as we also saw in section~\ref{subsec:examples-1d-waveguide}.
For both charge densities, the number of energy levels rapidly grows as the radius $R$ increases. When new energy levels appear, this may be either due to the radial quantum number $n$, or to the angular quantum number $l$.
In accordance with our physical intuition, there are no levels above the energy $E_L$ corresponding to $n^{(0)}_\text{max}$. On the other hand, there are levels below the plasma energy $E_P$ corresponding to $n^{(0)}_\text{max}$. These states arise from the decay of the potential and are located close to the outer radius $R$.
We have thus computed the energy levels for a spherically symmetric potential, and have shown that many different states can occur depending on the radius $R$.

\section{Conclusion and outlook}
\label{sec:conclusion}

In this article, we have systematically developed a semiclassical theory for plasmons in spatially inhomogeneous media at the level of the RPA. This is not only interesting at the fundamental level, but also has potential applications for plasmonic devices, for which the quantum regime can now be reached experimentally.
We considered two different sources of inhomogeneity, namely a position-dependent charge density $n^{(0)}(x)$ (which leads to a position-dependent Fermi momentum $p_F(x)$) and a position-dependent background dielectric constant $\varepsilon_b(x)$.
From an experimental point of view, these local variations can, for instance, be created by considering combinations of different materials, by locally doping a material, or by applying local strain. 

In section~\ref{sec:derivation}, we derived our semiclassical formalism. We showed that the secular equation~(\ref{eq:secular-eq-home}), which describes plasmons in a homogeneous medium, becomes the Hamilton-Jacobi equation~(\ref{eq:Hamilton-Jacobi-epsilon}) when we consider an inhomogeneous medium. From this equation, we extracted the effective classical Hamiltonian~(\ref{eq:def-L0-epsilon}), which describes the motion of plasmons in an inhomogeneous medium. 
This Hamiltonian can be obtained from the Lindhard function for the homogeneous case by replacing $p_F$ by $p_F(x)$ and $\varepsilon_b$ by $\varepsilon_b(x)$. We also obtained the leading-order term~(\ref{eq:solution-sc-potential}) of the induced potential, and showed that it does not have a geometric or Berry phase.

In section~\ref{sec:investigation}, we investigated the equations of motion at zero temperature. For an effectively one-dimensional waveguide, we constructed the trajectories in classical phase space and found that there are two types of bound states, shown schematically in figures~\ref{fig:regular-bound-states-phase-space} and~\ref{fig:Landau-bound-states-phase-space}. The energies of these bound states are determined by the quantization conditions~(\ref{eq:quantization-cond-regular-bound}) and~(\ref{eq:quantization-cond-Landau-bound}) that we constructed. Since the allowed energies depend on the momentum $q_\parallel$ along the waveguide, we obtain a plasmon spectrum.
In section~\ref{sec:examples}, we computed this spectrum explicitly, see figure~\ref{fig:dispersion-metal-regions}. We investigated how the spectrum depends on the different parameters, considering charge densities corresponding to both metals and semiconductors, and using realistic length scales.

In a similar way, we constructed the bound states for a system with a spherically symmetric Fermi momentum. We established the quantization condition~(\ref{eq:quantization-r-final}) by studying the motion in classical phase space.
Using a model for the charge density of an atom, we subsequently showed that the atomic plasmon does not exist within our model. Finally, we computed the spectrum of bound states within a spherically symmetric nanoparticle with a linear electric field.
Once again, we considered both metallic and semiconductor charge densities.

Although we presented a comprehensive theory in this article, several open questions still remain.
The first of these is related to the Landau-type bound states that we discussed in section~\ref{sec:investigation}. These bound states include a peculiar Landau-type turning point, which we investigated in detail in section~\ref{subsec:Landau-damping-threshold}. 
By studying the behavior of the effective classical Hamiltonian near this turning point, captured by expression~(\ref{eq:L0-approx-qL-reduced}), we were able to establish that there are no allowed plasmon states in the Landau-damped region when we consider real energies.
However, as we concluded in section~\ref{subsec:complex-energies-consequences-inhomogeneous}, we should consider complex energies whenever the trajectory in phase space includes a Landau-type turning point. A proper study therefore requires that we consider a complex phase space in the vicinity of the turning point. Moreover, the jump in the value of the momentum also requires a modification of the standard semiclassical construction. Since these two modifications require more advanced techniques, they were not considered in this paper. 

Nevertheless, we believe that a more accurate description of the Landau-type turning point would be very important. First of all, it would allow us to accurately determine the spectrum of the Landau-type bound states, as well as their damping, as discussed in section~\ref{subsec:complex-energies-consequences-inhomogeneous}. The damping is directly related to the broadening of the plasmon resonances that are measured in experiments. So far, we only speculated that the damping is larger for smoother potentials, but were not able to derive an expression for the damping.
Second, a more accurate description of the Landau-type turning point would enable us to study the case of finite temperature. When the temperature is not equal to zero, Landau damping is always present, although it may be small, as can be seen in appendix~\ref{app:analytic-cont}. An accurate description of the bound states at finite temperature therefore requires a concise description of the Landau-type turning point. Similarly, Landau damping is always present for a classical plasma~\cite{Landau46,Vlasov38}, and its description therefore also requires a more detailed study of the Landau-type turning point.

Another interesting research direction would be to extend our results to two-dimensional systems. 
Plasmons in these systems have been studied extensively over the past years, both numerically, for example in  Refs.~\cite{Jiang21,Westerhout21} and experimentally, see e.g. Ref.~\cite{Grigorenko12}. Two-dimensional systems would therefore provide an ideal platform to verify our predictions.
These predictions would not only concern the spectrum of bound states, but could also involve scattering theory. In two-dimensional systems, the differential cross section and total cross section can be determined experimentally. A semiclassical approximation for these quantities can be constructed using the scattering phases or phase shifts~\cite{Fluegge94}.

When one considers scattering in real space, special attention is often paid to the regions where the intensity is maximal. This happens near so-called caustics, which are composed of the points where the density of classical trajectories diverges~\cite{Poston78,Arnold82,Reijnders18}. As a result, the Jacobian $J(x)$ vanishes at these points, and our asymptotic solution~(\ref{eq:solution-sc-potential-index}) diverges. It therefore does not accurately represent the true solution of the original differential equations in the vicinity of the caustic. In fact, the simple turning point that we discussed in section~\ref{subsec:simple-turning-point} is the simplest example of a caustic. As we already mentioned there, an accurate solution in the vicinity of this point requires a new and lengthy derivation in our case. We nevertheless believe that this derivation would be important from a fundamental point of view. It would not only rigorously prove that the Maslov index of the simple turning point equals $-1$, but would also enable us to obtain an accurate asymptotic solution in the vicinity of caustics. The latter is also important in problems where the charge density depends on more that one spatial coordinate.

In section~\ref{subsec:derivation-induced-potential}, we saw that the induced potential $V(x,t)$ does not contain a geometric or Berry phase. This led to the quantization condition~(\ref{eq:quantization-condition-general}), in which no geometric or Berry phase is present either.
We can therefore raise the question whether this always happens for our system of differential equations, or whether it is a consequence of our choice of the Hamiltonian $\hat{H}_0$, given by expression~(\ref{eq:Ham-0}). An additional Berry phase might lead to interesting new effects in the quantization condition and in the equations of motion, as discussed e.g. in Refs.~\cite{Littlejohn91,Xiao10}. It would be useful to understand, both from a fundamental point of view and from an experimental point of view, under which circumstances these additional phases might arise for plasmons in inhomogeneous media.

Finally, we may think of research directions that are more important from the point of view of applications. These include applying our theory to more realistic density profiles, and comparing the outcomes with numerical results. As we mentioned in the introduction, the numerical examples given in section~\ref{sec:examples} should be considered as a proof of concept of our theory, and not as model calculations for realistic materials. 
We remark that it would probably be more useful, from a practical point of view, to make this comparison for two-dimensional systems. We may also think about considering periodically modulated density profiles, which should lead to the appearance of a gap in the plasmon spectrum. This system could serve as a model for plasmonic crystals.

In summary, we have presented a novel semi-analytical approach for bulk plasmons in inhomogeneous media based on the semiclassical  approximation. We have derived effective equations of motion for plasmons in inhomogeneous media, and have computed the spectrum of a plasmonic waveguide and a spherically symmetric nanoparticle. Although there are still several open questions related to our approach, we believe that our method provides a theoretical basis to describe different setups in quantum plasmonics.

\section*{Acknowledgements}

We are grateful to Tjacco Koskamp, Malte R\"osner, Sergey Dobrokhotov, Vladimir Nazaikinskii, Erik van Loon and Tim de Laat for helpful discussions.
One of us (M.I.K.) started to work in this field long ago under the supervision of Boris Ishmukhametov (1929--2020). We dedicate this paper to his memory.
The work of K.J.A.R. and M.I.K. was supported by the ERC Synergy Grant, Project No. 854843 FASTCORR.

\clearpage

\appendix

\section{Details of the semiclassical derivation}
\label{app:details-derivation}

In this appendix, we discuss several technical aspects of the derivation in section~\ref{sec:derivation}. We start with a brief review of pseudodifferential operators in appendix~\ref{subapp:review-pseudo-diff-operators}: we define the symbol and summarize several of its properties. Appendix~\ref{subapp:operator-eq-to-symbol} supplements the discussion in section~\ref{subsec:derivation-density-matrix}, where we determine the symbol of the operator $\hat{w}$. We present a detailed derivation of the principal and subprincipal symbols, given by expressions~(\ref{eq:relation-u0-A0}) and~(\ref{eq:u1-final}).
In appendices~\ref{subapp:symbol-action-sandwich} and~\ref{subapp:symbol-rho0-functional}, we zoom in on two elements of this derivation.
Appendices~\ref{subapp:trace-general} and~\ref{subapp:trace-precise} contain additional details on the derivation in section~\ref{subsec:derivation-charge-density}. We review the expression for the trace of a pseudodifferential operator and prove that the operator in expression~(\ref{eq:def-density}) is trace-class.
In appendix~\ref{subapp:pd-pol-Ham}, we finally discuss the general structure of our outcomes and their relation with the general formulation of the semiclassical approximation. This allows us to introduce two new pseudodifferential operators $\hat{\Pi}$ and $\hat{L}$.

\subsection{Brief review of pseudodifferential operators and their symbols}  \label{subapp:review-pseudo-diff-operators}

As we discussed in the main text, the symbol $a(x,p,\hbar)$ of the operator $\hat{a}$ is a function on classical phase space. In order to obtain a unique relation between the operator $\hat{a}$ and the symbol $a(x,p,\hbar)$, we have to determine the operator ordering. In section~\ref{sec:derivation} of the main text, we use standard quantization, in which the operator $\hat{p}$ acts first and the operator $x$ acts second.
Within standard quantization the operator $\hat{a}$ and its symbol $a(x,p,\hbar)$ are related by the expression~\cite{Maslov81,Martinez02,Zworski12}
\begin{equation}  \label{eq:standard-quantization}
  \hat{a} \, f(x) 
  = \frac{1}{(2\pi \hbar)^d} \int e^{i \langle p , x-y \rangle/\hbar} a\big( x, p, \hbar \big) f(y) \text{d}y \text{d}p ,
\end{equation}
where $f(x)$ is a function and $d$ is the dimensionality of space. Using the Fourier transform and its inverse, defined by equation~(\ref{eq:fourier-transform}), we can also express this relation as
\begin{equation}  \label{eq:standard-quantization-fourier}
  \hat{a} \, f(x) = \mathcal{F}^{-1}_{p \to x} a(x,p,\hbar) \mathcal{F}_{y \to p} f(y) .
\end{equation}
With a slight abuse of notation, we may also write $\hat{a}=a(x,\hat{p},\hbar)$ for the operator $\hat{a}$.

We denote the operator of taking the symbol of an operator by $\sigma$, that is, we write $a(x,p,\hbar)=\sigma(\hat{a})(x,p,\hbar)$.  
Within standard quantization, this operation is particularly simple and is given by~\cite{Martinez02,Zworski12}
\begin{equation} \label{eq:symbol-from-operator}
  a(x, p, \hbar) = \sigma(\hat{a}) = e^{-i \langle p, x \rangle/\hbar} ( \hat{a} e^{i \langle p, x \rangle/\hbar} ) .
\end{equation}
As we saw in the main text, the symbol may explicitly depend on $\hbar$. Throughout this article, we consider symbols which have an asymptotic expansion in powers of $\hbar$, that is,
\begin{equation}
  a(x, p, \hbar) = \sum_j a_j(x,p) \hbar^j .
\end{equation}
We call $a_0(x,p)$ the principal part of the symbol $a(x,p,\hbar)$ and $a_1(x,p)$ its subprincipal part.

From a physical point of view, standard quantization is not the most natural scheme to use. For instance, the Hermitian operator $\tfrac{1}{2}( \langle x, \hat{p} \rangle + \langle \hat{p}, x \rangle )$ has the complex-valued symbol $\langle x, p \rangle - \tfrac{d}{2} i \hbar$ within standard quantization. If we want a self-adjoint operator to correspond to a real symbol, we thus have to choose a different quantization scheme. 
We therefore briefly consider the more general $t$-quantization.
Compared to the symbol $a(x, p, \hbar) = \sigma(\hat{a})(x,p,\hbar)$ in standard quantization, the $t$-symbol $a^{(t)}(x, p, \hbar) = \sigma_t(\hat{a})(x,p,\hbar)$ in $t$-quantization is defined by~\cite{Martinez02}
\begin{equation}  \label{eq:symbol-t}
  \hat{a} \, f(x) = \frac{1}{(2\pi \hbar)^d} \int e^{i \langle p , x-y \rangle/\hbar} a^{(t)}\big( (1-t)x + t y, p, \hbar\big) f(y) \text{d}y \text{d}p .
\end{equation}
Note that we have implicitly introduced the notation $a^{(0)}(x,p,\hbar) = a(x,p,\hbar)$.
One can pass from the symbol $a^{(t)} = \sigma_t(\hat{a})$ in $t$-quantization to the symbol $a^{(t')} = \sigma_{t'}(\hat{a})$ in $t'$-quantization using the formula~\cite{Martinez02}
\begin{equation}  \label{eq:symbol-change-quant}
  a^{(t')}(x, p, \hbar) 
  = \exp\left(i \hbar (t'-t) \left\langle \frac{\partial}{\partial x} , \frac{\partial}{\partial p} \right\rangle \right) a^{(t)}(x, p, \hbar) 
  = \sum_\beta \frac{(i \hbar (t'-t))^{|\beta|}}{\beta!} \frac{\partial^\beta}{\partial x^\beta} \frac{\partial^\beta}{\partial p^\beta} a^{(t)}(x,p,\hbar) ,
\end{equation}
where $\beta=(\beta_1,\ldots,\beta_d)$ is a multi-index, and $p^\beta = \prod_j p_j^{\beta_j}$. Furthermore, the last equality should be understood as an asymptotic equivalence, as defined in Ref.~\cite{Martinez02}. This formula implies that, from a mathematical point of view, the different quantization schemes are equivalent, as one can pass from one type of quantization to another using well defined formulas. In particular, this formula shows that the principal part of the symbol does not depend on the quantization and is the same for all $t$, i.e., $a_0^{(t)}(x,p) = a_0(x,p)$.

At this point, we would like to establish a relation between the symbol of an operator and the symbol of its adjoint. Following Ref.~\cite{Martinez02}, the $t$-symbol of an operator $\hat{a}$ is related to the $(1-t)$-symbol of its adjoint $\hat{a}^\dagger$ by the formula
\begin{equation}  \label{eq:symbol-adjoint-t}
  \sigma_{1-t}(\hat{a}^\dagger) = (\sigma_t(\hat{a}))^* ,
\end{equation}
where the star denotes complex conjugation. When $t=\tfrac{1}{2}$, we have $\sigma_{1/2}(\hat{a}^\dagger) = (\sigma_{1/2}(\hat{a}))^*$, meaning that the symbol of a self-adjoint operator is real. The quantization scheme with $t=\tfrac{1}{2}$ is known as Weyl quantization, and is the most logical choice from a physical point of view. Unfortunately, it makes the calculations much more complicated. We therefore use the simpler standard quantization in section~\ref{sec:derivation}.
Within the latter quantization scheme, we have
\begin{equation}  \label{eq:symbol-adjoint-standard}
  \sigma(\hat{a}^\dagger)(x,p,\hbar) = \left(\sigma_{1}(\hat{a})\right)^*(x,p,\hbar) = \left(\sum_\beta \frac{(i \hbar)^{|\beta|}}{\beta!} \frac{\partial^\beta}{\partial x^\beta} \frac{\partial^\beta}{\partial p^\beta} \sigma(\hat{a})(x,p,\hbar) \right)^* ,
\end{equation}
where the last equality follows from equation~(\ref{eq:symbol-change-quant}). When we consider a self-adjoint operator, with $\hat{a}^\dagger = \hat{a}$, we can expand both sides of this equation in powers of $\hbar$ and ensure equality for each order separately. This gives
\begin{equation} \label{eq:constraints-symbol-self-adjoint}
  a_0(x,p) = a_0^*(x,p) , \qquad \text{Im} \, a_1(x,p) + \frac{1}{2} \sum_j \frac{\partial^2 a_0(x,p)}{\partial x_j \partial p_j} = 0 .
\end{equation}
So although self-adjoint operators can have complex-valued symbols within standard quantization, there are clear constraints on the symbol.

Finally, we briefly discuss how to compute the symbol of an operator product. Within standard quantization, we have~\cite{Martinez02,Zworski12}
\begin{equation}
  \begin{aligned} \label{eq:symbol-product}
    \sigma( \hat{a} \hat{b} )(x,p,\hbar)
    &= \left. \exp\left( - i \hbar \left\langle \frac{\partial}{\partial q} , \frac{\partial}{\partial y} \right\rangle \right)  a(x, q, \hbar) b(y, p, \hbar) \right|_{\substack{y=x \\ q=p}} \\
    &= \sum_{\beta} \frac{(-i \hbar)^{|\beta|}}{\beta!} 
    \left( \frac{\partial^\beta}{\partial p^\beta} a(x, p, \hbar) \right) \left( \frac{\partial^\beta}{\partial x^\beta}  b(x, p, \hbar) \right) ,
  \end{aligned}
\end{equation}
where the second equality shows the asymptotic expansion of the symbol in powers of $\hbar$. One can derive similar expressions for different $(t)$-quantizations~\cite{Martinez02}. Generally, these expressions define a product on the space of symbols, known as the star product. For Weyl quantization, this product is known as the Moyal product.

\subsection{Detailed derivation of the symbols $w_0(x,p)$ and $w_1(x,p)$} 
\label{subapp:operator-eq-to-symbol}

In this appendix, we show in detail how to convert equation~(\ref{eq:operator-eq-u}) into an equation for symbols.
Since the right-hand side of equation~(\ref{eq:operator-eq-u}) contains several operator products, we first use equation~(\ref{eq:symbol-product}) to express the symbols of these operator products in terms of the symbols of the individual operators. We find that
\begin{align}  \label{eq:intermediate-1-operator-eq-u-to-symbol}
  E w(x,p,\hbar) = \;& \sigma\left( \exp\left(-\frac{i}{\hbar} S(x) \right) \hat{H}_0 \exp\left(\frac{i}{\hbar} S(x) \right) \right)(x,p,\hbar) w(x,p,\hbar) \nonumber \\
  & - i\hbar \left\langle \frac{\partial}{\partial p} \sigma\left( \exp\left(-\frac{i}{\hbar} S(x) \right) \hat{H}_0 \exp\left(\frac{i}{\hbar} S(x) \right) \right)(x,p,\hbar) , \frac{\partial w}{\partial x}(x,p,\hbar)\right\rangle \nonumber \\
  & - w(x,p,\hbar) H_0(x,p) + i\hbar \left\langle \frac{\partial w}{\partial p}(x,p,\hbar) , \frac{\partial H_0}{\partial x}(x,p) \right\rangle + \varphi(x,\hbar) \sigma\left( \hat{\rho}_0 \right)(x,p,\hbar) \nonumber \\
  & - \sigma\left( \exp\left(-\frac{i}{\hbar} S(x) \right) \hat{\rho}_0 \exp\left(\frac{i}{\hbar} S(x) \right) \right)(x,p,\hbar) \varphi(x,\hbar) \nonumber \\
  & + i\hbar \left\langle \frac{\partial}{\partial p} \sigma\left( \exp\left(-\frac{i}{\hbar} S(x) \right) \hat{\rho}_0 \exp\left(\frac{i}{\hbar} S(x) \right) \right)(x,p,\hbar) , \frac{\partial \varphi}{\partial x}(x,\hbar) \right\rangle + \mathcal{O}(\hbar^2) .
\end{align}
The second step is to compute the symbol of the quantity $\exp(-i S(x)/\hbar)\hat{a}\exp(i S(x)/\hbar)$, where $\hat{a}$ is a certain operator.
In appendix~\ref{subapp:symbol-action-sandwich}, we show that the asymptotic expansion of this symbol is given by
\begin{multline} \label{eq:symbol-sandwich-asymptotic}
  \sigma\left( \exp\left(-\frac{i}{\hbar} S(x) \right) \hat{a} \exp\left(\frac{i}{\hbar} S(x) \right) \right)(x,p,\hbar) \\
  = a\left(x, p+\frac{\partial S}{\partial x}, \hbar \right) 
  -\frac{i \hbar}{2} \sum_{j,k} \frac{\partial^2 a}{\partial p_j \partial p_k}\left( x, p+\frac{\partial S}{\partial x}, \hbar \right) \frac{\partial^2 S}{\partial x_j \partial x_k} + \mathcal{O}(\hbar^2) .
\end{multline}
Inserting this result into equation~(\ref{eq:intermediate-1-operator-eq-u-to-symbol}), and absorbing corrections of $\mathcal{O}(\hbar)$ to terms that are already $\mathcal{O}(\hbar)$ into $\mathcal{O}(\hbar^2)$, we find that
\begin{align}
  E w(x,p,\hbar) = \;& H_0\left(x,p+\frac{\partial S}{\partial x} \right) w(x,p,\hbar) - \frac{i\hbar}{2} \sum_{j,k} \frac{\partial^2 H_0}{\partial p_j \partial p_k}\left(x,p+\frac{\partial S}{\partial x}\right) \frac{\partial^2 S}{\partial x_j \partial x_k} w(x,p,\hbar) \nonumber \\
  & - i\hbar \left\langle \frac{\partial H_0}{\partial p}\left(x,p+\frac{\partial S}{\partial x} \right) , \frac{\partial w}{\partial x}(x,p,\hbar) \right\rangle - w(x,p,\hbar) H_0(x,p) \nonumber \\
  & + i\hbar \left\langle \frac{\partial w}{\partial p}(x,p,\hbar) , \frac{\partial H_0}{\partial x}(x,p) \right\rangle + \varphi(x,\hbar) \sigma\left( \hat{\rho}_0 \right)(x,p,\hbar) \nonumber \\
  & - \sigma\left( \hat{\rho}_0 \right) \left(x,p+\frac{\partial S}{\partial x},\hbar \right) \varphi(x,\hbar) \nonumber \\
  & + \frac{i\hbar}{2} \sum_{j,k} \frac{\partial^2 \sigma\left( \hat{\rho}_0 \right)}{\partial p_j \partial p_k} \left(x,p+\frac{\partial S}{\partial x},\hbar \right) \frac{\partial^2 S}{\partial x_j \partial x_k} \varphi(x,\hbar) \nonumber \\
  & + i\hbar \left\langle \frac{\partial \sigma\left( \hat{\rho}_0 \right)}{\partial p} \left(x,p+\frac{\partial S}{\partial x},\hbar \right) , \frac{\partial \varphi}{\partial x}(x,\hbar) \right\rangle + \mathcal{O}(\hbar^2) .
  \label{eq:intermediate-2-operator-eq-u-to-symbol}
\end{align}

Our third step is to compute the symbol of the equilibrium part $\hat{\rho}_0$ of the density operator. Until now, we only stated that it has to commute with $\hat{H}_0$. In order to determine its precise form, we need to consider additional physical arguments. As we discussed in the main text, we can make a comparison with the homogeneous case, where the operator $\hat{\rho}_0$ evaluated on one of the eigenstates becomes the Fermi-Dirac distribution~(\ref{eq:Fermi-Dirac-dist}) evaluated at the energy eigenvalue, i.e. $\rho_0(E_{p})$. Since the homogeneous case should emerge from the inhomogeneous case when we consider a constant potential, we naturally arrive at
\begin{equation}  \label{eq:rho0-operator-function}
  \hat{\rho}_0 = \rho_0(\hat{H}_0) ,
\end{equation}
where $\rho_0(z)$ is the Fermi-Dirac distribution~(\ref{eq:Fermi-Dirac-dist}).
Although this expression seems rather intuitive, giving it a precise meaning requires the introduction of a functional calculus for pseudodifferential operators, see e.g. Ref.~\cite{Dimassi99}.
Looking at equation~(\ref{eq:rho0-operator-function}), it seems natural that the commutator $ [\hat{\rho}_0, \hat{H}_0 ]$ vanishes, as required by the Liouville-von Neumann equation. This statement can be rigorously proven using the spectral theorem, see e.g. Ref.~\cite{Hall13}.
In appendix~\ref{subapp:symbol-rho0-functional}, we discuss why expression~(\ref{eq:rho0-operator-function}) implies that the asymptotic expansion of the symbol of $\hat{\rho}_0$ is given by
\begin{equation} \label{eq:symbol-rho0-asymptotic}
  \sigma(\hat{\rho}_0)(x,p,\hbar) = \rho_0(H_0(x,p)) - \frac{i\hbar}{2} \bigg\langle \frac{\partial H_0}{\partial p}(x,p) , \frac{\partial H_0}{\partial x}(x,p) \bigg\rangle \rho_0''(H_0(x,p)) + \mathcal{O}(\hbar^2) .
\end{equation}

When we insert the result~(\ref{eq:symbol-rho0-asymptotic}) into equation~(\ref{eq:intermediate-2-operator-eq-u-to-symbol}), we obtain
\begin{align}
  E w(x,p,\hbar) = \;& H_0\left(x,p+\frac{\partial S}{\partial x} \right) w(x,p,\hbar) - \frac{i\hbar}{2} \sum_{j,k} \frac{\partial^2 H_0}{\partial p_j \partial p_k}\left(x,p+\frac{\partial S}{\partial x}\right) \frac{\partial^2 S}{\partial x_j \partial x_k} w(x,p,\hbar) \nonumber \\
  & - i\hbar \left\langle \frac{\partial H_0}{\partial p}\left(x,p+\frac{\partial S}{\partial x} \right) , \frac{\partial w}{\partial x}(x,p,\hbar) \right\rangle - w(x,p,\hbar) H_0(x,p) \nonumber \\ 
  & + i\hbar \left\langle \frac{\partial w}{\partial p}(x,p,\hbar) , \frac{\partial H_0}{\partial x}(x,p) \right\rangle + \varphi(x,\hbar) \rho_0( H_0(x,p)) \nonumber \\
  & - \frac{i\hbar}{2} \varphi(x,\hbar) \rho_0''(H_0(x,p)) \left\langle \frac{\partial H_0}{\partial p}(x,p) , \frac{\partial H_0}{\partial x}(x,p) \right\rangle \nonumber \\
  & - \rho_0\left(H_0\left(x,p+\frac{\partial S}{\partial x} \right) \right) \varphi(x,\hbar) \nonumber \\
  & + \frac{i\hbar}{2} \rho_0''\left(H_0\left(x,p+\frac{\partial S}{\partial x} \right) \right) \left\langle \frac{\partial H_0}{\partial p}\left(x,p+\frac{\partial S}{\partial x} \right) , \frac{\partial H_0}{\partial x}\left(x,p+\frac{\partial S}{\partial x} \right) \right\rangle \varphi(x,\hbar)
  \nonumber \\
  & + \frac{i\hbar}{2} \sum_{j,k} \frac{\partial^2 \rho_0\left(H_0\left(x,p+\frac{\partial S}{\partial x} \right) \right) }{\partial p_j \partial p_k} \frac{\partial^2 S}{\partial x_j \partial x_k} \varphi(x,\hbar) \nonumber \\
  & + i\hbar \left\langle \frac{\partial \rho_0\left(H_0\left(x,p+\frac{\partial S}{\partial x} \right)\right)}{\partial p} , \frac{\partial \varphi}{\partial x}(x,\hbar) \right\rangle + \mathcal{O}(\hbar^2) .  \label{eq:intermediate-operator-eq-u-to-symbol}
\end{align}
We now note that
\begin{equation}
  \frac{\partial \rho_0\left(H_0(x,p)\right)}{\partial p} 
  = \rho_0'\left(H_0(x,p)\right) \frac{\partial H_0}{\partial p}(x,p)
\end{equation}
and that
\begin{align}
  \frac{\partial^2 \rho_0(H_0(x,p))}{\partial p_j \partial p_k} 
  &= \frac{\partial}{\partial p_j} \left( \rho_0'\left(H_0(x,p)\right) \frac{\partial H_0}{\partial p_k}(x,p) \right) \nonumber \\
  &= \rho_0''(H_0(x,p)) \frac{\partial H_0}{\partial p_j}(x,p) \frac{\partial H_0}{\partial p_k}(x,p) + \rho_0'(H_0(x,p)) \frac{\partial^2 H_0}{\partial p_j \partial p_k}(x,p) .
\end{align}
Inserting these results into equation~(\ref{eq:intermediate-operator-eq-u-to-symbol}), we find that
\begin{align}
    E w(x,p,\hbar) &= H_0\left(x,p+\frac{\partial S}{\partial x} \right) w(x,p,\hbar) - \frac{i\hbar}{2} \sum_{j,k} \frac{\partial^2 H_0}{\partial p_j \partial p_k}\left(x,p+\frac{\partial S}{\partial x}\right) \frac{\partial^2 S}{\partial x_j \partial x_k} w(x,p,\hbar) \nonumber \\
  & \;\; - i\hbar \left\langle \frac{\partial H_0}{\partial p}\left(x,p+\frac{\partial S}{\partial x} \right) , \frac{\partial w}{\partial x}(x,p,\hbar) \right\rangle - w(x,p,\hbar) H_0(x,p) \nonumber \\ 
  & \;\; + i\hbar \left\langle \frac{\partial w}{\partial p}(x,p,\hbar) , \frac{\partial H_0}{\partial x}(x,p) \right\rangle + \varphi(x,\hbar) \rho_0( H_0(x,p)) \nonumber \\
  & \;\; - \frac{i\hbar}{2} \varphi(x,\hbar) \rho_0''(H_0(x,p)) \left\langle \frac{\partial H_0}{\partial p}(x,p) , \frac{\partial H_0}{\partial x}(x,p) \right\rangle \nonumber \\
  & \;\; - \rho_0\left(H_0\left(x,p+\frac{\partial S}{\partial x} \right) \right) \varphi(x,\hbar) \nonumber \\
  & \;\; + \frac{i\hbar}{2} \rho_0''\left(H_0\left(x,p+\frac{\partial S}{\partial x} \right) \right) \left\langle \frac{\partial H_0}{\partial p}\left(x,p+\frac{\partial S}{\partial x} \right) , \frac{\partial H_0}{\partial x}\left(x,p+\frac{\partial S}{\partial x} \right) \right\rangle \varphi(x,\hbar)
  \nonumber \\
  & \;\; + \frac{i\hbar}{2} \rho_0''\left(H_0\left(x,p+\frac{\partial S}{\partial x} \right) \right) \sum_{j,k} \frac{\partial H_0}{\partial p_j} \! \left(x,p+\frac{\partial S}{\partial x} \right) \frac{\partial H_0}{\partial p_k} \! \left(x,p+\frac{\partial S}{\partial x} \right) \frac{\partial^2 S}{\partial x_j \partial x_k} \varphi(x,\hbar) \nonumber \\
  & \;\; + \frac{i\hbar}{2} \rho_0'\left(H_0\left(x,p+\frac{\partial S}{\partial x} \right) \right) \sum_{j,k} \frac{\partial^2 H_0}{\partial p_j \partial p_k}\left(x,p+\frac{\partial S}{\partial x} \right)  \frac{\partial^2 S}{\partial x_j \partial x_k} \varphi(x,\hbar) \nonumber \\
  & \;\; + i\hbar \rho_0'\left(H_0\left(x,p+\frac{\partial S}{\partial x} \right) \right) \left\langle \frac{\partial H_0}{\partial p} \left(x,p+\frac{\partial S}{\partial x} \right) , \frac{\partial \varphi}{\partial x}(x,\hbar) \right\rangle + \mathcal{O}(\hbar^2) , \label{eq:final-operator-eq-u-to-symbol-app}
\end{align}
which corresponds to equation~(\ref{eq:final-operator-eq-u-to-symbol}) in section~\ref{subsec:derivation-density-matrix} of the main text.

As discussed in the main text, we expand $\varphi(x,\hbar)$ and $w(x,p,\hbar)$ in $\hbar$ to solve this equation order by order in $\hbar$. Gathering the terms of order $h^0$, we obtain equation~(\ref{eq:relation-u0-A0-eq}), which has the solution~(\ref{eq:relation-u0-A0}). 
Our next step is to collect the terms of order $\hbar$ in equation~(\ref{eq:final-operator-eq-u-to-symbol-app}). Rearranging them slightly, we find
\begin{align}
  E w_1(x,p) = \;& H_0\left(x,p+\frac{\partial S}{\partial x} \right) w_1(x,p) - w_1(x,p) H_0(x,p) + \varphi_1(x) \rho_0( H_0(x,p)) \nonumber \\
  & - \rho_0\left(H_0\left(x,p+\frac{\partial S}{\partial x} \right) \right) \varphi_1(x)- \frac{i}{2} \sum_{j,k} \frac{\partial^2 H_0}{\partial p_j \partial p_k}\left(x,p+\frac{\partial S}{\partial x}\right) \frac{\partial^2 S}{\partial x_j \partial x_k} w_0(x,p) \nonumber \\
  & - i \left\langle \frac{\partial H_0}{\partial p}\left(x,p+\frac{\partial S}{\partial x} \right) , \frac{\partial w_0}{\partial x}(x,p) \right\rangle + i \left\langle \frac{\partial w_0}{\partial p}(x,p) , \frac{\partial H_0}{\partial x}(x,p) \right\rangle  \nonumber \\
  & - \frac{i}{2} \varphi_0(x) \rho_0''(H_0(x,p)) \left\langle \frac{\partial H_0}{\partial p}(x,p) , \frac{\partial H_0}{\partial x}(x,p) \right\rangle \nonumber \\
  & + \frac{i}{2} \rho_0''\left(H_0\left(x,p+\frac{\partial S}{\partial x} \right) \right) \left\langle \frac{\partial H_0}{\partial p}\left(x,p+\frac{\partial S}{\partial x} \right) , \frac{\partial H_0}{\partial x}\left(x,p+\frac{\partial S}{\partial x} \right) \right\rangle \varphi_0(x) \nonumber \\
  & + \frac{i}{2} \rho_0''\left(H_0\left(x,p+\frac{\partial S}{\partial x} \right) \right) \sum_{j,k} \frac{\partial H_0}{\partial p_j}\left(x,p+\frac{\partial S}{\partial x} \right) \frac{\partial H_0}{\partial p_k}\left(x,p+\frac{\partial S}{\partial x} \right) \frac{\partial^2 S}{\partial x_j \partial x_k} \varphi_0(x) \nonumber \\
  & + \frac{i}{2} \rho_0'\left(H_0\left(x,p+\frac{\partial S}{\partial x} \right) \right) \sum_{j,k} \frac{\partial^2 H_0}{\partial p_j \partial p_k}\left(x,p+\frac{\partial S}{\partial x} \right)  \frac{\partial^2 S}{\partial x_j \partial x_k} \varphi_0(x) \nonumber \\
  & + i \rho_0'\left(H_0\left(x,p+\frac{\partial S}{\partial x} \right) \right) \left\langle \frac{\partial H_0}{\partial p} \left(x,p+\frac{\partial S}{\partial x} \right) , \frac{\partial \varphi_0}{\partial x} \right\rangle .  
  \label{eq:u1-intermediate}
\end{align}
By using the relation $w_0(x,p) = \zeta(x,p,\partial S/\partial x) \varphi_0(x)$, where $\zeta(x,p,q)$ is given by expression~(\ref{eq:def-eta}) in the main text, we can eliminate $w_0(x,p)$ from this equation.
To this end, we note that
\begin{equation}  \label{eq:eta-total-derivative-x}
  \frac{\partial w_0}{\partial x_j}(x,p) = \zeta\left(x,p,\frac{\partial S}{\partial x}\right) \frac{\partial \varphi_0}{\partial x_j}(x)
  + \frac{\partial \zeta}{\partial x_j}\left(x,p,\frac{\partial S}{\partial x}\right)\varphi_0(x)
  + \sum_k \frac{\partial \zeta}{\partial q_k}\left(x,p,\frac{\partial S}{\partial x}\right) \frac{\partial^2 S}{\partial x_j \partial x_k}\varphi_0(x) ,
\end{equation}
where
\begin{align}
  \frac{\partial \zeta}{\partial q_j}(x,p,q) &= \frac{\partial}{\partial q_j} \left( \frac{\rho_0( H_0(x,p)) - \rho_0\left(H_0\left(x,p+q \right)\right)}{H_0(x,p) - H_0\left(x,p+q \right) + E} \right) \nonumber \\
  &= \frac{\rho_0( H_0(x,p)) - \rho_0\left(H_0\left(x,p+q \right)\right)}{(H_0(x,p) - H_0\left(x,p+q \right) + E)^2}\frac{\partial H_0}{\partial p_j}(x,p+q) \nonumber \\
  & \hspace*{3cm}  - \frac{\rho_0'\left(H_0\left(x,p+q \right)\right)}{H_0(x,p) - H_0\left(x,p+q \right) + E}\frac{\partial H_0}{\partial p_j}(x,p+q) .
  \label{eq:d-eta-d-q}
\end{align}
We insert these results into equation~(\ref{eq:u1-intermediate}) and subsequently solve for $w_1(x,p)$.
Rearranging the various terms, we obtain
\begin{align}
  w_1(x,p) = \;& \frac{\rho_0(H_0(x,p))-\rho_0(H_0\left(x,p+\frac{\partial S}{\partial x}\right))}{H_0(x,p)-H_0\left(x,p+\frac{\partial S}{\partial x}\right)+E} \varphi_1(x) \nonumber \\
  & -i \frac{\rho_0(H_0(x,p))-\rho_0(H_0\left(x,p+\frac{\partial S}{\partial x}\right))}{(H_0(x,p)-H_0\left(x,p+\frac{\partial S}{\partial x}\right)+E)^2} \left\langle \frac{\partial H_0}{\partial p}\left(x, p+\frac{\partial S}{\partial x}\right), \frac{\partial \varphi_0}{\partial x} \right\rangle \nonumber \\
  & +i \frac{\rho_0'(H_0\left(x,p+\frac{\partial S}{\partial x}\right))}{H_0(x,p)-H_0\left(x,p+\frac{\partial S}{\partial x}\right)+E} \left\langle \frac{\partial H_0}{\partial p}\left(x, p+\frac{\partial S}{\partial x}\right), \frac{\partial \varphi_0}{\partial x} \right\rangle \nonumber \\
  & - i \sum_{j,k} B_{1,jk}\left(x,p,\frac{\partial S}{\partial x}\right) \frac{\partial^2 S}{\partial x_j \partial x_k} \varphi_0(x) 
  + B_2\left(x,p,\frac{\partial S}{\partial x}\right) \varphi_0(x)
  \label{eq:u1-intermediate-2}
\end{align}
where
\begin{align}
  B_{1,jk}\left(x,p,q\right) 
    =\,& \frac{1}{2} \frac{\rho_0( H_0(x,p)) - \rho_0\left(H_0\left(x,p+q \right)\right)}{(H_0(x,p) - H_0\left(x,p+q \right) + E)^2} \frac{\partial^2 H_0}{\partial p_j \partial p_k}\left(x,p+q\right) \nonumber \\
    &+ \frac{\rho_0( H_0(x,p)) - \rho_0\left(H_0\left(x,p+q \right)\right)}{(H_0(x,p) - H_0\left(x,p+q \right) + E)^3} \frac{\partial H_0}{\partial p_j}\left(x,p+q\right) \frac{\partial H_0}{\partial p_k}(x,p+q) \nonumber \\
    & - \frac{\rho_0'\left(H_0\left(x,p+q \right)\right)}{(H_0(x,p) - H_0\left(x,p+q \right) + E)^2} \frac{\partial H_0}{\partial p_j}\left(x,p+q\right) \frac{\partial H_0}{\partial p_k}(x,p+q) \nonumber \\
    & - \frac{1}{2} \frac{\rho_0''\left(H_0\left(x,p+q \right) \right)}{H_0(x,p) - H_0\left(x,p+q \right) + E} \frac{\partial H_0}{\partial p_j}\left(x,p+q \right) \frac{\partial H_0}{\partial p_k}\left(x,p+q \right) \nonumber \\
    & - \frac{1}{2} \frac{\rho_0'\left(H_0\left(x,p+q \right) \right)}{H_0(x,p) - H_0\left(x,p+q \right) + E} \frac{\partial^2 H_0}{\partial p_j \partial p_k}\left(x,p+q \right) ,
    \label{eq:B1-jk-def}
\end{align}
and
\begin{align}
  B_2\left(x,p,q\right) 
  =\,& \frac{-i}{{H_0(x,p)-H_0\left(x,p+\frac{\partial S}{\partial x}\right)+E}} \Bigg( \left\langle \frac{\partial H_0}{\partial p}\left(x,p+q\right) , \frac{\partial \zeta}{\partial x}\left(x,p,q\right) \right\rangle \nonumber \\
  & \hspace*{1.0cm} - \left\langle \frac{\partial H_0}{\partial x}(x,p) , \frac{\partial \zeta}{\partial p} \left(x,p,q\right) \right\rangle + \frac{1}{2} \rho_0''(H_0(x,p)) \left\langle \frac{\partial H_0}{\partial p}(x,p) , \frac{\partial H_0}{\partial x}(x,p) \right\rangle \nonumber \\
  & \hspace*{1.0cm} - \frac{1}{2} \rho_0''\left(H_0\left(x,p+q \right) \right) \left\langle \frac{\partial H_0}{\partial p}\left(x,p+q \right) , \frac{\partial H_0}{\partial x}\left(x,p+q \right) \right\rangle \Bigg) .
  \label{eq:B2-def}
\end{align}
We immediately recognize $\zeta(x,p,\partial S/\partial x)$ in the first term in the result~(\ref{eq:u1-intermediate-2}). 
Furthermore, comparing the second and the third term to our previous expression~(\ref{eq:d-eta-d-q}), we also recognize the derivative $\partial \zeta/\partial q$.
Computing the second derivative $\partial^2 \zeta/\partial q_j\partial q_k$, we find that
\begin{align}
  \frac{\partial^2 \zeta}{\partial q_j \partial q_k}(x,p,q) =\;
  & \frac{\rho_0( H_0(x,p)) - \rho_0\left(H_0\left(x,p+q \right)\right)}{(H_0(x,p) - H_0\left(x,p+q \right) + E)^2}\frac{\partial^2 H_0}{\partial p_j \partial p_k}(x,p+q) \nonumber \\
  & + 2 \frac{\rho_0( H_0(x,p)) - \rho_0\left(H_0\left(x,p+q \right)\right)}{(H_0(x,p) - H_0\left(x,p+q \right) + E)^3}\frac{\partial H_0}{\partial p_j}(x,p+q) \frac{\partial H_0}{\partial p_k}(x,p+q) \nonumber \\
  & - 2 \frac{\rho_0'\left(H_0\left(x,p+q \right)\right)}{(H_0(x,p) - H_0\left(x,p+q \right) + E)^2} \frac{\partial H_0}{\partial p_j}(x,p+q) \frac{\partial H_0}{\partial p_k}(x,p+q) \nonumber \\
  & -\frac{\rho_0''\left(H_0\left(x,p+q \right)\right)}{H_0(x,p) - H_0\left(x,p+q \right) + E} \frac{\partial H_0}{\partial p_j}(x,p+q) \frac{\partial H_0}{\partial p_k}(x,p+q) \nonumber \\
  & -\frac{\rho_0'\left(H_0\left(x,p+q \right)\right)}{H_0(x,p) - H_0\left(x,p+q \right) + E} \frac{\partial^2 H_0}{\partial p_j \partial p_k}(x,p+q)
  \label{eq:eta-second-derivative}
\end{align}
When we compare this result to our definition~(\ref{eq:B1-jk-def}), 
we immediately see that $B_{1,jk}(x,p,q)=\frac{1}{2}\frac{\partial^2 \zeta}{\partial q_j \partial q_k}(x,p,q)$. We can thus rewrite our previous result~(\ref{eq:u1-intermediate-2}) as
\begin{multline}
  w_1(x,p) = \zeta\left(x,p,\frac{\partial S}{\partial x}\right) \varphi_1(x) -i \left\langle \frac{\partial \zeta}{\partial q}\left(x, p,\frac{\partial S}{\partial x}\right), \frac{\partial \varphi_0}{\partial x} \right\rangle \\
   - \frac{i}{2} \sum_{j,k} \frac{\partial^2 \zeta}{\partial q_j \partial q_k}\left(x,p,\frac{\partial S}{\partial x}\right) \frac{\partial^2 S}{\partial x_j \partial x_k} \varphi_0(x) 
  + B_2\left(x,p,\frac{\partial S}{\partial x}\right) \varphi_0(x) .
  \label{eq:u1-intermediate-3}
\end{multline}
In the remainder of this appendix, we rewrite our expression~(\ref{eq:B2-def}) for $B_2(x,p,q)$ in a more elegant form.
To this end, we first compute the partial derivative
\begin{equation}
  \label{eq:zeta-deriv-x}
  \frac{\partial \zeta}{\partial x_j}\left(x,p,q\right) = \frac{\rho_0'( H_0(x,p)) \frac{\partial H_0}{\partial x_j}(x,p) - \rho_0'\left(H_0\left(x,p+q \right)\right) \frac{\partial H_0}{\partial x_j}(x,p+q) }{H_0(x,p) - H_0\left(x,p+q \right) + E} ,
\end{equation}
where we have used that the denominator $H_0(x,p) - H_0\left(x,p+q \right) + E$ does not depend on $x$.
We also have
\begin{multline}
  \frac{\partial \zeta}{\partial p_j}\left(x,p,q\right) = \frac{\rho_0'( H_0(x,p)) \frac{\partial H_0}{\partial p_j}(x,p) - \rho_0'\left(H_0\left(x,p+q \right)\right) \frac{\partial H_0}{\partial p_j}(x,p+q) }{H_0(x,p) - H_0\left(x,p+q \right) + E} \\
  - \frac{\rho_0( H_0(x,p)) - \rho_0\left(H_0\left(x,p+q \right)\right)}{(H_0(x,p) - H_0\left(x,p+q \right) + E)^2}\left(\frac{ \partial H_0}{\partial p_j}(x,p) - \frac{\partial H_0}{\partial p_j}\left(x,p+q \right) \right) ,
\end{multline}
Inserting these results into our expression~(\ref{eq:B2-def}) for $B_2$, we have
\begin{align}
  i B_2\left(x,p,q\right) 
  =\,& -\frac{\rho_0'(H_0(x,p))}{(H_0(x,p) - H_0\left(x,p+q \right) + E)^2} \left\langle \frac{\partial H_0}{\partial p}(x,p) - \frac{\partial H_0}{\partial p}(x,p+q), \frac{\partial H_0}{\partial x}(x,p) \right\rangle \nonumber \\
  & + \frac{\rho_0( H_0(x,p)) - \rho_0\left(H_0\left(x,p+q \right)\right)}{(H_0(x,p) - H_0\left(x,p+q \right) + E)^3}\left\langle \frac{\partial H_0}{\partial p}(x,p) - \frac{\partial H_0}{\partial p}(x,p+q), \frac{\partial H_0}{\partial x}(x,p) \right\rangle \nonumber \\
  & + \frac{1}{2} \frac{\rho_0''(H_0(x,p))}{H_0(x,p) - H_0\left(x,p+q \right) + E} \left\langle \frac{\partial H_0}{\partial p}(x,p) , \frac{\partial H_0}{\partial x}(x,p) \right\rangle \nonumber \\
  & - \frac{1}{2} \frac{\rho_0''\left(H_0\left(x,p+q \right) \right)}{H_0(x,p) - H_0\left(x,p+q \right) + E} \left\langle \frac{\partial H_0}{\partial p}\left(x,p+q \right) , \frac{\partial H_0}{\partial x}\left(x,p+q \right) \right\rangle ,
  \label{eq:B2-mod-1}
\end{align}
where we have used that $\frac{\partial H_0}{\partial x_j}(x,p+q) - \frac{\partial H_0}{\partial x_j}(x,p) = 0$.

For reasons that are explained in appendix~\ref{subapp:pd-pol-Ham}, we would like to relate $B_2(x,p,q)$ to the partial derivative $\sum_j \frac{\partial^2 \zeta}{\partial q_j \partial x_j}$. Using our previous result~(\ref{eq:zeta-deriv-x}) for the partial derivative of $\zeta$ with respect to $x_j$, we find that
\begin{multline}
  \sum_j \frac{\partial^2 \zeta}{\partial q_j \partial x_j}\left(x,p,q\right) 
  = -\frac{\rho_0''\left(H_0\left(x,p+q \right)\right)}{H_0(x,p) - H_0\left(x,p+q \right) + E} \left\langle \frac{\partial H_0}{\partial p}(x,p+q), \frac{\partial H_0}{\partial x}(x,p+q) \right\rangle \\
  + \frac{\rho_0'( H_0(x,p)) - \rho_0'\left(H_0\left(x,p+q \right)\right)}{(H_0(x,p) - H_0\left(x,p+q \right) + E)^2} \left\langle \frac{\partial H_0}{\partial p}\left(x,p+q \right),  \frac{\partial H_0}{\partial x}(x,p) \right\rangle ,
\end{multline}
where we have used once more that $\frac{\partial H_0}{\partial x_j}$ does not depend on $p$.
We therefore have
\begin{align}
  i B_2\left(x,p,q\right) 
  =\,& 
  \frac{1}{2} \sum_j \frac{\partial^2 \zeta}{\partial q_j \partial x_j}\left(x,p,q\right)
  -\frac{\rho_0'(H_0(x,p))}{(H_0(x,p) - H_0\left(x,p+q \right) + E)^2} \left\langle \frac{\partial H_0}{\partial p}(x,p), \frac{\partial H_0}{\partial x}(x,p) \right\rangle \nonumber \\
  & + \frac{1}{2} \frac{\rho_0'(H_0(x,p))}{(H_0(x,p) - H_0\left(x,p+q \right) + E)^2} \left\langle \frac{\partial H_0}{\partial p}(x,p+q), \frac{\partial H_0}{\partial x}(x,p) \right\rangle \nonumber \\
  & + \frac{1}{2} \frac{\rho_0'(H_0(x,p+q))}{(H_0(x,p) - H_0\left(x,p+q \right) + E)^2} \left\langle \frac{\partial H_0}{\partial p}(x,p+q), \frac{\partial H_0}{\partial x}(x,p) \right\rangle \nonumber \\
  & + \frac{\rho_0( H_0(x,p)) - \rho_0\left(H_0\left(x,p+q \right)\right)}{(H_0(x,p) - H_0\left(x,p+q \right) + E)^3}\left\langle \frac{\partial H_0}{\partial p}(x,p) - \frac{\partial H_0}{\partial p}(x,p+q), \frac{\partial H_0}{\partial x}(x,p) \right\rangle \nonumber \\
  & + \frac{1}{2} \frac{\rho_0''(H_0(x,p))}{H_0(x,p) - H_0\left(x,p+q \right) + E} \left\langle \frac{\partial H_0}{\partial p}(x,p) , \frac{\partial H_0}{\partial x}(x,p) \right\rangle .
  \label{eq:B2-mod-2}
\end{align}
We would like to show that we can write the other terms in this expression as the divergence (with respect to $p$) of a vector function. We start by considering
\begin{multline}
  \hspace*{-0.35cm}
  \sum_j \frac{\partial}{\partial p_j} \left( \frac{\rho_0'( H_0(x,p)) \frac{\partial H_0}{\partial x_j}(x,p)}{H_0(x,p) - H_0\left(x,p+q \right) + E} \right) 
  \!=\! \frac{\rho_0''( H_0(x,p))}{H_0(x,p) - H_0\left(x,p+q \right) + E} \! \left\langle \frac{\partial H_0}{\partial p}(x,p), \! \frac{\partial H_0}{\partial x}(x,p) \! \right\rangle \\
  - \frac{\rho_0'( H_0(x,p))}{(H_0(x,p) - H_0\left(x,p+q \right) + E)^2} \left\langle \frac{\partial H_0}{\partial p}(x,p) - \frac{\partial H_0}{\partial p}(x,p+q), \frac{\partial H_0}{\partial x}(x,p) \right\rangle ,
\end{multline}
which leads us to
\begin{align}
  i B_2\left(x,p,q\right) 
  =\,& 
  \frac{1}{2} \sum_j \frac{\partial^2 \zeta}{\partial q_j \partial x_j}\left(x,p,q\right) +\frac{1}{2} \sum_j \frac{\partial}{\partial p_j} \left( \frac{\partial H_0}{\partial x_j}(x,p) \frac{\rho_0'( H_0(x,p))}{H_0(x,p) - H_0\left(x,p+q \right) + E} \right) \nonumber \\
  & - \frac{1}{2} \frac{\rho_0'(H_0(x,p))}{(H_0(x,p) - H_0\left(x,p+q \right) + E)^2} \left\langle \frac{\partial H_0}{\partial p}(x,p), \frac{\partial H_0}{\partial x}(x,p) \right\rangle \nonumber \\
  & + \frac{1}{2} \frac{\rho_0'(H_0(x,p+q))}{(H_0(x,p) - H_0\left(x,p+q \right) + E)^2} \left\langle \frac{\partial H_0}{\partial p}(x,p+q), \frac{\partial H_0}{\partial x}(x,p) \right\rangle \nonumber \\
  & + \frac{\rho_0( H_0(x,p)) - \rho_0\left(H_0\left(x,p+q \right)\right)}{(H_0(x,p) - H_0\left(x,p+q \right) + E)^3}\left\langle \frac{\partial H_0}{\partial p}(x,p) - \frac{\partial H_0}{\partial p}(x,p+q), \frac{\partial H_0}{\partial x}(x,p) \right\rangle .
  \label{eq:B2-mod-3}
\end{align}
The last three terms in this expression are equal to
\begin{displaymath}
  -\frac{1}{2} \sum_j \frac{\partial}{\partial p_j} \left( \frac{\partial H_0}{\partial x_j} (x,p) \frac{\rho_0( H_0(x,p)) - \rho_0\left(H_0\left(x,p+q \right)\right)}{(H_0(x,p) - H_0\left(x,p+q \right) + E)^2} \right) ,
\end{displaymath}
as can be verified by explicitly computing the derivative.
We have therefore shown that
\begin{equation}
  B_2\left(x,p,q\right) =
    - \frac{i}{2} \sum_j \frac{\partial^2 \zeta}{\partial q_j \partial x_j}\left(x,p,q\right) 
    + \sum_j \frac{\partial B_{3,j}}{\partial p_j}(x,p,q)
\end{equation}
where
\begin{multline}
  B_{3,j}(x,p,q) = 
    - \frac{i}{2} \frac{\partial H_0}{\partial x_j}(x,p) \frac{\rho_0'( H_0(x,p))}{H_0(x,p) - H_0\left(x,p+q \right) + E} \\
    + \frac{i}{2} \frac{\partial H_0}{\partial x_j}(x,p) \frac{\rho_0( H_0(x,p)) - \rho_0\left(H_0\left(x,p+q \right)\right)}{(H_0(x,p) - H_0\left(x,p+q \right) + E)^2} .
  \label{eq:def-B3-app}
\end{multline}
Inserting this expression into equation~(\ref{eq:u1-intermediate-3}), we obtain our final result for $w_1(x,p)$, namely
\begin{multline}
  w_1(x,p) = 
    \zeta\left(x,p,\frac{\partial S}{\partial x}\right) \varphi_1(x) 
    - \frac{i}{2} \sum_j \frac{\partial^2 \zeta}{\partial q_j \partial x_j}\left(x,p,\frac{\partial S}{\partial x}\right) \varphi_0(x)
    + \sum_j \frac{\partial B_{3,j}}{\partial p_j}\left(x,p,\frac{\partial S}{\partial x}\right) \varphi_0(x) \\
    - i \left\langle \frac{\partial \zeta}{\partial q}\left(x, p,\frac{\partial S}{\partial x}\right), \frac{\partial \varphi_0}{\partial x} \right\rangle
    - \frac{i}{2} \sum_{j,k} \frac{\partial^2 \zeta}{\partial q_j \partial q_k}\left(x,p,\frac{\partial S}{\partial x}\right) \frac{\partial^2 S}{\partial x_j \partial x_k} \varphi_0(x) .
  \label{eq:u1-final-app}
\end{multline}
This expression corresponds to equation~(\ref{eq:u1-final}) in the main text.

\subsection{Computing the symbol of $\exp(-i S(x)/\hbar)\hat{a}\exp(i S(x)/\hbar)$}  \label{subapp:symbol-action-sandwich}

In this appendix, we consider the operator $\exp(-i S(x)/\hbar)\hat{a}\exp(i S(x)/\hbar)$ and discuss in detail how to compute its symbol. The proof is heavily inspired by a similar proof given in chapter 2.1 of Ref.~\cite{Guillemin77}. We note that the result can also be obtained as a special case of the stationary phase theorem, as discussed in Theorem 7.7.7 from Ref.~\cite{Hormander83}. For the general stationary phase theorem, we refer to Refs.~\cite{Maslov81,Hormander83}.

For the sake of brevity, we denote the symbol of the operator $\exp(-i S(x)/\hbar)\hat{a}\exp(i S(x)/\hbar)$ by $a_S$ throughout this section. According to equation~(\ref{eq:symbol-from-operator}), this symbol can be computed using the expression
\begin{equation}
  a_S(x,p,\hbar) = \sigma\!\left( \exp\left(-\frac{i}{\hbar} S(x) \right) \hat{a} \exp\left(\frac{i}{\hbar} S(x) \right) \right)\!(x,p)
  = e^{-i \langle p, x \rangle/\hbar} e^{-i S(x)/\hbar} \left( \hat{a} e^{i S(x)/\hbar} e^{i \langle p, x \rangle/\hbar} \right) \!.
\end{equation}
By definition~(\ref{eq:standard-quantization}), this equals
\begin{equation}
  a_S(x,p,\hbar) = e^{-i \langle p, x \rangle/\hbar} e^{-i S(x)/\hbar} \frac{1}{(2\pi h)^d} \int \text{d}y \text{d}p' a(x,p',\hbar) e^{i \langle p', x-y \rangle/\hbar} e^{i S(y)/\hbar} e^{i \langle p, y \rangle/\hbar} ,
  \label{eq:symbol-sandwich-intermediate}
\end{equation}
where $a(x,p,\hbar) =\sigma(\hat{a})(x,p,\hbar)$. When we define the function $g(z,x)$ by
\begin{equation} \label{eq:def-g-exp}
  g(z,x) = S(x) - S(z) - \left\langle \frac{\partial S}{\partial x}(z), x-z \right\rangle ,
\end{equation}
we can rewrite expression~(\ref{eq:symbol-sandwich-intermediate}) as
\begin{equation}
  a_S(x,p,\hbar) = \frac{1}{(2\pi h)^d} \int \text{d}y \text{d}p' a(x,p',\hbar) \exp\left(\frac{i}{\hbar}\left\langle p'-p-\frac{\partial S}{\partial x}(x),x-y\right\rangle \right) \exp\left(\frac{i}{\hbar} g(x,y) \right) .
\end{equation}
We can then perform a change of variables, from $p'$ to $\tilde{p} = p'-p-\partial S/\partial x(x)$, which gives
\begin{equation}
  a_S(x,p,\hbar) = \frac{1}{(2\pi h)^d} \int \text{d}y \text{d}\tilde{p} \, a\left(x,\tilde{p}+p+\frac{\partial S}{\partial x}(x),\hbar\right) e^{i\langle \tilde{p},x-y\rangle/\hbar}  \exp\left(\frac{i}{\hbar} g(x,y) \right) .
\end{equation}
At this point, we perform a Taylor expansion of $a(x,p,\hbar)$ in its second argument around $\tilde{p}=0$. We then obtain
\begin{equation}
  a_S(x,p,\hbar) = \frac{1}{(2\pi h)^d} \int \text{d}y \text{d}\tilde{p} \sum_\beta \frac{1}{\beta!} \frac{\partial^\beta a}{\partial p^\beta}\left(x,p+\frac{\partial S}{\partial x}(x),\hbar \right) \tilde{p}^\beta      
  e^{i\langle \tilde{p},x-y\rangle/\hbar} \exp\left(\frac{i}{\hbar} g(x,y) \right) ,
\end{equation}
which can be rewritten as
\begin{multline}
  a_S(x,p,\hbar) = \frac{1}{(2\pi h)^d} \sum_\beta \frac{1}{\beta!} \frac{\partial^\beta a}{\partial p^\beta}\left(x,p+\frac{\partial S}{\partial x}(x),\hbar \right) \\
  \times \int \text{d}y \text{d}\tilde{p} \, (-i\hbar)^{|\beta|} \frac{\partial^\beta}{\partial x^\beta} \left( e^{i\langle \tilde{p},x\rangle/\hbar} \right) e^{-i\langle \tilde{p},y\rangle/\hbar} \exp\left(\frac{i}{\hbar} g(x,y) \right) .
\end{multline}
At this point, we would like to take the derivative outside of the integral. Unfortunately, the function $g(x,y)$ also depends on $x$. We therefore replace $g(x,y)$ by $g(z,y)$ and set $z$ equal to $x$ at the very end of the calculation. This then allows us to take the derivative out of the integral, which gives
\begin{multline}
  a_S(x,p,\hbar) = \frac{1}{(2\pi h)^d} \sum_\beta \frac{1}{\beta!} \frac{\partial^\beta a}{\partial p^\beta}\left(x,p+\frac{\partial S}{\partial x}(x),\hbar \right)  \\
  \times (-i\hbar)^{|\beta|} \frac{\partial^\beta}{\partial x^\beta} \left. \int \text{d}y \text{d}\tilde{p} \, e^{i\langle \tilde{p},x-y\rangle/\hbar} \exp\left(\frac{i}{\hbar} g(z,y) \right) \right|_{z=x} ,
\end{multline}
The integral then corresponds to the Fourier transform, followed by the inverse Fourier transform, of $\exp(i g(z,x)/h)$. Alternatively, we can integrate over $\tilde{p}$ to obtain $(2\pi\hbar)^d \delta(x-y)$. We therefore arrive at
\begin{multline} \label{eq:symbol-sandwich-full}
  a_S(x,p,\hbar) = \sigma\left( \exp\left(-\frac{i}{\hbar} S(x) \right) \hat{a} \exp\left(\frac{i}{\hbar} S(x) \right) \right)(x,p,\hbar) \\
  = \sum_\beta \frac{1}{\beta!} \frac{\partial^\beta a}{\partial p^\beta}\left(x, p+\frac{\partial S}{\partial x}, \hbar \right) (-i \hbar)^{|\beta|} \frac{\partial^\beta}{\partial x^\beta} 
  \left.\left( \exp\left[\frac{i}{\hbar} g(z,x)\right] \right)\right|_{z=x} ,
\end{multline}
where $g(z,x)$ is given by equation~(\ref{eq:def-g-exp}). Note that the second argument of the symbol of $\hat{a}$ is shifted by $\partial S/\partial x$ by the presence of $\exp(i S(x)/\hbar)$.

Although the result(\ref{eq:symbol-sandwich-full}) looks like an asymptotic expansion, we have to be careful with its interpretation because of the factor $\hbar^{-1}$ in the exponent. We therefore reduce the power of $\hbar$ on the right-hand side by one every time we take the derivative of the exponent.
To obtain the asymptotic expansion, we first take a close look at the function $g(z,x)$. We immediately note that $g(z,x)$ equals zero at $z=x$. Furthermore, the gradient $\partial g(z,x)/\partial x$ also vanishes at $z=x$. The higher order derivatives $\partial^\beta g(z,x)/\partial x^\beta$ with $|\beta|\geq2$ are equal to $\partial^\beta S/\partial x^\beta$.
Thus, although taking the derivative of the exponent reduces the power of $\hbar$ by one, it also brings down a factor of $\partial g(x,z)/\partial x$, which vanishes when setting $z$ equal to $x$. We may nevertheless end up with nonzero terms when we subsequently take the derivative of $\partial g(x,z)/\partial x$, at the cost of increasing the total power of $h$ by one.
We therefore conclude that the first two terms in the asymptotic expansion of the symbol are given by
\begin{multline}
  \sigma\left( \exp\left(-\frac{i}{\hbar} S(x) \right) \hat{a} \exp\left(\frac{i}{\hbar} S(x) \right) \right)(x,p,\hbar) \\
  = a\left(x, p+\frac{\partial S}{\partial x}, \hbar \right) 
  -\frac{i \hbar}{2} \sum_{j,k} \frac{\partial^2 a}{\partial p_j \partial p_k}\left( x, p+\frac{\partial S}{\partial x}, \hbar \right) \frac{\partial^2 S}{\partial x_j \partial x_k} + \mathcal{O}(\hbar^2) ,
\end{multline}
which corresponds to equation~(\ref{eq:symbol-sandwich-asymptotic}) in appendix~\ref{subapp:operator-eq-to-symbol}.

\subsection{Computing the symbol of $\hat{\rho}_0$}   \label{subapp:symbol-rho0-functional}

In this appendix, we discuss how to determine the symbol $\sigma(\hat{\rho}_0) = \sigma(\rho_0(\hat{H}_0)$. We first derive equation~(\ref{eq:symbol-rho0-asymptotic}) using intuitive arguments, which are not entirely mathematically rigorous but may nevertheless help to improve the understanding of the reader. Afterwards, we discuss how to prove this result rigorously.

Naively, we can expand the function $\rho_0(\hat{H}_0)$ in a Taylor series, that is, $\hat{\rho}_0 = \sum_j \alpha_j \hat{H}_0^j$.
Since the operation of taking the symbol is additive, this immediately gives
\begin{equation}  \label{eq:rho0-Taylor-expansion-symbol}
  \sigma(\rho_0(\hat{H}_0))(x,p,\hbar) = \sum_j \alpha_j \sigma(\hat{H}_0^j)(x,p,\hbar) .
\end{equation}
The right-hand side of this expression can be simplified using the formula for the symbol of an operator product, see equation~(\ref{eq:symbol-product}). Using mathematical induction, we can then prove that
\begin{equation} \label{eq:symbol-power}
  \sigma(\hat{H}_0^j) = \big(\sigma(\hat{H}_0)\big)^j - i \hbar \frac{j(j-1)}{2} \big(\sigma(\hat{H}_0)\big)^{j-2} \left\langle \frac{\partial \sigma(\hat{H}_0)}{\partial p}, \frac{\partial \sigma(\hat{H}_0)}{\partial x} \right\rangle + \mathcal{O}(\hbar^2) .
\end{equation}
On an intuitive level, this result can be understood as follows. The only way to obtain a term of order $\hbar^0$ is to consider the product $\sigma(\hat{H}_0)^j$. The terms of order $\hbar$ are subsequently obtained by taking the derivative of one of the symbols $\sigma(\hat{H}_0)$ with respect to $p$ and then taking the derivative with respect to $x$ of one of the subsequent symbols. When we consider $\sigma(\hat{H}_0^j)$, there are a total of $j(j-1)/2$ ways to do this. Note that the Hamiltonian $\hat{H}_0$ does not have terms of order $\hbar$ in its symbol, as the operator does not contain products of $x$ and $\hat{p}$. If it had, this would have led to additional terms of order $\hbar$, as we will see shortly.

Applying equation~(\ref{eq:symbol-power}) to the expansion~(\ref{eq:rho0-Taylor-expansion-symbol}) for $\sigma(\rho(\hat{H}_0))$, and using that $\sigma(\hat{H}_0)(x,p)=H_0(x,p)$, we find that, up to terms of $\mathcal{O}(\hbar^2)$,
\begin{align}
  \sigma(\rho_0(\hat{H}_0))(x,p,\hbar)
  &= \sum_j \alpha_j \left( \big(H_0(x,p)\big)^j - i \hbar \frac{j(j-1)}{2} \big(H_0(x,p)\big)^{j-2} \bigg\langle \frac{\partial H_0}{\partial p}(x,p) , \frac{\partial H_0}{\partial x}(x,p) \bigg\rangle \right) \nonumber \\
  &= \sum_j \alpha_j \big(H_0(x,p)\big)^j - \frac{i\hbar}{2} \bigg\langle \frac{\partial H_0}{\partial p}(x,p) , \frac{\partial H_0}{\partial x}(x,p) \bigg\rangle \frac{\partial^2}{\partial z^2} \left.\left( \sum_j \alpha_j z^j \right)\right|_{z=H_0(x,p)}  \nonumber \\
  &= \rho_0(H_0(x,p)) - \frac{i\hbar}{2} \bigg\langle \frac{\partial H_0}{\partial p}(x,p) , \frac{\partial H_0}{\partial x}(x,p) \bigg\rangle \rho_0''(H_0(x,p))  ,
  \label{eq:symbol-rho0-asymptotic-app}
\end{align}
which corresponds to equation~(\ref{eq:symbol-rho0-asymptotic}) in appendix~\ref{subapp:operator-eq-to-symbol}.

In the remainder of this appendix, we discuss the rigorous justification of equation~(\ref{eq:symbol-rho0-asymptotic-app}). As we already mentioned in appendix~\ref{subapp:operator-eq-to-symbol}, giving a precise meaning to the expression $\hat{\rho}_0 = \rho_0(\hat{H}_0)$ requires the introduction of a functional calculus for pseudodifferential operators. This topic is discussed in detail in Ref.~\cite{Dimassi99}. In particular, Theorem 8.7 in Ref.~\cite{Dimassi99} demonstrates that if $\hat{a}$ is a pseudodifferential operator with Weyl symbol $\sigma_{1/2}(\hat{a}) = a^{(1/2)}(x,p,\hbar) = a^{(1/2)}_0(x,p) + \hbar a^{(1/2)}_1(x,p) + \mathcal{O}(\hbar^2)$ and $f(z)$ is a function that tends to zero at infinity, then $f(\hat{a})$ is again a pseudodifferential operator. Morevover, the asymptotic expansion of its Weyl symbol is given by
\begin{equation}
  \sigma_{1/2}(f(\hat{a}))(x,p) = f(a_0^{(1/2)}(x,p)) + \hbar a_1^{(1/2)}(x,p) f'(a_0^{(1/2)}(x,p)) + \mathcal{O}(\hbar^2) .
\end{equation}
We can now use equation~(\ref{eq:symbol-change-quant}) to obtain the standard symbol $\sigma(f(\hat{a}))(x,p)$ from the Weyl symbol $\sigma_{1/2}(f(\hat{a}))(x,p)$. Noting that we can use the same formula to obtain $a_0(x,p)$ and $a_1(x,p)$ from $a_0^{(1/2)}(x,p)$ and $a_1^{(1/2)}(x,p)$, we find that, up to terms of $\mathcal{O}(\hbar^2)$,
\begin{align}
  \sigma(f(\hat{a}))(x,p) &= f(a_0^{(1/2)}(x,p)) + \hbar a_1^{(1/2)}(x,p) f'(a_0^{(1/2)}(x,p)) -\frac{i \hbar}{2} \sum_j \frac{\partial^2 f(a_0^{(1/2)}(x,p))}{\partial x_j \partial p_j} \nonumber \\
  &= f(a_0(x,p)) + \hbar \bigg( \! a_1(x,p)+\frac{i}{2}\sum_j \frac{\partial^2 a_0(x,p)}{\partial x_j \partial p_j}  \bigg) f'(a_0(x,p)) -\frac{i \hbar}{2} \sum_j \frac{\partial^2 f(a_0(x,p))}{\partial x_j \partial p_j} \nonumber \\
  &= f(a_0(x,p)) + \hbar a_1(x,p) f'(a_0(x,p)) -\frac{i \hbar}{2} \left\langle \frac{\partial a_0(x,p)}{\partial p} , \frac{\partial a_0(x,p)}{\partial x} \right\rangle f''(a_0(x,p)) 
  \label{eq:symbol-rho0-asymptotic-app-extended}
\end{align}
Note that this formula is more general than equation~(\ref{eq:symbol-rho0-asymptotic-app}), as it also covers the case of a nonzero subprincipal symbol $a_1(x,p)$. Since in our case the role of the operator $\hat{a}$ is played by $\hat{H}_0$, which does not have a subprincipal symbol, the second term in equation~(\ref{eq:symbol-rho0-asymptotic-app-extended}) vanishes. Our previous result~(\ref{eq:symbol-rho0-asymptotic-app}) is therefore proven rigorously.

\subsection{The Hilbert space trace of a pseudodifferential operator} \label{subapp:trace-general}

In this appendix, we review the definition of the Hilbert space trace for pseudodifferential operators. Our exposition is mainly based on Refs.~\cite{Dimassi99,Jimenez10,Lesch10}. We do not limit ourselves to standard quantization, but consider the more general $t$-quantization discussed in appendix~\ref{subapp:review-pseudo-diff-operators}.

For a general integral operator $\hat{A}$ with kernel $K_A(x,y)$, we have
\begin{equation} \label{eq:def-kernel}
  \big(\hat{A} f\big)(x) = \int K_A(x,y) f(y) \text{d}y .
\end{equation}
Comparing this expression to equation~(\ref{eq:symbol-t}), we see that the kernel $K_a(x,y)$ of a pseudodifferential operator $\hat{a}$ is related to its $t$-symbol $a^{(t)}(x,p,\hbar)$ by
\begin{equation}  \label{eq:kernel-pd-op}
  K_a(x,y) = \frac{1}{(2\pi \hbar)^d} \int e^{i \langle p , x-y \rangle/\hbar} a^{(t)}\big( (1-t)x + t y, p,\hbar\big) \text{d}p .
\end{equation}
If an integral operator $\hat{A}$ is trace class, its trace is expressed in terms of the kernel $K_A(x,y)$ as
\begin{equation} \label{eq:def-trace}
  \text{Tr}(\hat{A}) = \int K_A(x,x) \text{d}x .
\end{equation}
For a pseudodifferential operator $\hat{a}$, whose kernel is given by equation~(\ref{eq:kernel-pd-op}), this implies that
\begin{equation}  \label{eq:trace-pd-op}
  \text{Tr}(\hat{a}) = \frac{1}{(2\pi \hbar)^d} \int a^{(t)}( x, p, \hbar ) \text{d}p \text{d}x ,
\end{equation}
see also Refs.~\cite{Martinez02,Hall13}. For a rigorous justification of this formula, we refer to Refs.~\cite{Dimassi99,Jimenez10,Lesch10} and references therein.
Note that this formula holds regardless of the value of $t$, that is, regardless of the quantization. It is therefore in particular valid for standard quantization, which we discuss in the main text. However, it also holds for Weyl quantization, which corresponds to $t=\tfrac{1}{2}$ in the above formulas.

\subsection{Precise considerations for the trace and the induced density}  \label{subapp:trace-precise}

In this appendix, we review the precise conditions for a pseudodifferential operator to be of trace-class. We subsequently discuss how to deal with the delta function in our definition~(\ref{eq:def-density}) of the charge density and finally prove that the operator in this expression satisfies the conditions of a trace-class operator.

Following Ref.~\cite{Martinez02}, a pseudodifferential operator $\hat{a}$ is trace-class if its (standard) symbol $a(x,p)$ satisfies
\begin{equation}  \label{eq:trace-symbol-condition}
  \left| \frac{\partial^\beta}{\partial p^\beta} \frac{\partial^\gamma}{\partial x^\gamma} a(x,p) \right| \leq C_{\beta \gamma} (1+ |x|^2)^{m/2} (1+ |p|^2)^{m/2} ,
\end{equation}
and the order $m$ satisfies $m<-d$, where $d$ is the dimensionality of space. As the notation indicates, the constant $C_{\beta \gamma}$ may depend on the multi-indices $\beta$ and $\gamma$.
In words, we may say that the symbol and its derivatives have to be bounded on the entire integration interval and have to decay faster than $|x|^{-d} |p|^{-d}$ as $x$ and $p$ go to infinity. These conditions ensure that the integral~(\ref{eq:trace-pd-op}) converges.
We remark that many other definitions can be found in the literature, see e.g. Refs.~\cite{Jimenez10,Lesch10} for examples, which differ slightly from the one given above. These definitions usually explicate the relation between the trace-class property and the symbol classes introduced by H\"ormander~\cite{Hormander83}. Definition~(\ref{eq:trace-symbol-condition}) however suffices for our discussion.

Let us now consider the operator $\delta(x-x') \hat{\rho}_1$ in the trace~(\ref{eq:def-density}). Since the delta function is not an element of the Hilbert space $L^2(\mathbb{R}^d)$, we cannot directly consider the Hilbert space trace. This means that we have to modify our definition of the charge density in order to obtain a rigorous expression.
To this end, we consider the function 
\begin{equation} \label{eq:delta-rep-limit}
  \Delta_\varepsilon(x) = \frac{1}{\varepsilon^d} \eta\left(\frac{x}{\varepsilon}\right), \qquad 
  \eta(x) = \left\{ \begin{aligned} &\alpha_d \exp\left(-\frac{1}{1-x^2}\right)  && , \; |x| \leq 1 \\ & 0   && , \; |x| \geq 1 \end{aligned}  \right. \; ,
\end{equation}
where $\varepsilon > 0$ is a parameter and $\alpha_d$ is chosen in such a way that $\eta(x)$ is normalized, that is, $\int_{-\infty}^{\infty} \eta(x) \text{d}x = 1$. Note that the function $\eta(x)$ is compactly supported and infinitely differentiable. It is sometimes called the standard mollifier. We can then rigorously define the charge density as
\begin{equation}  \label{eq:def-density-precise}
  n(x,t) =\lim_{\varepsilon \to 0} g_s \text{Tr}(\Delta_\varepsilon(x-x')\hat{\rho}_1) .
\end{equation}
Let us first show that this definition agrees with our previous definition when we assume that the operator in equation~(\ref{eq:def-density-precise}) is trace-class.
Using our definitions~(\ref{eq:delta-rep-limit}) and~(\ref{eq:def-density-precise}), we have
\begin{align}
  n(x,t) &= \lim_{\varepsilon \to 0} g_s \int \frac{1}{\varepsilon^d} \eta\left(\frac{x-x'}{\varepsilon}\right) \rho_1(x',p) \, \text{d}x' \text{d}p , \nonumber \\
    &= \lim_{\varepsilon \to 0} g_s \int \eta\left(\tilde{x}\right) \rho_1(x+\varepsilon\tilde{x},p) \, \text{d}\tilde{x} \text{d}p ,
\end{align}
where we have made the substitution $\tilde{x} = (x' - x)/\varepsilon$. At this point we can take the limit into the integral and evaluate it, i.e.,
\begin{equation}
  n(x,t) = g_s \int \eta\left(\tilde{x}\right) \rho_1(x,p) \, \text{d}\tilde{x} \text{d}p 
    = g_s \int \rho_1(x,p) \, \text{d}p ,
  \label{eq:charge-density-precise-integral}
\end{equation}
where we have used the normalization of $\eta(x)$ in the final step. Comparing this result to our previous result~(\ref{eq:charge-density-integral-expression-general}), we see that they are identical. The rigorous definition~(\ref{eq:def-density-precise}) therefore leads to the same result as our previous definition~(\ref{eq:def-density}).

Now that we have understood how to deal with the delta function, we would like to show that the operator in the trace~(\ref{eq:def-density-precise}) satisfies the condition~(\ref{eq:trace-symbol-condition}) with $m<-d$. Strictly speaking, we have to do this for every order of $\hbar$ and for all derivatives. We start by considering the absolute value of the leading-order term, which we denote by $\left|\Delta_\varepsilon(x-x')\rho_1(x',p)\right|_{\text{l.o.}}$.
Using expressions~(\ref{eq:Ansatz-WKB-rho}),~(\ref{eq:relation-u0-A0}) and~(\ref{eq:def-eta}) for the leading-order term of $\rho_1(x,p)$, we have
\begin{equation}
  \left|\Delta_\varepsilon(x-x')\rho_1(x',p)\right|_{\text{l.o.}} 
    = \left| \frac{1}{(2\pi \varepsilon \hbar)^d} \eta\left(\frac{x-x'}{\varepsilon}\right) \frac{\rho_0( H_0(x',p)) - \rho_0\left(H_0\left(x',p+\frac{\partial S}{\partial x} \right)\right)}{H_0(x',p) - H_0\left(x',p+\frac{\partial S}{\partial x} \right) + E} \varphi_0(x') \right| . \label{eq:operator-trace-lo}
\end{equation}
By the mean value theorem, there exists a point $z$ in the interval $[ H_0(x',p), H_0\left(x',p+\frac{\partial S}{\partial x} \right) ]$ for which
\begin{equation} \label{eq:rho0-mean-value-theorem}
  \rho_0( H_0(x',p)) - \rho_0\left(H_0\left(x',p+\frac{\partial S}{\partial x} \right)\right) = \rho_0'(z) \left( H_0(x',p) - H_0\left(x',p+\frac{\partial S}{\partial x} \right) \right) .
\end{equation}
Let us denote the point in the interval $[ H_0(x',p), H_0\left(x',p+\frac{\partial S}{\partial x} \right) ]$ where $|\rho_0'(z)|$ assumes its maximum by $z_M(x',p)$. We can then estimate
\begin{align}
  \left| \frac{\rho_0( H_0(x',p)) - \rho_0\left(H_0\left(x',p+\frac{\partial S}{\partial x} \right)\right)}{H_0(x',p) - H_0\left(x',p+\frac{\partial S}{\partial x} \right) + E} \right| \!
    &\leq |\rho_0'(z_M(x',p))| \left| 1 - \frac{E}{H_0(x',p) - H_0\left(x',p+\frac{\partial S}{\partial x} \right) + E} \right| \nonumber \\
    &\leq |\rho_0'(z_M(x',p))| \! \left( \! 1 + \left| \frac{E}{H_0(x',p) - H_0\left(x',p+\frac{\partial S}{\partial x} \right) + E} \right| \right) \!\! . 
    \label{eq:eta-estimate}
\end{align}
Since $H_0(x',p) - H_0\left(x',p+\frac{\partial S}{\partial x} \right) + E \to 0$ as $|p| \to \infty$, the term between parentheses is bounded if there are no points for which $H_0(x',p) - H_0\left(x',p+\frac{\partial S}{\partial x} \right) + E \neq 0$.
Inserting the estimate~(\ref{eq:eta-estimate}) into our expression~(\ref{eq:operator-trace-lo}), we obtain
\begin{multline} \label{eq:operator-trace-lo-bound}
  \left|\Delta_\varepsilon(x-x')\rho_1(x',p)\right|_{\text{l.o.}} 
    \leq \frac{|\varphi_0(x')|}{(2\pi \varepsilon \hbar)^d} \eta\left(\frac{x-x'}{\varepsilon}\right) |\rho_0'(z_M(x',p))| \\
    \times \left( 1 + \left| \frac{E}{H_0(x',p) - H_0\left(x',p+\frac{\partial S}{\partial x} \right) + E} \right| \right) .
\end{multline}
Throughout this appendix, we assume that $|\varphi_0(x')|$ is bounded. We briefly discuss the background and the consequences of this assumption in section~\ref{subsec:simple-turning-point}. It is then clear that all terms on the right-hand side are bounded, as long as $H_0(x',p) - H_0\left(x',p+\frac{\partial S}{\partial x} \right) + E$ has no roots. We therefore only have to prove that the right-hand side decays faster than $|x|^{-d} |p|^{-d}$ as $|x|$, $|p|$ go to infinity.

Since $\eta(x)$ is compactly supported, it is zero outside of a certain interval. Our estimate~(\ref{eq:operator-trace-lo-bound}) therefore certainly decays sufficiently rapidly in $|x|$. Now consider the decay in $|p|$.
At zero temperature, the Fermi-Dirac distribution $\rho_0(z)$ is compactly supported, and we therefore directly see from equation~(\ref{eq:operator-trace-lo}) that $\left|\Delta_\varepsilon(x-x')\rho_1(x',p)\right|_{\text{l.o.}}$ also decays sufficiently rapidly in $|p|$.
At finite temperature, the derivative $\rho_0'(z)$ is given by
\begin{equation}  \label{eq:Fermi-Dirac-deriv}
  \rho_0'(z) = -\beta \frac{\exp(\beta(z-\mu))}{(\exp(\beta(z-\mu))+1)^2} \sim -\beta \exp(-\beta(z-\mu)) ,
\end{equation}
where the last relation holds as $z\to\infty$. For a given, fixed, value of $\frac{\partial S}{\partial x}$, the value of $z_m(x',p)$ increases as $|p|$ increases, since $z_m(x',p)$ lies within the interval $[ H_0(x',p), H_0\left(x',p+\frac{\partial S}{\partial x} \right) ]$ and $H_0(x',p)$ depends monotonically on $|p|$. Equation~(\ref{eq:Fermi-Dirac-deriv}) subsequently implies that $|\rho_0'(z_M(x',p))|$ decreases exponentially in $|p|$. Since the other terms on the right-hand side of our inequality~(\ref{eq:operator-trace-lo-bound}) are bounded, we conclude that $\left|\Delta_\varepsilon(x-x')\rho_1(x',p)\right|_{\text{l.o.}}$ decreases exponentially in $|p|$. It therefore certainly decays faster than any power of $|p|$, which finishes the proof.

Let us now consider the derivatives with respect to $x$ and $p$ in the condition~(\ref{eq:trace-symbol-condition}) and see what changes with respect to the leading-order term. When we take the derivative of $\eta(x)$ with respect to $x$, we still obtain a bounded and compactly supported function, so this does not change the arguments given above. When we instead apply the various derivatives to $\zeta\left(x,p,\frac{\partial S}{\partial x}\right)$, the expressions above change somewhat. 
In all cases we nevertheless obtain a part that exponentially decays in $|p|$, since we either end up with an expression containing $\rho_0( H_0(x',p)) - \rho_0\left(H_0\left(x',p+\frac{\partial S}{\partial x} \right)\right)$, to which we can apply the mean value theorem~(\ref{eq:rho0-mean-value-theorem}), or we directly obtain a derivative of $\rho_0(z)$, which decays exponentially. In equation~(\ref{eq:operator-trace-lo-bound}), the exponentially decaying term was multiplied by a bounded term. When we consider derivatives, we instead multiply by the rational function $R(x,p) = P(x,p)/(H_0(x',p) - H_0\left(x',p+\frac{\partial S}{\partial x} \right) + E)^m$, where $P(x,p)$ is polynomial in $p$ and $m$ is an integer. If the potential $U(x)$ and its derivatives do not have any singularities in a neighborhood of the point $x$ under consideration, the only singularities of the polynomial $R(x,p)$ lie at the points where $H_0(x',p) - H_0\left(x',p+\frac{\partial S}{\partial x} \right) + E=0$.
The product of the rational function $R(x,p)$ and the exponentially decaying term still decays faster than any power of $|p|$. The leading-order term therefore satisfies the condition~(\ref{eq:trace-symbol-condition}).

Looking at equation~(\ref{eq:u1-final}), we immediately see that the subleading term in the trace~(\ref{eq:def-density-precise}), and its derivatives, have the same structure as the derivatives of the leading-order term. They are therefore also compactly supported in $x$ and decay faster than any power of $|p|$. We do require that the potential $U(x)$ and its derivatives do not have any singularities in a neighborhood of the point $x$ under consideration, otherwise the condition~(\ref{eq:trace-symbol-condition}) does not hold. Summarizing, we conclude that the operator in the trace~(\ref{eq:def-density-precise}) satisfies the condition~(\ref{eq:trace-symbol-condition}) with $m<-d$.

Throughout this section, we only mentioned the possibility of zeroes of the denominator $H_0(x',p) - H_0\left(x',p+\frac{\partial S}{\partial x} \right) + E$, but did not investigate them. Let us have a look at them from a mathematical perspective here. We discuss the physical perspective in detail in section~\ref{subsec:Landau-damping-threshold}. First note that the difference $H_0(x',p) - H_0\left(x',p+\frac{\partial S}{\partial x} \right)$ does not depend on $x$. For a given value of $\frac{\partial S}{\partial x}$, the equation $H_0(x',p) - H_0\left(x',p+\frac{\partial S}{\partial x} \right) + E$ certainly has roots because $H_0(x,p)$ is quadratic in $|p|$.
When we consider the case of zero temperature, the numerator $\rho_0( H_0(x',p)) - \rho_0\left(H_0\left(x',p+\frac{\partial S}{\partial x} \right)\right)$ of $\zeta\left(x,p,\frac{\partial S}{\partial x}\right)$ is zero outside of a certain interval. The roots of $H_0(x',p) - H_0\left(x',p+\frac{\partial S}{\partial x} \right)$ therefore do not present any complication when they lie outside of this interval. In fact, this is the main example that we discuss in sections~\ref{sec:investigation} and~\ref{sec:examples}.
On the other hand, the condition~(\ref{eq:trace-symbol-condition}) is clearly violated when we consider finite temperatures, as the numerator of $\zeta\left(x,p,\frac{\partial S}{\partial x}\right)$ does not vanish. Nevertheless, we can still compute the integral~(\ref{eq:charge-density-precise-integral}), as the zeroes of the denominator lead to simple poles. However, in order to perform this computation, we have to consider complex momenta $p$, as we have to consider a tiny arc around the singularity. This could mean that we are not dealing with a fundamental problem, but rather with a technical problem that could possibly resolved by considering complex arguments. Further research is thus needed to clarify the consequences of this singularity.

\subsection{Commutation relations and new pseudodifferential operators} \label{subapp:pd-pol-Ham}

As we discussed at several points in section~\ref{sec:derivation}, the expression for $n(x,t)$ has a very particular structure. Similarly, the equations that are obtained when considering the leading and subleading orders for the Poisson equation have the same structure. In this section, we discuss this structure in more detail. We show why it regularly appears in the semiclassical approximation, and explain why it implies that we can introduce new pseudodifferential operators $\hat{\Pi}$ and $\hat{L}$.

To start our discussion, we consider a general pseudodifferential operator $\hat{\Gamma}$, with standard symbol $\Gamma(x,p,\hbar) = \Gamma_0(x,p) + \hbar \Gamma_1(x,p) + \mathcal{O}(\hbar^2)$. We consider its action on the semiclassical Ansatz $\varphi(x) \exp\left(\frac{i}{\hbar} S(x)\right)$, where the amplitude also has an asymptotic expansion: $\varphi(x) = \varphi_0(x) + \hbar \varphi_1(x) + \mathcal{O}(\hbar^2)$.
With the operator calculus, one can prove so-called commutation formulas for the operator and the semiclassical Ansatz. 
The commutation formula for the leading-order term of the asymptotic series reads
\begin{equation} \label{eq:commutation-formula-leading-order}
  \hat{\Gamma} \varphi(x) \exp\left(\frac{i}{\hbar} S(x)\right) = \exp\left(\frac{i}{\hbar} S(x)\right) \left( \Gamma_0\left(x,\frac{\partial S}{\partial x}\right) \varphi_0(x) + \mathcal{O}(\hbar) \right) ,
\end{equation}
where $\Gamma_0(x,p)$ is the principal symbol of $\hat{\Gamma}$. We refer to e.g. Refs.~\cite{Maslov81,Guillemin77} for the proof.
When we also include the subleading order, we obtain the commutation formula 
\begin{multline}  \label{eq:commutation-formula-subleading-order}
  \hat{\Gamma} \varphi(x) \exp\left(\frac{i}{\hbar} S(x)\right) = \exp\left(\frac{i}{\hbar} S(x)\right) \Bigg( \Gamma_0\left(x,\frac{\partial S}{\partial x}\right) (\varphi_0(x) + \hbar \varphi_1(x) ) + \hbar \Gamma_1\left(x,\frac{\partial S}{\partial x}\right) \varphi_0(x)  \\
  - i \hbar \left\langle \frac{\partial \Gamma_0}{\partial p}\left( x, \frac{\partial S}{\partial x} \right) , \frac{\partial \varphi_0}{\partial x} \right\rangle 
  - \frac{i \hbar}{2} \sum_{j,k} \frac{\partial^2 \Gamma_0}{\partial p_j \partial p_k} \left( x, \frac{\partial S}{\partial x} \right) \frac{\partial^2 S}{\partial x_j \partial x_k} \varphi_0 (x) + \mathcal{O}(\hbar^2) \Bigg) ,
\end{multline}
where $\Gamma_1(x,p)$ is called the subprincipal symbol.

With this is mind, let us look back at our expressions~(\ref{eq:n0}) and~(\ref{eq:n1}) for the leading and subleading order terms of the induced electron density $n(x,t)=n_0(x,t)+\hbar n_1(x,t)$. We immediately see that they have exactly the same structure as the right-hand sides of the commutation formulas~(\ref{eq:commutation-formula-leading-order}) and~(\ref{eq:commutation-formula-subleading-order}), provided that we introduce $\Pi_1(x,q)$, which takes the place of $\Gamma_1(x,q)$ above, as
\begin{equation} \label{eq:polarization-operator-subprincipal}
  \Pi_1(x,q) = -\frac{i}{2} \frac{\partial^2 \Pi_0}{\partial q \partial x}(x,q) ,
\end{equation}
and $\Pi_0(x,q)$ is defined by expression~(\ref{eq:def-polarization}).
It is curious and fairly unexpected that the expressions~(\ref{eq:n0}) and~(\ref{eq:n1}) have the same structure as the commutation formulas, as the procedure outlined in sections~\ref{subsec:derivation-density-matrix} and~\ref{subsec:derivation-charge-density} is far more complicated than letting a pseudodifferential operator act on the semiclassical Ansatz. However, now that we have obtained this result, it implies that we can introduce a pseudodifferential operator $\hat{\Pi}$, with symbol $\Pi(x,q,\hbar) = \Pi_0(x,q) + \hbar \Pi_1(x,q) + \mathcal{O}(\hbar^2)$ by reading equation~(\ref{eq:commutation-formula-subleading-order}) ``from the right to the left''. In other words, we can write the induced electron density as
\begin{equation}  \label{eq:charge-density-operator}
  n(x,t) = \hat{\Pi} \varphi(x) \exp\left(\frac{i}{\hbar} [S(x) - E t ]\right) = \hat{\Pi} V(x,t) ,
\end{equation}
where we have used the definition~(\ref{eq:Ansatz-WKB-potential}) of the semiclassical Ansatz.
This provides another interesting analogy with the homogeneous case, where the Fourier transform $\overline{n}(q)$ of the induced electron density satisfies the relation $\overline{n}(q) = \overline{\Pi}(q,E) \overline{V}(q)$. For the inhomogeneous case, this multiplicative relation has become an operator equation. 

Let us look a bit closer at the properties of the operator $\hat{\Pi}$. In appendix~\ref{subapp:review-pseudo-diff-operators}, we discussed that a necessary condition for a pseudodifferential operator to be self-adjoint is that it satisfies the constraints~(\ref{eq:constraints-symbol-self-adjoint}). The first constraint requires that $\Pi_0(x,q)$ is real, which is indeed the case for the examples we discuss in section~\ref{sec:examples}, see also the discussion in section~\ref{sec:investigation}. The second constraint is verified by virtue of expression~(\ref{eq:polarization-operator-subprincipal}). We therefore see that $\hat{\Pi}$ satisfies the necessary conditions to be a self-adjoint operator.

We wish to note that the similarity between our expressions and the commutation formulas is, strictly speaking, not sufficient to prove that we can introduce an operator $\hat{\Pi}$. To really prove this statement, one would have to show that all higher-order terms also have the correct structure, which would likely be very cumbersome. In particular, this means that the terms with higher-order derivatives of the action should appear in the charge density in the same way as they appear in the commutation formula, as these cannot be incorporated in the symbols $\Pi_n(x,q)$.
Likewise, the statement that the operator $\hat{\Pi}$ is self-adjoint also requires the consideration of all higher-order terms. Nevertheless, we consider our results a very strong indication that the operator $\hat{\Pi}$ truly exists. Either way, the results allow us to work with the commutation formula~(\ref{eq:commutation-formula-subleading-order}) as if the operator exists, since we have proven that the results coincide up to subleading order. This is why we could follow the standard semiclassical scheme in section~\ref{subsec:derivation-induced-potential} to obtain both the Hamilton-Jacobi equation and the transport equation, and to subsequently solve them.

In the same way, we can compare our results~(\ref{eq:comm-rel-h0}) and~(\ref{eq:comm-rel-h1}) with the commutation formulas~(\ref{eq:commutation-formula-leading-order}) and~(\ref{eq:commutation-formula-subleading-order}). As before, we see that their structure is exactly the same, provided that we define
\begin{equation} \label{eq:hamiltonian-operator-subprincipal}
  L_1(x,q) = -\frac{i}{2} \frac{\partial^2 L_0}{\partial q \partial x}(x,q) .
\end{equation}
This implies that we can introduce a second pseudodifferential operator $\hat{L}$, which has symbol $L(x,q,\hbar) = L_0(x,q) + \hbar L_1(x,q) + \mathcal{O}(\hbar^2)$, by reading equation~(\ref{eq:commutation-formula-subleading-order}) ``from the right to the left''. We can then rewrite equations~(\ref{eq:comm-rel-h0}) and~(\ref{eq:comm-rel-h1}) as an operator equation
\begin{equation}
  \label{eq:secular-eq-operator}
  \hat{L} \varphi(x) \exp\left(\frac{i}{\hbar} [S(x) - E t ]\right) = \hat{L} V(x,t) = 0 .
\end{equation}
The operator $\hat{L}$ is the effective Hamiltonian of the system.
In section~\ref{subsec:derivation-Poisson-SC}, we already noted the similarity between the leading-order result~(\ref{eq:comm-rel-h0}) and the secular equation for the homogeneous case, $\overline{\varepsilon}(q,E) \overline{V}(q) = 0$. 
Equation~(\ref{eq:secular-eq-operator}) generalizes this similarity, as it also incorporates the subleading order. We therefore have the same generalization as for the charge density: when going from the homogeneous case to the inhomogeneous case, the multiplicative relation in Fourier space becomes an operator equation. 

Similar to the operator~$\hat{\Pi}$, the operator $\hat{L}$ satisfies the necessary condition~(\ref{eq:constraints-symbol-self-adjoint}) for self-adjointness.
We remark that the constraint~(\ref{eq:constraints-symbol-self-adjoint}) by itself does not rule out that the subprincipal symbol $L_1(x,1)$ has a non-zero real part. However, in our system this real part is zero by virtue of expression~(\ref{eq:hamiltonian-operator-subprincipal}). If this real part were non-zero, it would lead to an additional term proportional to $(L_1(x,q) + (i/2) \sum_j \partial^2 L_0/\partial x_j \partial q_j) \varphi_0 = \text{Re} (L_1(x,q)) \varphi_0$ in equation~(\ref{eq:amplitude-eq-pre-final}). In turn, this gives rise to an additional phase factor in the amplitude: the geometric or Berry phase. Expression~(\ref{eq:hamiltonian-operator-subprincipal}) shows precisely that the induced potential $V(x,t)$ does not contain any geometric or Berry phase. For a more general treatment of this topic, we refer the interested reader to Refs.~\cite{Maslov81,Guillemin77,Reijnders18}.

In this appendix, we have thus shown that we can introduce two self-adjoint pseudodifferential operators, $\hat{\Pi}$ and $\hat{L}$, which represent the polarization and the effective Hamiltonian, respectively. With these operators, we established two operator equations~(\ref{eq:charge-density-operator}) and~(\ref{eq:secular-eq-operator}) for the induced electron density and the effective equation of motion. As we previously mentioned, we did not rigorously prove these operator equations, since we did not show that all higher-order terms also have the correct structure. Moreover, we based our derivation on the semiclassical Ansatz~(\ref{eq:Ansatz-WKB-potential}), which is only valid when the Jacobian $J(x)$ does not vanish. In other words, we have only made these equations plausible for those parts of the Lagrangian manifold that can be (locally) projected onto the configuration space, where the points are parametrized by the position vector $x$, cf. the discussion in section~\ref{subsec:simple-turning-point}. A complete proof of the operator equations would therefore also involve the points that do not have this property, such as the classical turning point discussed in section~\ref{subsec:simple-turning-point}. As we mention in the same section, a proof for these points involves a lengthy derivation, and will therefore be considered in a separate publication.

Nevertheless, the operator equations~(\ref{eq:charge-density-operator}) and~(\ref{eq:secular-eq-operator}) provide a deeper insight into the mathematical structure of the problem. These operator equations are the generalizations of the multiplicative relations that are obtained in the homogeneous case after the Fourier transform. It is important to note that the pseudodifferential operators that we introduced are single-particle operators, acting on the induced (local) potential $V(x,t)$. Nevertheless, these single-particle operators capture the electron-electron interaction, which is an many-body effect. This is due to the nature of the RPA, which is the only approximation in which you have a closed system of local equations for the single-particle density operator.

\section{Alternative derivation of the Maslov index for a simple turning point}
\label{app:derivation-Maslov-ind-alternative}

In section~\ref{subsec:simple-turning-point}, we showed that $q_1^2 \propto x_1 - x_{1,c}$ in the vicinity of a simple turning point. In this appendix, we use the method discussed in Refs.~\cite{Zwaan29,Heading62,Froeman65,Berry72,Reijnders13} to derive that the Maslov index $\mu(x_{1,0},x_{1,p})$ equals $-1$ for this turning point.
We only briefly summarize the main aspects of the method and refer the reader to the aforementioned references for more information.

In the vicinity of the turning point $x_{1,c}$, we have two solutions for $q_1$: one positive solution and one negative solution, see expression~(\ref{eq:L0-E-expansion-small-q-Taylor-fold}).
For a monotonically decreasing $E_P(x)$, shown in the example in figure~\ref{fig:simple-turning-point-phase-space}(a), we can write $q_1(x) = \pm\tfrac{3}{2} c_1 (x_1 - x_{1,c})^{1/2}$ in the vicinity of $x_{1,c}$, where $c_1 > 0$ is a positive constant. A similar expression can be obtained when $q_\parallel \neq 0$.
We can thus write the part $S_1(x_1)$ of the action~(\ref{eq:action-separated-cartesian}) that involves $x_1$ as
\begin{equation}
  S_1^\pm(x_1) = \pm \int_{x_{1,c}}^{x_1} \frac{3}{2} c_1 (X_1 - x_{1,c})^{1/2} \text{d} X_1 
  = \pm c_1 (x_1 - x_{1,c})^{3/2} ,
\end{equation}
where we have taken $x_{1,0}$ as $x_{1,c}$.
We therefore have two asymptotic solutions~(\ref{eq:solution-sc-potential-index}).
For $x_1 > x_{1,c}$ these correspond to right-moving $(q_1<0: S_1^-)$ and left-moving $(q_1 > 0: S_1^+)$ waves. For $x_1 < x_{1,c}$ the action becomes purely imaginary and we have exponentially increasing and decreasing waves.

\begin{figure}[tb]
  \begin{center}
    \includegraphics[width=0.6\textwidth]{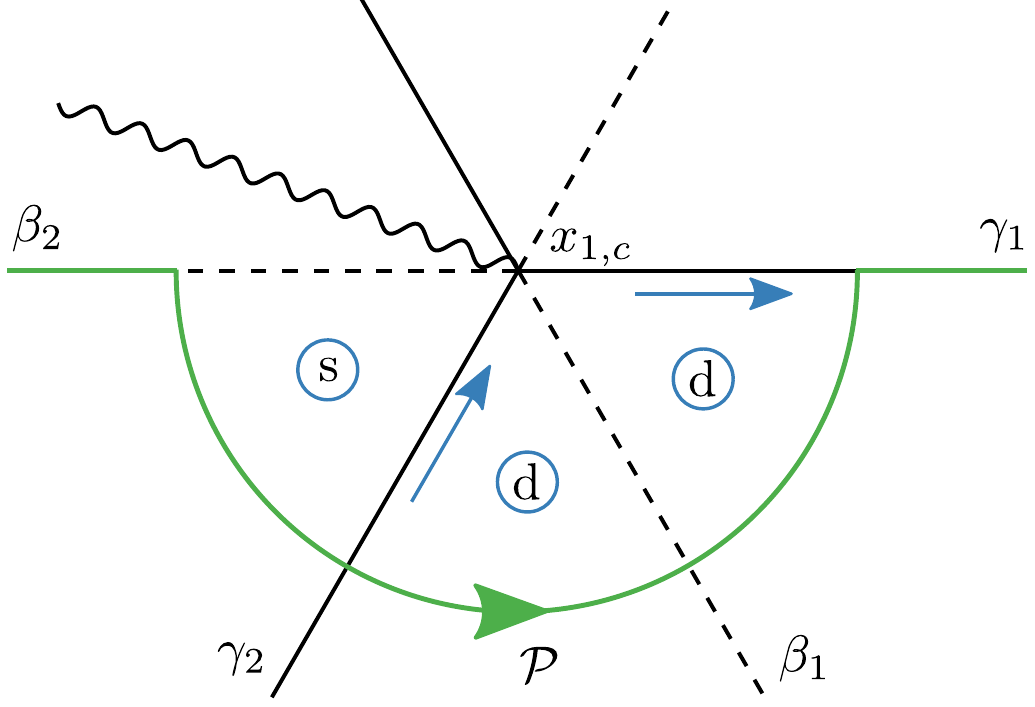}
  \end{center}
  \caption{Stokes diagram. The solid lines indicate the anti-Stokes lines $\gamma_i$, the dashed lines the Stokes lines $\beta_i$. The wavy line shows our choice of the branch cut. The path $\mathcal{P}$ is shown in green. The blue arrows on the anti-Stokes lines indicate the direction in which $S^+$ increases. The encircled letters show whether $V^+$ is dominant or subdominant in that region.}
  \label{fig:stokes-diagram}
\end{figure}

The key ingredient of Zwaan's approach is to consider the behavior of the two asymptotic solutions in the complex plane~\cite{Zwaan29,Heading62,Froeman65}. We therefore consider a complex coordinate $z_1$. Using the polar representation, $z_1 = x_{1,c} + r e^{i \phi}$, we have  $S_1^\pm(x_1) = c_1 r^{3/2} e^{3 i \phi/2}$. This shows that there are three so called anti-Stokes lines, shown as solid lines in figure~\ref{fig:stokes-diagram}, on which the action is purely real. Similarly, there are three so-called Stokes lines, shown as dashed lines in figure~\ref{fig:stokes-diagram}, on which the action is purely imaginary. We indicate our choice of the branch cut of the square root by the wavy line. Finally, the arrows on the anti-Stokes lines in figure~\ref{fig:stokes-diagram} indicate the direction in which $S^+$ increases. The goal of this section is to establish a connection between the exponentially decreasing and increasing waves for $x_1 < x_{1,c}$ and the traveling waves for $x_1 > x_{1,c}$. We do this by moving along a circular contour in the complex plane that stays sufficiently far away from the turning point.

As long as we stay sufficiently far away from the turning points, the semiclassical expression~(\ref{eq:solution-sc-potential}) is a good approximation to the true solution.
On the anti-Stokes lines the two asymptotic solutions
\begin{equation}  \label{eq:solution-asymptotic-app}
  V^\pm(z) = \frac{A_0^0}{\sqrt{|J(z)|}} \exp\left( \frac{i}{\hbar} [ S_1^\pm(z_1) + q_2 x_2 + q_3 x_3 - E t ] \right)
\end{equation}
have the same size, as the actions $S_1^{\pm}(z_1)$ are purely real and $|J(z)|$ is the same for both values of $q_1$. When we move away from an anti-Stokes line, the actions acquire a complex part and one of the functions $V^{\pm}$ becomes exponentially increasing, viewed from the direction of the turning point. We call this function the dominant term. Similarly, we call the function that becomes exponentially decreasing the subdominant term.
When we consider the anti-Stokes line $\gamma_1$, on which $S_1^+$ increases as we move away from the turning point. By the Cauchy-Riemann relations, $S_1^+$ acquires a negative imaginary part as we move away from $\gamma_1$ in the clockwise direction~\cite{Heading62,Reijnders13}, meaning that $V^+$ is dominant in the region between $\gamma_1$ and $\gamma_2$, see figure~\ref{fig:stokes-diagram}.

Let us now consider the transition between two anti-Stokes lines. For definiteness, let us say that on the Stokes line $\gamma_i$ the solution of the differential equation is accurately represented by the asymptotic solution
\begin{equation}  \label{eq:solution-asymptotic-expansion-app}
  V(z) = C_{\gamma_i}^+ V^+(z) + C_{\gamma_i}^- V^-(z) ,
\end{equation}
where $C_{\gamma_i}^\pm$ are two coefficients.
At this point, it is important to note that $V^\pm$ are asymptotic solutions that were derived up to terms of order $\hbar$: it is more accurate to write $V^\pm(z) \big(1 + \mathcal{O}(\hbar)\big)$. 
Let us consider what happens when we move from an anti-Stokes line $\gamma_i$ to a neighboring anti-Stokes line $\gamma_{i+1}$. In the region between the two anti-Stokes lines, the dominant term is the leading-order term of the asymptotic solution $V(z)$. Hence, its coefficient does not change when we move from $\gamma_i$ to $\gamma_{i+1}$~\cite{Heading62,Froeman65}. 
On the other hand, the subdominant term quickly becomes much smaller than the $\mathcal{O}(\hbar)$ correction to the dominant term as we move away from $\gamma_i$. We therefore lose information on the coefficient of the subdominant term due to the accuracy of the method~\cite{Heading62,Froeman65,Reijnders13}. This is the origin of the so-called connection problem~\cite{Heading62,Froeman65,Berry72,Reijnders13}. For simple turning points, like the one we consider here, the coefficient of the subdominant term can be reconstructed by noting that the solution should be uniquely defined when we make a full turn around the turning point.
After some calculations, which are explained in detail in Refs.~\cite{Heading62,Froeman65,Berry72,Reijnders13}, one finds that upon passing from $\gamma_{i}$ to $\gamma_{i+1}$, the coefficient $C_{\gamma_{i+1}}^s$ of the subdominant term is given by $C_{\gamma_{i}}^s + \alpha_S C_{\gamma_{i}}^d$, where $C_{\gamma_{i}}^d$ is the coefficient of the dominant term and $\alpha_S$ is known as the Stokes constant. It equals $-i$ when moving in the clockwise direction, and $i$ when moving in the counterclockwise direction~\cite{Heading62,Froeman65,Berry72,Reijnders13}.

After this brief explanation of the method, we return to the problem at hand. We already established that $V^+$ is dominant in the region between $\gamma_1$ and $\gamma_2$. With our choice of the branch cut, the action for $x < x_{1,c}$ is given by
\begin{equation}
  S_1^\pm(x_1) = \pm c_1 (x_1 - x_{1,c})^{3/2} = \pm e^{-3 i\pi/2} |x_1 - x_{1,c}|^{3/2} =  \pm i |x_1 - x_{1,c}|^{3/2} .
\end{equation}
This shows that $V^+$ is subdominant for $x < x_{1,c}$, i.e. on the Stokes line $\beta_2$.
We now consider the path $\mathcal{P}$ shown in figure~\ref{fig:stokes-diagram}. Because we require the wavefunction to be normalizable, we start with an exponentially decaying solution for $x_1 < x_{1,c}$, i.e. $C_{\beta_2}^+ = C_2$, $C_{\beta_2}^- = 0$. Because we only have a subdominant term, we can continue this solution to the anti-Stokes line $\gamma_2$ without loss of accuracy, i.e. $C_{\gamma_2}^+ = C_2$, $C_{\gamma_2}^- = 0$. The final step is passing from $\gamma_2$ to $\gamma_1$ using the Stokes constant, as we just explained. In matrix notation, we have
\begin{equation}
  \begin{pmatrix} C_{\gamma_1}^+ \\ C_{\gamma_1}^- \end{pmatrix}
  = \begin{pmatrix} 1 & 0 \\ i & 0 \end{pmatrix} \begin{pmatrix} C_{\gamma_2}^+ \\ C_{\gamma_2}^- \end{pmatrix}
  = \begin{pmatrix} 1 & 0 \\ i & 0 \end{pmatrix} \begin{pmatrix} C_2 \\ 0 \end{pmatrix}
  = \begin{pmatrix} C_2 \\ i C_2 \end{pmatrix} .
\end{equation}
On the anti-Stokes line $\gamma_i$, we have an incoming (left-moving) wave $V^+$ and a reflected (right-moving) wave $V^-$.
Comparing expressions~(\ref{eq:solution-asymptotic-app}), (\ref{eq:solution-sc-potential-index}) and~(\ref{eq:solution-asymptotic-expansion-app}), we see that the Maslov index $\mu(x_{1,0},x_{1,p})$ is related to the coefficients $C_{\gamma_1}^{\pm}$ by
\begin{equation}
  \exp\left( -\frac{i \pi}{2} \mu(r_{1,0},r_{1,p}) \right) = \frac{C_{\gamma_1}^-}{C_{\gamma_1}^+} = i .
\end{equation}
This confirms that $\mu(r_{1,0},r_{1,p}) = -1$ for the simple turning point discussed in section~\ref{subsec:simple-turning-point}.

\section{Analytic continuation of the dielectric function}
\label{app:analytic-cont}

In this appendix, we discuss the analytic continuation of the polarization $\overline{\Pi}(q,E)$ and the dielectric function $\overline{\varepsilon}(q,E)$ for a three-dimensional homogeneous quantum plasma. In appendix~\ref{subapp:analytic-cont-procedure}, we review the formalism explained in Refs.~\cite{Bonitz93,Hamann20}. Subsequently, we consider its application to the dielectric function at finite temperature in appendix~\ref{subapp:analytic-cont-finite-T}. We compare various analytic continuations and discuss which one leads to a well-defined limit as the temperature goes to zero. In appendix~\ref{subapp:analytic-cont-zero-T}, we explicitly consider the case of zero temperature. We show numerically that the analytic continuation gives rise to well-defined plasmons with complex energies when we enter the Landau damped region. In appendix~\ref{subapp:analytic-cont-proof-roots-zero-T}, we finally give an explicit proof that the analytic continuation of the dielectric function has roots in the complex plane.

\subsection{Analytic continuation using the spectral representation}
\label{subapp:analytic-cont-procedure}

Before we discuss the analytic continuation of the polarization, let us first consider it in slightly more detail.
As explained in section~\ref{subsec:damping-homogeneous-systems}, we consider the retarded polarization throughout the main text.
This retarded polarization vanishes for $t<t_0$, where $t_0$ is the time at which a stimulus is applied to the system. By the Titchmarsh theorem, the retarded polarization is analytic for $E$ in the upper half plane. On the real axis, one therefore often uses the notation $E+i\eta$, as we briefly discussed below equation~(\ref{eq:potential-time-dependence}).
In a similar way, one can define the advanced polarization, which vanishes for $t>t_0$. Retracing the steps in the proof of the Titchmarsh theorem, one naturally sees that this advanced polarization is analytic for $E$ in the lower half plane.
On the real axis, the retarded ($R$, plus sign) and advanced ($A$, minus sign) polarization are given by
\begin{equation}  \label{eq:homo-pol-RA}
  \overline{\Pi}^{R/A}(q,E) = \frac{g_s}{(2 \pi \hbar)^d} \int \frac{\rho_0(E_p) - \rho_0(E_{p+q})}{E_p - E_{p+q} + E \pm i \eta} \text{d}p ,
\end{equation}
where $\eta \to 0^+$ is small and positive.

We would like to rewrite these expressions in a way that facilitates the analytic continuation. Following the derivation in Refs.~\cite{Bonitz93,Hamann20}, we therefore define the spectral function $\widetilde{\Pi}(q,E)$ by
\begin{equation}  \label{eq:homo-spectral-def}
  \widetilde{\Pi}(q,E) = \frac{1}{i} \lim_{\eta \to 0} \left( \overline{\Pi}^{R}(q,E)-\overline{\Pi}^{A}(q,E) \right) .
\end{equation}
Note that the spectral function is proportional to the discontinuity in $\overline{\Pi}^{R/A}(q,E)$ when we cross the real axis.
With the help of the Sokhotski–Plemelj formula
\begin{equation}  \label{eq:sokhotsky-plemelj}
  \lim_{\eta \to 0^+} \frac{1}{x \pm i\eta} = \mathcal{P}\left(\frac{1}{x}\right) \mp i \pi \delta(x) , \; 
  \text{or, equivalently} \lim_{\eta \to 0^+} \int_{-\infty}^{\infty} \frac{1}{x \pm i\eta} \text{d} x = \mathcal{P} \int_{-\infty}^{\infty} \frac{\text{d}x}{x} \mp i \pi ,
\end{equation}
where $\mathcal{P}$ denotes the Cauchy principal value, we find
\begin{equation}  \label{eq:homo-spectral-res}
  \widetilde{\Pi}(q,E) = -2\pi \frac{g_s}{(2\pi \hbar)^3} \int \left(\rho_0(E_p) - \rho_0(E_{p+q})\right) \delta\left(E_p - E_{p+q} + E \right) \text{d}p .
\end{equation}
The spectral representation for the polarization then reads~\cite{Bonitz93,Hamann20}
\begin{equation}  \label{eq:homo-spectral-representation}
  \overline{\Pi}^{R/A}(q,z) = \int_{-\infty}^{\infty} \frac{\text{d} E}{2\pi} \frac{\widetilde{\Pi}(q,E)}{E - z},
\end{equation}
where we obtain $\overline{\Pi}^{R}$ for $z$ in the upper half plane, and $\overline{\Pi}^{A}$ for $z$ in the lower half plane.
This can be easily verified by inserting expression~(\ref{eq:homo-spectral-res}) for the spectral function, that is,
\begin{equation}
\begin{aligned}
  \overline{\Pi}^{R/A}(q,z) 
    &= \frac{g_s}{(2\pi \hbar)^3} \int_{-\infty}^{\infty} \frac{\text{d} E}{2\pi} \int \text{d}p (-2\pi) \frac{\rho_0(E_p) - \rho_0(E_{p+q})}{E - z} \delta\left(E_p - E_{p+q} + E \right) \\
    &= \frac{g_s}{(2\pi \hbar)^3} \int \text{d}p \frac{\rho_0(E_p) - \rho_0(E_{p+q})}{E_p - E_{p+q} + z} ,
\end{aligned}
\end{equation}
where we performed the integration over $E$ with the help of the delta function.

Expression~(\ref{eq:homo-spectral-representation}) allows us to perform the analytic continuation. When $z$ lies in the upper half plane, the integrand has a pole at $E=z$. When we decrease the imaginary part of $z$, the location of the pole shifts towards the real axis, until it eventually crosses the real axis. We can then perform the analytic continuation to the lower half plane in the same way as Landau did for the classical plasma~\cite{Landau46}: we deform the integration contour in such a way that it always lies below the pole. This leads to an additional contribution to the polarization, which equals the residue at the pole. In other words, the analytic continuation~$\check{\Pi}(q,E)$ reads~\cite{Bonitz93,Hamann20}
\begin{equation}  \label{eq:homo-polarization-analytical-continuation}
  \check{\Pi}^{R}(q,z) =
  \begin{cases}
    \overline{\Pi}^R(q,z) , & \text{Im} \, z > 0 \\
    \overline{\Pi}^A(q,z) + 2\pi i \frac{\widetilde{\Pi}(q,z)}{2\pi} = \overline{\Pi}^A(q,z) + i \widetilde{\Pi}(q,z), & \text{Im} \, z < 0
  \end{cases} .
\end{equation}
This expression is continuous across the real $z$-axis, since
\begin{equation}
  \lim_{\eta \to 0} \left( \check{\Pi}^{R}(q,E+i\eta) - \check{\Pi}^{R}(q,E-i\eta) \right) 
    = \lim_{\eta \to 0} \left( \overline{\Pi}^R(q,E+i\eta) - \overline{\Pi}^A(q,E-i\eta) - i \widetilde{\Pi}(q,E-i\eta) \right) ,
\end{equation}
which vanishes due to the definition~(\ref{eq:homo-spectral-def}) of the spectral function.
It is important to note that the result~(\ref{eq:homo-polarization-analytical-continuation}) actually contains a second analytic continuation. In expression~(\ref{eq:homo-spectral-res}), we defined the spectral function for real values of the energy $E$. However, in the result~(\ref{eq:homo-polarization-analytical-continuation}) we use it with a complex argument $z$, which means that we have to analytically continue the spectral function in the complex plane. We come back to this point in the next subsections.

We finally obtain the analytic continuation $\check{\varepsilon}^R(q,z)$ of the retarded dielectric function with the help of expression~(\ref{eq:dielectric-function-homo}), that is,
\begin{equation}  \label{eq:homo-epsilon-analytical-continuation}
  \check{\varepsilon}^R(q,z) = \varepsilon_b - \frac{4 \pi e^2 \hbar^2}{q^2} \check{\Pi}^R(q, z) .
\end{equation}
We analyze this function in detail in the following subsections. To simplify the notation, we will drop the superscript $R$ in the remainder of this appendix, and simply write $\check{\varepsilon}(q,z)$. Furthermore, we use $\varepsilon_b=1$ in all numerical calculations, but this is not a fundamental limitation.

\subsection{Calculations for finite temperature}
\label{subapp:analytic-cont-finite-T}

Before we consider the analytic continuation $\check{\varepsilon}(q,z)$ for the zero temperature case, which we discuss in the main text, we first consider what happens at finite temperature. The latter case was also discussed in Ref.~\cite{Hamann20}, where a specific analytic continuation was considered. In this subappendix, we show that the analytic continuation $\check{\varepsilon}(q,z)$ is not unique, and present a physical argument that helps to choose the most suitable analytic continuation.

Let us calculate the spectral function $\widetilde{\Pi}(q,E)$ on the real axis before we analytically continue it into the complex plane. When we split the integral in expression~(\ref{eq:homo-spectral-def}) into two separate integrals, and change variables $p+q\to p$ in the second integral, we find
\begin{equation}
  \widetilde{\Pi}(q,E) = -2\pi \frac{g_s}{(2\pi \hbar)^3} \left( 
      \int \rho_0(E_p) \delta\left(E_p - E_{p+q} + E \right) \text{d}p 
      - \int \rho_0(E_{p}) \delta\left(E_{p-q} - E_{p} + E \right) \text{d}p \right) .
\end{equation}
Since $E_p$ only depends on $|p|$, we introduce polar coordinates, aligning the $z$-axis of the coordinate system with the vector $q$. Setting $E_p=p^2/(2m)$, performing the integral over the polar angle $\phi$ and introducing a new variable $\varrho=|p|/|p_F|$, we find
\begin{multline}
  \widetilde{\Pi}(q,E) = -(2\pi)^2 p_F^3 \frac{g_s}{(2\pi \hbar)^3} \bigg( 
    \int_0^\infty \text{d}\varrho \, \varrho^2 \int_0^\pi \text{d}\theta \sin\theta \, \rho_0\left( \frac{(\varrho p_F)^2}{2m} \right) \delta\left(\frac{|q|p_F}{m} \left( \nu_- - \varrho\cos\theta \right) \right) \\
    - \int_0^\infty \text{d}\varrho \, \varrho^2 \int_0^\pi \text{d}\theta \sin\theta \, \rho_0\left( \frac{(\varrho p_F)^2}{2m} \right) \delta\left(\frac{|q|p_F}{m} \left( \nu_+ - \varrho\cos\theta \right) \bigg)
  \right) ,
\end{multline}
where we used our previous definition~(\ref{eq:nu-pm}) for $\nu_\pm$. The next step is to pull the factor $|q|p_F/m$ out of the delta function. Inside the delta function, we then have a function $f(\theta) = \nu_\pm - \varrho \cos\theta$. Its roots are given by the values $\theta_0^\pm$ that satisfy $f(\theta_0^\pm)=0$. On the interval $(0,\pi)$, $\cos\theta$ takes all values between -1 and 1 exactly once. Therefore, $f(\theta)$ only has roots whenever $-\varrho < \nu_\pm < \varrho$, i.e. when $\varrho^2 - \nu_\pm^2 > 0$.
Moreover, the absolute value of the derivative equals $|f'(\theta_0^\pm)| = \varrho \sin\theta_0^\pm$, since $\sin\theta$ only takes positive values on $(0,\pi)$. We thus obtain
\begin{multline}
  \widetilde{\Pi}(q,E) = -(2\pi)^2 p_F^3 \frac{m}{|q|p_F} \frac{g_s}{(2\pi \hbar)^3} \bigg( 
  \int_0^\infty \text{d}\varrho \, \varrho^2 \int_0^\pi \text{d}\theta \rho_0\left( \frac{(\varrho p_F)^2}{2m} \right) \frac{\sin\theta}{\varrho\sin\theta_0^-} \Theta(\varrho^2-\nu_-^2) \delta(\theta-\theta_0^-) \\
  - \int_0^\infty \text{d}\varrho \, \varrho^2 \int_0^\pi \text{d}\theta \rho_0\left( \frac{(\varrho p_F)^2}{2m} \right) \frac{\sin\theta}{\varrho\sin\theta_0^+} \Theta(\varrho^2-\nu_+^2) \delta(\theta-\theta_0^+) ,
  \bigg) .
\end{multline}
where $\Theta(x)$ is the Heaviside step function.
At this point we can easily perform the integral over $\theta$. Introducing the new variable $\chi=\varrho^2$, we find
\begin{equation}  \label{eq:homo-spectral-pre-int}
  \widetilde{\Pi}(q,E) = - \frac{g_s}{(2\pi \hbar)^3} \frac{2 \pi^2 m p_F^2}{|q|} \bigg( 
  \int_0^\infty \text{d}\chi \, \rho_0\left( \frac{\chi p_F^2}{2m} \right) \Theta(\chi -\nu_-^2) 
  - \int_0^\infty \text{d}\chi \, \rho_0\left( \frac{\chi p_F^2}{2m} \right) \Theta(\chi-\nu_+^2) 
  \bigg) .
\end{equation}
The Fermi-Dirac distribution is given by expression~(\ref{eq:Fermi-Dirac-dist}). Noting that $\mu = p_F^2/(2m)$ and setting $\tau=2 m k_B T/p_F^2$, we obtain
\begin{equation}
  \rho_0\left( \frac{\chi p_F^2}{2m} \right) = \frac{1}{1+\exp\left(\frac{1}{k_B T}\left(\frac{\chi p_F^2}{2 m}-\mu\right)\right)} = \frac{1}{1+\exp((\chi-1)/\tau)} .
\end{equation}
The step functions now turn the lower limits of the integrals into $\nu_\pm^2$, and we have
\begin{equation}  \label{eq:homo-spectral-int}
  \widetilde{\Pi}(q,E) = - \frac{g_s}{(2\pi \hbar)^3} \frac{2 \pi^2 m p_F^2}{|q|} \bigg( 
  \int_{\nu_-^2}^\infty \frac{\text{d}\chi}{1+\exp((\chi-1)/\tau)} 
  - \int_{\nu_+^2}^\infty \frac{\text{d}\chi}{1+\exp((\chi-1)/\tau)}
  \bigg) .
\end{equation}
Since we calculate the spectral function for real energies $E$, the quantities $\nu_\pm$ are real.

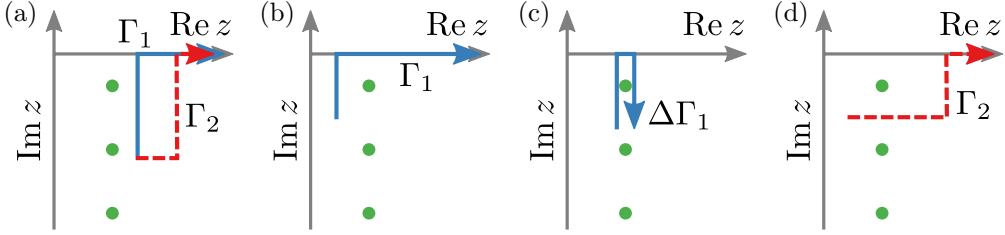
\begin{figure}[tb]
  \hfill
  \begin{tikzpicture}
    \node at (0cm,0cm)
    {\includegraphics[width=0.9\textwidth]{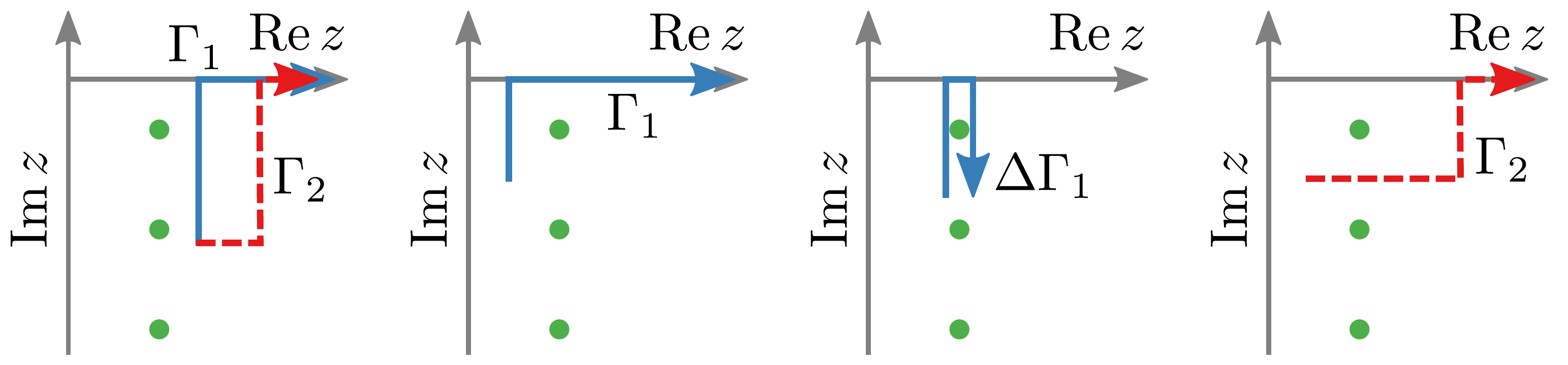}};
    \node at (-6.48cm,1.4cm) {(a)};
    \node at (-3.1cm,1.4cm) {(b)};
    \node at (0.28cm,1.4cm) {(c)};
    \node at (3.66cm,1.4cm) {(d)};
  \end{tikzpicture}
  \hfill\hfill
  \caption{Different integration paths. (a) For $\text{Re}(\nu_\pm^2) > 1$, the paths $\Gamma_1$ and $\Gamma_2$ lead to the same result. (b) The path $\Gamma_1$ for $\text{Re}(\nu_\pm^2) < 1$. (c) Path $\Delta\Gamma_1$ for the difference between two points left and right of the lines $\text{Re}(\nu_\pm^2) = 1$. (d) The path $\Gamma_2$ for $\text{Re}(\nu_\pm^2) < 1$.
  }
  \label{fig:poles-complex-plane}
\end{figure}

Expression~(\ref{eq:homo-spectral-int}) gives us important information about the analytic continuation. We therefore first consider it in more detail before we compute it. Both integrands in expression~(\ref{eq:homo-spectral-int}) have poles at $\chi=1+(2n+1)\tau \pi i$, where $n$ is an integer. These poles arise due to the Fermi function and are shown in figure~\ref{fig:poles-complex-plane}. When we consider complex values $z$ for the energy, and therefore complex $\nu_\pm$, these poles lead to branch cuts for the function~$\widetilde{\Pi}(q,z)$. However, we have a certain freedom in the choice of the branch cuts, which comes from the freedom to choose the path of integration in the analytic continuation of expression~(\ref{eq:homo-spectral-int}). Let us consider this argument in more detail. First of all, when $\text{Re}(\nu_\pm^2) > 1$ or when $\text{Re}(\nu_\pm^2) \leq 1$ and $-\tau\pi < \text{Im}(\nu_\pm^2) < 0$, we can choose both integration paths shown in figure~\ref{fig:poles-complex-plane}(a). Since both paths $\Gamma_1$ and $\Gamma_2$ go around the poles in the same way, they give the same result for the analytic continuation~$\widetilde{\Pi}(q,z)$.

On the other hand, when $\text{Re}(\nu_\pm^2) \leq 1$ and $-3\tau\pi < \text{Im}(\nu_\pm^2) < -\tau\pi$, the choice of the integration path becomes more important. When we use the path $\Gamma_1$ shown in figure~\ref{fig:poles-complex-plane}(b), the resulting analytic continuation $\widetilde{\Pi}_{\Gamma_1}(q,z)$ has a branch cut along the line $\text{Re}(\nu_\pm^2) = 1$. This can be seen by evaluating $\widetilde{\Pi}_{\Gamma_1}(q,z)$ at a point just to the left of this line and at a point just to the right of this line. The difference between the two function values can be written as an integral along the contour $\Delta\Gamma_1$, shown in figure~\ref{fig:poles-complex-plane}(c). Since this contour wraps around the pole at $\chi = 1 - \tau\pi i$, and possibly also others, this difference is nonzero and we have a branch cut. On the other hand, we can choose the path $\Gamma_2$ shown in figure~\ref{fig:poles-complex-plane}(d). The resulting analytic continuation $\widetilde{\Pi}_{\Gamma_2}(q,z)$ has branch cuts along the lines $\text{Im}(\nu_\pm^2) = -(2n+1)\pi\tau$. Other choices for the integration path lead to more complicated placements of the branch cuts.

\begin{figure}[tb]
  \hfill
  \includegraphics[width=0.45\textwidth]{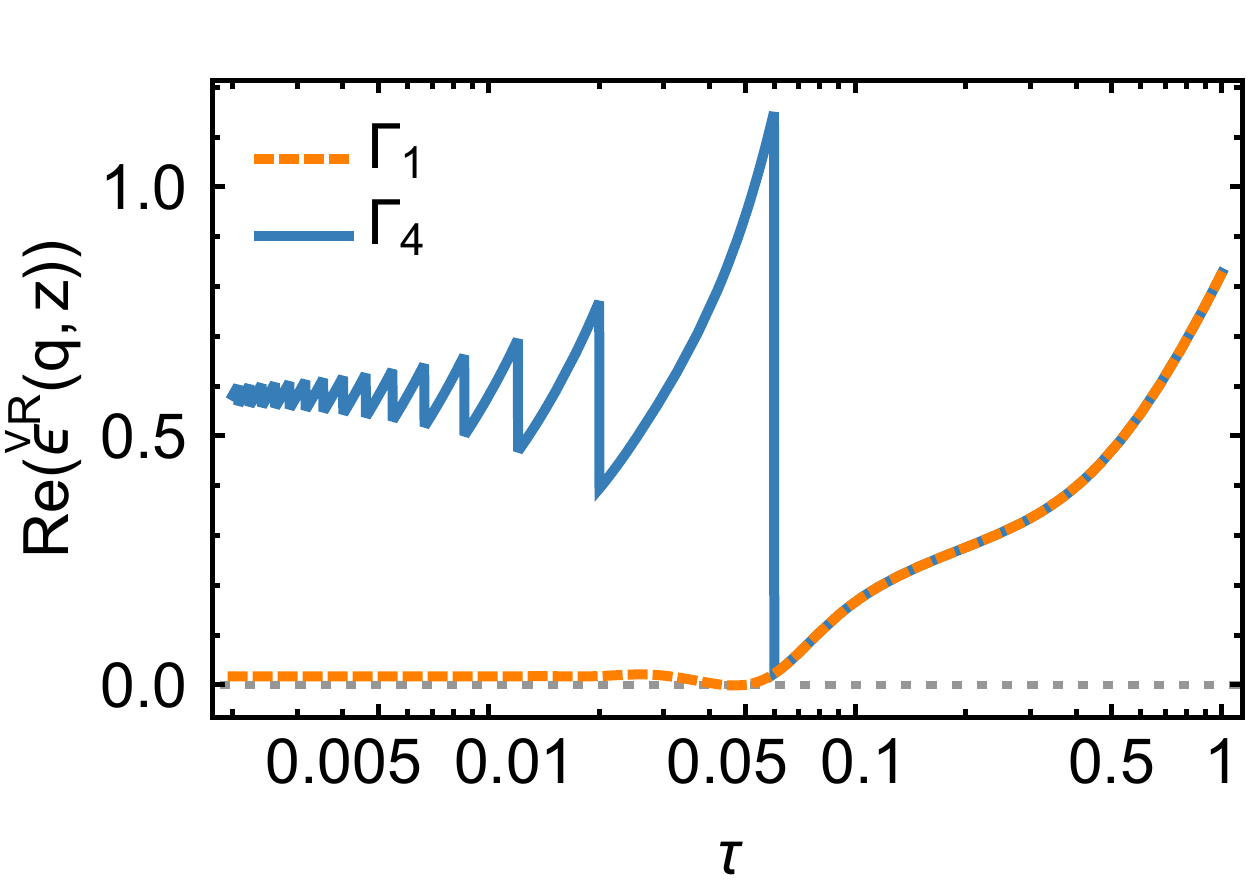}
  \hfill\hfill
  \caption{The real part of the analytic continuations $\check{\varepsilon}_{\Gamma_1}(q,z)$ (orange dashed line) and $\check{\varepsilon}_{\Gamma_4}(q,z)$ (solid blue line) as a function of $\tau$ for $n^{(0)}=1/a_0^3$ and $q=0.6 p_F$. The value of $z$ is chosen such that $\check{\varepsilon}_{\Gamma_1}(q,z)=0$ at $\tau=0.05$. The discontinuities in $\check{\varepsilon}_{\Gamma_4}$ arise from the branch cuts.
    \label{fig:epsilon-limit-T-zero}
  }
\end{figure}

Is there a physical argument that could help us to make a choice between the different integration paths when $\text{Re}(\nu_\pm^2) \leq 1$ and $-3\tau\pi < \text{Im}(\nu_\pm^2) < -\tau\pi$? The argument that we would like to present here is that $\widetilde{\Pi}(q,z)$ should be well-behaved when we consider the limit $T \to 0$. Let us evaluate the path $\Gamma_2$ in light of this argument. In the limit $T\to 0$, the spacing between the poles in figure~\ref{fig:poles-complex-plane} goes to zero. As we said before, the integration path $\Gamma_2$ gives rise to a function $\widetilde{\Pi}_{\Gamma_2}(q,z)$ with branch cuts along the lines $\text{Im}(\nu_\pm^2) = -(2n+1)\pi\tau$. When we fix $q$ and $z$ and decrease the temperature, the branch cuts pass through the point $z$, leading to a function $\widetilde{\Pi}_{\Gamma_2}(q,z)$ with many discontinuities. We show this explicitly for a slightly different path in figure~\ref{fig:epsilon-limit-T-zero}. Moreover, in the limit $T=0$, it is not clear how to choose an integration path $\Gamma_2$ that goes between two poles, as the poles lie on a continuous line.
On the other hand, the path $\widetilde{\Pi}_{\Gamma_1}(q,z)$ leads to two branch cuts along the lines $\text{Re}(\nu_\pm^2) = 1$, regardless of the temperature $T$. This analytic continuation therefore does not exhibit the previously described discontinuities when we consider the limit $T\to 0$, see also figure~\ref{fig:epsilon-limit-T-zero}.
The choice $\Gamma_1$ therefore seems more sensible on physical grounds. We come back to this shortly.

On the other hand, we may also take the approach considered in Ref.~\cite{Hamann20}, where the authors computed the integral~(\ref{eq:homo-spectral-int}) explicitly on the real axis, and subsequently considered its analytic continuation.
Performing the integration, we obtain
\begin{align}  \label{eq:homo-spectral-int-calculated-v1}
  \widetilde{\Pi}_{\Gamma_3}(q,E) &= - \frac{g_s}{(2\pi \hbar)^3} \frac{2 \pi^2 m p_F^2}{|q|} \tau \left( 
  \log( 1+ \exp[ -(\nu_-^2-1)/\tau ] )
  - \log( 1+ \exp[ -(\nu_+^2-1)/\tau ] )
  \right) , \\
  \widetilde{\Pi}_{\Gamma_4}(q,E) &= - \frac{g_s}{(2\pi \hbar)^3} \frac{2 \pi^2 m p_F^2}{|q|} \tau 
  \log\left( \frac{1+ \exp( -(\nu_-^2-1)/\tau )}{1+ \exp( -(\nu_+^2-1)/\tau )} \right) .
  \label{eq:homo-spectral-int-calculated}
\end{align}
Expression~(\ref{eq:homo-spectral-int-calculated}) was found in Ref.~\cite{Hamann20}, where its analytic continuation into the complex plane was studied using the principal branch of the logarithm. 
It is important to note that although expressions~(\ref{eq:homo-spectral-int-calculated-v1}) and~(\ref{eq:homo-spectral-int-calculated}) are equivalent on the real axis, they lead to a different placement of the branch cuts once they are viewed as functions of a complex variable $z$.
In fact, the branch cuts of $\widetilde{\Pi}_{\Gamma_3}(q,E)$ coincide with those of $\widetilde{\Pi}_{\Gamma_2}(q,z)$ and lie along the lines $\text{Im}(\nu_\pm^2) = -(2n+1)\pi\tau$, see also figure~\ref{fig:poles-complex-plane}(d). Numerically evaluating both functions, we find that they coincide. This can also be seen analytically be noting that, by Cauchy's theorem, we can always deform the integration path in the complex plane as long as we do not encircle any poles.
Although it is not that straightforward to relate expression~(\ref{eq:homo-spectral-int-calculated}) to an integration path in the complex plane, we use the label $\Gamma_4$ to denote it.

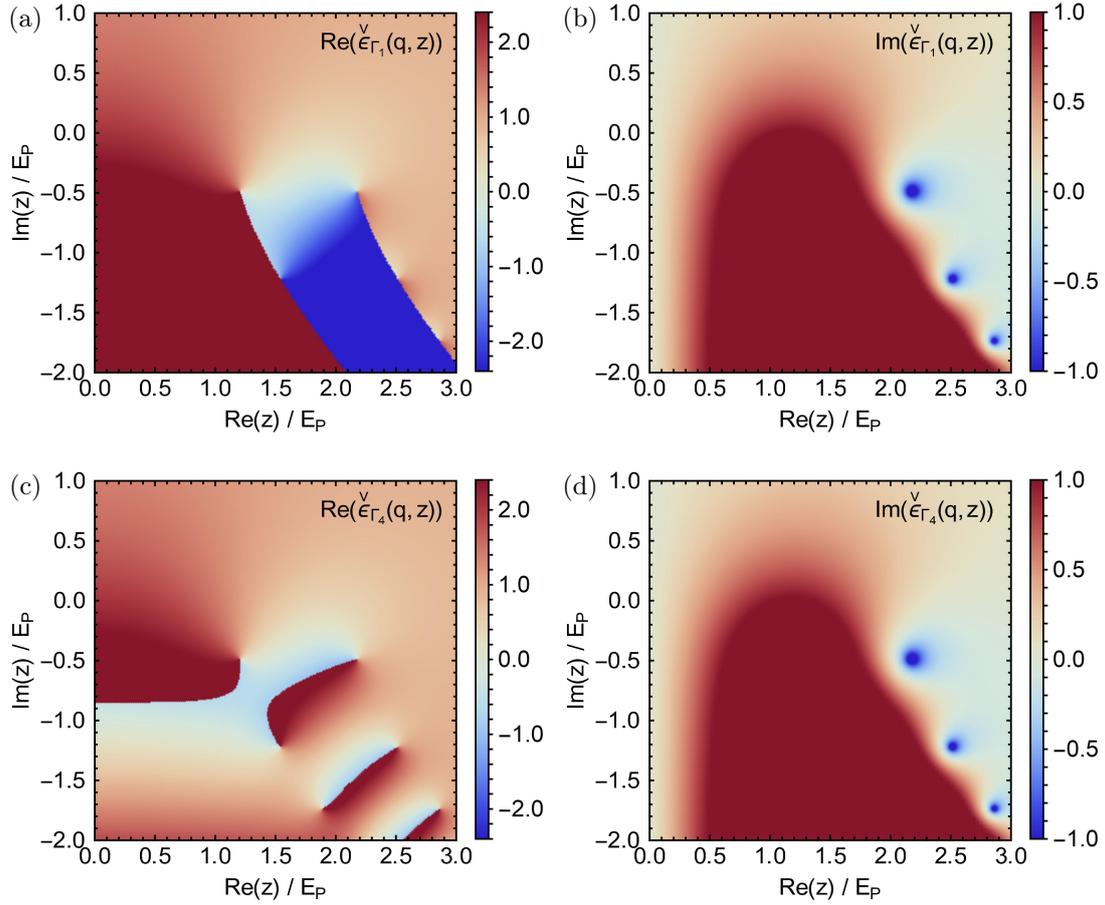
\begin{figure}[!p]
  \hfill
  \begin{tikzpicture}
    \node at (0cm,0cm)
    {\includegraphics*[width=0.48\textwidth]{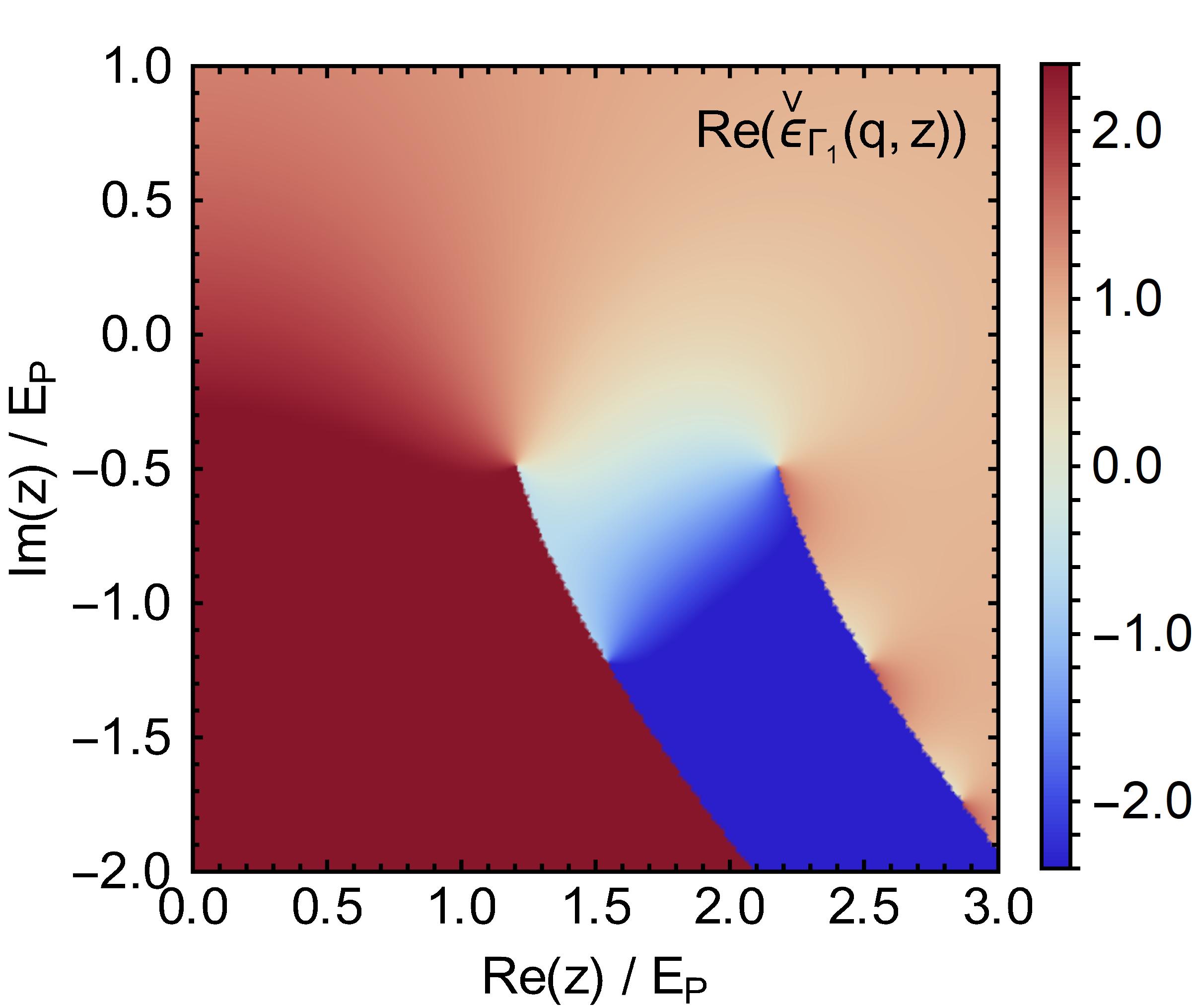}};
    \node at (-3.3cm,2.5cm) {(a)};
    
    \node at (7.3cm,0cm)
    {\includegraphics[width=0.48\textwidth]{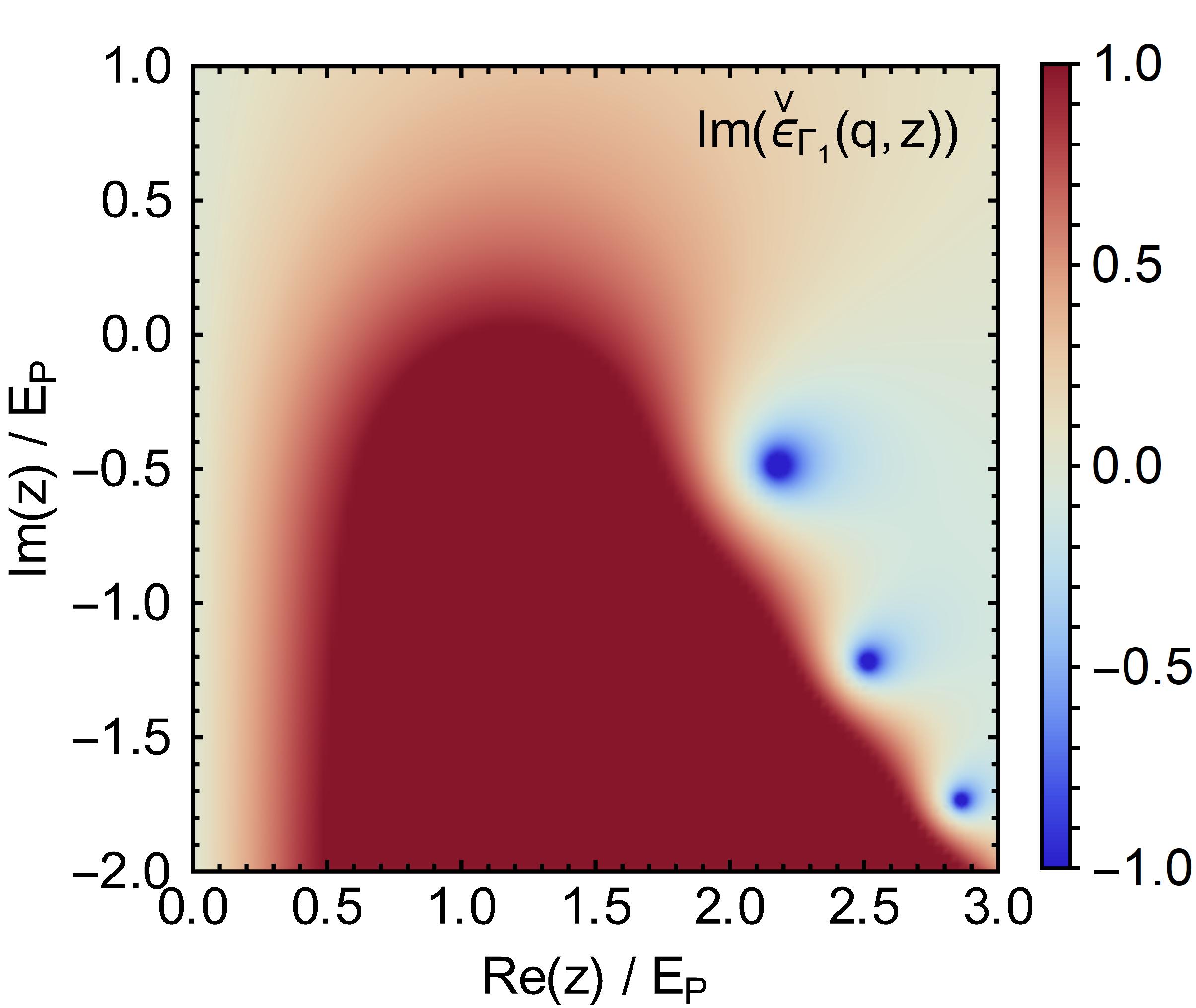}}; 
    \node at (4.0cm,2.5cm) {(b)};
    
    \node at (0cm,-6.2cm)
    {\includegraphics*[width=0.48\textwidth]{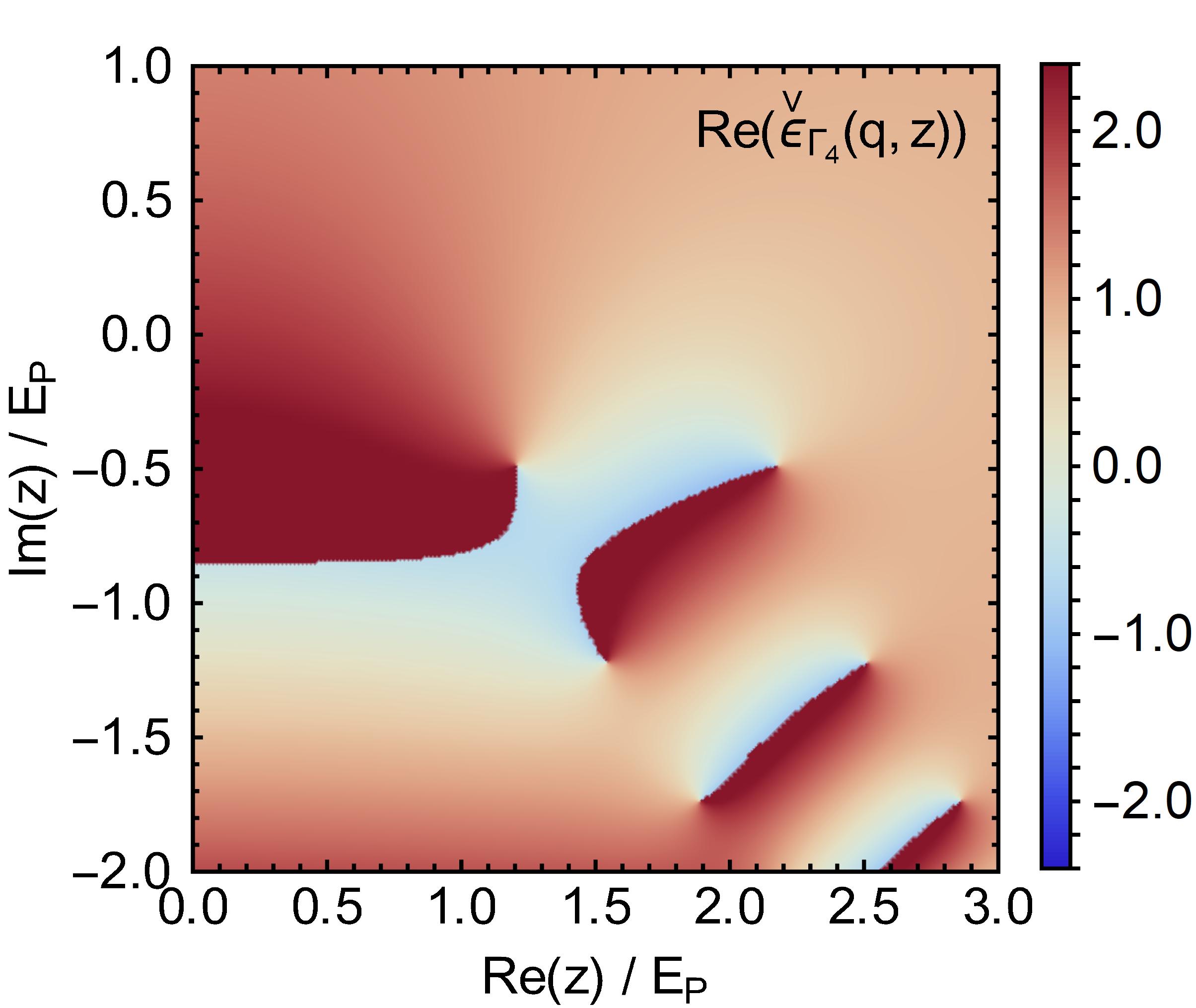}};
    \node at (-3.3cm,-3.7cm) {(c)};
    
    \node at (7.3cm,-6.2cm)
    {\includegraphics[width=0.48\textwidth]{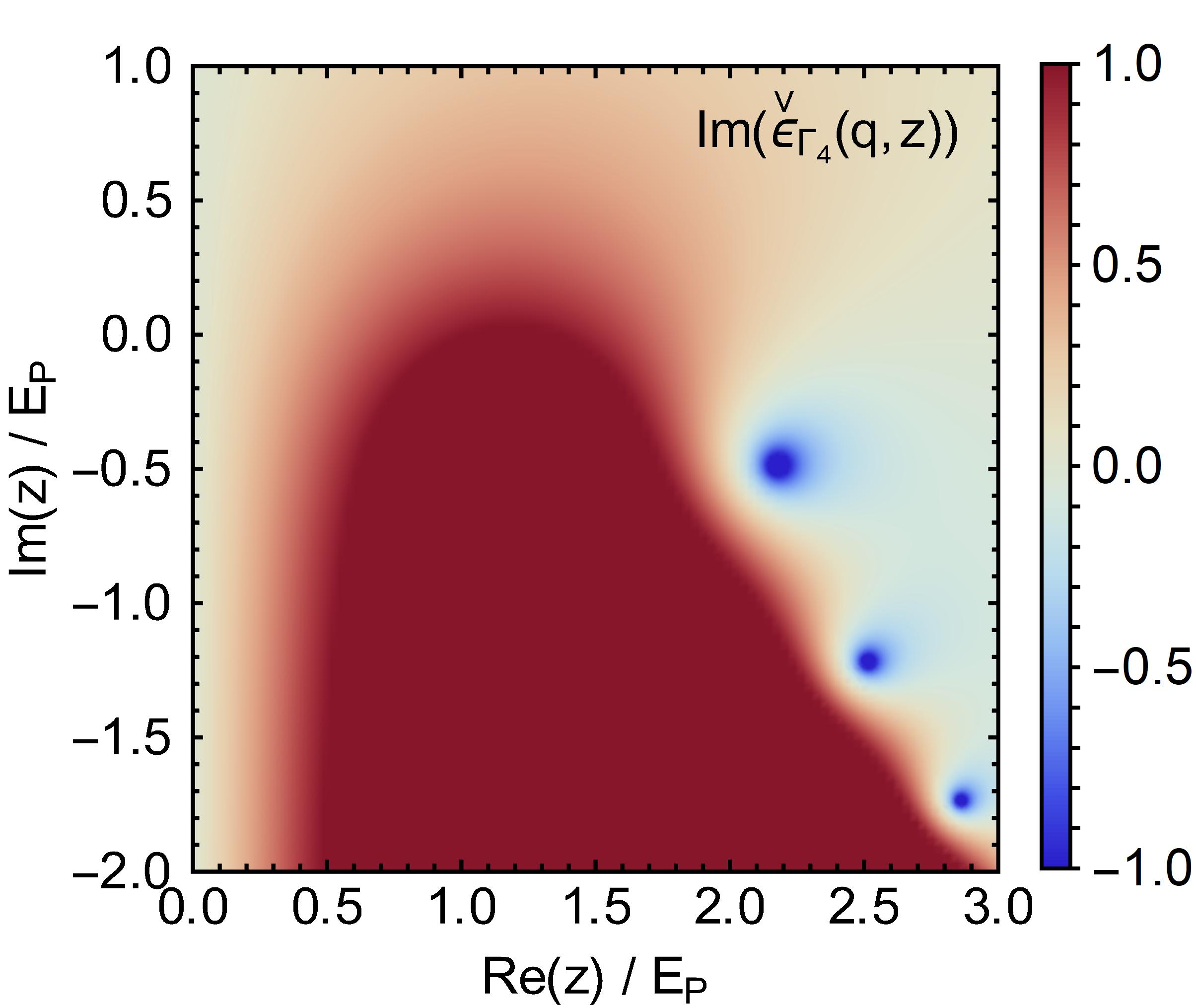}}; 
    \node at (4.0cm,-3.7cm) {(d)};
  \end{tikzpicture}
  \hfill\hfill
  \caption{The analytic continuation $\check{\varepsilon}(q,z)$ in the complex plane for $n^{(0)}=1/a_0^3$, $q=0.6 p_F$ and $\tau=0.2$.
    (a) and (b): real and imaginary parts, respectively, of $\check{\varepsilon}_{\Gamma_1}(q,z)$. The branch cuts in the real part lie along the lines $\text{Re}(\nu_\pm^2) = 1$. (c) and (d): real and imaginary parts, respectively, of $\check{\varepsilon}_{\Gamma_4}(q,z)$. The branch cuts in the real part have a complicated placement.
  }
  \label{fig:density-epsilon-comparison-T}
\end{figure}

\begin{figure}[!p]
  \hfill
  \begin{tikzpicture}
    \node at (0cm,0cm)
    {\includegraphics*[width=0.45\textwidth]{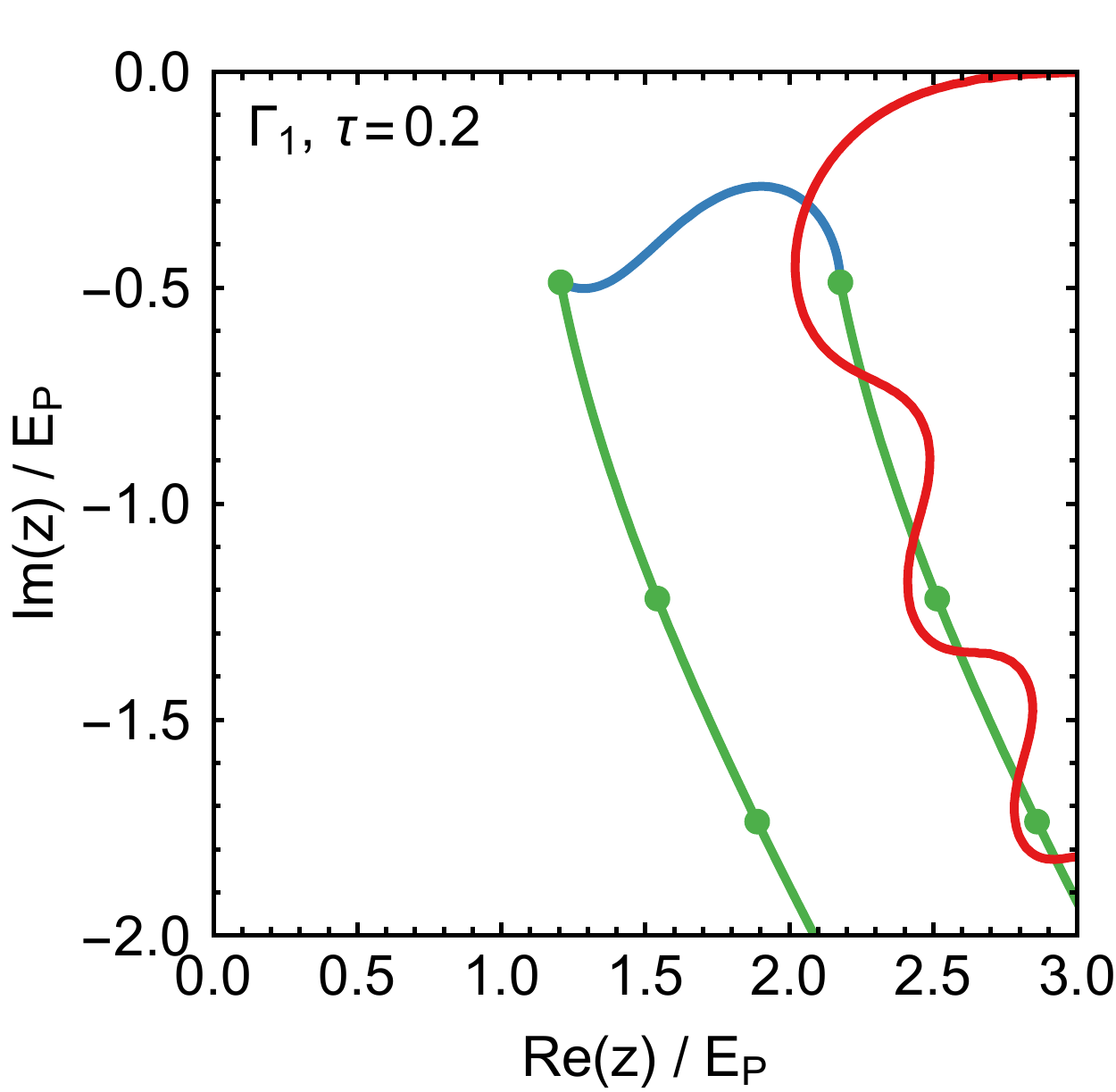}};
    \node at (-3.2cm,2.7cm) {(a)};
    
    \node at (7.3cm,0cm)
    {\includegraphics[width=0.45\textwidth]{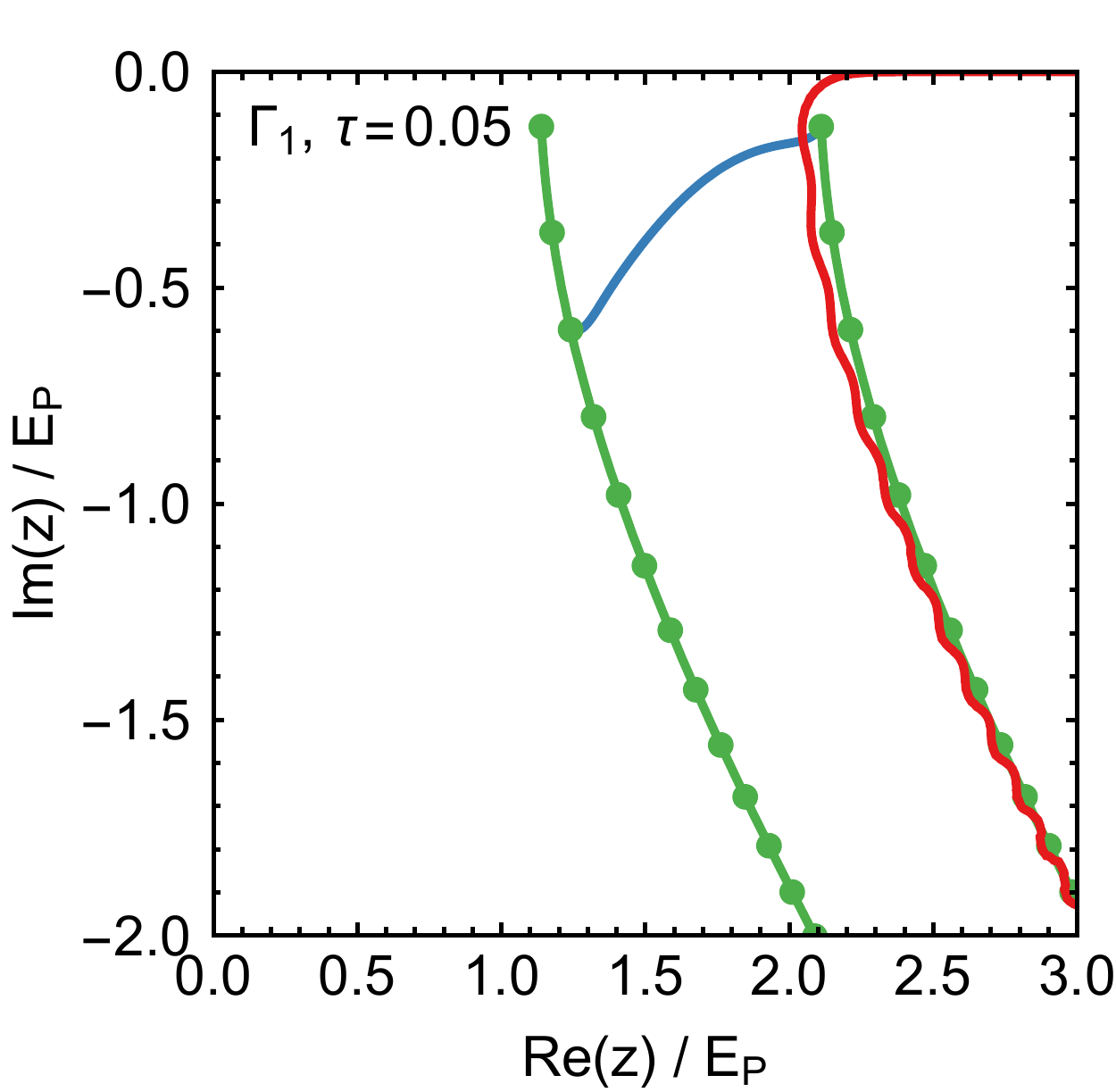}}; 
    \node at (4.1cm,2.7cm) {(b)};
    
    \node at (0cm,-6.5cm)
    {\includegraphics*[width=0.45\textwidth]{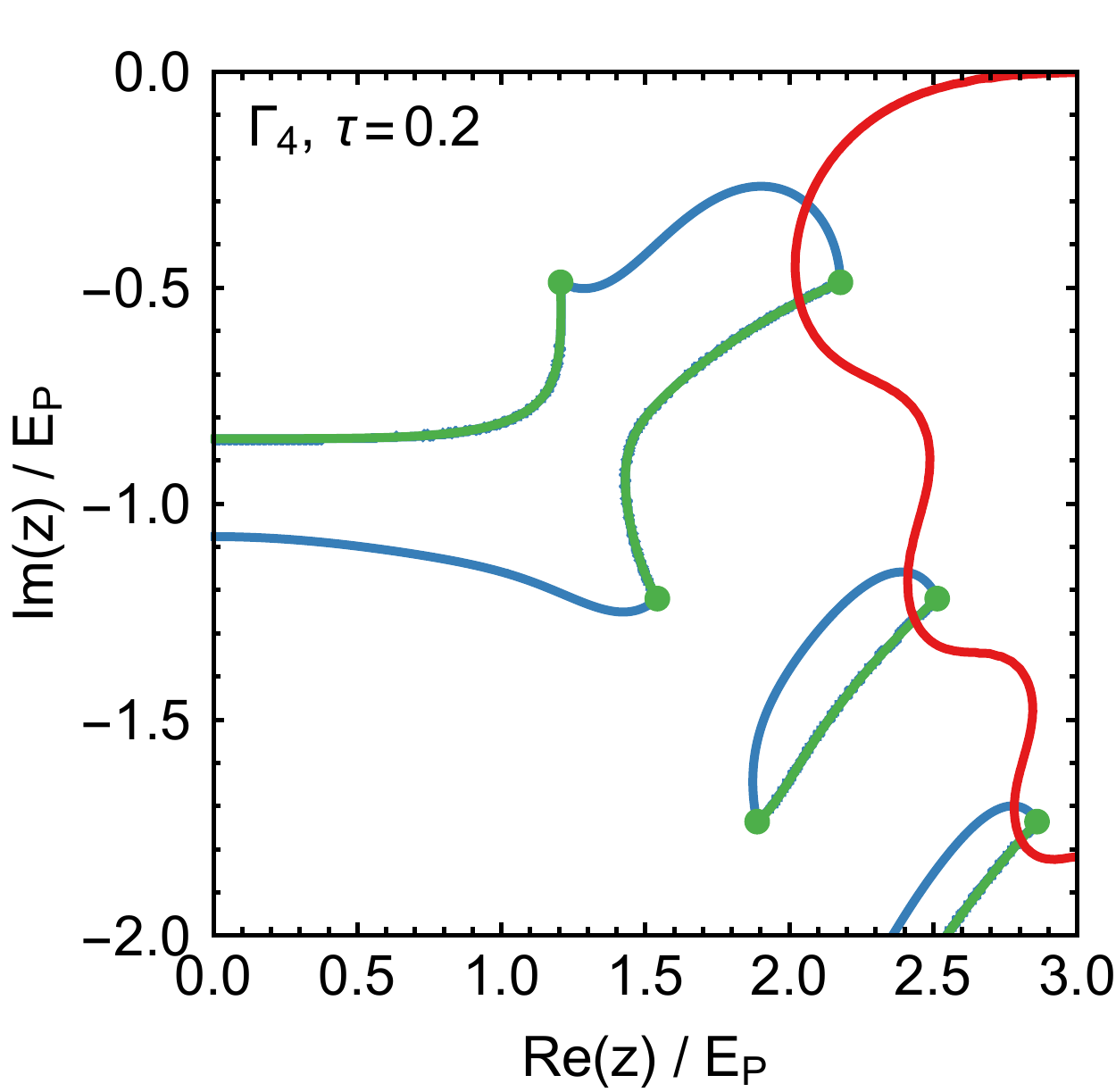}};
    \node at (-3.2cm,-3.8cm) {(c)};
    
    \node at (7.3cm,-6.5cm)
    {\includegraphics[width=0.45\textwidth]{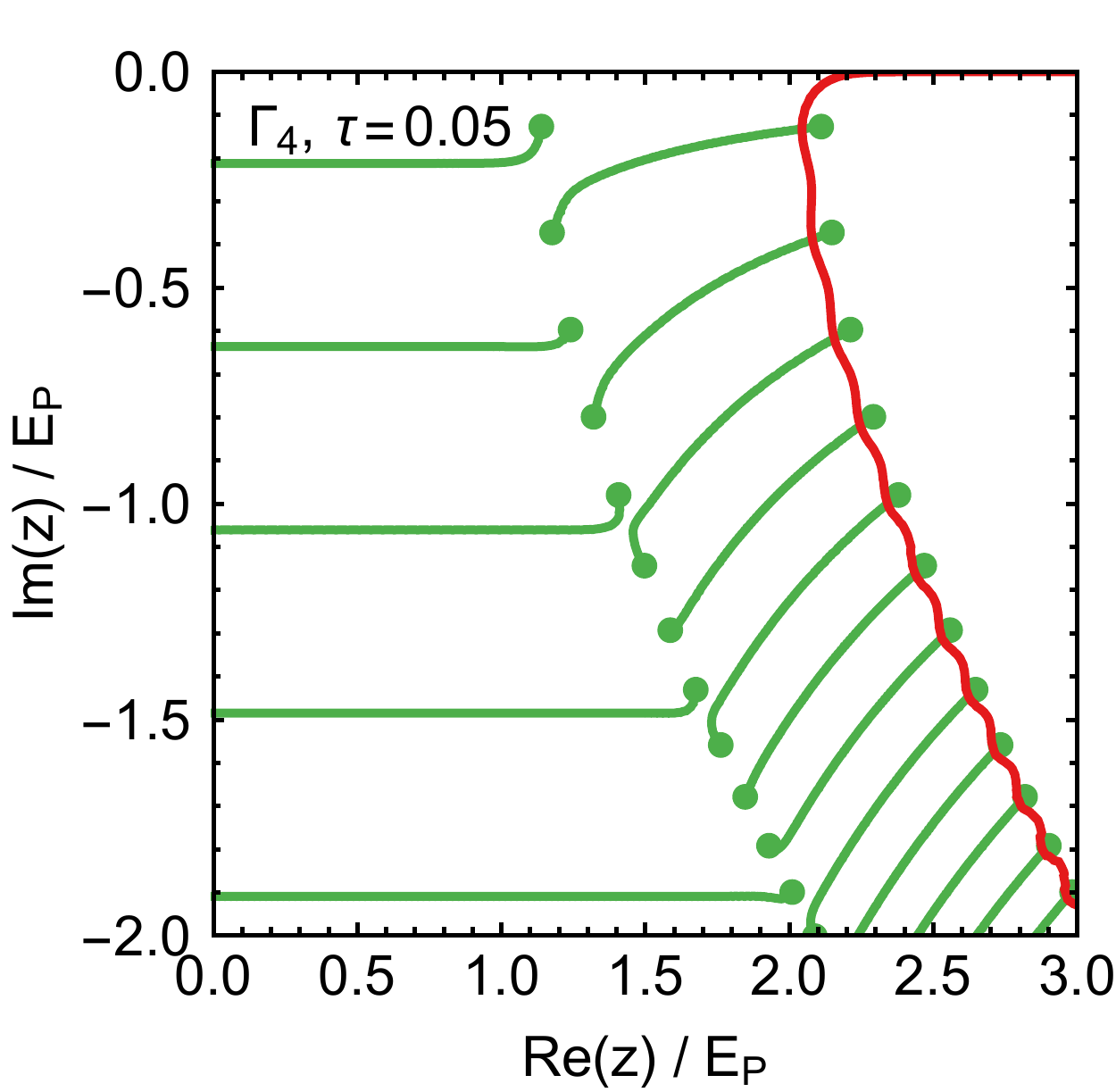}}; 
    \node at (4.1cm,-3.8cm) {(d)};
  \end{tikzpicture}
  \hfill\hfill
  \caption{Various contours for the analytic continuation $\check{\varepsilon}(q,z)$ in the complex plane, with parameters $n^{(0)}=1/a_0^3$ and $q=0.6 p_F$. The green lines show the branch cuts, and the green points the branch points. The blue lines show where $\text{Re}(\check{\varepsilon}(q,z))=0$, and the red lines where $\text{Im}(\check{\varepsilon}(q,z))=0$.
  (a) and (b): $\check{\varepsilon}_{\Gamma_1}(q,z)$ for $\tau=0.2$ and $\tau=0.05$, respectively. In both cases we have a well-defined root, since the blue and red lines cross.
  (c) and (d): $\check{\varepsilon}_{\Gamma_4}(q,z)$ for $\tau=0.2$ and $\tau=0.05$, respectively. For $\tau=0.05$ we no longer have a root.
  }
  \label{fig:contours-epsilon-comparison-T}
\end{figure}

Figure~\ref{fig:density-epsilon-comparison-T} shows the real and the imaginary parts of the analytic continuations $\check{\varepsilon}_{\Gamma_1}(q,z)$ and $\check{\varepsilon}_{\Gamma_4}(q,z)$. 
We consider $n^{(0)}=1/a_0^3$, which translates to $r_s = (3/(4\pi))^{1/3} \approx 0.62$, $\tau=0.2$ and $|q| = 0.6 p_F$.
Comparing figures~\ref{fig:density-epsilon-comparison-T}(b) and~\ref{fig:density-epsilon-comparison-T}(d), we immediately see that the imaginary parts of $\check{\varepsilon}_{\Gamma_1}(q,z)$ and $\check{\varepsilon}_{\Gamma_4}(q,z)$ coincide. The difference in the branch cuts is clearly visible in the real parts shown in figures~\ref{fig:density-epsilon-comparison-T}(a) and~\ref{fig:density-epsilon-comparison-T}(c).
For $\Gamma_1$, these branch cuts lie along the lines $\text{Re}(\nu_\pm^2) = 1$, whilst they have a more complicated placement for $\Gamma_4$.
Figure~\ref{fig:contours-epsilon-comparison-T} shows various contour plots for the same parameters, where the branch cuts are indicated by green lines, and the branch points by green dots. The blue lines show where the real part of $\check{\varepsilon}(q,z)$ vanishes, and the red lines show where its imaginary part vanishes. A root of $\check{\varepsilon}(q,z)$ exists at a point where the red and blue lines cross. Figures~\ref{fig:contours-epsilon-comparison-T}(a) and~\ref{fig:contours-epsilon-comparison-T}(c) show that for $\tau=0.2$ the roots of $\check{\varepsilon}_{\Gamma_1}(q,z)$ and $\check{\varepsilon}_{\Gamma_4}(q,z)$ coincide. This makes sense in light of our previous discussion, since $-\tau\pi < \text{Im}(\nu_-^2) < 0$ even though $\text{Re}(\nu_-^2) \leq 1$.

For $\tau=0.05$, $\check{\varepsilon}_{\Gamma_1}(q,z)$, shown in figure~\ref{fig:contours-epsilon-comparison-T}(b), still has a root, whereas the root of $\check{\varepsilon}_{\Gamma_4}(q,z)$ has disappeared, see figure~\ref{fig:contours-epsilon-comparison-T}(d).
To understand why this happens, we come back to our previous discussion of $\Gamma_2$. Although we cannot precisely infer the integration path $\Gamma_4$ from the placement of the branch cuts, we can extract some information by noting that an integration path does not pass through the points of a branch cut, see also figure~\ref{fig:poles-complex-plane}(c). We therefore conclude that $\Gamma_4$ passes in between the poles depicted in figure~\ref{fig:poles-complex-plane}, qualitatively somewhat similar to $\Gamma_2$. As for $\Gamma_2$, this leads to a function with many discontinuities when we fix $q$ and $z$ and lower the temperature.

We show these discontinuities explicitly in figure~\ref{fig:epsilon-limit-T-zero}, where we plot $\check{\varepsilon}_{\Gamma_1}(q,z)$ and $\check{\varepsilon}_{\Gamma_4}(q,z)$ as a function of temperature. We set $|q|=0.6 p_F$ and choose $z$ in such a way that $\check{\varepsilon}_{\Gamma_1}(q,z)$ has a root for $\tau=0.05$. We observe that $\check{\varepsilon}_{\Gamma_4}(q,z)$ does not have a root at the same point, because it has a large discontinuity at a slightly larger temperature. This discontinuity, as well as the subsequent ones, arises when a branch cut passes through the point $z$ as we lower the temperature, cf. figures~\ref{fig:contours-epsilon-comparison-T}(c) and~\ref{fig:contours-epsilon-comparison-T}(d). Figure~\ref{fig:epsilon-limit-T-zero} shows explicitly that this leads to erratic behavior as $\tau \to 0$. The analytic continuations $\check{\varepsilon}_{\Gamma_2}(q,z)$ and $\check{\varepsilon}_{\Gamma_3}(q,z)$ show exactly the same erratic behavior.
On the other hand, $\check{\varepsilon}_{\Gamma_1}(q,z)$ gives rise to a well-defined limit for $\tau\to 0$, and therefore for $T\to 0$.
Because we require the analytic continuation to be well-behaved as $T \to 0$, we conclude that $\check{\varepsilon}_{\Gamma_1}(q,z)$ is the most suitable analytic continuation.

Although we believe that, on physical grounds, $\Gamma_1$ is more suitable than $\Gamma_4$, we remark that this does not affect the results obtained in Ref.~\cite{Hamann20}. Since they studied the dispersion and damping for higher temperatures than we consider, the complex roots that they find mostly satisfy $\text{Re}(\nu_-^2) > 1$, or $-\tau\pi < \text{Im}(\nu_-^2) < 0$ when $\text{Re}(\nu_-^2) \leq 1$. In these regions, all analytic continuations give the same results, see figure~\ref{fig:poles-complex-plane}(a). Nevertheless, we think that their finding that complex roots of $\check{\varepsilon}(q,z)$ stop to exist close to the pair continuum may be influenced by their choice of the analytic continuation.

\begin{figure}[tb]
  \hfill
    \begin{tikzpicture}
    \node at (0cm,0cm)
    {\includegraphics[width=0.45\textwidth]{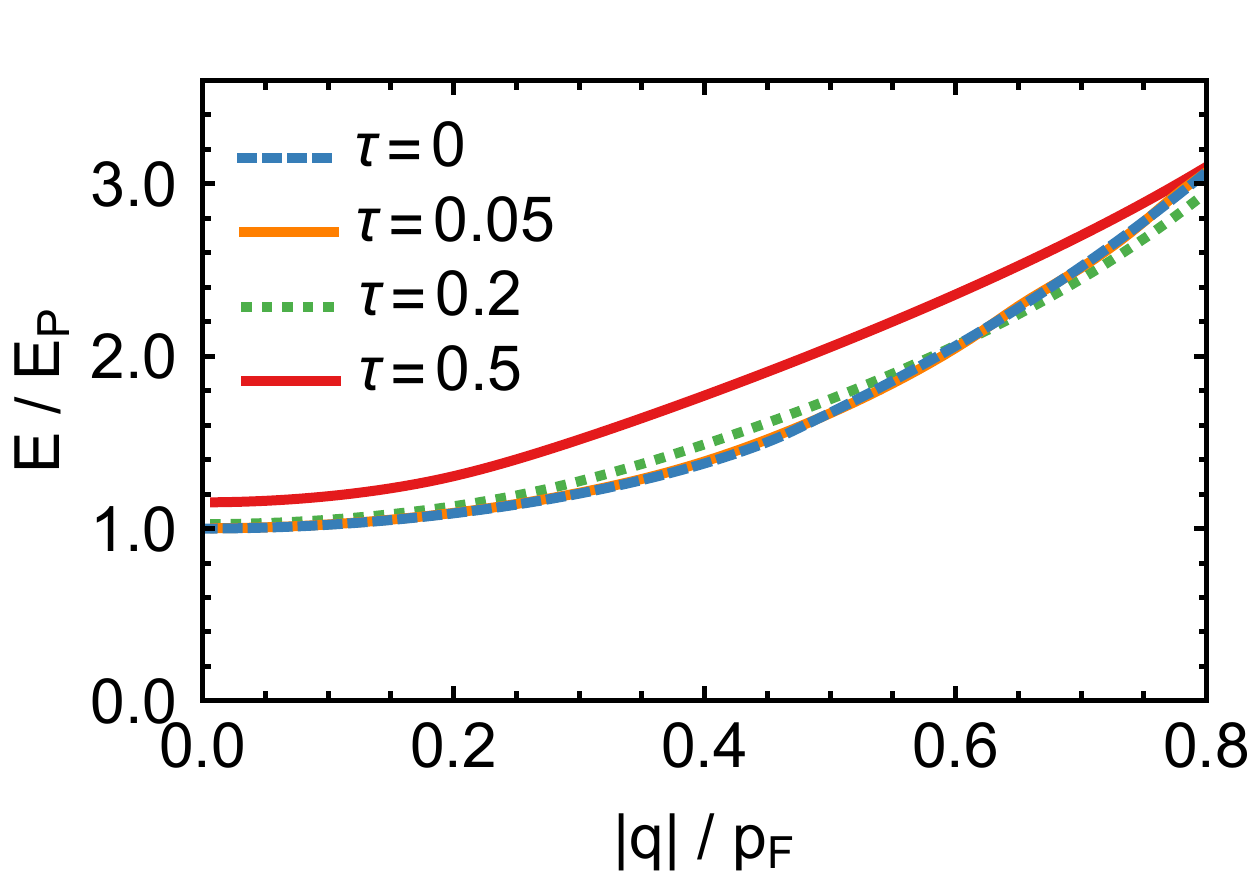}};
    \node at (-3.3cm,1.85cm) {(a)};
    
    \node at (7.1cm,0cm)
    {\includegraphics[width=0.45\textwidth]{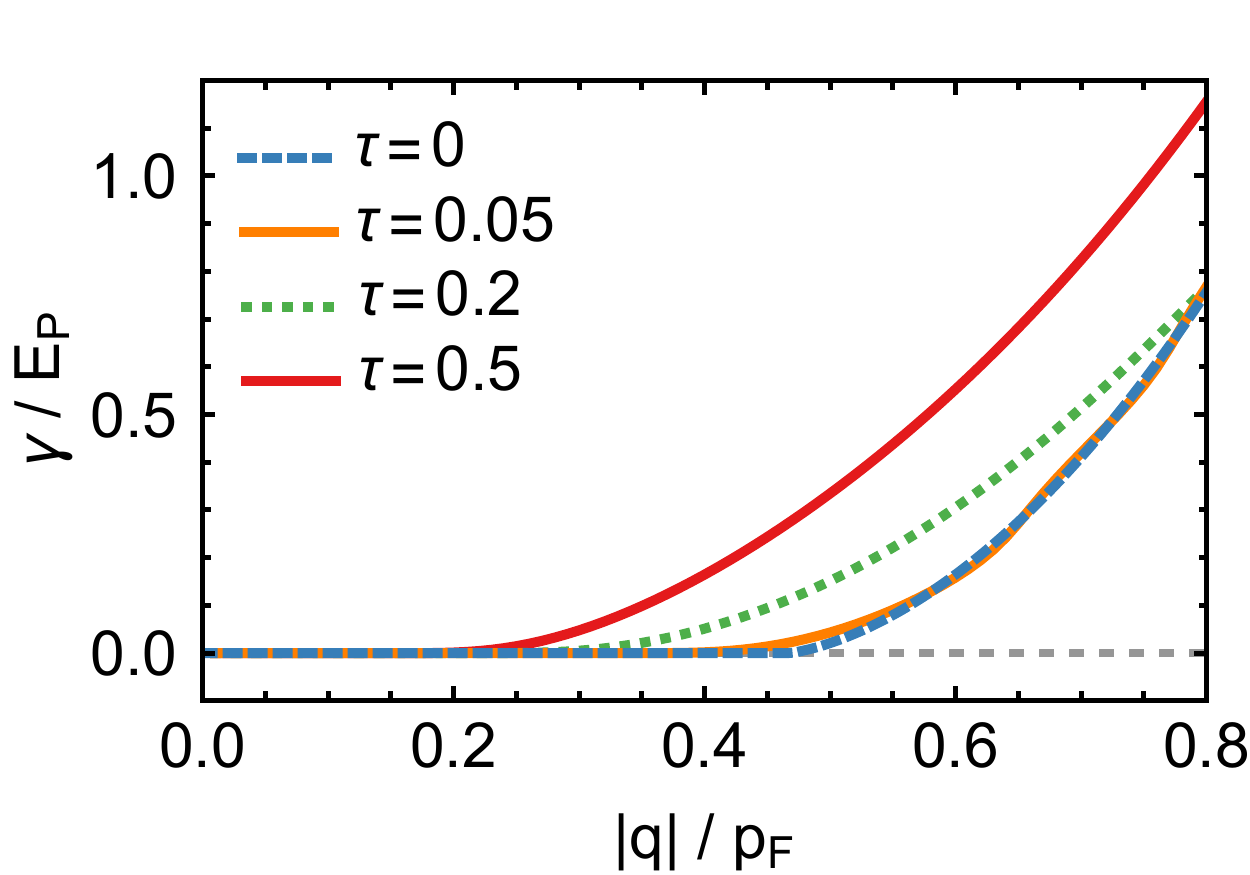}}; 
    \node at (3.8cm,1.85cm) {(b)};
  \end{tikzpicture}
  \hfill\hfill
  \caption{Solution of $\check{\varepsilon}_{\Gamma_1}(q,E-i\gamma)=0$ for $n^{(0)}=1/a_0^3$. Panels (a) and (b) show the dispersion $E$ and the damping $\gamma$, respectively, for different temperatures. There is good convergence to the result for $T=0$, computed in appendix~\ref{subapp:analytic-cont-zero-T}.}
  \label{fig:dispersion-damping-several-T}
\end{figure}

Finally, we discuss how the position of the complex root of $\check{\varepsilon}_{\Gamma_1}(q,z)$ changes when we alter the temperature. We set $z=E-i\gamma$, and compute the position of the root as a function of $|q|$ for various temperatures. Figure~\ref{fig:dispersion-damping-several-T} shows the dispersion $E$ and damping $\gamma$. 
For $T>0$, the root is always complex, so Landau damping is always present as $\gamma$ is never zero. However, as figure~\ref{fig:dispersion-damping-several-T}(b) shows, the damping can be very small and practically negligible.
We observe that there is a clear convergence to the result for zero temperature, which we compute in the next section. Note that the results for finite temperature oscillate around the zero temperature result, as can be most clearly seen in the damping for $\tau=0.05$. We ascribe this effect to the oscillations in the red lines in figure~\ref{fig:contours-epsilon-comparison-T}, which show where $\text{Im}(\check{\varepsilon}(q,z)) = 0$. The size of these oscillations indeed decreases with temperature.
We therefore conclude that the analytic continuation $\check{\varepsilon}_{\Gamma_1}(q,z)$ gives rise to a well-defined limit as $T\to 0$. We compute this limit explicitly in the next subappendix.

\subsection{Calculations for zero temperature}
\label{subapp:analytic-cont-zero-T}

Let us consider the spectral function~(\ref{eq:homo-spectral-def}) for zero temperature. For real values of $E$ (and $\nu_\pm$), we can follow the steps from the previous subappendix to arrive at expression~(\ref{eq:homo-spectral-int}).
At zero temperature, the Fermi-Dirac function is a step function. Since it is not immediately clear how to evaluate this function for complex arguments, let us first consider the integral expression~(\ref{eq:homo-spectral-int}) on the real axis.
We have
\begin{equation}  \label{eq:homo-spectral-int-zero-T}
  \widetilde{\Pi}(q,E) = - \frac{g_s}{(2\pi \hbar)^3} \frac{2 \pi^2 m p_F^2}{|q|} \bigg( 
  \int_0^\infty \text{d}\chi \, \Theta(1-\chi) \Theta(\chi -\nu_-^2) 
  - \int_0^\infty \text{d}\chi \, \Theta(1-\chi) \Theta(\chi-\nu_+^2) 
  \bigg) .
\end{equation}
The product of these two step functions is only non-zero when $\nu_\pm^2 < \chi < 1$, which immediately implies that the integral vanishes for $\nu_\pm^2 > 1$, or $1 - \nu_\pm^2 < 0$. Since the integrand equals one when $1 - \nu_\pm^2 > 0$, we obtain
\begin{equation}  \label{eq:homo-spectral-int-zero-T-real-axis}
  \widetilde{\Pi}(q,E) = - \frac{g_s}{(2\pi \hbar)^3} \frac{2 \pi^2 m p_F^2}{|q|} \bigg( 
  (1 - \nu_-^2) \Theta(1 - \nu_-^2) 
  - (1 - \nu_+^2) \Theta(1 - \nu_+^2) 
  \bigg) .
\end{equation}
This expression clearly has two discontinuities at $\nu_+^2=1$.

The question that remains is how we can analytically continue the result~(\ref{eq:homo-spectral-int-zero-T-real-axis}) into the complex plane. Since the step function is not analytic and only defined for real arguments, we should regularize it.
In order to do this in the correct way, we start from the integral expression~(\ref{eq:homo-spectral-int}). Just as in the case of finite temperature, the step functions in this expression constrain the lower limits of the integrals, and set them to $\nu_\pm^2$. We therefore have
\begin{equation}  \label{eq:homo-spectral-int-zero-T-with-lower-limit}
  \widetilde{\Pi}(q,E) = - \frac{g_s}{(2\pi \hbar)^3} \frac{2 \pi^2 m p_F^2}{|q|} \bigg( 
  \int_{\nu_-^2}^\infty \text{d}\chi \, \rho_0\left( \frac{\chi p_F^2}{2m} \right) 
  - \int_{\nu_+^2}^\infty \text{d}\chi \, \rho_0\left( \frac{\chi p_F^2}{2m} \right) 
  \bigg) .
\end{equation}
The second step function, coming from the Fermi-Dirac distribution, has to be regularized by using a limiting procedure. In Ref.~\cite{Bonitz94b}, where the analytic continuation was studied for a one-dimensional plasma, the regularization
\begin{equation}  \label{eq:FD-regularization-Bonitz}
  \Theta(1-y) = \lim_{\Delta \to 0} \Theta_\Delta(1-y) , \qquad \text{with } \Theta_\Delta(1-y) = \frac{1}{2} - \frac{1}{\pi} \arctan\left( \frac{y-1}{\Delta} \right)  .
\end{equation}
was used. As noted in Ref.~\cite{Bonitz94b}, the imaginary of $\Theta_\Delta$ goes to zero as $\Delta$ goes to zero. In the complex plane, the limiting function can be written as $\Theta(1-\text{Re}(z))$. Although this function does not look analytic, it should really be understood in the sense of the regularization.
However, the regularization~(\ref{eq:FD-regularization-Bonitz}) is not unique. Since the step function under consideration is actually the limit of the Fermi-Dirac distribution when we decrease the temperature, a more natural regularization seems to be
\begin{equation}  \label{eq:FD-regularization-temperature}
  \Theta(1-y) = \lim_{\tau \to 0} \rho_\tau(1-y) , \qquad \text{with } \rho_\tau(1-y) = \frac{1}{1+\exp((y-1)/\tau)}  .
\end{equation}
Note that this leads to the same limiting function $\Theta(1-\text{Re}(z))$ in the complex plane.
The regularization~(\ref{eq:FD-regularization-temperature}) implies that we can study the case of zero temperature as a limit of the finite temperature case. In the previous subappendix, we already established that integrating along the path $\Gamma_1$ shown in figure~\ref{fig:poles-complex-plane}(b) leads to a well-defined limit as the temperature goes to zero. We can therefore write down the spectral function for complex arguments $z$ as
\begin{multline} \label{eq:homo-spectral-int-pre-calculated-zero-T}
  \widetilde{\Pi}_{\Gamma_1}(q,z) = - \frac{g_s}{(2\pi \hbar)^3} \frac{2 \pi^2 m p_F^2}{|q|} \lim_{\tau\to 0} \bigg( 
  \int_{\nu_-^2}^{\text{Re}(\nu_-^2)} \text{d}\chi \, \rho_\tau(1-\chi)
  + \int_{\text{Re}(\nu_-^2)}^\infty \text{d}\chi \, \rho_\tau(1-\chi) \\
  - \int_{\nu_+^2}^{\text{Re}(\nu_+^2)} \text{d}\chi \, \rho_\tau(1-\chi)
  - \int_{\text{Re}(\nu_+^2)}^{\nu_+^2} \text{d}\chi \, \rho_\tau(1-\chi)
  \bigg) .
\end{multline}
A straightforward computation then gives
\begin{equation} \label{eq:homo-spectral-int-calculated-zero-T}
  \widetilde{\Pi}_{\Gamma_1}(q,z) = - \frac{g_s}{(2\pi \hbar)^3} \frac{2 \pi^2 m p_F^2}{|q|} \Big(
  (1-\nu_-^2) \Theta\left(1-\text{Re}(\nu_-^2)\right) - (1-\nu_+^2) \Theta\left(1-\text{Re}(\nu_+^2)\right) \Big)
\end{equation}
The step function in this result should once again be understood in terms of a proper regularization. Expression~(\ref{eq:homo-spectral-int-calculated-zero-T}) explicitly shows that the limit $T\to 0$ is well defined, since its discontinuities exactly coincide with the branch cuts that emerged for finite temperatures.

Alternatively, we can obtain the same result by taking the limit $\tau\to 0$ in the analytic continuation of expression~(\ref{eq:homo-spectral-int-calculated-v1}). Indeed, when $\text{Re}(1-\nu_\pm^2) < 0$, the exponent within the logarithm goes to zero in the limit $\tau\to 0$, and we obtain zero. On the other hand, when $\text{Re}(1-\nu_\pm^2) > 0$, the exponent becomes very large and we have
\begin{equation}
  \tau \log( 1 + \exp[ (1-\nu_\pm^2)/\tau ]) = \tau (1-\nu_\pm^2)/\tau + \tau \log( \exp[ -(1-\nu_\pm^2)/\tau] + 1) \to 1-\nu_\pm^2
\end{equation}
We can therefore also obtain the result~(\ref{eq:homo-spectral-int-calculated-zero-T}) by taking the limit $\tau\to 0$ in the analytic continuation of expression~(\ref{eq:homo-spectral-int-calculated-v1}).

With the result~(\ref{eq:homo-spectral-int-calculated-zero-T}) we can compute the analytic continuation~$\check{\Pi}(q,z)$ of the spectral function. Using expressions~(\ref{eq:homo-polarization-analytical-continuation}) and~(\ref{eq:chi-homogenegous-computed}) we have, in the lower half plane,
\begin{multline}  \label{eq:homo-Pi-analytic-cont-bc}
  \check{\Pi}_{\Gamma_1}^{R}(q,z) 
    = \frac{g_s m}{2 \pi^2 \hbar^3} \frac{p_F^2}{|q|} \bigg( 
      -\frac{|q|}{2 p_F} 
      + \frac{1-\nu_-^2}{4} \left( \log\left(\nu_- + 1\right) - \log\left(\nu_- - 1\right) - 2\pi i \Theta[\text{Re}(1-\nu_-^2)] \right) \\
      - \frac{1-\nu_+^2}{4} \left( \log\left(\nu_+ + 1 \right) - \log\left(\nu_+ - 1 \right) - 2\pi i \Theta[\text{Re}(1-\nu_+^2)] \right) \bigg) .
\end{multline}
The only difference with our previous expression~(\ref{eq:chi-homogenegous-computed}) is the presence of the step function, which should be understood in terms of the regularization. Upon closer inspection, we see that this step function actually shifts the branch cut of the logarithm $\log\left(\nu_\pm - 1\right)$. For our previous expression~(\ref{eq:chi-homogenegous-computed}), which is valid on the real axis, the branch cuts lie along the negative real axis, that is, at an angle $\pm\pi$. This comes from the fact that we consider the principal branch of the logarithm and is a direct consequence of the fact that we are discussing the retarded response function, see also the discussion in the main text.
In the new expression~(\ref{eq:homo-Pi-analytic-cont-bc}), these branch cuts are shifted due to the presence of the step function and lie along the lines~$\text{Re}(\nu_\pm^2)=1$. In more mathematical terms, one could say that the analytic continuation tells us to consider a different Riemann sheet of the multi-valued logarithm function.

\begin{figure}[tb]
  \hfill
    \begin{tikzpicture}
    \node at (0cm,0cm)
    {\includegraphics*[width=0.48\textwidth]{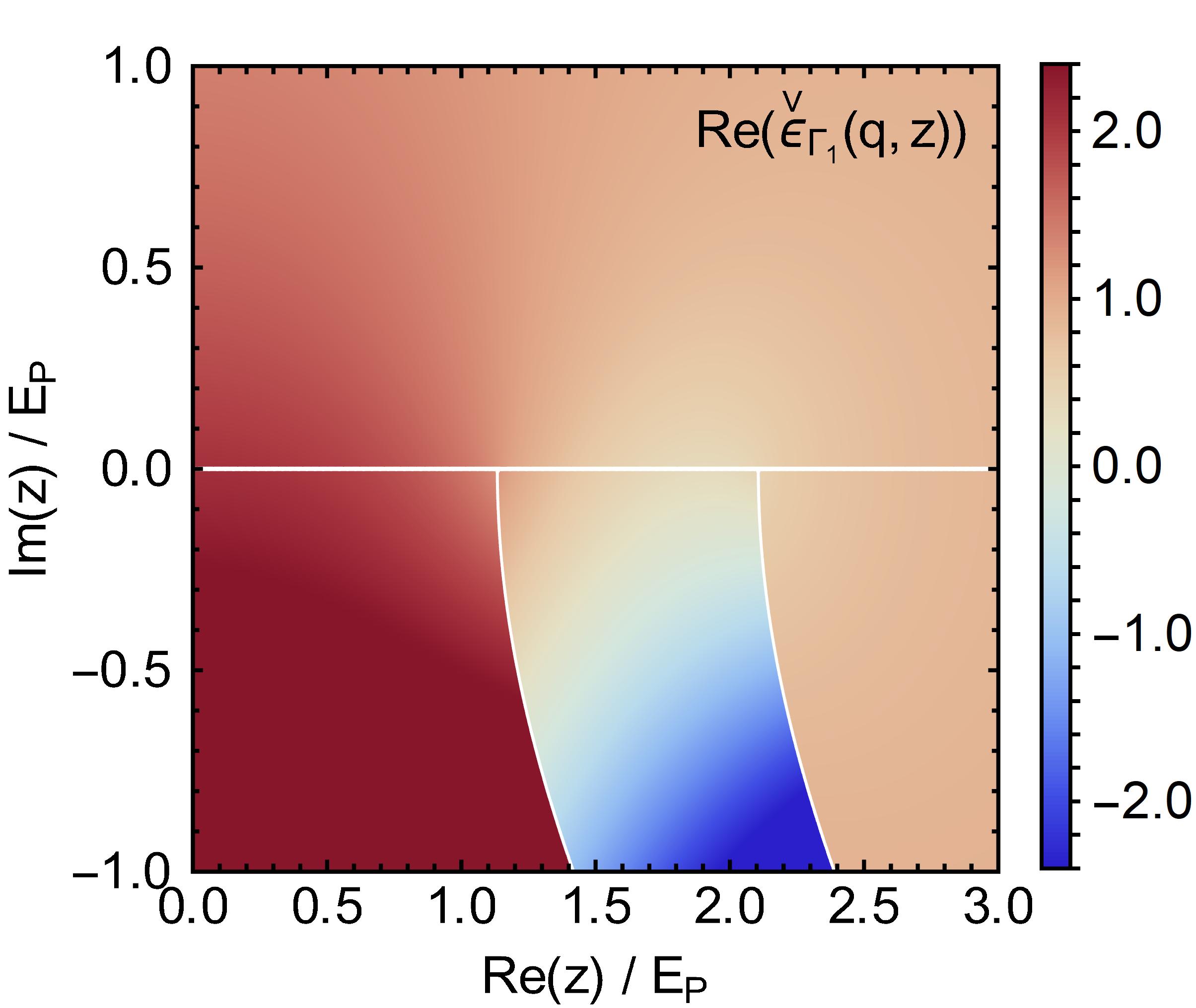}};
    \node at (-3.3cm,2.5cm) {(a)};
    
    \node at (7.3cm,0cm)
    {\includegraphics[width=0.48\textwidth]{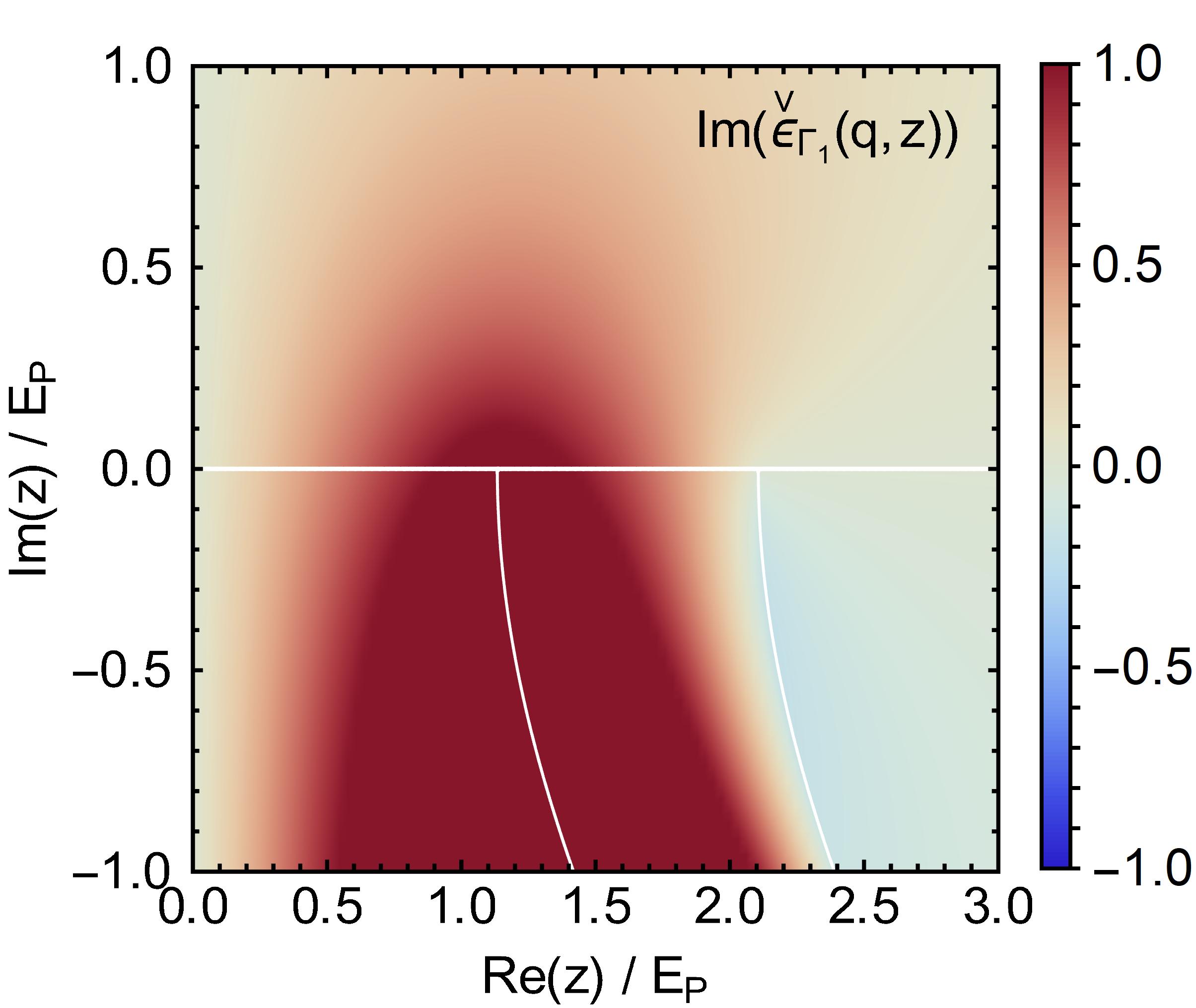}}; 
    \node at (4.0cm,2.5cm) {(b)};
  \end{tikzpicture}
  \hfill
  \caption{The analytic continuation $\check{\varepsilon}_{\Gamma_1}(q,z)$ in the complex plane at zero temperature for $n^{(0)}=1/a_0^3$ and $q=0.6 p_F$. Panels (a) and (b) show the real and imaginary parts, respectively. The branch cuts of the logarithm clearly lie along the lines $\text{Re}(\nu_\pm^2) = 1$.
  }
  \label{fig:density-epsilon-T-zero}
\end{figure}

\begin{figure}[tb]
  \hfill
  \includegraphics[width=0.45\textwidth]{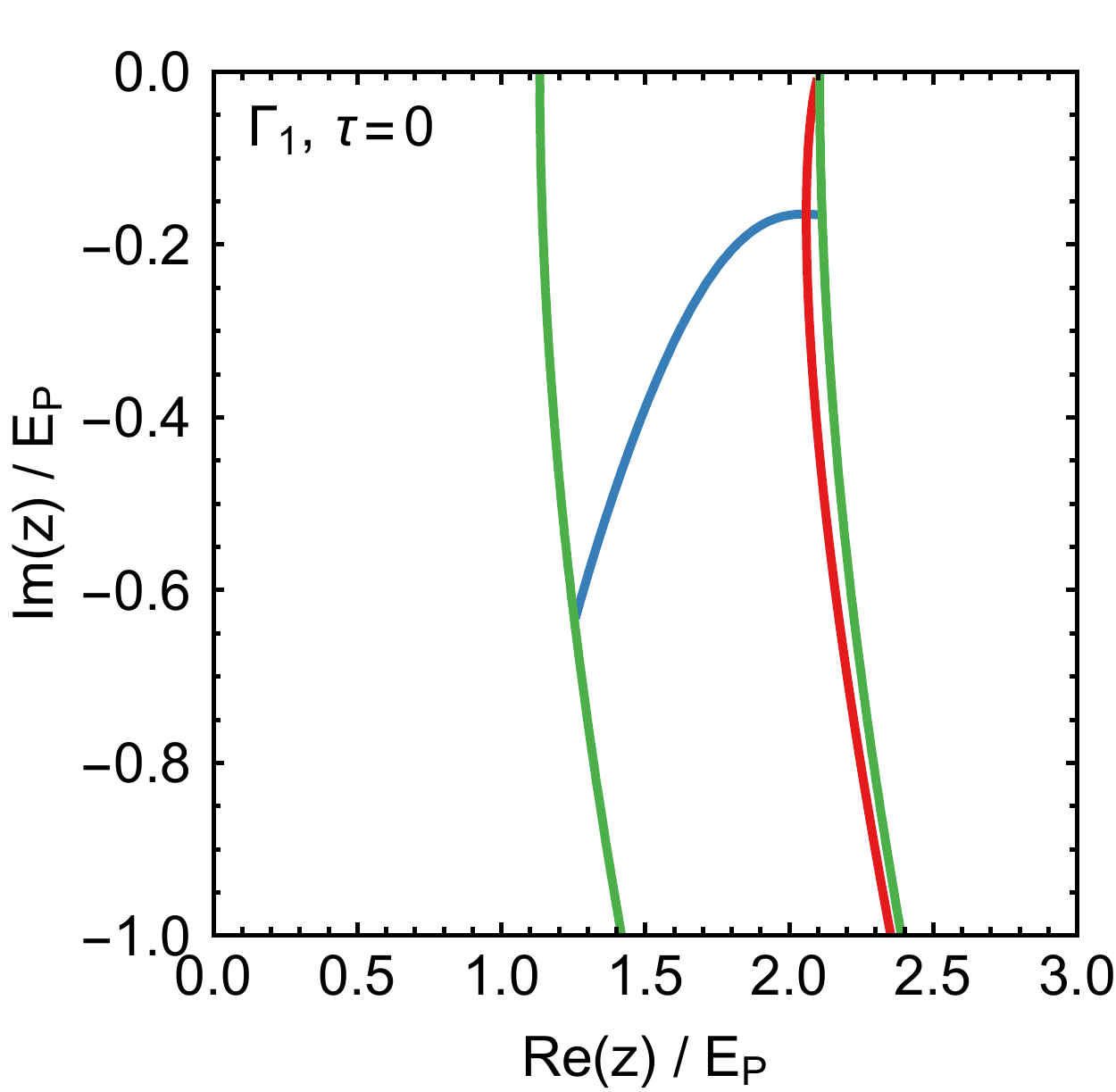}
  \hfill\hfill
  \caption{Various contours for the analytic continuation $\check{\varepsilon}_{\Gamma_1}(q,z)$ in the complex plane, with parameters $n^{(0)}=1/a_0^3$ and $q=0.6 p_F$. The green lines show the branch cuts, whilst the blue and red lines show where the real and imaginary parts of $\check{\varepsilon}_{\Gamma_1}(q,z)=0$ vanish, respectively. Since the blue and red lines cross, we find that $\check{\varepsilon}_{\Gamma_1}(q,z)$ has a root.
  }
  \label{fig:contours-epsilon-T-zero}
\end{figure}

In figure~\ref{fig:density-epsilon-T-zero}, we show the real and the imaginary parts of the analytic continuation $\check{\varepsilon}_{\Gamma_1}(q,z)$ that results from our expression for $\check{\Pi}_{\Gamma_1}^{R}(q,z)$ through definition~(\ref{eq:homo-epsilon-analytical-continuation}). We consider $n^{(0)}=1/a_0^3$ and $|q| = 0.6 p_F$. In figure~\ref{fig:density-epsilon-T-zero}, we immediately see that the branch cuts do not lie along the real axis. Instead, the analytic continuation $\check{\varepsilon}_{\Gamma_1}(q,z)$ is continuous across the entire real axis. 
The contour plot corresponding to these graphs is shown in figure~\ref{fig:contours-epsilon-T-zero}. Once again, the branch cuts are indicated by green lines, and the blue and red lines show where the real and imaginary parts of $\check{\varepsilon}_{\Gamma_1}(q,z)$ vanish, respectively. One clearly sees that there is a point $z$ with $\check{\varepsilon}_{\Gamma_1}(q,z)=0$ in the region between the two branch cuts, as the red and blue lines cross. Since $|q_L(E_L)| = 0.466 p_F$ for our choice of parameters, we also expected a complex solution.

We now set $z=E-i\gamma$, and compute the position of the root as a function of $|q|$. 
We show the dispersion $E$ and the damping $\gamma$ in figure~\ref{fig:dispersion-damping-several-T}, and also in figure~\ref{fig:dispersion-damping-homo} in the main text. For zero temperature, the damping is zero until the point $|q_L(E_L)| = 0.466 p_F$, which means that the solution lies on the real axis.
In the next subappendix, we give an explicit proof that the analytic continuation $\check{\varepsilon}_{\Gamma_1}(q,z)$ has roots for complex $z$ when $|q|>|q_L(E_L)|$, that is, in the Landau damped region.

\subsection{Existence of complex roots of the dielectric function}
\label{subapp:analytic-cont-proof-roots-zero-T}

In this subappendix, we continue the analysis of the dielectric function at zero temperature.
In section~\ref{sec:investigation}, we defined $q_L(E)$ as the momentum for which $\nu_-=1$. We now invert this relation, and define $E_{LD}(|q|)$ by the requirement that $\nu_-=1$. In figure~\ref{fig:plasmon-dispersion-homo}, both $|q_L(E)|$ and $E_{LD}(|q|)$ correspond to the boundary of the red region.
Previously, we also defined $E_L$ by $\overline{\varepsilon}(q_L(E_L),E_L) = 0$. Equivalently, we can define $|q_{LD}|$ by $\overline{\varepsilon}(|q_{LD}|,E_{LD}(|q_{LD}|)) = 0$. Of course, we have $|q_{LD}| = |q_L(E_L)|$ and $E_{LD}(|q_{LD}|) = E_L$.
The momentum $|q_{LD}|$ therefore lies at the boundary of the Landau damped region: for $|q|>|q_{LD}|$ the equation $\overline{\varepsilon}(q,E) = 0$ has no solutions for real energies $E$. In figure~\ref{fig:dispersion-damping-several-T}, we already saw numerically that the equation $\check{\varepsilon}_{\Gamma_1}(q,z) = 0$ has solutions for complex $z$ when $|q|>|q_L(E_L)|$. In this subappendix, we prove analytically that these solutions exist. The proof is similar to the proof given in section~\ref{subsec:no-solutions-real-energies}.
Its main ingredient is once again the behavior of the function $z \log z$ in the complex plane around the origin.

In the upper half plane, $\check{\varepsilon}_{\Gamma_1}(q,z)$ is given by $\overline{\varepsilon}_{\Gamma_1}(q,z)$ according to expressions~(\ref{eq:homo-epsilon-analytical-continuation}) and~(\ref{eq:homo-polarization-analytical-continuation}).
According to the derivation in Ref.~\cite{Bonitz94}, the equation $\overline{\varepsilon}(q,z)=0$ has no roots in the upper half plane for any $|q|$. We may therefore focus our attention on the lower half plane, where $\check{\varepsilon}_{\Gamma_1}(q,z)$ is defined by the analytic continuation.

In the first step of the proof, we expand the analytic continuation $\check{\varepsilon}_{\Gamma_1}(q,E)$, which can be computed explicitly with the help of expressions~(\ref{eq:homo-epsilon-analytical-continuation}) and~(\ref{eq:homo-Pi-analytic-cont-bc}), in the vicinity of $E_{LD}(|q|)$. We define the variable $\delta E$ by $\delta E = E - E_{LD}(|q|)$. Since, by definition, $\nu_-=1$ at $E_{LD}(|q|)$, we can approximate $\nu_- - 1 \approx c_1(|q|)(E - E_{LD}(|q|)) = c_1(|q|) \delta E$ for $E$ close to $E_{LD}(|q|)$. We remark that $c_1(|q|) > 0$, since $\nu_- - 1 > 0$ for $E > E_{LD}(|q|)$. The term $(\nu_- - 1)(\log(\nu_- - 1)+2\pi i \Theta[\text{Re}(1-\nu_-^2)])$ in $\check{\varepsilon}_{\Gamma_1}$ then becomes
\begin{equation}
  c_1(q) \delta E \big( \log \delta E + \log c_1(|q|) + 2\pi i \Theta[\text{Re}(- (\nu_- + 1) c_1(|q|) \delta E)]\big)
\end{equation} 
for $E$ in the vicinity of $E_{LD}(|q|)$. We can use a first-order Taylor expansion for the other terms in $\check{\varepsilon}_{\Gamma_1}(q,E)$, as they do not show singular behavior. In the vicinity of $E_{LD}(|q|)$, the dielectric function $\check{\varepsilon}_{\Gamma_1}(q,E)$ therefore behaves as
\begin{equation}
  \check{\varepsilon}_{\Gamma_1}(q,E) = c_2(|q|) + c_3(|q|) \delta E \big( \log \delta E + 2\pi i \Theta[\text{Re}(-\delta E)] \big) + c_4(|q|) \delta E
\end{equation}
for certain coefficients $c_j(|q|)$. As before, we can eliminate the term $c_4(|q|) \delta E$ by introducing a new variable $z = \exp(c_4(|q|)/c_3(|q|)) \delta E$. This gives
\begin{equation}
  \check{\varepsilon}_{\Gamma_1}(q,E) = c_2(|q|) + c_5(|q|) z \big( \log z + 2\pi i \Theta[\text{Re}(-z)] \big) ,
\end{equation}
where $c_5(|q|) = c_3(|q|) \exp(-c_4(|q|)/c_3(|q|))$.
We now specialize to the situation where $q$ is close to $q_{LD}$. At the point $|q| = |q_{LD}|$, we have a plasmon with momentum $E_{LD}(|q_{LD}|)$, so $\check{\varepsilon}_{\Gamma_1}(q,E)=0$ and $z=0$. This means that $c_2(|q_{LD}|)=0$, and that we can write $c_2(|q|) \approx c_6 (|q| - |q_{LD}|)$ when $|q|$ is close to $|q_{LD}|$.

We now move on to the second step of the proof, in which we consider the solutions of the equation
\begin{equation}
  z \log z + 2\pi i z \Theta[\text{Re}(-z)] = - c_2(|q|)/c_5(|q|) \equiv \lambda ,
\end{equation}
where $\lambda$ is real and small. We allow $z$ to be complex, and use the polar form $z=r e^{i\phi}$. Separating real and imaginary parts, we obtain a set of equations, namely
\begin{equation}  \label{eq:homo-singularity-near-qLD}
  \begin{aligned}
    r \cos\phi \log r - r \phi \sin \phi - 2 \pi r \sin\phi \Theta(-\cos\phi) &= \lambda , \\
    r \sin\phi \log r + r \phi \cos \phi + 2 \pi r \cos\phi \Theta(-\cos\phi) &= 0 .
  \end{aligned}
\end{equation}
As we mentioned in the previous subappendix, we are considering the lower half plane. Moreover, the logarithm in the above equations comes from the advanced response function, which means that we have to consider its principal branch. The variable $\phi$ therefore lies in the interval $[-\pi,0]$.
The equations for the upper half plane can be derived from equations~(\ref{eq:homo-singularity-near-qLD}) by eliminating the step functions, and letting $\phi$ lie in the interval $[0, \pi]$. Note that both prescriptions lead to the same equations on the real axis, and that we do not have a discontinuity when we cross it.

We start with the case $|q|<|q_{LD}|$, where the set of equations~(\ref{eq:homo-singularity-near-qLD}) should have a solution for real $z$. From the definition of $z$, we extract $E = E_{LD}(|q|) + r e^{i\phi} \exp(-c_4(|q|)/c_3(|q|))$. Since the plasmon is not damped for $|q|<|q_{LD}|$, we should have $E > E_{LD}(|q|)$, so $r>0$ and $\phi=0$. For $\phi=0$, the second equation is automatically satisfied, and the first equations becomes $r \log r = \lambda$. We previously established that $\lambda$ is small, and proportional to $|q| - |q_{LD}|$ in first order. Since $r \log r < 0$ for small $r$, we see that the constant of proportionality has to be positive.

When $|q|>|q_{LD}|$, the solution with $\phi=0$ no longer exists, since the right-hand side of the first equation~(\ref{eq:homo-singularity-near-qLD}) is positive, whereas the left-hand side is negative. The first equation however implies that $r$ is small, since $\lambda$ is still small. 

We therefore consider the second equation, and distinguish two cases. The first case is when $\phi$ lies in the interval $(-\pi/2,0)$, so the step function is zero. In fact, we can include the upper half plane into the description by allowing $\phi$ to lie in the interval $(-\pi/2,\pi]$. This can be seen by noting that the step functions are absent there, as the rest of the derivation is analogous. Because we ruled out $\phi=0$, we can now divide the second equation~(\ref{eq:homo-singularity-near-qLD}) by $r \phi \cos\phi \log r$, to find
\begin{equation}  \label{eq:homo-singularity-second-rewritten}
  \frac{\tan\phi}{\phi} = -\frac{1}{\log r} .
\end{equation}
For small $r$, the right-hand side is positive and small. The left-hand side, however, lies between one and infinity when $\phi$ lies in the interval $(-\pi/2,\pi/2)$, and between negative infinity and zero when $\phi$ lies in the interval $(\pi/2,\pi]$. This implies that the system of equations~(\ref{eq:homo-singularity-near-qLD}) does not have a solution when $\phi$ lies in the interval $(-\pi/2,\pi]$.
This can be regarded as a different way to prove the result of Ref.~\cite{Bonitz94}, i.e., that the equation $\overline{\varepsilon}(q,z)=0$ has no roots in the upper half plane for any $|q|$. However, our proof does not rule out solutions that cannot be continuously connected to $E_{LD}(q_{LD})$, i.e. that appear as a sudden jump in the value of $E$. The proof given in Ref.~\cite{Bonitz94} shows that such solutions does not exist. Alternatively, this can be seen from the figures in the previous subappendix.

The second case is when $\phi$ lies in the interval $[-\pi,-\pi/2)$. This time, the step functions are not zero, and we have
\begin{equation}  \label{eq:homo-singularity-near-qLD-specialized}
  \begin{aligned}
    r \cos\phi \log r - r (\phi + 2\pi) \sin \phi &= \lambda , \\
    r \sin\phi \log r + r (\phi + 2\pi) \cos \phi &= 0 .
  \end{aligned}
\end{equation}
Since the sine and cosine are $2\pi$ periodic, we can change variables to $\tilde{\phi} = \phi+2\pi$ to obtain the system of equations
\begin{equation}  \label{eq:homo-singularity-near-qLD-specialized-changed-vars}
  \begin{aligned}
    r \cos\tilde{\phi} \log r - r \tilde{\phi} \sin \tilde{\phi} &= \lambda , \\
    r \sin\tilde{\phi} \log r + r \tilde{\phi} \cos \tilde{\phi} &= 0 ,
  \end{aligned}
\end{equation}
where $\tilde{\phi}$ now lies in the interval $[\pi,3\pi/2)$. As before, the first equation shows that $r$ is small, whilst the second can be rewritten as $\tan\tilde{\phi}/\tilde{\phi} = -1/\log r$. However, this time $\tan\tilde{\phi}/\tilde{\phi}$ takes all values between zero and infinity, since $\tilde{\phi}$ lies between $\pi$ and $3\pi/2$. We can therefore find a solution $r(\tilde{\phi})$, which can be inserted into the first equation. This first equation now also has a solution, since both terms on the left-hand side are positive, $(r \log r)\cos\tilde{\phi} > 0$ and $- r \tilde{\phi} \sin \tilde{\phi} > 0$ and small, and on the right-hand side $\lambda >0$ is also positive and small. The set of equations~(\ref{eq:homo-singularity-near-qLD-specialized-changed-vars}) therefore has a solution when $\tilde{\phi}$ lies in the interval $[\pi,3\pi/2)$. Note that this solution only exists for the analytically continued dielectric function $\check{\varepsilon}_{\Gamma_1}(q,z) = 0$. When we consider the original dielectric function $\overline{\varepsilon}(q,z)$, we arrive at equation~(\ref{eq:homo-singularity-second-rewritten}), where $\phi$ lies in the interval $(-\pi,-\pi/2)$. This equation has no solutions, as the left-hand side lies between negative infinity and zero, whereas the right-hand side is positive.

We have thus proven analytically that $\check{\varepsilon}_{\Gamma_1}(q,z) = 0$ has solutions for complex $z$ when $|q|>|q_{LD}|$. Moreover, we showed that these solutions lie in the lower half plane, to the left of the line $\text{Re}(1-\nu_-^2) = 0$. This is in exact agreement with our numerical calculations in the previous subappendix, see figure~\ref{fig:contours-epsilon-T-zero}.

\section{Review of a prior derivation}
\label{app:prior-derivation}

The derivations presented in section~\ref{sec:derivation} of this article arose after studying  Refs.~\cite{Ishmukhametov71,Ishmukhametov75}. The formula $L_0(x,q)=0$ was first presented in Ref.~\cite{Ishmukhametov71}, in a slightly different notation. A justification of this formula was given in Ref.~\cite{Ishmukhametov75}. Unfortunately, the latter article is not available digitally and very hard to come by.
In order to sketch the context of the present work and to show its inspiration, we repeat the arguments presented in Ref.~\cite{Ishmukhametov75} in this appendix.
We emphasize that they are not completely rigorous, but may give the reader a more intuitive idea of the formal transformations presented in the main text.

This appendix consists of two parts. In the first part, we briefly repeat the formulation of the RPA in real space, as presented in the textbook~\cite{Vonsovsky89}. More precisely, we follow the three steps outlined in section~\ref{subsec:derivation-outline} to obtain an integro-differential equation for the induced potential $V(x,t)$. This discussion serves as an introduction for the second part, where we present the arguments given in Ref.~\cite{Ishmukhametov75}. 
We use various manipulations to rewrite the integro-differential equation in a more convenient form and then apply the semiclassical approximation to obtain the effective Hamiltonian $L_0(x,p)$.
The notation in this appendix corresponds to the notation in the main text, instead of to the notation in Refs.~\cite{Ishmukhametov71,Ishmukhametov75}. Although the original article does not include a varying background dielectric constant $\varepsilon_b(x)$,
we include it in the arguments presented here, in analogy with the main text.

As discussed in the main text, we assume that $\hat{H}_0$ does not depend on time. This allows us to split $V(x,t)$ in a spatially varying part $V(x)$ and a time-dependent factor $\exp(-i(E+i\eta)t/\hbar)$, see equation~(\ref{eq:potential-time-dependence}) in the main text. As discussed there, we will write $E$ instead of $E+i \eta$, and assume that $E$ lies slightly above the real axis ($\eta \to 0^+$).

We start by considering
the linearized Liouville-von Neumann equation for the density operator, that is,
\begin{equation} \label{eq:liouville-vonNeumann-lin-app}
  i \hbar \frac{\partial \hat{\rho}_1}{\partial t} = [ \hat{H}_0, \hat{\rho}_1 ] + [ V, \hat{\rho}_0 ] ,
\end{equation}
cf. equation~(\ref{eq:liouville-vonNeumann-lin}) in the main text.
Let $\psi_\nu(x)$ be the eigenfunctions of $\hat{H}_0$, where $\nu$ is a combined momentum and spin index.
We can use these eigenfunctions as a basis for our Hilbert space, and compute the matrix elements of the operators in equation~(\ref{eq:liouville-vonNeumann-lin-app}) in this basis.
Using bra-ket notation, with $|\nu\rangle$ indicating the eigenstate with index $\nu$ and $\langle \nu | \nu'\rangle$ indicating the inner product on Hilbert space, we have
\begin{equation}
\begin{aligned}
  \langle \nu | \, [ \hat{H}_0, \hat{\rho}_1 ] \, | \nu' \rangle &= (E_\nu - E_{\nu'} ) (\rho_1)_{\nu \nu'} , \\
  \langle \nu | \, [ \hat{V}, \hat{\rho}_0 ] \, | \nu' \rangle &= \big( \rho_0(E_{\nu'}) - \rho_0(E_{\nu}) \big) V_{\nu \nu'} , \\
  \langle \nu | \, i \hbar  \frac{\partial \hat{\rho}_1}{\partial t} \, | \nu' \rangle &= E (\rho_1)_{\nu \nu'} ,
\end{aligned}
\end{equation}
where $E_\nu$ is the eigenvalue of $\hat{H}_0$ corresponding to the eigenfunction $\psi_\nu$. Furthermore, the matrix elements are denoted by $A_{\nu \nu'} = \langle \nu | \hat{A} | \nu'\rangle$ and we have used that $\hat{\rho}_1$ has the same time dependence as the potential $V(x)$. Finally, $\hat{\rho}_0 | \nu \rangle = \rho_0(E_\nu) | \nu \rangle$.
We can now solve the Liouville-von Neumann equation~(\ref{eq:liouville-vonNeumann-lin-app}) for the matrix element $(\rho_1)_{\nu \nu'}$ and find
\begin{equation}  \label{eq:rho1-basis-app}
   (\rho_1)_{\nu \nu'} = \frac{\rho_0(E_{\nu}) - \rho_0(E_{\nu'})}{E_\nu - E_{\nu'} - E} V_{\nu \nu'}
     = e^{-i E t/\hbar} \frac{\rho_0(E_{\nu}) - \rho_0(E_{\nu'})}{E_\nu - E_{\nu'} - E} \int \text{d}x' \psi_{\nu}^*(x') V(x') \psi_{\nu'}(x') ,
\end{equation}
where the star denotes complex conjugation.

We proceed by computing
the induced electron density. Since the electron density operator is given by $\hat{n}(x) = \delta(x - x')$, its matrix elements equal
\begin{equation}  \label{eq:n-basis-app}
  n_{\nu' \nu}(x) = \int \text{d} x' \psi_{\nu'}^*(x') \delta(x - x') \psi_{\nu}(x') = \psi_{\nu'}^*(x) \psi_{\nu}(x) .
\end{equation}
Combining the results~(\ref{eq:rho1-basis-app}) and~(\ref{eq:n-basis-app}), we obtain an expression for the induced electron density, namely
\begin{align}  \label{eq:n-full-basis-app}
  n(x,t) &= \text{Tr}( \hat{n}(x) \hat{\rho}_1 )
       = \sum_{\nu \nu'} n_{\nu' \nu}(x) (\rho_1)_{\nu \nu'} \nonumber \\
       &= e^{-i E t/\hbar} \sum_{\nu \nu'} \frac{\rho_0(E_{\nu}) - \rho_0(E_{\nu'})}{E_\nu - E_{\nu'} - E} \psi_{\nu'}^*(x) \psi_{\nu}(x) \int \text{d}x' \psi_{\nu}^*(x') V(x') \psi_{\nu'}(x') .
\end{align}
This naturally leads to the third step, where we relate the induced electron density $n(x,t)$ to the induced potential $V(x,t)$ by means of the Poisson equation.
We have
\begin{equation}
  \nabla \cdot \big( \varepsilon_b(x) \nabla V(x,t) \big)
    = -4 \pi e^2 n(x, t) ,
\end{equation}
where the dot denotes the Cartesian inner product.
Inserting the result~(\ref{eq:n-full-basis-app}) and dividing by the time-dependent exponential on both sides, we obtain an integro-differential equation for $V(x)$, namely
\begin{equation} \label{eq:integro-diff-v1}
  \nabla \cdot \big( \varepsilon_b(x) \nabla V(x) \big)
    = -4 \pi e^2 \sum_{\nu \nu'} \frac{\rho_0(E_{\nu}) - \rho_0(E_{\nu'})}{E_\nu - E_{\nu'} - E} \psi_{\nu'}^*(x) \psi_{\nu}(x) \int \text{d}x' \psi_{\nu}^*(x') V(x') \psi_{\nu'}(x') .
\end{equation}
For a homogeneous system, where the potential $U(x)$ in $\hat{H}_0$ vanishes, the eigenstates $\psi_\nu(x)$ are plane waves. When we insert these in the above expression, together with the Fourier transform~(\ref{eq:V-fourier-expansion}), we naturally find the secular equation~(\ref{eq:secular-eq-home}), with the dielectric function~(\ref{eq:dielectric-function-homo}) for a homogeneous system, see also Ref.~\cite{Vonsovsky89}.

For an inhomogeneous system, where $U(x)$ does not vanish, the eigenstates have a more complicated form and 
inserting them into expression~(\ref{eq:integro-diff-v1}) does not lead to tractable equations.
In Ref.~\cite{Ishmukhametov75}, the authors therefore perform a series of manipulations and subsequently apply the semiclassical approximation. We present their arguments in the second part of this appendix.

The first step of the approach is to split the right-hand side into two terms, corresponding to each of the terms in the numerator $\rho_0(E_{\nu}) - \rho_0(E_{\nu'})$. We subsequently interchange the role of the indices $\nu$ and $\nu'$ in the second term. This leads to
\begin{multline} \label{eq:integro-diff-v2}
  \nabla \cdot \big( \varepsilon_b(x) \nabla V(x) \big)
    = -4 \pi e^2 \sum_{\nu \nu'} \rho_0(E_{\nu}) \left( \frac{1}{E_\nu - E_{\nu'} - E} \psi_{\nu'}^*(x) \psi_{\nu}(x) \int \text{d}x' \psi_{\nu}^*(x') V(x') \psi_{\nu'}(x') \right. \\
    \left. - \frac{1}{E_{\nu'} - E_{\nu} - E} \psi_{\nu}^*(x) \psi_{\nu'}(x) \int \text{d}x' \psi_{\nu'}^*(x') V(x') \psi_{\nu}(x') \right) .
\end{multline}
In the next step, we first pull the fraction $1/(E_\nu - E_{\nu'} - E)$ into the first integral. We then expand this fraction in a Taylor series, tacitly assuming that we are within the radius of convergence. We subsequently replace $E_{\nu'}$ by the operator $\hat{H}_0$, which is possible because it acts on $\psi_{\nu'}(x')$. We then apply the geometric series expansion to the resulting operator product and finally apply the same procedure to the second integral. We obtain
\begin{multline} \label{eq:integro-diff-v3}
  \nabla \cdot \big( \varepsilon_b(x) \nabla V(x) \big)
    = -4 \pi e^2 \sum_{\nu \nu'} \rho_0(E_{\nu}) \left( \psi_{\nu'}^*(x) \psi_{\nu}(x) \int \text{d}x' \psi_{\nu}^*(x') V(x') \frac{1}{E_\nu - \hat{H}_0 - E} \psi_{\nu'}(x') \right. \\
    \left. - \psi_{\nu}^*(x) \psi_{\nu'}(x) \int \text{d}x' \psi_{\nu}(x') V(x') \frac{1}{\hat{H}_0 - E_{\nu} - E} \psi_{\nu'}^*(x') \right) .
\end{multline}
We now rewrite the integral in the first term as an inner product. When we use that the operator $\hat{H}_0$ is Hermitian and does not contain $i$, and that $V^\dagger = V^*$, we find
\begin{multline}
  \int \text{d}x' \psi_{\nu}^*(x') V(x') \frac{1}{E_\nu - \hat{H}_0 - E} \psi_{\nu'}(x') 
    = \left\langle \psi_{\nu} \middle| V \frac{1}{E_\nu - \hat{H}_0 - E} \psi_{\nu'} \right\rangle \\
    = \left\langle \frac{1}{E_\nu - \hat{H}_0 - E} V^\dagger \psi_{\nu} \middle| \psi_{\nu'} \right\rangle
    = \int \text{d}x' \left( \frac{1}{E_\nu - \hat{H}_0 - E} V(x') \psi_{\nu}^*(x') \right) \psi_{\nu'}(x') .
\end{multline}
We apply a similar procedure to the second term. After rearranging, we obtain the expression
\begin{multline} \label{eq:integro-diff-v4}
  \nabla \cdot \big( \varepsilon_b(x) \nabla V(x) \big)
    = -4 \pi e^2 \sum_{\nu \nu'} \rho_0(E_{\nu}) \left( \psi_{\nu'}^*(x) \psi_{\nu}(x) \int \text{d}x' \psi_{\nu'}(x') \frac{1}{E_\nu - \hat{H}_0 - E} \psi_{\nu}^*(x') V(x') \right. \\
    \left. - \psi_{\nu}^*(x) \psi_{\nu'}(x) \int \text{d}x' \psi_{\nu'}^*(x') \frac{1}{\hat{H}_0 - E_{\nu} - E} \psi_{\nu}(x') V(x') \right) .
\end{multline}
At this point we use the completeness of the set of eigenfunctions $\psi_\nu(x)$, that is,
\begin{equation}
  \sum_{\nu'} \psi_{\nu'}^*(x) \psi_{\nu'}(x') = \delta(x-x')
\end{equation}
The delta function allows us to easily perform the integration over $x'$, yielding
\begin{multline} \label{eq:integro-diff-v5}
  \nabla \cdot \big( \varepsilon_b(x) \nabla V(x) \big)
    = -4 \pi e^2 \sum_{\nu} \rho_0(E_{\nu}) \left( \psi_{\nu}(x) \frac{1}{E_\nu - \hat{H}_0 - E} \psi_{\nu}^*(x) V(x) \right. \\
    \left. + \psi_{\nu}^*(x) \frac{1}{E_{\nu} - \hat{H}_0 + E} \psi_{\nu}(x) V(x) \right) .
\end{multline}
Thus, we managed to eliminate both the sum over $\nu'$ and the integration over $x'$.

The next step in the derivation aims to pull $\psi_\nu^*$ in the first term to the front, and similarly $\psi_\nu$ in the second term. To this end, we look at $\hat{H}_0 \psi_\nu V$. We have
\begin{equation}  \label{eq:comm-H0-psi-V}
  \begin{aligned}
    \hat{H}_0 \psi_\nu V &= -\frac{\hbar^2}{2 m} \nabla^2(\psi_\nu V) + U \psi_\nu V \\
      &= \left( -\frac{\hbar^2}{2 m} \nabla^2(\psi_\nu) + U \psi_\nu \right) V - \frac{\hbar^2}{m} (\nabla \psi_\nu) \cdot (\nabla V) - \frac{\hbar^2}{2 m} \psi_\nu \nabla^2 V \\
      &= \psi_\nu \left( E_\nu - \frac{\hbar^2}{m} (\nabla \log \psi_\nu) \cdot \nabla - \frac{\hbar^2}{2 m} \nabla^2 \right) V ,
  \end{aligned}
\end{equation}
where we used that $\hat{H}_0 \psi_\nu = E_\nu \psi_\nu$ in the final step.
We can prove a similar identity for $\hat{H}_0^n \psi_\nu V$; the only difference in the result being that the term in parentheses should be raised to the $n$-th power. Since we previously understood $(E_\nu - \hat{H}_0 - E)^{-1}$ as a power series, we can apply the above identity~(\ref{eq:comm-H0-psi-V}) to this function.
We therefore find
\begin{multline}  \label{eq:integro-diff-v6}
  \nabla \cdot \big( \varepsilon_b(x) \nabla V(x) \big)
    = -4 \pi e^2 \sum_{\nu} \rho_0(E_{\nu}) |\psi_{\nu}(x)|^2 \left(
    \left(\frac{\hbar^2}{m} (\nabla \log \psi_\nu^*) \cdot \nabla + \frac{\hbar^2}{2 m} \nabla^2 - E\right)^{-1} \right. \\
    \left. + \left(\frac{\hbar^2}{m} (\nabla \log \psi_\nu) \cdot \nabla + \frac{\hbar^2}{2 m} \nabla^2 + E\right)^{-1} \right) V(x) .
\end{multline}
At this point we apply the semiclassical approximation. We assume that the eigenfunctions $\psi_\nu(x)$ have the form of the semiclassical Ansatz, with amplitude $\chi_\nu(x)$ and action $K_\nu(x)$, that is,
\begin{equation}
  \psi_\nu(x) = \chi_\nu(x) \exp\left( \frac{i}{\hbar} K_\nu(x) \right) , \qquad \frac{\partial K_\nu}{\partial x} = p_\nu (x) .
\end{equation}
Similarly, we assume that $V(x)$ has the form of the semiclassical Ansatz, that is,
\begin{equation}
  V(x) = \varphi(x) \exp\left( \frac{i}{\hbar} S(x) \right), \qquad \varphi(x) = \varphi_0(x) + \hbar \varphi_1(x) + \ldots
\end{equation}
which coincides with the Ansatz~(\ref{eq:Ansatz-WKB-potential}) in the main text.

From this point onward, we only consider the leading-order term. We have
\begin{equation}
  \nabla \log \psi_\nu = \frac{i}{\hbar} \nabla K_\nu + \mathcal{O}(1) = \frac{i}{\hbar} p_\nu + \mathcal{O}(1); 
  \qquad \nabla V = \frac{i}{\hbar} (\nabla S) \varphi_0 \exp\left( \frac{i}{\hbar} S \right) + \mathcal{O}(1).
\end{equation}
Because we understand the functions of operators as power series, we can use the above expressions in equation~(\ref{eq:integro-diff-v6}). Note that we neglect additional terms of $\mathcal{O}(\hbar)$ in this process, 
since we only consider the action of the operators on the exponents in $\psi_\nu$ and $V$, and we do not consider the action of the operators on the amplitudes or on terms containing $\nabla S$.
The leading-order term of the asymptotic expansion of equation~(\ref{eq:integro-diff-v6}) therefore becomes
\begin{multline}  \label{eq:integro-diff-v8}
  \hspace*{-0.12cm}
  - \varepsilon_b(x) \frac{|\nabla S|^2}{\hbar^2} \varphi_0(x) \exp\left( \frac{i}{\hbar} S(x) \right) = - 4 \pi e^2 \sum_{\nu} \rho_0(E_{\nu}) |\psi_{\nu}(x)|^2 \left(
  \left(\frac{p_\nu(x) \cdot \nabla S}{m}  - \frac{| \nabla S|^2}{2 m} - E\right)^{-1} \right. \\
  \left. + \left(-\frac{p_\nu(x) \cdot \nabla S}{m} - \frac{|\nabla S|^2 }{2 m} + E\right)^{-1} \right) \varphi_0(x) \exp\left( \frac{i}{\hbar} S(x) \right) .
\end{multline}
In the deep semiclassical limit, the spacing between the different energy levels $E_\nu$ is very small, and the spectrum becomes quasi-continuous.
In the leading order, we can then replace the sum over all states by an integral over the continuous momentum variable $p$, that is,
\begin{equation}  \label{eq:continuum-limit-summation}
  \sum_\nu |\psi_\nu(x)|^2 \to \frac{g_s}{(2\pi \hbar)^3} \int \text{d}p .
\end{equation}
We now give some arguments to illustrate this limiting procedure for the one-dimensional case.

Let us consider a particle trapped in a potential well, whose behavior is governed by the one-dimensional Schr\"odinger equation. The index $\nu$ then consists of the spin $\sigma$ and the index $n$ that determines the energy level.
In the classically allowed region, between the two turning points $b_1$ and $b_2$, the wavefunction is a standing wave~\cite{Heading62}, which arises as the sum of a right-moving and a left-moving wave, i.e.
\begin{equation} \label{eq:psi-sc-app}
\begin{aligned}
  \psi_n(x) &= \frac{c_n}{\sqrt{p_n(x)}} \left( \exp\left( \frac{i}{\hbar} \int_{b_2}^{x} p_n(x) \text{d}x \right) - i \exp\left( -\frac{i}{\hbar} \int_{b_2}^{x} p_n(x) \text{d}x \right) \right) \\
    &= \frac{2 c_n e^{-i\pi/4}}{\sqrt{p_n(x)}} \cos\left( \frac{i}{\hbar} \int_{b_2}^{x} p_n(x) \text{d}x + \frac{\pi}{4} \right) ,
\end{aligned}
\end{equation}
where $p_n = \sqrt{2 m (E - U(x) )}$ is the momentum of a particle with energy $E$ in the potential $U(x)$. The constant $c_n$ needs to be chosen such that the wavefunction is normalized. The normalization procedure is discussed in detail in, e.g., Ref.~\cite{Heading62}, and leads to
\begin{equation}
  c_n = \left( 2 \int_{b_1}^{b_2} \frac{\text{d}x}{p_n(x)} \right)^{-1/2} .
\end{equation}
The quantization condition for a bound state with energy $E_n$ is given by
\begin{equation}
  \frac{1}{\hbar} \int_{b_1}^{b_2} p_n(x) \text{d}x 
    = \frac{1}{\hbar} \int_{b_1}^{b_2} \sqrt{2 m (E_n - U(x)) } \, \text{d}x
    = \left(n + \tfrac{1}{2} \right) \pi .
\end{equation}
Considering $n$ as a continuous variable and taking the derivative of this condition with respect to $n$, we find
\begin{equation}
  \frac{1}{\hbar} \left( p_n(b_2) \frac{\partial b_2}{\partial E_n} - p_n(b_1) \frac{\partial b_1}{\partial E_n} + m \int_{b_1}^{b_2}  \frac{\text{d}x}{p_n(x)} \right) \frac{\partial E_n}{\partial n}
    = \pi .
\end{equation}
Because $b_1$ and $b_2$ are turning points, we have $p_n(b_1)=p_n(b_2)=0$. The amplitude $c_n$ can therefore be expressed as $c_n^2 = (m/(2\pi\hbar)) \partial E_n/\partial n$ and we have 
\begin{equation}
  \psi_n(x) = \sqrt{\frac{m \frac{\partial E_n}{\partial n}}{2\pi \hbar}} \frac{2 e^{-i\pi/4}}{\sqrt{p_n(x)}} \cos\left( \frac{i}{\hbar} \int_{b_2}^{x} p_n(x) \text{d}x + \frac{\pi}{4} \right)
\end{equation}
in the classically allowed region.
Since we would like to make the limiting procedure~(\ref{eq:continuum-limit-summation}) plausible, let us consider $|\psi_n|^2$.
Replacing the square of the cosine by its average value $\tfrac{1}{2}$, we have
\begin{equation}
  |\psi_n|^2 = \frac{m}{2\pi \hbar} \frac{\partial E_n/\partial n}{p_n(x)} = \frac{1}{2\pi \hbar} \frac{\partial E_n/\partial n}{\partial E_n/\partial p_n} = \frac{1}{2\pi \hbar} \frac{\partial p_n}{\partial n} ,
\end{equation}
where the second equality follows either from Hamilton's equations or directly from the definition of $p_n$.
In the limit where the spectrum becomes quasi-continuous, we therefore have
\begin{equation}  \label{eq:continuum-limit-summation-1d}
  \sum_\nu |\psi_\nu(x)|^2 = g_s \sum_n |\psi_n(x)|^2 =  \frac{g_s}{2\pi \hbar} \sum_n \frac{\partial p_n}{\partial n} \Delta n \to \frac{g_s}{2\pi \hbar} \int \text{d}p .
\end{equation}
These arguments illustrate how one can think of the limiting procedure in one dimension.

Expression~(\ref{eq:continuum-limit-summation}) is the three-dimensional analog of the limiting procedure~(\ref{eq:continuum-limit-summation-1d}). It is important to note that at the same time the occupation function $\rho_0(E_\nu)$ in expression~(\ref{eq:integro-diff-v8}) becomes $\rho_0(H_0(x,p))$.
Taken together, these equalities are a reflection of the requirement that in the classical limit the averaging over quantum mechanical states passes to integration over phase space.
We emphasize once more that we only consider the leading-order term, as taking the limit~(\ref{eq:continuum-limit-summation}) may introduce additional terms of subleading order.
We also remark that the above process somewhat resembles the way the Wigner quasiprobability distribution (or Wigner function) is introduced in the formulation of quantum mechanics on phase space~\cite{Case08}.

After taking the limit, we can rewrite the result in terms of $H_0(x,p)= \frac{p^2}{2 m} + U(x)$. This gives
\begin{multline}  \label{eq:integro-diff-v10}
  - \varepsilon_b(x) \frac{|\nabla S|^2}{\hbar^2} \varphi_0(x) \exp\left( \frac{i}{\hbar} S(x) \right) = - 4 \pi e^2 \frac{g_s}{(2\pi\hbar)^3} \int \text{d}p \left(
  \frac{\rho_0(H_0(x, p))}{- H_0(x, p - \nabla S) + H_0(x,p) - E} \right. \\
  \left. + \frac{\rho_0(H_0(x, p))}{H_0(x,p) - H_0(x, p + \nabla S) + E}
  \right) \varphi_0(x) \exp\left( \frac{i}{\hbar} S(x) \right)
\end{multline}
At this point, we shift the integration variable $p \to p + \nabla S$ in the first term in the integral. Using the notation $\partial S/\partial x$ instead of $\nabla S$, we arrive at our final result
\begin{equation}
  \frac{1}{\hbar^2} \left(
  \varepsilon_b(x) \left\langle \frac{\partial S}{\partial x} , \frac{\partial S}{\partial x} \right\rangle 
  - 4 \pi e^2 \hbar^2 \Pi_0\left(x, \frac{\partial S}{\partial x} \right)
  \right) \varphi_0(x) \exp\left( \frac{i}{\hbar} S(x) \right) = 0 ,
\end{equation}
with the polarization
\begin{equation}
  \Pi_0\left(x, q \right) = \frac{g_s}{(2\pi\hbar)^3} \int \text{d}p \frac{\rho_0(H_0(x, p)) - \rho_0(H_0(x, p + q))}{H_0(x,p) - H_0(x, p + q) + E} .
\end{equation}
These last two expressions exactly coincide with our previous results~(\ref{eq:comm-rel-h0}) and~(\ref{eq:def-polarization}).

The arguments presented in this appendix thus provide an alternative way to obtain the leading-order term of the results obtained in section~\ref{sec:derivation}. Although these arguments do not have the same level of rigor as the derivations presented in the main text, they may be more intuitive for some readers and may therefore help to understand the formal transformations in section~\ref{sec:derivation}. However, these arguments are not able to reproduce the terms of subleading order.



\begin{thebibliography}{10}
  \expandafter\ifx\csname url\endcsname\relax
  \def\url#1{\texttt{#1}}\fi
  \expandafter\ifx\csname urlprefix\endcsname\relax\def\urlprefix{URL }\fi
  \expandafter\ifx\csname href\endcsname\relax
  \def\href#1#2{#2} \def\path#1{#1}\fi
  
  \bibitem{Tonks29}
  L.~Tonks, I.~Langmuir, Oscillations in ionized gases, Phys. Rev. 33 (1929)
  195--210.
  
  \bibitem{Vlasov38}
  A.~A. Vlasov, On the vibrational properties of an electron gas, J. Exp. Theor.
  Phys. 8 (1938) 291.
  
  \bibitem{Landau46}
  L.~Landau, On the vibration of the electronic plasma, J. Exp. Theor. Phys. 16
  (1946) 574.
  
  \bibitem{Pines52}
  D.~Pines, D.~Bohm, A collective description of electron interactions: Ii.
  collective vs individual particle aspects of the interactions, Phys. Rev. 85
  (1952) 338--353.
  
  \bibitem{Vonsovsky89}
  S.~V. Vonsovsky, M.~I. Katsnelson, Quantum solid-state physics,
  Springer-Verlag, Berlin Heidelberg, 1989.
  
  \bibitem{Giuliani05}
  G.~F. Giuliani, G.~Vignale, Quantum theory of the electron liquid, Cambridge
  University Press, Cambridge, 2005.
  
  \bibitem{Platzman73}
  P.~M. Platzman, P.~A. Wolff, Waves and Interactions in Solid State Plasmas,
  Academic Press, 1973.
  
  \bibitem{Nozieres99}
  P.~Nozi\`eres, D.~Pines, Theory Of Quantum Liquids, Hachette, United Kingdom,
  1999.
  
  \bibitem{Kittel05}
  C.~Kittel, Introduction to Solid State Physics, Wiley, 2005.
  
  \bibitem{Tame13}
  M.~S. Tame, K.~R. McEnery, {\relax \c{S}}.~K. \"Ozdemir, J.~Lee, S.~A. Maier,
  M.~S. Kim, Quantum plasmonics, Nat. Phys. 9 (2013) 329--340.
  
  \bibitem{Fitzgerald16}
  J.~M. Fitzgerald, P.~Narang, R.~V. Craster, S.~A. Maier, V.~Giannini, Quantum
  plasmonics, Proc. IEEE 104 (2016) 2307--2322.
  
  \bibitem{Fan14}
  X.~Fan, W.~Zheng, D.~J. Singh, Light scattering and surface plasmons on small
  spherical particles, Light Sci. Appl. 3 (2014) e179.
  
  \bibitem{Ritchie73}
  R.~Ritchie, Surface plasmons in solids, Surf. Sci. 34 (1973) 1--19.
  
  \bibitem{Scholl12}
  J.~A. Scholl, A.~L. Koh, J.~A. Dionne, Quantum plasmon resonances of individual
  metallic nanoparticles, Nature 483 (2012) 421--427.
  
  \bibitem{Eguiluz76}
  A.~Eguiluz, J.~J. Quinn, Hydrodynamic model for surface plasmons in metals and
  degenerate semiconductors, Phys. Rev. B 14 (1976) 1347--1361.
  
  \bibitem{Ciraci13}
  C.~Cirac\`i, J.~B. Pendry, D.~R. Smith, Hydrodynamic model for plasmonics: A
  macroscopic approach to a microscopic problem, ChemPhysChem 14 (2013)
  1109--1116.
  
  \bibitem{Mie08}
  G.~Mie, Beitr\"age zur optik tr\"uber medien, speziell kolloidaler
  metall\"osungen, Ann. Phys. 330 (1908) 377--445.
  
  \bibitem{Ishmukhametov81}
  B.~{\relax Kh}. Ishmukhametov, M.~I. Katsnelson, V.~N. Larionov, A.~M.
  Ustjuzhanin, On the existence of the atomic plasmon, Phys. Lett. A 82 (1981)
  387--388.
  
  \bibitem{Ishmukhametov81b}
  B.~{\relax Kh}. Ishmukhametov, M.~I. Katsnelson, V.~N. Larionov, A.~M.
  Ustyuzhanin, Spectrum of oscillations of the inhomogeneous electronic plasma,
  phys. stat. sol. (b) 104 (1981) K75--K78.
  
  \bibitem{Genzel75}
  L.~Genzel, T.~P. Martin, U.~Kreibig, Dielectric function and plasma resonances
  of small metal particles, Z. Phys. B 21 (1975) 339--346.
  
  \bibitem{Westerhout18}
  T.~Westerhout, E.~van Veen, M.~I. Katsnelson, S.~Yuan, Plasmon confinement in
  fractal quantum systems, Phys. Rev. B 97 (2018) 205434.
  
  \bibitem{Westerhout21}
  T.~Westerhout, M.~I. Katsnelson, M.~R\"osner, Quantum dot-like plasmonic modes
  in twisted bilayer graphene supercells, 2D Materials 9 (2021) 014004.
  
  \bibitem{Jiang21}
  Z.~Jiang, S.~Haas, M.~R\"osner, Plasmonic waveguides from coulomb-engineered
  two-dimensional metals, 2D Materials 8 (2021) 035037.
  
  \bibitem{Yannouleas92}
  C.~Yannouleas, R.~Broglia, Landau damping and wall dissipation in large metal
  clusters, Ann. Phys. 217 (1992) 105--141.
  
  \bibitem{Yannouleas93}
  C.~Yannouleas, E.~Vigezzi, R.~A. Broglia, Evolution of the optical properties
  of alkali-metal microclusters towards the bulk: The matrix
  random-phase-approximation description, Phys. Rev. B 47 (1993) 9849--9861.
  
  \bibitem{Reinhard90}
  P.-G. Reinhard, M.~Brack, O.~Genzken, Random-phase approximation in a local
  representation, Phys. Rev. A 41 (1990) 5568--5582.
  
  \bibitem{Brack93}
  M.~Brack, The physics of simple metal clusters: self-consistent jellium model
  and semiclassical approaches, Rev. Mod. Phys. 65 (1993) 677--732.
  
  \bibitem{Ishmukhametov71}
  B.~{\relax Kh}. Ishmukhametov, Collective oscillations of atomic electrons in
  the statistical theory of atoms, Phys. Status Solidi (B) 45 (1971) 669--678.
  
  \bibitem{Ishmukhametov75}
  B.~{\relax Kh}. Ishmukhametov, M.~I. Katsnelson, Collective oscillations of an
  inhomogeneous electron plasma in the semiclassical approximation, Fiz. Met.
  Metalloved. 40 (1975) 736.
  
  \bibitem{Bloch33}
  F.~Bloch, Bremsverm\"ogen von atomen mit mehreren elektronen, Z. Phys. 81
  (1933) 363--376.
  
  \bibitem{Jensen37}
  H.~Jensen, Eigenschwingungen eines fermi-gases und anwendung auf die blochsche
  bremsformel f\"ur schnelle teilchen, Z. Phys. 106 (1937) 620--632.
  
  \bibitem{Verkhovtseva76}
  {\relax \'E}.~T. Verkhovtseva, P.~S. Pogrebnyak, {\relax Ya}.~M. Fogel',
  Concerning the possibility of radiative decay of the collective levels of the
  argon atom, JETP lett. 24 (1976) 425--428.
  
  \bibitem{Torre17}
  I.~Torre, M.~I. Katsnelson, A.~Diaspro, V.~Pellegrini, M.~Polini,
  Lippmann-schwinger theory for two-dimensional plasmon scattering, Phys. Rev.
  B 96 (2017) 035433.
  
  \bibitem{Maslov81}
  V.~P. Maslov, M.~V. Fedoryuk, Semi-classical approximation in quantum
  mechanics, Reidel, Dordrecht, 1981.
  
  \bibitem{Guillemin77}
  V.~Guillemin, S.~Sternberg, Geometric asymptotics, American Mathematical
  Society, Providence, Rhode Island, 1977.
  
  \bibitem{Martinez02}
  A.~Martinez, An introduction to semiclassical and microlocal analysis,
  Springer-Verlag, New York, 2002.
  
  \bibitem{Zworski12}
  M.~Zworski, Semiclassical analysis, American Mathematical Society, Providence,
  Rhode Island, 2012.
  
  \bibitem{Hormander83}
  L.~H{\"o}rmander, The analysis of linear partial differential operators {I},
  Springer-Verlag, Berlin, 1983.
  
  \bibitem{Bir74}
  G.~L. Bir, G.~E. Pikus, Symmetry and Strain-induced Effects in Semiconductors,
  Wiley, New York, 1974.
  
  \bibitem{Blank78}
  A.~{\relax Ya}. Blank, V.~L. Berezinskii, Dispersion law of surface plasmons,
  Sov. Phys. JETP 48 (1978) 1170--1175.
  
  \bibitem{Curtis04}
  L.~J. Curtis, D.~G. Ellis, Use of the {E}instein-{B}rillouin-{K}eller action
  quantization, Am. J. Phys. 72 (2004) 1521--1523.
  
  \bibitem{Langer37}
  R.~E. Langer, {O}n the {C}onnection {F}ormulas and the {S}olutions of the
  {W}ave {E}quation, Phys. Rev. 51 (1937) 669--676.
  
  \bibitem{Heading62}
  J.~Heading, An Introduction to Phase-Integral Methods, Methuen, London, 1962.
  
  \bibitem{Froeman65}
  N.~Fr{\"o}man, P.~O. Fr{\"o}man, {JWKB} Approximation, Contributions to the
  Theory, North-Holland, Amsterdam, 1965.
  
  \bibitem{Lieb81}
  E.~H. Lieb, Thomas-{F}ermi and related theories of atoms and molecules, Rev.
  Mod. Phys. 53 (1981) 603--641.
  
  \bibitem{Tietz55}
  T.~Tietz, Comparison of the approximate solutions for free neutral atoms, J.
  Chem. Phys. 23 (1955) 1167--1167.
  
  \bibitem{Fluegge94}
  S.~Fl\"ugge, Practical Quantum Mechanics, Springer-Verlag, Berlin Heidelberg,
  1994.
  
  \bibitem{Mayer55}
  M.~Goeppert~Mayer, J.~H.~D. Jensen, Elementary theory of nuclear shell
  structure, John Wiley, New York, 1955.
  
  \bibitem{Toscano15}
  G.~Toscano, J.~Straubel, A.~Kwiatkowski, C.~Rockstuhl, F.~Evers, H.~Xu,
  N.~Asger~Mortensen, M.~Wubs, Resonance shifts and spill-out effects in
  self-consistent hydrodynamic nanoplasmonics, Nat. Commun. 6 (2015) 7132.
  
  \bibitem{David14}
  C.~David, F.~J. Garc{\'i}a~de Abajo, Surface plasmon dependence on the electron
  density profile at metal surfaces, ACS Nano 8 (2014) 9558--9566.
  
  \bibitem{Lang70}
  N.~D. Lang, W.~Kohn, Theory of metal surfaces: Charge density and surface
  energy, Phys. Rev. B 1 (1970) 4555--4568.
  
  \bibitem{Li13}
  X.~Li, D.~Xiao, Z.~Zhang, Landau damping of quantum plasmons in metal
  nanostructures, New J. Phys. 15 (2013) 023011.
  
  \bibitem{Mahan90}
  G.~D. Mahan, Many-Particle Physics, Springer, New York, 2000.
  
  \bibitem{Abrikosov65}
  A.~A. Abrikosov, L.~P. Gor'kov, I.~{\relax Ye}. Dzyaloshinskii, Quantum field
  theoretical methods in statistical physics, Pergamon Press, Oxford, 1965.
  
  \bibitem{Ceperley80}
  D.~M. Ceperley, B.~J. Alder, Ground state of the electron gas by a stochastic
  method, Phys. Rev. Lett. 45 (1980) 566--569.
  
  \bibitem{Arnold89}
  V.~I. Arnold, Mathematical methods of classical mechanics, second Edition,
  Springer, New York, 1989.
  
  \bibitem{Reijnders18}
  K.~J.~A. Reijnders, D.~S. Minenkov, M.~I. Katsnelson, S.~{\relax {Yu}}.
  Dobrokhotov, Electronic optics in graphene in the semiclassical
  approximation, Ann. Phys. 397 (2018) 65--135.
  
  \bibitem{Poston78}
  T.~Poston, I.~N. Stewart, Catastrophe theory and its applications, Pitman,
  Boston, 1978.
  
  \bibitem{Arnold82}
  V.~I. Arnold, S.~M. Gusein-Zade, A.~N. Varchenko, Singularities of
  differentiable maps, Vol.~1, Birkh\"auser, Basel, 1982.
  
  \bibitem{Griffiths05}
  D.~J. Griffiths, Introduction to quantum mechanics, second Edition, Pearson
  Prentice Hall, Upper Saddle River, 2005.
  
  \bibitem{Berry72}
  M.~V. Berry, K.~E. Mount, Semiclassical approximations in wave mechanics, Rep.
  Progr. Phys. 35 (1972) 315--397.
  
  \bibitem{Chester57}
  C.~Chester, B.~Friedman, F.~Ursell, An extension of the method of steepest
  descents, Proc. Camb. Phil. Soc. 53 (1957) 599--611.
  
  \bibitem{Berry21}
  M.~V. Berry, C.~J. Howls, Integrals with coalescing saddles, in: F.~W.~J.
  Olver, A.~B. {Olde Daalhuis}, D.~W. Lozier, B.~I. Schneider, R.~F. Boisvert,
  C.~W. Clark, B.~R. Miller, B.~V. Saunders, H.~S. Cohl, M.~A. McClain (Eds.),
  NIST Digital Library of Mathematical Functions, \url{http://dlmf.nist.gov/},
  Release 1.1.3, 2021.
  
  \bibitem{Anikin19}
  A.~{\relax {Yu}}. Anikin, S.~{\relax {Yu}}. Dobrokhotov, V.~E. Nazaikinskii,
  A.~V. Tsvetkova, Uniform asymptotic solution in the form of an airy function
  for semiclassical bound states in one-dimensional and radially symmetric
  problems, Theor. Math. Phys. 201 (2019) 1742--1770.
  
  \bibitem{Arnold67}
  V.~I. Arnold, Characteristic class entering in quantization conditions, Funct.
  Anal. Appl. 1 (1967) 1--14.
  
  \bibitem{Dobrokhotov03}
  S.~{\relax Yu}. Dobrokhotov, P.~N. Zhevandrov, Asymptotic expansions and the
  maslov canonical operator in the linear theory of water waves. i. main
  constructions and equations for surface gravity waves, Russ. J. Math. Phys.
  10 (2003) 1--31.
  
  \bibitem{Zwaan29}
  A.~Zwaan, Intensit{\"a}ten im {C}a-funkenspektrum, Ph.D. thesis, Utrecht
  university (1929).
  
  \bibitem{Reijnders13}
  K.~J.~A. Reijnders, T.~Tudorovskiy, M.~I. Katsnelson, Semiclassical theory of
  potential scattering for massless dirac fermions, Ann. Phys. 333 (2013)
  155--197.
  
  \bibitem{Bonitz94}
  M.~Bonitz, Impossibility of plasma instabilities in isotropic quantum plasmas,
  Physics of Plasmas 1 (1994) 832--833.
  
  \bibitem{Bonitz94b}
  M.~Bonitz, R.~Binder, D.~C. Scott, S.~W. Koch, D.~Kremp, Theory of plasmons in
  quasi-one-dimensional degenerate plasmas, Phys. Rev. E 49 (1994) 5535--5545.
  
  \bibitem{Bonitz93}
  M.~Bonitz, R.~Binder, D.~Scott, S.~Koch, D.~Kremp, Plasmons and instabilities
  in quantum plasmas, Contrib. Plasma Phys. 33 (1993) 536--539.
  
  \bibitem{Hamann20}
  P.~Hamann, J.~Vorberger, T.~Dornheim, Z.~A. Moldabekov, M.~Bonitz, Ab initio
  results for the plasmon dispersion and damping of the warm dense electron
  gas, Contrib. Plasma Phys. 60 (2020) e202000147.
  
  \bibitem{Nazaikinskii14}
  V.~E. Nazaikinskii, The maslov canonical operator on lagrangian manifolds in
  the phase space corresponding to a wave equation degenerating on the
  boundary, Math. Notes 96 (2014) 248--260.
  
  \bibitem{Anikin18}
  A.~{\relax Yu}. Anikin, S.~{\relax Yu}. Dobrokhotov, V.~E. Nazaikinskii, Simple
  asymptotics for a generalized wave equation with degenerating velocity and
  their applications in the linear long wave run-up problem, Math. Notes 104
  (2018) 471--488.
  
  \bibitem{Dobrokhotov21}
  S.~{\relax Yu}. Dobrokhotov, D.~S. Minenkov, V.~E. Nazaikinskii, Representation
  of bessel functions by the maslov canonical operator, Theor. Math. Phys. 208
  (2021) 1018--1037.
  
  \bibitem{Goldstein02}
  H.~Goldstein, C.~P. Poole, J.~L. Safko, Classical Mechanics, third Edition,
  Addison Wesley, San Fransisco, 2002.
  
  \bibitem{Mathematica}
  Wolfram {M}athematica, \url{www.wolfram.com/mathematica}. The computations in
  this article were performed with version 12.1.0.0 (1988--2022).
  
  \bibitem{Grigorenko12}
  A.~N. Grigorenko, M.~Polini, K.~S. Novoselov, Graphene plasmonics, Nat.
  Photonics 6 (2012) 749--758.
  
  \bibitem{Littlejohn91}
  R.~G. Littlejohn, W.~G. Flynn, Geometric phases in the asymptotic theory of
  coupled wave equations, Phys. Rev. A 44 (1991) 5239--5256.
  
  \bibitem{Xiao10}
  D.~Xiao, M.-C. Chang, Q.~Niu, Berry phase effects on electronic properties,
  Rev. Mod. Phys. 82 (2010) 1959--2007.
  
  \bibitem{Dimassi99}
  M.~Dimassi, J.~Sj{\"o}strand, Spectral asymptotics in the semi-classical limit,
  Cambridge University Press, Cambridge, 1999.
  
  \bibitem{Hall13}
  B.~C. Hall, Quantum theory for mathematicians, Springer, New York, 2013.
  
  \bibitem{Jimenez10}
  C.~N. Jim\'enez, Cohomology of classes of symbols and classification of traces
  on corresponding classes of operators with non positive order, PhD Thesis,
  Rheinischen Friedrich-Wilhelms-Universit\"at Bonn, 2010.
  
  \bibitem{Lesch10}
  M.~Lesch, Pseudodifferential operators and regularized traces, Clay Math. Proc.
  12 (2010) 37--72.
  
  \bibitem{Case08}
  W.~B. Case, Wigner functions and weyl transforms for pedestrians, Am. J. Phys.
  76 (2008) 937--946.
  
\end{thebibliography}
\end{document}